\documentclass[english,twoside,openright]{report}
\usepackage{amssymb}
\usepackage{amsfonts}
\usepackage{amsmath}
\usepackage{mathrsfs}
\usepackage{eufrak}
\usepackage{fancyhdr}
\usepackage{graphicx}
\usepackage{babel}
\begin{document} 
\ifx\href\undefined\else\hypersetup{linktocpage=true}\fi 
\topmargin 0pt
\oddsidemargin 30mm
\evensidemargin 10mm

\newcommand{\ve}{\varepsilon}

\def\h{\eta}
\def\vt{\vartheta}

\def\r{\rho}
\def\vp{\varphi}

\def\sfrac#1#2{{\textstyle\frac{#1}{#2}}}
\def\rd#1{\buildrel{_{_{\hskip 0.01in}\rightarrow}}\over{#1}}
\def\ld#1{\buildrel{_{_{\hskip 0.01in}\leftarrow}}\over{#1}}
\def\+{\dagger}
\def\={\ =\ }
\def\tr{\mathrm{tr}}
\def\det{\mathrm{det}}
\def\res{\mathrm{res}}
\def\ch{\mathrm{ch}}
\def\sh{\mathrm{sh}}
\def\th{\mathrm{th}}

\newcommand{\mbf}[1]{{\boldsymbol{#1}}}
\newcommand{\im}{\,\mathrm{i}\,}
\newcommand{\diff}{\mathrm{d}}
\newcommand{\intd}{\mathrm{d}}
\newcommand{\R}{{\mathbb{R}}}
\newcommand{\C}{{\mathbb{C}}}
\newcommand{\Ecal}{{\cal E}}
\newcommand{\mb}{\bar{\mu}}
\newcommand{\Lb}{\bar{L}^{A,B}}
\newcommand{\Lbt}{\widetilde{\bar{L}}^{A,B}}
\newcommand{\Pht}{\widetilde{\Phi}}
\newcommand{\Pt}{\widetilde{P}}
\newcommand{\St}{\widetilde{S}}
\newcommand{\Tt}{\widetilde{T}}
\newcommand{\gt}{\widetilde{g}}
\newcommand{\Psh}{\widehat{\Psi}}
\newcommand{\Sh}{\widehat{S}}

\newcommand{\EQ}{\begin{equation}}
\newcommand{\beq}{\begin{equation}}
\newcommand{\EN}{\end{equation}}
\newcommand{\eeq}{\end{equation}}
\newcommand{\bea}{\begin{eqnarray}}
\newcommand{\ena}{\end{eqnarray}}
\newcommand{\eea}{\end{eqnarray}}
\newcommand{\vs}[1]{\vspace{#1 mm}}
\renewcommand{\a}{\alpha}
\renewcommand{\b}{\beta}
\renewcommand{\c}{\gamma}
\renewcommand{\d}{\delta} 
\renewcommand{\th}{\theta}
\renewcommand{\TH}{\Theta}
\newcommand{\pa}{\partial}
\newcommand{\g}{\gamma}
\newcommand{\G}{\Gamma}
\newcommand{\A}{\Alpha}
\newcommand{\B}{\Beta}
\newcommand{\D}{\Delta}
\newcommand{\e}{\epsilon}
\newcommand{\E}{\Epsilon}
\newcommand{\z}{\zeta}
\newcommand{\Z}{\Zeta}
\renewcommand{\l}{\lambda}
\renewcommand{\L}{\Lambda}
\newcommand{\m}{\mu}
\newcommand{\M}{\Mu}
\newcommand{\n}{\nu}
\newcommand{\N}{\Nu}
\newcommand{\x}{\chi}
\newcommand{\X}{\Chi}
\newcommand{\p}{\pi}
\renewcommand{\P}{\Pi}
\renewcommand{\r}{\rho}
\newcommand{\s}{\sigma}
\renewcommand{\S}{\Sigma}
\renewcommand{\t}{\tau}
\newcommand{\T}{\Tau}
\newcommand{\y}{\upsilon}
\newcommand{\Y}{\Upsilon}
\renewcommand{\o}{\omega}
\renewcommand{\O}{\Omega}
\newcommand{\ad}{{\dot{\alpha}}}
\newcommand{\bd}{{\dot{\beta}}}
\newcommand{\gd}{{\dot{\gamma}}}
\newcommand{\dd}{{\dot{\delta}}}
\newcommand{\ed}{{\dot{\epsilon}}}
\newcommand{\Qb}{{\bar{Q}}}
\newcommand{\thb}{{\bar{\theta}}}
\newcommand{\Db}{{\bar{D}}}
\newcommand{\dl}{{\overleftarrow{\partial}}}
\newcommand{\dr}{{\overrightarrow{\partial}}}
\newcommand{\Dl}{{\overleftarrow{D}}}
\newcommand{\Dr}{{\overrightarrow{D}}}
\newcommand{\Dbr}{{\overrightarrow{\bar{D}}}}
\newcommand{\Dbl}{{\overleftarrow{\bar{D}}}}
\newcommand{\non}{\nonumber}
\newcommand{\pab}{\bar\pa}
\newcommand{\half}{\frac{1}{2}}
\newcommand {\cd}  {\newpage{\pagestyle{empty}\cleardoublepage}}

\thispagestyle{empty}
\begin{center}
\begin{Large}
UNIVERSITY OF PAVIA\\
\vskip 0.5cm
Department of Nuclear and Theoretical Physics
\end{Large}
\end{center}
\vskip 3cm
\begin{center}
\begin{huge}
Noncommutative\\
Supersymmetric/Integrable Models\\
and\\[0.4cm]
String Theory
\end{huge}
\end{center}
\vskip 3cm
\begin{flushleft}
\begin{normalsize}
Supervisor: Prof. Annalisa Marzuoli\\
External Referee: Prof. Ulf Lindstr\"om
\end{normalsize}
\end{flushleft}
\vskip 2cm
\begin{flushright}
{\normalsize Doctoral Thesis of}\\[0.2cm]
{\Large Laura Tamassia}
\end{flushright}
\vskip 2cm
\begin{center}
Dottorato di Ricerca in Fisica XVII Ciclo
\end{center}

\cd
\pagenumbering{roman}
\tableofcontents
\cd
\addcontentsline{toc}{chapter}{Introduction}
\newpage
{\Large\bf{Introduction}}
\vskip 22pt
Quantum field theory, i.e. quantum mechanics with the basic observables living at spacetime points, is the principle underlying our current understanding of natural phenomena.
The incredibly successful Standard Model explains all the forces of nature, except gravity, in terms of local symmetry principles. 

At the energies we are currently able to test, gravity is well described by general relativity.
However, it is expected that at energies of the order of the Planck mass $M_P\sim10^{19}$ GeV, a quantum description of gravity will be necessary.
The union of gravity with quantum field theory leads to a nonrenormalizable theory. 
History of physics suggests that this should be interpreted as a sign that new physics will appear at higher energies, comparing for instance to the Fermi theory case.  

A natural way to modify quantum mechanics to describe a ``quantum spacetime" is to generalize Heisenberg commutation relations to let the space coordinates be noncommuting. In a field theory setting, in principle one could include time in noncommutativity and consider a quantum spacetime where 
\beq
\left[X^\m,X^\n\right]=i\th^{\m\n}(X)
\nonumber
\eeq
This can be realized in a functional formulation by deforming the algebra of functions on spacetime, replacing the ordinary, commutative product with a noncommutative one that, when applied to coordinates themselves, gives $[x^\m,x^\n]_\ast=x^\m\ast x^\n-x^\n\ast x^\m=i\th^{\m\n}$.
The simplest case, with a constant $\th^{\m\n}$, is related to spacetime translation invariance. The corresponding (Moyal) product can be uniquely identified by requiring that the deformed algebra of functions remains associative. 

When a field theory is deformed by replacing ordinary products with Moyal products in the action, 
vertices in the Feynman rules are multiplied by a phase factor with a dependence on the momenta, that generically acts as a regulator. For this reason, spacetime noncommutativity was first introduced with the hope that it might reduce the degree of divergence of a given theory \cite{regular}. However, since the contributions from a certain diagram also include the case where the phase factor does not depend on loop momenta, but only on external momenta, this does not work \cite{filk}.

At the moment, we only know one way to consistently cut off the divergences appearing in a quantum field theory of gravity. This is string theory \cite{pol,gsstring}, where pointlike interactions are replaced by splitting and joining of one-dimensional objects, so that the interaction itself does not happen at a certain point in spacetime but is described by a smooth two-dimensional surface.

String theory provides a consistent quantum theory of gravity. The number of spacetime dimensions is fixed to ten, the theory is consistent only in the presence of supersymmetry, relating bosonic and fermionic degrees of freedom, and leads to gauge groups that include the Standard Model one.
String theories are unique in the sense that there are no free parameters and no freedom in choosing gauge groups. There is more than one possible string theory (actually five), but they happen to be all related to each other by dualities. Therefore, a unique underlying theory is expected to exist, from which the various string theories can be obtained as limits in a certain space of parameters.

It is striking that the somehow natural proposal for a quantum spacetime where Heseinberg commutation relations are generalized to include a nontrivial coordinate algebra can be obtained from string theory.  
The first works in this direction showed that noncommutative tori are solutions to the problem of compactifying M(atrix) theory \cite{torinc}.
Noncommutative field theories were shown to emerge as low energy limits of open string dynamics in the presence of a constant Neveu-Schwarz Neveu-Schwarz (NS-NS) background and D-branes (i.e. p-dimensional manifolds where open string ends are attached) \cite{presb,seibwit}. Since the background selects preferred directions in spacetime, the resulting effective theory breaks Lorentz invariance \cite{lorentz}. 
Therefore, noncommutative field theory must not be thought of as a fundamental theory, but as an effective description to be used when certain backgrounds are present. 

The discovery that noncommutative geometry is somehow embedded in string theory induced a growing interest in the field. Various aspects of noncommutative field theory, such as renormalization properties \cite{micu,roibanren}, solitons (see for instance the lectures \cite{ictpgopak,nikitasol,harvey}) and instantons \cite{insta} have been investigated. Furthermore, it was shown that when time is involved in noncommutativity the resulting field theories display awkward features, such as acausality \cite{causal} and nonunitarity \cite{unitar}, that spoil the consistency of the theory. However,  it was also shown that these ill-defined noncommutative field theories cannot be obtained as a low energy limit of string theory \cite{unitar}. String corrections conspire to cancel out the inconsistencies \cite{causal,timenc}.

Applications that are not directly related to string theory have been considered, for instance a model to describe the quantum Hall effect was proposed in \cite{susskind}, based on noncommutative Chern-Simons theory.  Moreover, a mechanism to drive inflation without an inflaton field was studied in \cite{inflation}, based on noncommutativity of spacetime. Particle physics phenomenology based on a noncommutative version of the Standard Model has been studied \cite{standard} and possible experimental tests have been proposed \cite{standardtest}. In these applications also noncommutative deformations that have not yet been shown to arise from string theory have been considered \cite{kappamink}. 
A natural question would be at which energy one expects to observe effects of noncommutativity. If one considers this as an effective description in the presence of a background, the energy is related to the background scale, and therefore cannot be determined a priori. 

As I said, string theory is consistent only in the presence of supersymmetry. Supersymmetric theories are better described in superspace formalism  \cite{wessbagger,superspace,sohnius}, where spacetime is enlarged by adding fermionic coordinates. Superspace geometry is not flat, torsion is present. 
Since noncommutative geometry arises in a string theory context and strings require supersymmetry, it is compelling to study the possible deformations of supersymmetric theories. The natural way to do that is to investigate the possible consistent deformations of superspace.
This I did in my first paper \cite{mypaper1}, in collaboration with D. Klemm and S. Penati.
We found that consistent deformations of superspace are possible in both Minkowski and Euclidean signatures in four dimensions. In the Euclidean case more general structures are allowed because of the different spinor reality conditions.

After the appearance of my paper \cite{mypaper1}, deformed superspaces were found to emerge in superstring theory \cite{vafaooguri2,seiberg,antonio}. However, to see this effect, a particular formulation of superstring theory has to be considered, where all the symmetries of the theory are manifest. This superPoincar\'e covariant formulation, found in 2000 by Berkovits \cite{berkovits}, is still ``work in progress". However, it has already proven to be superior in handling certain string backgrounds (Ramond-Ramond), that lead to the superspace deformations I found, and to prove general theorems regarding string amplitudes, like for instance a theorem  stating the ``finiteness"  of perturbative string theory \cite{multiloop}.
I've been studying this formalism for the superstring and contributed, in \cite{mypaper4}, written together with P.A. Grassi, by giving an iterative procedure to simplify the computation of superstring vertex operators, that are the main ingredient for the evaluation of superstring amplitudes. Some applications of our analysis are also discussed in the paper. For instance, the vertices associated to a R-R field stength  with a linear dependence on the bosonic coordinates have been computed. These play a role in the study of superspace deformation with a Lie-algebraic structure such that bosonic coordinates of superspace are obtained as the anticommutator of the fermionic ones. This can be interpreted by saying that spacetime has a fermionic substructure \cite{Schwarz:pf}.

Going back to noncommutative field theory, a puzzling aspect is that the noncommutative generalization of a given ordinary theory is not unique. A selection principle can be based on trying to preserve the nice properties of a model in its noncommutative deformation. The symmetry structure of a theory is its most important characteristic and actually defines the theory when the field content is given. Therefore, preserving the symmetry structure of a theory should be the first criterium to consider when constructing its deformation. This is also the guiding principle we used in studying deformations of supersymmetric theories in \cite{mypaper1}.

A very special case is given by systems endowed with an infinite number of local conserved charges. These systems, known as integrable by analogy to the case with a finite number of degrees of freedom, display very special features. The presence of such a strong symmetry structure underlying the system puts constraints on the dynamics, for instance causing the S-matrix to be factorized and preventing particle production processes to occur. Moreover, integrable systems usually display localized classical solutions, known as solitons, that do not change shape or velocity after collisions.
Therefore, when constructing the noncommutative generalization of an integrable system, one would like to preserve both the symmetry structure underlying the system and the consequences that this has in ordinary geometry. 
The two-dimensional case is particularly problematic in noncommutative field theory, because time must be necessarily involved in noncommutativity and thus it is expected that acausality and nonunitarity will spoil the consistency of the theory. One may hope that integrability can cure these inconsistencies.

I studied the problem of constructing the generalization of two-dimensional integrable systems by focusing on a very special example, the sine-Gordon theory, describing the dynamics of a scalar field governed by an oscillating potential. Integrability is a property of the equations of motion and in most cases an action generating them is not known. The sine-Gordon model is one of the cases where an action is also known. Therefore it is possible to move on to a quantum description of the model.

The project about the noncommutative sine-Gordon was started by M.T. Grisaru and S. Penati in \cite{us}, where they constructed a noncommutative version of the sine-Gordon system by making use of a bidifferential calculus approach, so that it was guaranteed that the system possessed an infinite number of conserved currents. In the resulting noncommutative model the dynamics of a single, scalar field is described by two equations. The second equations has the form of a conservation law and becomes trivial in the noncommutative limit. This pattern is somehow unavoidable, since it is related to the necessary extension of $SU(2)$ to $U(2)$ in noncommutative geometry \cite{closure}. 

Because of the unusual structure of the equations, in \cite{us} an action for the model was not found.
This was the first goal of my paper \cite{mypaper2}, written in collaboration with M.T. Grisaru, L. Mazzanti and S. Penati. There we found an action for the model and studied its tree-level scattering amplitudes. We discovered that they are plagued by acausality and that particle production occurs.
Therefore this model had to be discarded and a different one had to be constructed.

In \cite{mypaper3}, in collaboration with O. Lechtenfeld, L. Mazzanti, S. Penati and A.D. Popov, we proposed another model obtained by a two-step reduction from selfdual Yang-Mills theory, that proved to possess a well-defined, causal and factorized S-matrix. Moreover, soliton solutions of this model were constructed.

The two different noncommutative versions of the sine-Gordon model I studied differ in the generalization of the oscillating interaction term. They can both be obtained by dimensional reduction from noncommutative selfdual Yang-Mills equations, with a different parametrization of the gauge group. 
The second model is obtained by reducing $U(2)$ to its $U(1)\times U(1)$ subgroup, which seems the most natural choice and indeed leads to a successful integrable theory.
\vskip 11pt
This thesis contains four chapters. In the first one, I will give an introduction to Moyal deformation of bosonic spacetime. I will review the main results in noncommutative field theory and in particular I will discuss the problems arising when time is involved in noncommutativity. I will summarize Kontsevich results about coordinate-dependent noncommutativity. I will then move on to supersymmetric theories, discuss the results I obtained in \cite{mypaper1} concerning superspace non(anti)commutative deformations in four dimensions and their relation with following works. In the rest of the chapter I will review the string theory results concerning the low energy description of D-brane dynamics in terms of non(anti)commutative field theory. In particular I will consider the bosonic, RNS, GS and $N=2$ strings in the presence of constant NS-NS backgrounds, the covariant superstring compactified on a Calabi-Yau three-fold in the presence of a R-R selfdual field strength and the generalization to the uncompactified case. Finally, I will discuss the generalization to nonconstant NS-NS backgrounds, leading to Kontsevich deformations, and I will speculate about a possible similar analysis of nonconstant R-R backgrounds.

In the second chapter, I will first give an introduction to selected topics concerning two-dimensional integrable systems, such as their derivation from zero curvature conditions for a bidifferential calculus, the possibility of obtaining them as dimensional reductions from selfdual Yang-Mills equations, the factorization of the S-matrix and the presence of solitons solutions. I will only consider the sine-Gordon model as an example. Then I will introduce the first attempt to construct an integrable noncommutative version of the sine-Gordon model. I will report my results in \cite{mypaper2}, concerning the construction of an action for the model and the computation of tree-level scattering amplitudes, that proved to be nonfactorized and acausal. Finally, I will discuss the results I obtained in \cite{mypaper3}, regarding the construction of a different noncommutative generalization of the sine-Gordon model that proved to possess all the nice properties one expects from a two-dimensional integrable theory,  such as factorization of the S-matrix.

In the third chapter, I will first review the pure spinor approach to superstring theory. In particular, I will focus on the construction of open and closed superstring vertex operators in this formalism. I will then present the results I obtained in \cite{mypaper4}, where I constructed an iterative procedure to compute the vertices. I will then discuss an application of this analysis, concerning the computation of the vertex operators for nonconstant R-R field strengths, that are expected to be related to coordinate-dependent superspace deformations. In particular, I will show how to compute vertices for a R-R field strength that is linear in the bosonic spacetime coordinates and is expected to be associated to a Lie algebraic superspace deformation providing a fermionic substructure for bosonic coordinates.

In the fourth chapter I will summarize my results and discuss open problems and possible applications.

\chapter[An introduction to non(anti)commutative geometry and\\ superstring theory]{An introduction to non(anti)commutative geometry and superstring theory}
\setcounter{page}{1}
\setcounter{chapter}{1} 
\pagenumbering{arabic}
\section[A brief introduction to noncommutative and non(anti)commutative\\ field theory]{A brief introduction to noncommutative and non(anti)commutative field theory}
\subsection{The Moyal product}
\subsubsection{Weyl transform definition} 
Noncommutative geometry deals with manifolds whose coordinates do not commute. This kind of manifold appeared in physics much before noncommutative geometry itself was born as a branch of mathematics \cite{connes}. A well-known example is quantum phase space. This is a manifold described by $2n$ operator coordinates $\hat{X}_1$,...,$\hat{X}_n$,$\hat{P}_1$,...,$\hat{P}_n$, satisfying nontrivial commutation relations
\bea
&&\left[\hat{X}_i,\hat{P}_j\right]=i\hbar\d_{ij}\cr
&&\left[\hat{X}_i,\hat{X}_j\right]=\left[\hat{P}_i,\hat{P}_j\right]=0
\label{qm}
\ena
It is expected that the geometric nature of spacetime will be modified at very short distances. The physical idea underlying modern noncommutative geometry is that  a quantum spacetime will be uncovered then, where the usual trivial commutation relations among coordinates are no longer valid and noncommutativity emerges as
\EQ
\left[\hat{X}^{\m},\hat{X}^{\n}\right]=i\th^{\m\n}\left(\hat{X}\right)
\label{ncalgebra}
\EN
with $\th^{\m\n}=-\th^{\n\m}$.
In the limit $\th^{\m\n}\rightarrow 0$ ordinary commutative geometry must be recovered.
Coordinate algebra (\ref{ncalgebra}) gives rise to the spacetime uncertainty relations
\beq
\Delta X^\m \Delta X^\n \geq \frac{1}{2} \vert \th^{\m\n} \vert 
\eeq
In (\ref{ncalgebra}) the possibility is left open that time may be involved in noncommutativity. 
In this case (\ref{ncalgebra}) would not be simply a generalization of the quantum mechanical commutation relations (\ref{qm}) and it is somehow expected that it may clash with quantum mechanics. 
For a while I will not worry about this, I will reconsider the case of noncommuting time later on.

The connection between noncommutative geometry and quantum mechanics is easily seen when the latter is discussed in the Weyl formalism.  In this formalism an explicit map between  functions $f(x,p)$ of the phase space variables $x$, $p$ and corresponding operators $\hat{O}_f(\hat{X},\hat{P})$ is constructed, where $\hat{X}$, $\hat{P}$ are noncommuting operators corresponding to classical variables $x$, $p$.\\
I will briefly discuss this formalism, in the case of two variables $x^1$, $x^2$, with corresponding operators satisfying $\left[\hat{X}^1,\hat{X}^2\right]=\th^{12}$ with constant $\th$.
After this discussion the natural embedding of noncommutative geometry in quantum mechanics will be clear to the reader \cite{harvey}.

On ``phase space" described by coordinates $x_1,x_2$ let us consider a function $f(x^1,x^2)$.  Given its Fourier transform 
\EQ
\tilde{f}(\a_1,\a_2)=\int d^2 x ~e^{i(\a_1 x^1 +\a_2 x^2)}~ f(x^1,x^2)
\EN
we can define an operator $\hat{O}_f(\hat{X^1},\hat{X^2})$ as follows
\EQ
\hat{O}_f(\hat{X^1},\hat{X^2})=\frac{1}{(2\p)^2}\int d^2\a~ U(\a_1,\a_2)\tilde{f}(\a_1,\a_2)
\EN
where
\EQ
U(\a_1,\a_2)=e^{-i(\a_1 \hat{X^1}+\a_2 \hat{X^2})}
\EN
Making use of Baker-Campbell-Hausdorff  formula \cite{wilcox} we find
\EQ
U(\a_1,\a_2)U(\b_1,\b_2)=e^{-\frac{i}{2}(\a_1\b_2-\a_2\b_1)\th^{12}}U(\a_1+\b_1,\a_2+\b_2)
\EN
The map $f\longrightarrow \hat O_f$ defines the Weyl-Moyal correspondence between functions on phase space and operators.

Now we would like to determine which function corresponds to the operator $\hat{O}_f \hat{O}_g$.
We know that in general  $\hat{O}_f \hat{O}_g\not= \hat{O}_g \hat{O}_f$, so we expect some noncommutative deformation of the ordinary product to arise by the Weyl-Moyal correspondence. 
\bea
&&\hat{O}_f \hat{O}_g=\frac{1}{(2\p)^4}\int d^2\a~ d^2\b~ U(\a_1,\a_2) U(\b_1,\b_2)\tilde{f}(\a_1,\a_2)\tilde{g}(\b_1,\b_2)=\cr
&&=\frac{1}{(2\p)^4}\int d^2\a~ d^2\b~ U(\a_1+\b_1,\a_2+\b_2)e^{-\frac{i}{2}(\a_1\b_2-\a_2\b_1)\th^{12}}\tilde{f}(\a_1,\a_2)\tilde{g}(\b_1,\b_2)\cr
&&~~~~~~
\ena
By making the change of variables $\g_1=\a_1+\b_1$, $\d_1=\frac{1}{2}(\a_1-\b_1)$, $\g_2=\a_2+\b_2$, $\d_2=\frac{1}{2}(\a_2-\b_2)$ we obtain
\bea
&&\hat{O}_f \hat{O}_g=\frac{1}{(2\p)^4}\int d^2\g d^2 \d~ U(\g_1,\g_2)e^{\frac{i}{2}\th^{12}(\g_1\d_2-\d_1\g_2)}\tilde{f}\left(\frac{\g_1}{2}+\d_1,\frac{\g_2}{2}+\d_2\right)\cdot\cr
&&~~~~~~~~~~~~~~~~~~~~~~~~~~~~~~~~~~~~~~~~~~~~~~\cdot\tilde{g}\left(\frac{\g_1}{2}-\d_1,\frac{\g_2}{2}-\d_2\right)
\label{1}
\ena
Let us define Moyal product between two functions  $f$ and $g$ in $R^{2n}$ as
\EQ
\left(f\ast g\right)(x)=e^{\frac{i}{2}\th^{ij}\pa_i \pa'_j}f(x)g(x')\vert_{x=x'}
\label{moyal1}
\EN
In our simple two-dimensional case it becomes
\EQ
(f\ast g)(x)=e^{\frac{i}{2}\th^{12}(\pa_1\pa'_2-\pa_2\pa'_1)}f(x)g(x')\vert_{x=x'}
\EN
In momentum space we can obtain the following formula for the Fourier transform of $f\ast g$
\EQ
\widetilde{f\ast g}(\g_1,\g_2)=\frac{1}{(2\p)^2}\int d^2\d~ e^{\frac{i}{2}\th^{12}(\g_1\d_2-\g_2\d_1)} \tilde{f}(\frac{\g_1}{2}+\d_1,\frac{\g_2}{2}+\d_2)\tilde{g}(\frac{\g_1}{2}-\d_1,\frac{\g_2}{2}-\d_2)
\label{3}
\EN
From this and (\ref{1}) it is clear that
\EQ
\hat{O}_f \hat{O}_g=\frac{1}{(2\p)^2}\int d^2\g~ U(\g_1,\g_2)\widetilde{f\ast g}(\g_1,\g_2)=\hat{O}_{f \ast g}
\EN
So Moyal product (\ref{moyal1}) naturally emerges in the context of quantum mechanics, when the latter is expressed in the Weyl-Moyal formalism. It is the functional product corresponding to operator product between quantum observables.  Applying (\ref{moyal1}) to the special case $f=x^i$, $g=x^j$, we obtain the coordinate algebra
\EQ
x^i\ast x^j-x^j\ast x^i=[x^i,x^j]_{\ast}=i\th^{ij}
\label{algmoyal}
\EN
Therefore the quantum commutation relations we began with are reproduced in the functional formalism as $\ast$ commutators.
\subsubsection{Translation covariance and associativity as a definition}
Moyal product (\ref{moyal1}) can also be obtained from  a general discussion concerning the algebraic requirement of associativity and the geometric requirement of covariance with respect to translations. 
These two properties uniquely determine Moyal product.
Before I discuss this, I will introduce the general ideas concerning a field called deformation quantization and its connections with modern noncommutative geometry.

Ordinary geometry is based on the concept of point. This is not true anymore for noncommutative geometry, since a noncommutative manifold is completely defined in terms of the properties of the algebra of functions on it \cite{castellani}. 
In ordinary geometry many sets of points can be completely described when the algebra $A$ of functions on them with values in $R$ or $C$ is known. A finite dimensional vector space $V$ is a familiar example of this, since the space of functions $f:V\longrightarrow R$ (or $C$) is the dual space $V^*$, which is isomorphic to $V$. In this case studying the algebra of functions on the manifold or the manifold itself is the same thing.

We can consider the more general case of a $C^*$ algebra $A$, i.e. an algebra endowed with a norm and an involution. Every $C^*$ algebra is isomorphic to the algebra $A'$ of complex continuos functions on a certain compact space $V$. When $A'$ is commutative we can go back to the space $V$, that can be described as a set of points in ordinary geometry. When $A'$ is noncommutative, instead, going back to the space $V$ can be very complicated and in some cases impossible. However, this is irrelevant for the purpose of studying a physical theory, since all the needed information are encoded in $A'$.

A recipe to obtain a theory on noncommutative space from a given one on ordinary space is the following. Consider the algebra of functions with values in R (or C), deform its product to a new, noncommutative one, that I will call $\ast$, defined in terms of a parameter $\hbar$. In the limit $\hbar\rightarrow 0$ one must recover ordinary, commutative case. Now rewrite the old theory replacing all ordinary products with $\ast$ products, and think of the new theory as a deformation of the ordinary one, defined on noncommutative space. This I will call the natural deformation of a theory. It is not the only possible definition of a noncommutative generalization of an ordinary theory and I will discuss this point in more detail in section 1.1.3. 

Given two functions $f$, $g$, I will denote their ordinary, commutative product as $fg$.
I will deform it in the following way
\EQ
f\ast g \equiv fg~+~\hbar P(f,g)~+~\mathcal{O}(\hbar^2)
\label{2}
\EN
where $P(f,g)$ is a bilinear operator in the two functions $f$, $g$ and $\hbar$ is the parameter governing noncommutativity.
The example of quantum phase space discussed before suggests a good candidate for the bilinear operator $P$. When the manifold we are considering is endowed with a Poisson structure  $\left\{~~,~\right\}_P$, we will choose
\EQ
P(f,g)=\left\{f,g\right\}_P=P^{\m\n}~\pa_{\m}~ \pa'_{\n}~f(x)~g(x')\vert_{x=x'}=f\overleftarrow{\pa}_{\m}~P^{\m\n}~\overrightarrow{\pa}_{\n}g
\label{poisson}
\EN
(The last equality is just to present a different notation. It will be preferred since it is more suitable to  superspace extension, where the presence of fermionic indices makes different orderings inequivalent).

In the 70's the deformation of Poisson manifolds was studied in a completely different context. In the paper by Bayen et al. \cite{bayen} a different approach to quantization was proposed. Quantization had to be understood as a deformation of the algebraic structure of functions and not as a radical change in the nature of physical observables. Moyal product $\ast$ was then introduced with the goal of reinterpreting quantum mechanics in the context of algebraic deformations.
In particular, in \cite{bayen} an analysis of the possible noncommutative but associative products that can be obtained as a perturbative series in the parameter $\hbar$ was performed. The results obtained there are very interesting when they are reread in the light of the new ideas of noncommutative geometry. 
 
Consider a manifold $\O$ endowed with a Poisson structure $P$, where a set of derivatives $\nabla_\m$ is defined such that $\nabla_\m P=0$.
We will also assume that this set of derivatives is torsion free and without curvature.
We define a generic product $\ast$ on $\O$ by the smooth function
\EQ
u(z)=\sum_{r=0}^{\infty}a_r\left(\frac{z^r}{r!}\right)
\EN
with $a_0=a_1=1$, as
\EQ
f\ast g =\sum_{r=0}^{\infty}~\hbar^r ~\frac{a_r}{r!}~ P^r(f,g)
\label{svil1}
\EN
where 
\EQ
P^r(f,g)=P^{\m_1 \n_1}.......P^{\m_r \n_r}~\nabla_{\m_1}...\nabla_{\m_r}~f ~\nabla_{\n_1}...\nabla_{\n_r}~g
\label{svil2}
\EN
One can show that the exponential function is the only possible choice for $u$ leading to an associative product, i. e. satisfying
\EQ
(f \ast g) \ast h=f \ast (g \ast h)
\label{moyalasso}
\EN
To show this one imposes (\ref{moyalasso}), writing every $\ast$ product explicitly as in (\ref{svil1}, \ref{svil2}). Order by order in  $\hbar$ one gets constraints on the coefficients  $a_r$. The proof makes a strong use of the assumptions on the derivative $\nabla$, since one needs to exchange derivatives and to pass the Poisson tensor $P^{\m\n}$ through derivatives without getting extra terms from commutators. Finally one obtains that (\ref{moyalasso}) is satisfied if and only if $a_r=1~\forall r$ and this uniquely identifies the function $u$ with the exponential.

Summarizing, under the assumptions made for $\O$, $\nabla$ and $P$, the unique associative $\ast$ product has the form:
\EQ
f \ast g= e^{(\hbar P)}(f,g)
\EN
(modulo a constant overall factor and linear changes of variables).
If at least one of the three hypotesis is not satisfied ($P$ constant with respect to $\nabla$, $\nabla$ without torsion and curvature), then Moyal product is not associative anymore. This can be easily seen by considering second and third order terms in $\hbar$.\\
Once the product  $\ast$ is known, the commutation relations among coordinates are determined by considering the special case of two coordinates themselves as functions $f$ and $g$ .
If $\Omega$ is flat spacetime described by coordinates $\left\{x^{\m}\right\}$ and ordinary derivatives, we can take as a Poisson structure the one associated to a constant antisymmetric matrix $P^{\m\n}$. In this case we obtain the coordinate algebra
\EQ
\left[x^{\m}, x^{\n}\right]_{\ast}\equiv x^{\m}\ast x^{\n}-x^{\n}\ast x^{\m}=2\hbar P^{\m\n}.
\EN
Usually in the definition of the commutator an $i$ is factorized so that the matrix $P$ is hermitian. Moreover, the parameter $\hbar$ is sometimes absorbed into the definition of the matrix $P$.

In flat spacetime the choice of a constant $P^{\m\n}$ is deeply related to translation invariance. In fact, if we want to deform a theory with this symmetry, the only deformation of the coordinate algebra that preserves it is the one associated to a constant symplectic matrix.  

Consider the commutation relations $\left[x^{\m},x^{\n}\right]=i\th^{\m\n}(x)$ that we would like to implement in a certain theory originally defined in terms of commuting coordinates $x^{\m}$.  Suppose the original theory to be symmetric with respect to the transformation $x\longrightarrow x'$.
For the symmetry to be preserved in the deformed theory the new coordinate algebra must be invariant with respect to that trasformation. When we say invariant we mean that the functional dependence on $x$ variable must not change under the transformation, that is
\bea 
\left[x^{\m},x^{\n}\right]=i\th^{\m\n}(x) \Longrightarrow_{x\rightarrow x'}\left[x'^{\m},x'^{\n}\right]=i\th^{\m\n}(x')
\ena
Note that the matrix $\th^{\m\n}$ only transforms punctually and does not become a new matrix $\th'^{\m\n}$.
The matrix $\th^{\m\n}$ is arbitrarily chosen, defines the noncommutative manifold and must be same for all $x$ on the manifold.

Again, let us consider the case of flat spacetime. We would like to deform a Poincar\'e invariant theory. We will first consider translations $x\longrightarrow x+a$ to see which conditions must be imposed on the matrix $\th^{\m\n}$ for the deformed algebra not to break this symmetry.
\bea
\left[x'^{\m},x'^{\n}\right]=\left[x^{\m}+a^{\m},x^{\n}+a^{\n}\right]=\left[x^{\m},x^{\n}\right]
\label{invariance}
\ena
So $\th$ must satisfy the constraint
\EQ
\th^{\m\n}(x+a)=\th^{\m\n}(x)
\label{invariance2}
\EN
Since $\th^{\m\n}$ must be a local function, it has to be constant. Therefore, in flat spacetime the only nontrivial deformation preserving translation invariance is the one with constant commutators. 

Now we will consider Lorentz invariance \cite{lorentz}. 
Two different kinds of Lorentz transformations can be considered, the ones where the observer moves while the particle stands still (``observer" Lorentz transformations) and the ones where the particle is boosted or rotated and the observer is fixed (``particle" Lorentz transformations). In the first case it is sufficient for the physics of the system not to change that the matrix $\th^{\m\n}$ transforms covariantly.
In the second case instead the matrix $\th^{\m\n}$ must not transform, since the coordinate algebra must remain unaltered while moving from $x$ to $x'$. Thus in this case physics changes under the transformation and the symmetry is broken.

Let us explicitly consider the ``particle" Lorentz transformation $x^{\m}\longrightarrow x'^{\m}=\L^{\m}_{~\n}x^{\n}$.
It happens that
\EQ
\left[x'^{\m},x'^{\n}\right]=\left[\L^{\m}_{~\r}~x^{\r},\L^{\n}_{~\s}~x^{\s}\right]=\L^{\m}_{~\r}~\left[x^{\r},x^{\s}\right]~\L^{\n}_{~\s}=\L^{\m}_{~\r}~\th^{\r\s}~\L^{\n}_{~\s}\not= _{D>2}\th^{\m\n}
\EN
We conclude that noncommutative theories in $D>2$ dimensions cannot preserve ``particle" Lorentz transformation, while ``observer" Lorentz transformation are not broken by the deformation. An exception to this general rule is the two-dimensional case, where every antisymmetric matrix is a number times the Ricci tensor $\e^{\m\n}$, which is Lorentz invariant. 

In most papers concerning noncommutative field theory the following choice is made
 \bea
 \th^{0i}=0,~~~~~~~ \th^{ij}\not=0
 \label{restriction}
 \ena
Time is ``isolated" with respect to spacial directions and Lorentz symmetry is manifestly broken. As anticipated in the beginning of this section, time-space noncommutativity is likely to cause a breakdown of the usual framework of quantum mechanics. 
Actually, it has been shown that time-space noncommutativity is responsible for unitarity \cite{unitar} and causality \cite{causal} problems in noncommutative field theory (see section 1.1.2). To avoid this, the restriction (\ref{restriction}) is applied in most work concerning noncommutative field theory. 

Finally I would like to point out that the discussion about symmetries I have presented here is based on the assumption that the symmetry group is undeformed (i.e. it is a classical symmetry group and not a quantum group). This means that parameters of symmetry transformations are commuting. This is not the only possible way to proceed.  There is a branch of mathematics called Quantum Algebra that studies the deformation of symmetry groups. An interesting example is the $\kappa$-deformation of Minkowski space \cite{polacchi}, where parameter and coordinate algebras have an identical structure. Since Minkowski spacetime can be defined as the quotient between Poincar\'e and Lorentz groups, it may seem natural to take also into consideration deformations of Minkowski space that are accompanied by an analogous deformation in the translation symmetry group.
\vskip 24pt
I briefly summarize the results obtained, in the special case $\O=R^{2n}$ (the extension to Minkowski signature is straightforward).
The only product $\ast$ defined as in (\ref{2}, \ref{poisson}) that is associative and that preserves translation invariance is Moyal product 
\EQ
(f\ast g)(x)=e^{\frac{i}{2}\hbar\th^{ij}\pa_{i}\pa'_{j}}~f(x)~g(x')\vert_{x=x'}
\label{moyal}
\EN
that generates the coordinate algebra 
\EQ
\left[x^i,x^j\right]_{\ast}=i~\hbar~\th^{ij}
\EN
As we have seen before, Moyal product is also naturally obtained in quantum mechanics through the Weyl-Moyal correspondence defined between quantum operators and functions on ``phase-space". 
\subsubsection{Properties of Moyal product}
Here I will summarize some useful properties of Moyal product $\ast$.
Associativity and covariance with respect to translations have been already discussed.
\begin{enumerate}
\item{The $\ast$ product between exponential functions reflects from the functional point of view the well-known  Baker-Campbell-Hausdorff formula
\bea
&&e^{ikx}\ast e^{iqx}=e^{i(k+q)x} e^{-\frac{i}{2}(k\th q)}\cr
&&k\th q\equiv k^{\m} q^{\n}\th_{\m\n}
\label{exp}
\ena}
\item{ By making use of the previous formula we can obtain the representation of  $\ast$ in momentum space
\EQ
\left(f\ast g\right)(x)=\frac{1}{(2\p)^8}\int d^4k d^4q \tilde{f}(k) \tilde{g}(q) e^{-\frac{i}{2}(k\th q)} e^{i(k+q)x}
\label{momenti}
\EN}
\item{Commutativity is recovered under integration
\EQ
\int(f\ast g)(x) d^4x=\int(g \ast f)(x)d^4x=\int(f\cdot g)(x) d^4x
\label{int}
\EN
since all the corrections in (\ref{moyal}) with respect to the ordinary product are total derivatives, because of the antisymmetry of $\th^{\m\n}$.}
\item{A cyclicity property can be deduced from the previous relation
\EQ
\int(f_1 \ast f_2 \ast...\ast f_n)(x)d^4x=\int(f_n\ast f_1\ast...\ast f_{n-1})(x)d^4x
\label{cicli}
\EN}
\item{Finally, $\ast$ has the following behavior with respect to complex conjugation
\EQ
\left(f\ast g\right)^*=g^* \ast f^*
\EN
because of the antisymmetry of  $\th^{\m\n}$. Clearly $f\ast f$ is real when $f$ is real, but when both $f$ and  $g$ are real, $ f \ast g$ is generally complex.}
\end{enumerate}
\subsection{The natural Moyal deformation of a field theory}
In this section I will discuss the main properties of noncommutative field theories that are obtained from ordinary ones by replacing ordinary products with Moyal $\ast$ products (\ref{moyal}) in the action.
This is what I will call the natural deformation of a given field theory. I will first discuss the simple case of scalar field theory with $\Phi^4$ interaction \cite{micu}. This is chosen for simplicity and most of the features we will find in this case can be easily generalized to more complicated situations. I will then move to gauge theories, to see how gauge invariance is modified in noncommutative space.
In this first two subsections I will only take into consideration the restricted case (\ref{restriction}).
In the last subsection I will instead discuss unitarity and causality  problems arising when time is involved in noncommutativity. 

\subsubsection{A simple example: The scalar $\Phi^4$ theory}
Let us consider the natural noncommutative deformation of a given ordinary field theory, for instance the scalar theory with $\Phi^4$ interaction. We have already seen that we can obtain the deformed theory by replacing ordinary products with $\ast$ products everywhere in the action. We choose Moyal product because we want to preserve translation invariance in the deformed theory and we want associativity. The action for the noncommutative theory is 
\EQ
S\left[\Phi\right]=\int d^4x\left[\frac{1}{2}\pa_{\m}\Phi \ast \pa^{\m}\Phi-\frac{m^2}{2}\Phi\ast\Phi-\frac{\l}{4!}\Phi\ast\Phi\ast\Phi\ast\Phi\right]
\EN
Property (\ref{int}) implies that the quadratic part of the action does not receive corrections from the star products. Only the interaction term is modified, so the free theory is the same as the ordinary one. The noncommutative theory is built on the same Fock space as the commutative one, but it has different interactions. This feature is common to all theories obtained as deformation of ordinary ones by implementing Moyal product, since it just relies on property (\ref{int}).

We can easily deduce Feynman rules from (\ref{momenti}). Introducing the Fourier components $\phi(k)$ of $\Phi(x)$
\EQ
\Phi(x)=\frac{1}{(2\p)^4}\int d^4k e^{ikx}\phi(k)
\EN
We obtain
\bea
&&S_{\rm{int}}=\frac{\l}{4!}\int d^4x~ \Phi\ast\Phi\ast\Phi\ast\Phi\cr
&&=\frac{1}{(2\p)^{16}}\frac{\l}{3\cdot4!}\int d^4k_1...d^4k_4~\phi(k_1)\phi(k_2)\phi(k_3)\phi(k_4)\cdot(2\p)^4\d^{(4)}(\sum_{i=1}^4 k_i)\cr
&&\cdot\left[\cos \frac{k_1\th k_2}{2}\cos\frac{k_3\th k_4}{2}+\cos\frac{k_1\th k_3}{2}\cos\frac{k_2\th k_4}{2}+\cos\frac{k_1\th k_4}{2}\cos\frac{k_2\th k_3}{2}\right]\cr
&&~~~~~~~
\label{feynman}
\ena
We conclude that the only difference between the natural deformation of a field theory and the field theory itself is a phase factor depending on momenta and noncommutativity parameter $\th$, appearing in front of every vertex in the Feynman rules. This procedure can be clearly generalized to other field theories.

Now I'm going to discuss how this phases modify perturbation theory, in particular ultraviolet behaviour and renormalization. 
Since the phases appearing in front of vertices depend on the momenta, when we compute the contribution coming from a certain diagram we have to distinguish between two different situations. If the phase is only depending on external momenta, it does not affect loop integrations and thus it does not modify the degree of divergence. This case we will call planar. Instead, when the phase factor depends on internal, loop momenta, it generally modifies the ultraviolet behavior of the diagram. This case we will call nonplanar. So a single diagram in the ordinary theory decomposes in various planar and nonplanar contributions, depending on the ordering of momenta in the vertices. 

A nice feature of natural Moyal deformations of ordinary field theories is that nonplanar graphs always display a better ultraviolet behavior with respect to the corresponding planar ones, since the phase acts as a regulator. So one can say that such deformation of a renormalizable theory will also be renormalizable. It will display the same degree of divergence in planar diagrams and a lower degree of divergence in nonplanar ones \cite{filk}. A general discussion of renormalizability properties of noncommutative field theory can be found in \cite{roibanren}.

Now I would like to discuss a typical feature of noncommutative field theories called UV/IR mixing. To this purpose I will present the result of the 1-loop computation for the renormalized two-point function $\G^{(2)}_{\rm{ren}}$ in the case of  $\Phi^4$ theory. I will not give any detail about the computation. The interested reader should refer to \cite{micu}.

Let $\l$ be the coupling constant of the theory, $\L$ the ultraviolet cutoff, $M$ the renormalized mass, $p$ the incoming momentum. Moreover we will define  $p\circ k\equiv p\th\th k=p_{\m}\th^{\m\r}\th_{\r}^{~\n}k_{\n}$, where $\th^{\m\n}$ is the noncommutativity matrix characterizing the theory.\\ One finds that the renormalized  $\G^{(2)}$ in the limit $\L_{\rm{eff}}\equiv\frac{1}{\left(p\circ p+\frac{1}{\L^2}\right)^{\frac{1}{2}}}\rightarrow 0$ takes the form
\EQ
\G^{(2)}_{\rm{ren}}(p,M,\L)=p^2+M^2+\frac{\l}{96(2\p)^2(p\circ p+\frac{1}{\L^2})}-\frac{\l M^2}{96\p^2}\ln\frac{1}{M^2(p\circ p+\frac{1}{\L^2})}+\mathcal{O}(\l^2)
\label{gamma2}
\EN
If we then take the limit $\L\rightarrow \infty$ we observe that an infrared divergence appears when we take $p\rightarrow 0$.
If we instead take the limit $p\rightarrow 0$ first, we discover that the cutoff does not appear explicitly anymore and the two-point function diverges for  $\L\rightarrow \infty$. So we observe an interesting connection between the UV and IR behaviors in the extra terms appearing in the two-point function because of noncommutativity. This is known in the literature as UV/IR mixing.

Finally, we will consider the limit $\th\rightarrow 0$. In this limit we expect to obtain the standard result for the renormalized $\G^2$ of ordinary $\Phi^4$ theory. We have already discussed the fact that a diagram in the ordinary theory splits in planar and nonplanar parts in the noncommutative theory. 
In (\ref{gamma2}) the subtraction made to obtain the renormalized mass only took into consideration the divergences coming from the planar graphs. The last two terms represent the contribution coming from nonplanar diagrams. In the limit $\th\rightarrow 0$ one may verify that by adding to the planar contributions the nonplanar ones in the computation of the renormalized mass, one obtains the well-known result for  ordinary $\Phi^4$.
\subsubsection{Gauge theories}
Up to now I have considered scalar theory for simplicity. Now I would like to discuss Yang-Mills theories. In ordinary geometry these theories are constructed by promoting a global invariance to a local one. In general the gauge group is nonabelian.

First of all a remark is needed regarding the choice of the gauge group. In noncommutative geometry described by Moyal product it is easy to show that $SU(n)$, $SO(n)$ and $Sp(n)$ are not closed any more and the same is valid for the corresponding Lie algebras \cite{closure}. This is due to the fact that a nontrivial trace part appears in the product of two traceless matrices. So we will consider $U(n)$ gauge theory as our example\footnote{The cases $SO(n)$, $Sp(n)$ seem to allowed from the subtle string theory discussion in \cite{tomasiello}}. 

The ordinary gauge theory is described by an $n\times n$  hermitian matrix of vectors, $A_{\m}$. This transforms as follows under the local gauge symmetry with parameter $\l$ (also an $n\times n$ matrix)
\EQ
\d_{\l}A_{\m}=\pa_{\m}\l+i[\l,A_{\m}]
\EN
 The field strength $F_{\m\n}$ and its transformation law are given by
\bea
&&F_{\m\n}=\pa_{\m}A_{\n}-\pa_{\n}A_{\m}-i[A_{\m},A_{\n}]\cr
&&\d_{\l}F_{\m\n}=i[\l,F_{\m\n}]
\label{gauge}
\ena
The pure gauge action is given by
\EQ
\int Tr\left(F^{\m\n}F_{\m\n}\right)
\label{gaugeaction}
\EN
where the trace acts on gauge indices and the integral is taken over spacetime variables.
The invariance of (\ref{gaugeaction}) under (\ref{gauge}) follows from cyclicity of the trace.

We are going to construct the natural Moyal deformation of the theory by substituting ordinary products with $\ast$ products in the action. This time we have matrix-valued fields, so the product between two of them will be the tensor product between $\ast$ and the matrix product. We obtain the action
\EQ
\int Tr\left(F^{\m\n}\ast F_{\m\n}\right)
\label{actionnc}
\EN
If we also substitute ordinary products with Moyal products in the gauge transformation and definition of the field strength 
\bea
&&\d_{\l}A_{\m}=\pa_{\m}\l+i\l\ast A_{\m}-iA_{\m}\ast \l\cr
&&F_{\m\n}=\pa_{\m}A_{\n}-\pa_{\n}A_{\m}-iA_{\m}\ast A_{\n}+iA_{\n}\ast A_{\m}\cr
&&\d_{\l}F_{\m\n}=i\l\ast F_{\m\n}-iF_{\m\n}\ast \l
\label{gaugenc}
\ena
we find that the gauge transformation is still a symmetry of the action.
In the noncommutative case the invariance of (\ref{actionnc}) under (\ref{gaugenc}) is more subtle, though. Cyclicity of the trace is valid for ordinary matrix product, but not for the tensor product of the latter with $\ast$. However, the cyclicity property of Moyal product under integral (\ref{cicli}) can be extended to the case of matrix-valued fields and used to prove the invariance of the action. In the limit $\th\rightarrow 0$ one recovers the ordinary theory.
 
It is very important to observe that  while in the ordinary case $n=1$ corresponds to an abelian theory  with field strength and transformation laws 
\bea
&&\d_{\l}A_{\m}=\pa_{\m}\l\cr
&&F_{\m\n}=\pa_{\m}A_{\n}-\pa_{\n}A_{\m}\cr
&&\d_{\l}F_{\m\n}=0
\ena
in the noncommutative case the commutator of two gauge transformations with parameters $\l_1$ and $\l_2$ is the gauge transformation with parameter  $\l_1\ast \l_2-\l_2\ast\l_1$ \cite{seibwit}. This is nontrivial even in the case $n=1$, so this is a nonabelian theory and its features perfectly mimic the case with $n>1$.
\subsubsection{Unitarity and causality problems}
As anticipated, problems arise when time-space noncommutativity is considered.
The structure of Moyal product (\ref{moyal}) leads to terms in the action with an infinite number of derivatives of fields. This renders a Moyal-deformed field theory non local. In particular, when time is involved in noncommutativity, nonlocality in time appears and the usual framework of quantum mechanics breaks down. 

In \cite{unitar} unitarity of noncommutative field theory with time-space noncommutativity has been studied. Scalar field theory deformed with time-space noncommutativity has been considered and several one loop amplitudes have been shown not to be unitary. In particular, the two point function in noncommutative $\Phi^3$ theory has been shown not to satisfy the usual cutting rules when $\th^{0i}\neq 0$, while these rules are satisfied when only spatial noncommutativity is present. Moreover, $2\rightarrow2$ scattering in noncommutative $\Phi^4$ has been considered and again unitarity of the S-matrix is satisfied only when $\th^{0i}=0$.

Recently in a series of papers \cite{doplicher} a different approach to perturbative noncommutative field theories with a noncommuting time variable has been proposed. It has been argued that time-ordering is nontrivial when time is involved in noncommutativity and so a new prescription for the computation of Green functions must be given. This is different with respect to the naive Feynman rules obtained by multiplying the usual vertices by a phase factor (see (\ref{feynman})). It has been shown that in this framework unitarity is preserved when the lagrangian of the theory is hermitian.

In \cite{causal} causality of scattering processes in noncommutative field theory with time-space noncommutativity has been investigated. In particular, $2\rightarrow 2$ tree-level scattering amplitudes for massless scalars with $\Phi^4$ interaction in a two-dimensional noncommutative spacetime have been computer there.
The ordinary result for the $2\rightarrow 2$ amplitude is 
\beq
i{\cal M}=-ig
\eeq
where $g$ is the coupling constant of the theory.
In the noncommutative case with $[t,x]=i\th$, because of the phases appearing in front of the vertices, one obtains instead
\beq
i{\cal M}\sim g[\cos(p_1\wedge p_2)\cos(p_3\wedge p_4)+ 2\leftrightarrow 3 + 2\leftrightarrow 4]
\label{ampli}
\eeq
where $p_1,\dots,p_4$ are the two-momenta of the particles satisfying the conservation law $\sum_{i=1}^4 p_i =0$ (all momenta are incoming) and the wedge product is defined as $p\wedge q\equiv \th(p^0 q^1 -p^1 q^0)$.
In the center of mass frame (\ref{ampli}) becomes
\beq
i{\cal M}\sim g[\cos(4 p^2 \th) +2]
\eeq
Given an incoming gaussian wave packet
\beq
\phi_{in}(p)\sim \left(e^{-\frac{1}{\l}(p-p_0)^2} + e^{-\frac{1}{\l} (p+p_0)^2}\right)
\eeq
the outgoing wave packet, in the limit $p_0\gg \l^{\frac{1}{2}}\gg \frac{1}{p_0 \th}$, $\l\th\gg 1$, is expressed as follows
\beq
\Phi_{out}(x)\sim g\left[F(x;-\th,\l,p_0) + 4\sqrt \l e^{-\l\frac{x^2}{4}} e^{i p_0 x} + F(x;\th,\l,p_0) \right] +(p_0\rightarrow -p_0)
\eeq
where $F(x;\th,\l,p_0)$ represents a packet concentrated at $x=-8 p_0 \th$.
So the incoming wave-packet, in the special high energy limit considered, splits into three parts, one concentrated at $x=-8 p_0 \th$, one at $x=0$ and one at $x=8 p_0 \th$. All three propagate towards $x\rightarrow \infty$.

 In the center of mass frame scattering can be seen as bouncing on a wall. The first packet is an advanced one, which means that it leaves the wall much before the arrival of the incoming packet. The third term instead corresponds to a delayed wave, appearing well after the arrival of the incoming packet. These two terms suggest an interpretation of the noncommutative particle as a rigid rod. Both terms originate from the phase factor due to noncommutativity.
The second term is not interesting, since it is neither significantly delayed or advanced.

The advance by itself is not an indication of acausality. A nonrelativistic example of this is the reflection of a rigid rod of length L oriented along the direction of motion. The center of mass of the rod appears to reflect before reaching the wall, but the event is not acausal. However, there is a problem when both causality and Lorentz invariance are considered. In our case the advance increases with energy, therefore the rod seems to expand instead of Lorentz-contract at growing energies. 
This bizarre behavior is a sign of the inconsistency of a field theory with time-space noncommutativity.

When only space-space noncommutativity is considered (in a 2+1 dimensional field theory), the effect of the phase in front of the vertices is to let outgoing scattered waves originate from the diplaced position $y=\frac{1}{2} \th p_x $. This again suggests the interpretation of the incident particles as extended rods of size $\th p$, but orthogonally oriented with respect to their momentum. In this case there is no violation of causality.

In \cite{timenc} it has been shown that in noncommutative field theories with time-space noncommutativity tachyonic particles are produced. This gives a physical interpretation of the perturbative breakdown of unitarity. Moreover, in this paper a quantitative study of various locality and causality properties of noncommutative field theories at the quantum level has been performed.

In collaboration with M.T. Grisaru, O. Lechtenfeld, L. Mazzanti, S. Penati and A. Popov I have also addressed the problem of acausality in noncommutative field theory in \cite{mypaper2,mypaper3}.
We have conjectured that in a noncommutative two-dimensional field theory that is classically integrable, i.e. it has an infinite number of conserved charges, acausality may disappear. In \cite{mypaper3} we have shown that this is indeed the case for the noncommutative integrable sine-Gordon model, whose S-matrix is factorized, as expected for an integrable system, and causal. 
These results will be discussed in detail in chapter 2.\\
The noncommutative generalization of the sine-Gordon system characterized by a well-defined S-matrix is not the natural one, though.  In the next section I will begin to explore some possible noncommutative versions of the scalar theory that differ from the natural one considered in this section. 

In the last part of section 1.2.1 I will discuss time-space noncommutativity from the string theory point of view. There we will see that the ill-defined field theories with time-space noncommutativity do not arise as consistent limits of string theory. However, there exists a limit where one obtains a theory of open strings living in a noncommutative spacetime (NCOS).

\subsection[Other possible deformations: The free scalar field theory\\ example]{Other possible deformations: The free scalar field theory example}
Up to now I have considered the natural deformation of a given field theory, obtained by simply replacing ordinary products with $\ast$ products everywhere in the action (and in the definitions of the field strength and gauge transformations in the Yang-Mills case). I have discussed some of the peculiar properties of noncommutative field theories obtained in such a way and noted that some of these are not welcome in a reasonable field theory.

The natural deformation is not the only way to proceed. Given an ordinary field theory we can more generally define a noncommutative deformation of it as a theory written in terms of $\ast$ products that reproduces the original commutative theory in the limit $\th\rightarrow 0$. Of course the natural deformation is included in this definition, but different deformations can be constructed, just by adding new terms that vanish in the limit $\th\rightarrow 0$.

I would like to discuss a very simple example, the two-dimensional free massless scalar field theory.
We have previously noted that quadratic terms in the action are not modified by Moyal product.
So the natural deformation of a free scalar field theory is trivial.
There are more possible deformations, though, that are highly nontrivial and very interesting indeed.

In ordinary geometry we can consider the element $g$ of a nonabelian group $\cal {G}$. With this we can  construct the principal chiral model action
\bea
S_{PC}=\int d^2 x~ \pa_\m g^{-1} \pa^\m g
\ena
The corresponding equation of motion is given by
\bea
\pa_\m\left(g^{-1}\pa^\m g\right)=0
\ena
In the case of the abelian group ${\cal{G}}=U(1)$ we can parametrize $g=e^{i\a\phi}$ and see that the action reduces to the free massless scalar field action. In the nonabelian case instead the model is nontrivial.

It is possible to add to the principal chiral action a new term, called Wess-Zumino (WZ) term, that is written in terms of a commuting parameter $\rho\in [0,1]$. The resulting theory is called Wess-Zumino-Witten (WZW) \cite{wittenwzw} model and its action in a $(+,-)$ signature is
\bea
&&S_{WZW}={1\over 2}\int d^2 x ~ \pa_\m g^{-1} \pa^\m g -{1 \over 3}\int d^2x ~d\rho ~\e^{\m\n\sigma} ~\hat g^{-1}\pa_\m \hat g~ \hat g^{-1}\pa_\n \hat g~ \hat g^{-1}\pa_\sigma \hat g\cr
&&~~~~~~~~~~~
\ena
where we have introduced the homotopy path $\hat{g}(\rho)$ such that $\hat{g}(0)=1$,
$\hat{g}(1) = g$. 
The variation of the WZ term is a total derivative in $\rho$, so the equations of motion obtained from this action are truly two-dimensional and given by
\bea
\bar \pa\left(g^{-1}\pa g\right)=0
\ena
where we have defined $\pa\equiv\pa_0 +\pa_1$ and $\bar \pa\equiv \pa_0-\pa_1$.
For an abelian $U(1)$ group the WZ term vanishes and the WZW model reduces to the free massless scalar theory. 

We have previously observed that in noncommutative geometry the $U(1)$ group is no longer abelian. Parametrizing the element of noncommutative $U(1)$ as
\bea
g=e_\ast^{i\a\phi}
\ena
we can define noncommutative $U(1)$ principal chiral and WZW models \cite{wzwnc}, just substituting $\ast$ products everywhere in the given actions and considering $g$ in the noncommutative $U(1)$ group.
Both models reduce to the free massless scalar theory in the commutative limit, so they are nontrivial noncommutative generalizations of it.

Since many possible noncommutative generalizations of a single ordinary field theory can be constructed, it is natural to wonder whether one of these may be ``better" than the others so that it could be chosen as ``the" noncommutative version of the original theory.

First of all, criteria should be given to decide whether one generalization is better than the other.
There are a certain number of properties that render a field theory a ``good" theory.
For instance symmetries, classical integrability, causality, unitarity, renormalizability, quantum integrability, absence of anomalies, dualities are properties that, when present in the ordinary theory, we would like to preserve in its noncommutative version. So a good criteria could be to find a deformation that preserves the good properties of the ordinary theory, or some of them at least.

The case of a free scalar field theory is very simple and cannot give us any hint. However, it is possible to add a potential term to the scalar theory to obtain an interacting theory enjoying nice properties.
In my papers \cite{mypaper2,mypaper3}, in collaboration with M. Grisaru, O. Lechtenfeld, L. Mazzanti, S. Penati and A. Popov, I have studied the possible generalizations of a very special scalar theory, the sine-Gordon system, describing a scalar field autointeracting by an oscillating potential. As will be explained in detail in chapter 2, from our analysis of noncommutative sine-Gordon the WZW-like generalization of the kinetic term for scalar fields seems to be preferred.
This agrees with bosonization considerations in \cite{MS, schiappa}.
 
\subsection{The Kontsevich product}
In section 1.1.1 we have seen that, in a Poisson manifold with a derivative without torsion and curvature, such that $\nabla P=0$,the only associative product is Moyal $\ast$ (\ref{moyal}). If one of the three assumptions is not verified, then this product is no longer associative.
Moreover, we have noticed that in flat spacetime with ordinary derivatives the assumption of constant $P$ is a restriction needed for the noncommutative deformation of a translation invariant theory to preserve this property.

In this section I would like to consider more general situations, where spacetime may be curved and translations may not be a symmetry anymore. In particular, as I will show in detail in section 1.2.1, noncommutative geometry naturally emerges in the context of string theory, that is naturally embedded in curved backgrounds. Therefore, spacetimes with torsion and curvature should be considered. Different symmetries underlying the theory may require Poisson structures $P$ with a particular dependence on the coordinates.
For this reasons it is interesting to consider the case when the Poisson structure $P$ is not constant with respect to a certain set of derivatives. One may think about relaxing the other two constraints regarding the set of derivatives chosen. Actually, at least in some cases it is possible to rewrite a Poisson structure with nonflat derivatives and a covariantly constant Poisson tensor in terms of a nonconstant Poisson tensor and flat derivatives. The superspace case we will study in the next section is an example of this.

M. Kontsevich in \cite{kontsevich} generalized the results of Bayen et al. to the case where derivatives $\nabla$ are without torsion and curvature but the Poisson structure is not covariantly constant $\nabla{P}\not=0$. In this case Moyal product (\ref{moyal}) is not associative anymore. However, it is possible to modify Moyal product order by order in the deformation parameter $\hbar$ to obtain associativity.

Let $\O$ be a Poisson manifold, with coordinates $\{x^{\m}\}$ and flat derivatives $\{\nabla_{\m}\}$. Let $P^{\m\n}=P^{\m\n}(x)$ be the Poisson structure of this manifold, written in terms of the chosen coordinates as follows
\EQ
[x^{\m},x^{\n}]=P^{\m\n}(x)
\EN 
First of all we recall that the definition of a Poisson structure requires associativity
\EQ
P^{\m\r}\nabla_{\r} P^{\n\s}+P^{\n\r}\nabla_{\r} P^{\s\m}+P^{\s\r}\nabla_{\r} P^{\m\n}=0
\label{assoc}
\EN
that is completely equivalent to the validity of Jacobi identity for the coordinate algebra
\EQ
\left[[x^{\m},x^{\n}],x^{\s}\right]+\left[[x^{\s},x^{\m}],x^{\n}\right]+\left[[x^{\n},x^{\s}],x^{\m}\right]=0
\label{jacobi}
\EN  
When $P^{\m\n}$ is constant, this requirement is trivially satisfied and does not give any further constraint on  the matrix $P$. When a general coordinate dependence is allowed, associativity constrains the functional dependence of the matrix. 
When $P^{\m\n}$ is invertible, i.e. it exists $P^{-1}_{\m\n}$ satisfying $P^{\m\n}P^{-1}_{\n\rho}=\d^\m_\rho$, (\ref{assoc}) can easily be rewritten in terms of the vanishing of the three-form $H$
\beq
H_{\m\n\rho}=\nabla_\m P_{\n\rho} +{\rm cycl.}=0
\label{assocH}
\eeq

Let us now consider Moyal product (\ref{moyal}), expanded up to second order in the parameter $\hbar$
\bea
&&f \ast g=fg+\hbar P^{\m\n}(x)\nabla_{\m}f~\nabla_{\n}g+\hbar^2 P^{\m\n}(x)P^{\r\s}(x)\nabla_{\m}\nabla_{\r}f~ \nabla_{\n}\nabla_{\s}g+\mathcal{O}(\hbar^3)\cr
&&~~~~~~
\label{moyal2}
\ena
We then evaluate the quantity $(f\ast g)\ast h-f \ast(g\ast h)$ up to second order in $\hbar$ (note that associativity is trivially satisfied at first order):
\EQ
(f\ast g)\ast h-f \ast(g\ast h)=-\hbar^2\left(P^{\s\r}\nabla_{\r}P^{\m\n}+P^{\m\r}\nabla_{\r}P^{\n\s}\right)\nabla_{\m}f~\nabla_{\n}g~\nabla_{\s}h+\mathcal{O}(\hbar^3)
\EN
So nonvanishing terms arise, because of the $x$ dependence of the Poisson structure. Kontsevich observed that once the trilinear dependence on the functions $f$, $g$, $h$ is factorized, one obtains terms with an identical structure with respect to the ones emerging in the associativity equation for $P$ (\ref{assoc}). If one could modify the definition of the product $\ast$ (\ref{moyal2}) in such a way to obtain exactly the quantity that is constrained to be zero in (\ref{assoc}), one would obtain a product associative up to order $\hbar^2$.

We have to add new terms of order $\hbar^2$, since associativity is trivially satisfied at first order. 
Therefore let us define a new product $\star$ by adding to $\ast$ a new term of order $\hbar^2$ as follows
\bea
f\star g\equiv&& fg+\hbar P^{\m\n}(x)\nabla_{\m}f~\nabla_{\n}g+\hbar^2 P^{\m\n}(x)P^{\r\s}(x)\nabla_{\m}\nabla_{\r}f~ \nabla_{\n}\nabla_{\s}g\cr
&&+A\hbar^2 P^{\m\r}\nabla_{\r}P^{\n\s}\left(\nabla_{\m}\nabla_{\n}f~\nabla_{\s}g-\nabla_{\n}f~\nabla_{\m}\nabla_{\s}g\right)+\mathcal{O}(\hbar^3)\cr
&&~~~~~
\label{kont}
\ena
where $A$ is a coefficient to be determined in such a way that  $(f\star g)\star h-f \star(g\star h)$ is proportional to the constraint (\ref{assoc}). One obtains
\EQ
A=\frac{1}{3}
\label{kont2}
\EN

The product $\star$ that we have defined up to second order in $\hbar$ is associative if and only if the Jacobi identity for the coordinate algebra (\ref{jacobi}) is satisfied. Kontsevich showed that order by order in $\hbar$ it is always possible to modify Moyal product $\ast$ in order to make extra terms vanish when the Jacobi identity for the coordinates is valid. So Kontsevich $\star$ product is uniquely defined at any order in the deformation parameter $\hbar$ by the requirement of associativity.
\subsection{Deforming superspace}
Up to now,  we have only considered noncommutativity of  bosonic coordinates, in the form
\beq
[x^{\mu}, x^{\nu}] = i\Theta^{\mu\nu}(x)
\label{bosonic}
\eeq
where $\Theta^{\mu\nu}(x)$ is antisymmetric.
Since, as it will be explained in section 2, noncommutative field theories emerge naturally in the context of string theory and string theory is only consistent in the presence of supersymmetry, it is natural to consider the problem of deforming a supersymmetric theory.

It is well-known that the natural setting for discussing supersymmetric theories is a nontrivial extension of bosonic space, known as superspace, where bosonic coordinates $x$ are accompanied by fermionic ones, that I will generally denote with $\th$. 
So it seems natural and
compelling to ask what happens if we deform also the anticommutators
between fermionic coordinates of superspace. Exactly as in the bosonic case discussed before, we would like the deformation to preserve the symmetries of our original theory, described by the group of supertranslations. Moreover, we would like the deformed algebra to be associative. In the bosonic case we have seen that this two properties in flat space were identifying Moyal product. 

In collaboration with D. Klemm and S. Penati, I have addressed the problem of deforming superspace in \cite{mypaper1}. 
First steps had been taken in this direction before the appearance of this paper. 
Nonvanishing anticommutators of fermionic coordinates have been
considered in \cite{Schwarz:pf} in the context of a
possible fermionic substructure of spacetime.
In \cite{kosinski}, quantum deformations of the Poincar\'{e}
supergroup were considered. 
In a modern noncommutative geometry context, trivial superspace deformation of supersymmetric field theories have been analysed, where only the bosonic sector of the coordinate algebra is modified \cite{zanon}. In \cite{ferrara}, a Moyal-like deformation of $d=4$ $N=1$ superspace was proposed involving fermionic coordinates, that is associative and covariant with respect to supersymmetry, but does not preserve the complex conjugation rules that characterize Majorana-Weyl spinors in four dimensions. In this paper it was also
shown that in general the set of chiral superfields is not closed under
star products that involve fermionic coordinates. I will discuss the results in \cite{ferrara} in detail in the second part of this section.

In \cite{mypaper1} we were mainly concerned with the conditions imposed
on the possible deformations of superspace by requirements such
as covariance under {\em classical} translations and supertranslations,
Jacobi identities, associativity of the star product and closure of the set of
chiral superfields under the star product, but we wanted superspace conjugation relations to be still valid in the deformation. The motivation for this requirement relies in the fact that in a theory with $N$ supersymmetries formulated in superspace, the number of fermionic degrees of freedom is $N$ times the bosonic one. As I will show in detail, by relaxing spinor conjugation relations in the deformation, one in fact does not preserve the number of supersymmetries.

The main results in \cite{mypaper1} are that it is possible to fulfill all the requirements in a $d=4$ $N=1$ Minkowski superspace, even if the contraints imposed on the supercoordinate algebra are strong and only  $[x,\th]$ and $[x,x]$ can be turned on. Moreover, in the same paper we have shown that euclidean signature is less restrictive and allows for a nonanticommutative superspace with $\{\th,\th\}$ different from zero. The results obtained in my paper will be discussed in section 1.1.6.

Since then a lot of progress has been done in understanding superspace deformations.
Non(anti)commutative superspaces have been shown to emerge in a superstring theory context, in the presence of Ramond-Ramond (RR) backgrounds \cite{vafaooguri2,seiberg,antonio,seibergberkovits}. I will discuss these results in section 1.2.4. Furthermore, in the paper by N. Seiberg \cite{seiberg}, a deformed superspace that only preserves $N={1\over 2}$ supersymmetric of the original $N=1$ has been introduced. I will review the properties of this deformation in section 1.1.7 and compare with the ones obtained in my paper.
Many deformations of superspace field theories have been studied and their quantum properties have been discussed. I will give a brief summary of the main results obtained in the second part of section 1.1.7.

I will not give an introduction to supersymmetry and its superspace formulation. For an introduction to this topics, I suggest the books \cite{wessbagger,superspace} and the review paper \cite{sohnius}.
\subsubsection{The Ferrara-Lled\`o proposal}
In \cite{ferrara} the authors consider the problem of generalizing Moyal product (\ref{moyal}) to $d=4$ $N=1$ superspace. 
This is described by the set of superspace coordinates $Z^A = (x^{\a \ad}, \th^\a, \thb^\ad)$,
where $x^{\a \ad}$ are four real bosonic coordinates and $\th^\a$,
$\thb^\ad$ are two--component complex Weyl fermions.
The conjugation rule $\thb^\ad = (\th^\a)^{\dag}$ follows from the
requirement to have a four component Majorana fermion
(we use conventions of {\em Superspace} \cite{superspace}).
In the standard (anti)commutative superspace the algebra of the coordinates is
\bea
&& \{ \th^\a , \th^\b \} ~=~\{ \thb^\ad , \thb^\bd \} ~=~
\{ \th^\a , \thb^\ad \} ~=~ 0 \non\\
&& [ x^{\a \ad} , \th^\b ] ~=~ [ x^{\a \ad} , \thb^\bd ] ~=~ 0 \non\\
&& [ x^{\a \ad} , x^{\b \bd} ] ~=~ 0
\label{coord}
\eea
and it is trivially covariant under the superPoincar\'e group.
The subgroup of the classical (super)translations (spacetime translations and
supersymmetry transformations)
\bea
&& \th'^\a ~=~ \th^\a ~+~ \epsilon^\a \non \\
&& \thb'^\ad ~=~ \thb^\ad ~+~ \bar{\epsilon}^\ad \non \\
&& x'^{\a \ad} ~=~ x^{\a \ad} ~+~ a^{\a \ad}
~-~\frac{i}{2} \left( \epsilon^\a \thb^\ad ~+~
\bar{\epsilon}^\ad \th^\a \right)
\label{transf1}
\eea
is generated by two
complex charges $Q_\a$ ($\bar{Q}_\ad = Q_\a^\dag$) and the four--momentum
$P_{\a \ad}$ subjected to
\beq
\{ Q_\a , Q_\b \} ~=~ \{ \bar{Q}_\ad , \bar{Q}_\bd \} ~=~ 0  \quad , \quad
\{ Q_\a , \bar{Q}_\ad \} ~=~ P_{\a \ad}
\eeq

Representations of supersymmetry are given by superfields
$V(x^{\a \ad}, \th^\a, \thb^{\ad})$ whose components are obtained by expanding
$V$ in powers of the spinorial coordinates. The set of superfields is closed
under the standard product of functions. The product of two superfields
is (anti)commutative, $V \cdot W = (-1)^{deg(V) \cdot deg(W)} W \cdot V$,
and associative, $(K \cdot V) \cdot W = K \cdot (V \cdot W)$.
The set of superspace covariant derivatives is given by $\pa_A=(\pa_{\a\ad}, D_\a, \bar D_\ad)$.
In both this section and the following one (so in the papers \cite{ferrara} and \cite{mypaper1}) the nonchiral representation of supersymmetry has been chosen, where
\bea
D_\a= \pa_\a +\frac{i}{2} \thb^\ad \pa_{\a\ad}~~~~~~~ ;~~~~~~~\bar D_\ad=\bar \pa_\ad +\frac{i}{2}\th^\a \pa_{\a\ad}
\ena
Superspace geometry is nontrivial, because of the presence of a nonvanishing torsion
\beq
\{D_\a,\bar D_\ad\}=i\pa_{\a\ad}
\label{torsion}
\eeq

To extend the construction of Moyal product to superspace, one has first to introduce a superspace Poisson structure generalizing (\ref{poisson}).  The authors propose
\beq
\left\{\Phi,\Psi\right\}=\Phi {\overleftarrow{\pa}}_{\a\ad} P^{\a\ad\b\bd} {\overrightarrow{\pa}}_{\b\bd} \Psi + \Phi {\overleftarrow {D}}_\a P^{\a\b} {\overrightarrow{D}}_\b \Psi
\eeq
where $P^{\a\ad\b\bd}$ and $P^{\a\b}$ are constant matrices. This Poisson structure is manifestly covariant with respect to supersymmetry. Moreover, it is associative since it involves only $\pa_{\a\ad}$ and $D_\a$ and not $\bar D_\ad$. Associativity would be broken if the whole set of superspace covariant derivatives had appeared, because of the nontrivial superspace torsion (\ref{torsion}). With this Poisson structure the authors construct a Moyal-like product in superspace as follows
\beq
\Phi \ast \Psi = \Phi~ \exp\left(\hbar\left(\overleftarrow{\pa}_{\a\ad} P^{\a\ad\b\bd} \overrightarrow{\pa}_{\b\bd} + \overleftarrow {D}_\a P^{\a\b} \overrightarrow{D}_\b\right)\right)\Psi
\label{ferrarastar}
\eeq
This is an associative product, as one can easily deduce by extending to superspace the discussion in section 1.1.1.

If we consider the special case when the superfields $\Phi$ and $\Psi$ are identified with the supercoordinates themselves, we obtain the following anticommutation relations
\bea
&& \left\{\th^\a,\th^\b\right\}= P^{\a\b}\cr
&& \left\{\thb^\ad, \thb^\bd\right\}= 0
\ena
They are not consistent with the $d=4$ $N=1$ superspace conjugation relation
\beq
\left(\th^\a\right)^\dagger=\thb^\ad
\label{conjug}
\eeq
Even if in principle a noncommutative deformation is not required to respect the complex structure present on the original space, the relation (\ref{conjug}) is needed in the original space for the fermionic degrees of freedom to be equal to the bosonic ones. By relaxing it in the deformation, one modifies (doubles) the number of supersymmetries.

In the paper \cite{ferrara} it was also observed that the class of chiral superfields, defined by the relation $\bar D_\ad \Phi =0$, is not closed under the product (\ref{ferrarastar}). Already at first order in the $\hbar$ expansion nontrivial terms arise in $\bar D_\ad \left(\Phi \ast \Psi\right)$, where $\Phi$ and $\Psi$ are both chiral, because of the nontrivial superspace torsion (\ref{torsion}). Since the simplest superspace field theories are written in terms of chiral superfields, the lack of closure for the chiral class is a serious  obstruction in constructing deformations of known supersymmetric theories.
\subsection{Non(anti)commutative superspace}
In this section I will discuss the results that I obtained in \cite{mypaper1}, in collaboration with D. Klemm and S. Penati, regarding supersymmetric associative deformations of $d=4$ $N=1$ Minkowski and $N=2$ euclidean superspace.
\subsubsection{Supersymmetric deformations of $N=1$ $d=4$ superpace}
In \cite{mypaper1}, a more systematical approach with respect to \cite{ferrara} has been followed to determine 
 the most general non(anti)commutative geometry in $N=1$
four dimensional superspace, invariant under the classical
supertranslation group and associative. As I have anticipated before, the deformation will be required to preserve the complex conjugation relations that are valid in ordinary superspace.

We will consider the supercoordinates $Z^A$, generically satisfying the non\-(anti)\-com\-mu\-ta\-ti\-ve algebra 
\beq
\left[Z^A,Z^B\right\}=P^{AB}(Z)
\eeq
where we have introduced the (anti)commutator
\beq
\left[F_A,G_B\right\}\equiv F_A G_B - (-)^{ab} G_B F_A
\eeq
that is a commutator if at least one of the two indices $A$, $B$ is a vector and an anticommutator otherwise.

Let us consider the transformation $Z\rightarrow Z'$. This is the generic symmetry of the ordinary theory that we would like to preserve in the deformation. Exactly as in the bosonic case discussed in section 1.1.2, we will require that the functional dependence of the non(anti)commutative algebra is not modified under the transformation
\beq
\left[{Z'}^A, {Z'}^B\right\}= P^{AB}(Z')
\eeq
Since, as discussed in section 1.1.1, bosonic noncommutative deformations break ``particle" Lorentz invariance, we will not worry about this symmetry in our superspace generalization. We will only take into consideration the supertranslation group, containing ordinary bosonic translations and supersymmetry transformations.
For $N=1$ $d=4$ superspace conventions, we refer to \cite{superspace} (see summary in the previous section).

In order to define a non(anti)commutative superspace,
we consider the most
general structure of the algebra for a set of four bosonic real coordinates
and a complex two--component Weyl spinor with $(\th^\a)^\dag = \thb^\ad$
\bea
&&\left\{\th^{\a},\th^{\b}\right\} ~=~{\cal A}^{\a\b}(x,\th,
\bar{\th}) \qquad , \qquad
\left\{\bar{\th}^{\dot{\a}},\bar{\th}^{\dot{\b}}\right\} ~=~
\bar{\cal A}^{\dot{\a}\bd}(x,\th,\bar{\th})
\non\\
&&\left\{\th^{\a},\bar{\th}^{\dot{\a}}\right\}
~=~{\cal B}^{\a\dot{\a}}
(x,\th,\bar{\th})\non\\
&&\left[x^{\underline{a}},\th^{\b}\right]~=~i{\cal C}^{\underline{a}\b}
(x,\th,\bar{\th})\qquad , \qquad
\left[x^{\underline{a}},\bar{\th}^{\dot{\b}}\right]~=~
i\bar{\cal C}^{\underline{a} \dot{\b}}(x,\th,\bar{\th})\non \\
&&\left[x^{\underline{a}},x^{\underline{b}}\right]~=~
i{\cal D}^{\underline{a} \underline{b}}
(x,\th,\bar{\th})
\label{coord2}
\ena
Here, ${\cal A}, {\cal B}, {\cal C}, {\cal D}$ are local functions of the
superspace variables and we have defined $\bar{\cal A}^{\ad \bd}
\equiv ({\cal A}^{\a \b})^\dag$,
$\bar{\cal C}^{\underline{a} \bd} \equiv ({\cal C}^{\underline{a} \b})^\dag$.
From the conjugation rules for the coordinates it follows also
$\left({\cal B}^{\a\dot{\a}}\right)^{\dag}={\cal B}^{\a\dot{\a}}$ and
$\left({\cal D}^{\underline{a} \underline{b}}\right)^{\dag} =
{\cal D}^{\underline{a} \underline{b}}$.

To implement (\ref{coord2}) to be the algebra of the coordinates
of a non\-(anti)\-com\-mu\-ta\-ti\-ve $N=1$ superspace we require its invariance
under the group of space translations and supertranslations (\ref{transf1}).
As before, we restrict our analysis to the case of an
undeformed group where the parameters $a^{\a\ad}$, $\epsilon^\a$
and $\bar{\epsilon}^\ad$ in (\ref{transf1}) are kept
(anti)commuting \footnote{More general
constructions of non(anti)commutative geometries in grassmannian spaces
have been
considered, where also the algebra of the parameters is deformed
\cite{kosinski}.}.

As in the bosonic case discussed in section 1.1.1, we require the functional dependence of the
${\cal A} , {\cal B}, {\cal C} , {\cal D}$ in (\ref{coord2}) to be the
same at {\em any} point of the supermanifold.
To work out explicitly the constraints which follow,
we perform a (super)translation (\ref{transf1}) on the coordinates
and compute the algebra of the new coordinates in terms of
the old ones.
We find that the functions appearing in (\ref{coord2}) are constrained by 
the following set of independent equations
\beq
{\cal A}^{\a \b} (x',\th',\thb' ) ~=~
{\cal A}^{\a \b} (x,\th,\thb ) \quad , \quad
{\cal B}^{\a \ad} (x',\th',\thb' ) ~=~
{\cal B}^{\a \ad} (x,\th,\thb )
\label{first}
\eeq
\beq
{\cal C}^{\a\dot{\a}\b}(x',\th',\thb') ~=~
{\cal C}^{\a\dot{\a}\b}(x,\th,\thb )~-~\frac{1}{2}
\e^{\a}{\cal B}^{\b \ad}(x,\th,\thb )~-~\frac{1}{2}
\bar{\e}^{\ad}{\cal A}^{\a\b}(x,\th,\thb )
\label{second}
\eeq
\bea
&& {\cal D}^{\a\dot{\a}\b\dot{\b}}(x',\th',\thb') ~=~
{\cal D}^{\a\dot{\a}\b\dot{\b}}(x,\th,\overline{\th})
\non\\
&&-~
\frac{i}{2}\left(\e^{\b}\bar{\cal C}^{\a\dot{\a}\dot{\b}}
(x,\th,\thb )
+\bar{\e}^{\dot{\b}}{\cal C}^{\a\dot{\a}\b}(x,\th,\thb)
-\e^{\a}\bar{\cal C}^{\b\dot{\b}\ad}(x,\th,\thb )
-\bar{\e}^{\dot{\a}}{\cal C}^{\b\dot{\b}\a}(x,\th,\thb)\right)
\non\\
&& -\frac{i}{4}\left(\e^{\a}\bar{\cal A}^{\dot{\a}\dot{\b}}
(x,\th,\thb )\e^{\b}~+~\e^{\a} {\cal B}^{\b\dot{\a}}
(x,\th,\thb )\bar{\e}^{\dot{\b}} \right.
\non\\
&& ~~~~~~~~~~~~~~~~~~~~~~~~~~~~~\left. +\bar{\e}^{\dot{\a}}
{\cal B}^{\a\dot{\b}}(x,\th,\thb )\e^{\b}+
\bar{\e}^{\dot{\a}}{\cal A}^{\a\b}(x,\th,\thb )
\bar{\e}^{\dot{\b}}\right)
\non\\
&&~~~~~~~~
\label{third}
\eea
together with their hermitian conjugates.

Looking for the most general local solution brings us to the following algebra
for a non(anti)commutative
geometry in Minkowski superspace consistent with (super)translations
\bea
\left\{\th^{\a},\th^{\b}\right\} &&=~ A^{\a\b} \quad , \quad
\left\{ \thb^{\ad} , \thb^{\bd} \right\} ~=~ \bar{A}^{\ad \bd}
\quad , \quad
\left\{ \th^{\a},\thb^{\dot{\a}} \right\} ~=~ B^{\a\dot{\a}}
\non\\
\left[ x^{\a\dot{\a}},\th^{\b} \right] &&=~
i {\cal C}^{\a\dot{\a}\b} (\th, \thb)
\non\\
\left[x^{\a\dot{\a}},\thb^{\dot{\b}}\right] &&=~
i \bar{{\cal C}}^{\a\dot{\a}\dot{\b}}(\th, \thb)
\non\\
\left[x^{\a\dot{\a}},x^{\b\dot{\b}}\right]
&&=~ i {\cal D}^{\a\dot{\a}\b\dot{\b}}(\th,\thb)
\label{nonanti}
\eea
where
\bea
&& {\cal C}^{\a\dot{\a}\b}(\th,\thb ) ~=~
C^{\a\dot{\a}\b} ~-~\frac{1}{2}\th^{\a} B^{\b\dot{\a}}
~-~ \frac{1}{2}\thb^{\dot{\a}} A^{\a\b}\cr
&&{\cal D}^{\a\dot{\a}\b\dot{\b}}(\th,\thb )
=~ D^{\a\dot{\a}\b\dot{\b}}
~-~\frac{i}{2}\left(\th^{\b} \bar{C}^{\a \dot{\a}\dot{\b}}
~-~\thb^{\dot{\a}} C^{\b\dot{\b}\a}
~-~ \th^{\a} \bar{C}^{\b\dot{\b}\dot{\a}}
~+~ \thb^{\dot{\b}} C^{\a\dot{\a}\b}\right)
\non\\
&&-~\frac{i}{4}\left(\th^{\a} \bar{A}^{\dot{\a}\dot{\b}} \th^{\b}
~+~\th^{\a} B^{\b\dot{\a}} \thb^{\dot{\b}}
~+~ \thb^{\dot{\a}} B^{\a\dot{\b}}\th^{\b}
~+~ \thb^{\dot{\a}} A^{\a\b} \thb^{\dot{\b}} \right)
\label{choice}
\eea
and $A$, $B$, $C$ and $D$ are constant functions.

We note that, while invariance under spacetime
translations necessarily requires the non(anti)commutation functions to be
independent of the $x$ coordinates, as we have seen in section 1.1.1, invariance under supersymmetry is
less restrictive and allows for a particular dependence on the spinorial
coordinates. 

On the algebra of smooth functions of superspace variables we can formally
define a graded bracket which reproduces the fundamental
algebra (\ref{nonanti}) when applied to the coordinates.
In the case of bosonic Minkowski spacetime, the
noncommutative algebra (\ref{bosonic}) can be obtained by
interpreting the l.h.s. of this relation as the Poisson bracket of classical
commuting variables, where, for generic functions of spacetime, the Poisson
bracket is defined as $\{ f , g \}_{P} = i \Theta^{\m \n}\pa_\m f \pa_\n g$.
Generalizing to Minkowski superspace, the graded bracket
must be constructed as a bidifferential operator with respect to the
superspace variables. 
Using covariant derivatives $D_A \equiv (D_\a, \bar{D}_\ad,
\pa_{\a \ad})$, for generic functions $\Phi$ and $\Psi$ of the superspace
coordinates we define the bidifferential operator 
\beq
\{ \Phi , \Psi \}_P ~=~ \Phi \overleftarrow{D}_A \, P^{AB} \,
\overrightarrow{D}_B \Psi
\label{poisson1}
\eeq
where
\beq
P^{AB} ~\equiv~
\begin{pmatrix}
P^{\a \b}  & P^{\a \bd} & P^{\a \underline{b}} \cr
P^{\ad \b} & P^{\ad \bd} & P^{\ad \underline{b}}  \cr
P^{\underline{a} \b} & P^{\underline{a} \bd} &
P^{\underline{a} \underline{b}} \cr 
\end{pmatrix}
~=~ \begin{pmatrix}
-A^{\a \b}  & -B^{\a \bd} & iC^{\b \bd \a} \cr
-B^{\ad \b} & -\bar{A}^{\ad \bd} & i\bar{C}^{\b \bd \ad}  \cr
iC^{\a \ad \b} & i\bar{C}^{\a \ad \bd} & iD^{\a \ad \b \bd} \cr 
\end{pmatrix}
\label{supermatrix}
\eeq
is a constant graded symplectic supermatrix satisfying
$P^{BA} = (-1)^{(a+1)(b+1)} P^{AB}$, $a$ denoting the grading of $A$.
It is easy to verify that applying this
operator to the superspace coordinates we obtain (\ref{nonanti}).

Alternatively, one can express the graded brackets (\ref{poisson1}) 
in terms of torsion free, noncovariant spinorial derivatives
$\pa_A\equiv(\pa_\a, \bar{\pa}_\ad, \pa_{\a \ad})$ so obtaining a matrix $\tilde P^{A B}$
explicitly dependent on $(\th, \thb)$.
The bracket (\ref{poisson1}) is rewritten as follows
\beq
\{ \Phi , \Psi \}_P ~=~ \Phi \overleftarrow{\pa}_A \, \tilde P^{AB} \,
\overrightarrow{\pa}_B \Psi
\label{poissonnoncov}
\eeq
where 
\beq
\tilde P^{AB}(\th,\thb) ~\equiv~
\begin{pmatrix}
-A^{\a \b}  & -B^{\a \bd} & i{\cal C}^{\b\bd\a}(\th,\thb) \cr
-B^{\ad \b} & -{\bar A}^{\ad \bd} & i\bar {\cal C}^{\b\bd\ad} (\th,\thb) \cr
i{\cal C}^{\a\ad \b}(\th,\thb) & i\bar {\cal C}^{\a\ad \bd} (\th,\thb)&
i{\cal D}^{\a\ad \b\bd}(\th,\thb) \cr 
\end{pmatrix}
\label{supermatrixnoncov}
\eeq
and the functions ${\cal C}$, $\bar{\cal C}$ and $\cal {D}$ are given in (\ref{choice}).

The latter formulation of the graded bracket is not manifestly covariant, but it is however very useful, since  it makes clear that Kontsevich procedure outlined in section 1.1.4 can be generalized to superspace to construct an associative deformation of the product between superfields. 
All the results from now on can be written in both ways. In general the covariant formulation is preferred because it naturally leads to a geometrical interpretation of the results. 
\subsubsection{Deformation of the supersymmetry algebra}
It is important to note that the non(anti)commutative extension given in
(\ref{nonanti}) in general deforms the supersymmetry algebra. 
In the standard
case, defining $Q_A \equiv (Q_\a, \bar{Q}_\ad, -i\pa_{\a\ad})$, the
supersymmetry algebra can be written as
\bea
&& [Q_A,Q_B\} ~=~ i {T_{AB}}^C Q_C \quad , \quad
[D_A,D_B\} ~=~ {T_{AB}}^C D_C
\non\\
&& [Q_A,D_B\} ~=~ 0
\label{standard}
\eea
where ${T_{AB}}^C$ is the torsion of the flat superspace
(${T_{\a \bd}}^{\underline{c}} = {T_{\bd \a}}^{\underline{c}} = i \d^{~\g}_\a
\d_\bd^{~\dot{\g}}$ are the only nonzero components).
Turning on non(anti)commutativity in superspace leads instead to
\bea
&& [Q_A,Q_B\} ~=~ i {T_{AB}}^C Q_C ~+~ {R_{AB}}^{CD} Q_C Q_D
\non\\
&& [D_A,D_B\} ~=~ {T_{AB}}^C D_C ~+~ {R_{AB}}^{CD} D_C D_D
\non\\
&& [Q_A,D_B\} ~=~ {R_{AB}}^{CD} Q_C D_D
\label{modalg}
\eea
where ${T_{AB}}^C$ is still the torsion of the flat superspace, while
\beq
{R_{AB}}^{CD} ~=~ -\frac{1}{8} \, P^{MN} {T_{M[A}}^C {T_{B)N}}^D
\label{rtensor2}
\eeq
($[ab)$ means antisymmetrization when at least one of the indices
is a vector index, symmetrization otherwise)
is a curvature tensor whose presence is a direct consequence
of the non(anti)commutation of the grassmannian coordinates.
Its nonvanishing components are
\bea
&& {R_{\a \b}}^{\underline{c} \underline{d}} ~=~ \frac{1}{8}
P^{\dot{\gamma} \dot{\delta}} \d_{(\a}^{~\g} \d_{\b)}^{~\d}
\quad , \quad
{R_{\ad \bd}}^{\underline{c} \underline{d}} ~=~ \frac{1}{8}
P^{\gamma \delta} \d_{(\ad}^{~\dot{\g}} \d_{\bd)}^{~\dot{\d}}
\non\\
&& {R_{\a \bd}}^{\underline{c} \underline{d}} ~=~
{R_{\bd \a}}^{\underline{c} \underline{d}} ~=~ \frac{1}{8}
\left( P^{\gamma \dot{\d}} \d_\a^{~\d} \d_\bd^{~\dot{\g}}
~+~ P^{\d \dot{\g}} \d_\a^{~\g} \d_\bd^{~\dot{\d}} \right)
\label{rtensor}
\eea
I would like to stress that the curvature terms deforming the supersymmetry algebra are quadratic in bosonic derivatives and have no effect on the supercoordinate-algebra. Therefore their presence is not in disagreement with consistency of the supercoordinate algebra with supersymmetry.

Since the terms proportional to the curvature in (\ref{modalg})
are quadratic in supersymmetry charges and covariant derivatives,
we can define new graded brackets
\bea
&& [ Q_A , Q_B \}_q \equiv Q_A Q_B - (-1)^{ab} [ {\d_B}^C {\d_A}^D +
(-1)^{ab} {R_{AB}}^{CD} ] Q_C Q_D\cr
&&~~~~~~~~
\eea
and analogous ones for $[ D_A , D_B \}_q$ and $[ Q_A , D_B \}_q$, 
which satisfy the standard algebra (\ref{standard}).
The new brackets can be interpreted as a quantum deformation
associated to a $q$--parameter which in this case is a rank--four tensor
\beq
{q_{AB}}^{CD} ~\equiv~ {\d_B}^C {\d_A}^D ~+~ (-1)^{ab} {R_{AB}}^{CD}
\eeq
\subsubsection{Associativity and the geometry of deformed superspace}
Given the bidifferential operator (\ref{poisson1}) associated to 
the noncommutative supergeometry defined in (\ref{nonanti}) it is easy to prove
the following identities
\bea
&& \{ \Phi, \Psi \}_P ~=~ (-1)^{1+ deg(\Phi) \cdot deg(\Psi)} \, \{ \Psi ,
\Phi \}_P
\non \\
&& \{ c \Phi ,\Psi \}_P ~=~ c \, \{ \Phi, \Psi \}_P  \quad , \quad
\{ \Phi ,c \Psi \}_P ~=~ (-1)^{deg(c) \cdot deg(\Phi)} \,c \, \{ \Phi,
\Psi \}_P
\non\\
&& \{ \Phi + \Psi , \Omega \}_P ~=~ \{ \Phi , \Omega \}_P ~+~ \{\Psi,
\Omega\}_P
\eea
The operator $\{ \, , \}_P$ will then be promoted to a 
graded Poisson structure
on superspace if and only if Jacobi identities hold
\bea
&& \{ \Phi, \{ \Psi , \Omega \}_P \}_P +
(-1)^{deg(\Phi)\cdot [deg(\Psi) + deg(\Omega)]}
\{ \Psi, \{ \Phi, \Omega \}_P \}_P
\non\\
&& ~~~~~~~~~+ (-1)^{deg(\Omega)\cdot [deg(\Phi) + deg(\Psi)]}
\{ \Omega, \{ \Phi, \Psi \}_P \}_P
~=~ 0
\label{jacobi2}
\eea
for any triplet of functions of superspace variables.
This property is equivalent to associativity of the fundamental algebra (\ref{nonanti}). Since the latter is nontrivial (coordinate-dependent commutators appear), (\ref{jacobi2}) is not in general satisfied. 
Indeed, imposing (\ref{jacobi2}) yields the nontrivial
conditions
\begin{eqnarray}
&&P^{AR} P^{BS}T_{SR}^{\;\;\;\;C}(-1)^{c+b(c+a+r)} +
P^{BR}P^{CS}T_{SR}^{\;\;\;\;A}(-1)^{a+c(a+b+r)} \nonumber \\
&&~~~~~~~~~~~~~+\, P^{CR}P^{AS}T_{SR}^{\;\;\;\;B}(-1)^{b+a(b+c+r)} = 0
\label{Jac_cov}
\end{eqnarray}
\beq
(-1)^{bm} P^{AM} P^{BN} {R_{MN}}^{CD} ~=~ 0
\label{Jac_cov2}
\eeq
where the torsion ${T_{AB}}^C$ and the curvature ${R_{AB}}^{CD}$ have been
introduced in (\ref{standard}) and (\ref{modalg}).
Equation (\ref{Jac_cov}) is the covariant superspace generalization of bosonic associativity constraint  (\ref{assoc}) (the analogy with (\ref{assoc}) is clearer when it is rewritten in terms of the ``noncovariant" Poisson structure $\tilde P$ given in (\ref{poissonnoncov})). 

As in bosonic case, if $P^{AB}$ is invertible ($P_{AB}P^{BC}=\delta_A^C$), equation (\ref{Jac_cov}) is equivalent to
the vanishing of the contorsion tensor $H_{ABC}$ defined by
\begin{equation}
H_{ABC} = T_{AB}^{\;\;\;\;D}P_{DC}(-1)^{ac} +
T_{CA}^{\;\;\;\;D}P_{DB}(-1)^{cb} + T_{BC}^{\;\;\;\;D}P_{DA}(-1)^{ba}\,.
\end{equation}
This is the superspace generalization of (\ref{assocH}), in terms of manifestly covariant quantities.
The only nonvanishing components of $H$ are 
\bea
&& H_{\a \dot{\a} \b}=-i\left[ P_{\a \dot{\a}~ \b} + P_{\b \dot{\a}~ \a}
\right]
\non\\
&&H_{\a \dot{\a} \dot{\b}}=-i \left[ P_{\a\dot{\a}~\dot{\b}} +
P_{\a\dot{\b}~\dot{\a}} \right] 
\non\\
&&H_{\a \dot{\a} \underline{b}}=iP_{\a \dot{\a} ~\underline{b}}
\eea
We notice that its bosonic components $H_{\underline{a} \underline{b} 
\underline{c}}$ vanish due to the $x$--independence of the noncommutation
functions in (\ref{nonanti}). The nonvanishing of $H$ comes entirely from
the $\theta$--dependence of the functions in (\ref{choice}). As will be explained in section 1.3, it has been shown that bosonic coordinate-dependent deformations naturally emerge in string theory in the presence of curved backgrounds. Nonassociative deformations also emerge and the parameter governing nonassociativity is a bosonic three form $H$. Identifying the superspace analogue of this may help characterizing the superstring background where deformed superspaces may appear.

When $P$ is invertible, equation (\ref{Jac_cov2}) might seem to imply that the curvature $R$ is zero. This is not true in general because of the presence of the sign in front, which is dependent on the grading  of the summed index $M$. Moreover, being puzzled by the unusual pattern of equations found, we have been trying to prove that (\ref{Jac_cov2}) is algebraically implied by (\ref{Jac_cov}), but this doesn't seem to work.

We now search for the most general solutions of the conditions (\ref{Jac_cov}, \ref{Jac_cov2}). Writing them in terms of the $P^{AB}$ components we obtain
\bea
&& B^{\a\dot{\b}} A^{\b\g} ~+~ A^{\b\a}B^{\g\dot{\b}} ~=~ 0
\non\\
&& B^{\a\dot{\b}}B^{\b\dot{\g}} ~+~ A^{\b\a} \bar{A}^{\dot{\b}\dot{\g}}
~=~ 0
\non\\
&& \left( \bar{C}^{\b\dot{\b}\dot{\a}} A^{\a\g}
~+~ C^{\b\dot{\b}\a} B^{\g\dot{\a}} ~-~
\bar{C}^{\a\dot{\a}\dot{\b}} A^{\b\g}
~-~ C^{\a\dot{\a}\b} B^{\g\dot{\b}} \right) ~=~ 0
\non\\
&& {\cal I}{\rm m}
\left( \bar{C}^{\a\dot{\a}\dot{\b}} C^{\g\dot{\g}\b} ~+~
\bar{C}^{\g\dot{\g}\dot{\a}} C^{\b\dot{\b}\a}
~+~ \bar{C}^{\b\dot{\b}\dot{\g}} C^{\a\dot{\a}\g}\right) ~=~ 0\,.
\label{jacobi4}
\eea
The first two conditions necessarily imply the vanishing
of the constants $A$ and $B$.
Inserting this result in the third constraint we immediately realize
that it is automatically satisfied and the only nontrivial condition 
which survives is the last one. 
This equation has nontrivial solutions. For example, the matrix
\beq
C^{\a \ad \b} ~=~
\begin{pmatrix}
\psi^\b & \psi^\b \cr
\psi^\b & \psi^\b \cr 
\end{pmatrix}
\eeq
for any spinor $\psi^\b$, is a solution.
It would correspond to assume the
same commutations rules among any bosonic coordinate and the spinorial
variables.

We conclude that the most general {\em associative} and non(anti)commutative
algebra in Minkowski superspace has the form
\bea
\left\{\th^{\a},\th^{\b}\right\} &&=~
\left\{ \thb^{\ad} , \thb^{\bd} \right\}
~=~ \left\{ \th^{\a},\thb^{\dot{\b}} \right\} ~=~ 0
\non\\
\left[ x^{\a\dot{\a}},\th^{\b} \right] &&=~ i C^{\a\dot{\a}\b}
\non\\
\left[x^{\a\dot{\a}},\thb^{\dot{\b}}\right] &&=~ i
\bar{C}^{\a\dot{\a}\dot{\b}} \label{assnonanti} \\
\left[x^{\a\dot{\a}},x^{\b\dot{\b}}\right] &&=~
i D^{\a\dot{\a}\b\dot{\b}}
~+~\frac{1}{2} \left( \bar{C}^{\b\dot{\b}\dot{\a}}
\th^{\a} ~-~ \bar{C}^{\a\dot{\a}\dot{\b}} \th^\b
~+~ C^{\b\dot{\b}\a} \thb^{\dot{\a}}
~-~ C^{\a\dot{\a}\b} \thb^\bd \right)\,, \nonumber
\eea
where $C$ is subject to the last constraint in (\ref{jacobi4}).
Setting $C^{\a \ad \b}=0$ we recover the usual noncommutative
superspace considered so far in literature \cite{ferrara,chuzamora,zanon}.

Under conditions (\ref{jacobi4}) the graded brackets 
(\ref{poisson1}) satisfy the Jacobi identities (\ref{jacobi2}), as can be easily
proved by expanding the functions in power series.
In this case we have a well--defined super Poisson structure on superspace.

We note that a non(anti)commutative but associative geometry always
mantains the standard algebra (\ref{standard}) for the covariant derivatives. 
In fact,
in this case, from (\ref{rtensor}) it follows ${R_{AB}}^{CD} =0$. 
\subsubsection{Construction of a Kontsevich-like product on superspace}
We will now describe the first few steps towards the construction
of a star product defined on the class of general superfields.
By definition, this product must be associative, i.e. it has 
to satisfy the Jacobi identities (\ref{jacobi2}) when the fundamental
algebra is associative. 

In section 1.1.4 we have seen that in the 
nonsupersymmetric case the lack of associativity of the fundamental algebra
is signaled by the presence of a nonvanishing 3--form $H$.
A product has been constructed \cite{kontsevich} so that the terms
violating the Jacobi identities are proportional to $H$. 
The product is then automatically associative when 
the fundamental algebra is.

In the present case we have shown that the lack of associativity 
in superspace is related to a nonvanishing super 3--form.  
This suggests the possibility to construct 
a super star product by suitably generalizing Kontsevich 
construction \cite{kontsevich} to superspace.  
The supersymmetric Poisson structure we have constructed on superspace can be written in a manifestly covariant form, as in (\ref{poisson1}), in terms of a constant matrix and covariant derivatives that have nontrivial torsion, or as in (\ref{poissonnoncov}), in terms of a coordinate-dependent matrix and torsion free derivatives. The second formulation allows for a straightforward generalization of Kontsevich construction we outlined in section 1.1.4.  However, the same procedure can be performed in a manifestly covariant way, and I choose to give this second version, so that the geometric interpretation of the results will be clear.

We begin by considering Moyal--deformed product defined
in the usual way
\begin{equation}
\Phi \ast \Psi \equiv \Phi \exp(\hbar\overleftarrow{D}_A P^{AB}
                       \overrightarrow{D}_B) \Psi, \label{deformprod}
\end{equation}
where $\Phi$ and $\Psi$ are arbitrary superfields, and
$\hbar$ denotes a deformation parameter. In general, due to
the lack of (anti)commutativity among covariant derivatives (see eq.
(\ref{modalg})), it is easy to prove that the $\ast$--product is
not associative even when the Poisson brackets are.
However, inspired by Kontsevich procedure \cite{kontsevich}, 
we perturbatively define a modified product $\star$ with the 
property to
be associative up to second order in $\hbar$ when the Jacobi identities
are satisfied. Precisely, we find an explicit form for the product
by imposing the Jacobi identities (\ref{jacobi2})
to be violated at this order only by terms proportional to $H$. 
To this end we define
\begin{eqnarray}
\Phi \star \Psi &\equiv & \Phi\Psi ~+~ \hbar \Phi\overleftarrow{D}_A P^{AB}
\overrightarrow{D}_B \Psi ~+~ \frac{\hbar^2}{2}\Phi(\overleftarrow{D}_A P^{AB}
\overrightarrow{D}_B)(\overleftarrow{D}_C P^{CD}\overrightarrow{D}_D) \Psi
\nonumber \\
&& -\frac{\hbar^2}{3}\left(\overrightarrow{D}_A \Phi \, {\cal M}^{ABC}
\, \overrightarrow{D}_B\overrightarrow{D}_C \Psi - (-1)^c
\overrightarrow{D}_C\overrightarrow{D}_A \Phi \, {\cal M}^{ABC} \,
\overrightarrow{D}_B \Psi\right) \nonumber \\
&& + {\cal O}(\hbar^3)\,, \label{konts}
\end{eqnarray}
where
\begin{eqnarray}
{\cal M}^{ABC} &=& P^{AD}{T_{DE}}^C P^{EB} (-1)^{ce} +
            \frac12 P^{BD}{T_{DE}}^A P^{EC} (-1)^{ae + a + b + ab + bc}
            \nonumber \\
        & & + \frac12 P^{CD}{T_{DE}}^B P^{EA} (-1)^{be + a + c + ac + ab}.
\end{eqnarray}
Since it is straightforward to show that
\begin{eqnarray}
\lefteqn{(\Phi \star \Psi)\star \Omega ~-~ \Phi \star (\Psi \star \Omega)
~=~}
\nonumber \\
& & -\frac 23 \hbar^2(-1)^{(c+b)(e+1)+eg+cf}\overrightarrow{D}_A \Phi \,
    P^{AE}P^{BF}P^{CG} \, H_{GFE} \, \overrightarrow{D}_C \Omega \,
    \overrightarrow{D}_B \Psi \non\\
& & ~+~ {\cal O}(\hbar^3)
\end{eqnarray}
up to second order in $\hbar$ the product is associative if and only if $H=0$,
i.e. the fundamental algebra is associative. 
We note that at this order only the contorsion enters the breaking
of associativity, being the curvature tensor $R$ of order $\hbar$.

In \cite{mypaper1} we did not pursue the construction of the star product to
all orders in $\hbar$ but we believed that in principle there were no obstructions to the generalization of Kontsevich procedure to all orders. The extension of the superspace product we proposed to all orders in $\hbar$ has been obtained afterwards in \cite{chepelev}.

We now discuss the closure of the class of chiral superfields under the
deformed products we have introduced. For a generic choice
of the supermatrix $P^{AB}$ the star product of two chiral superfields
(satisfying $\bar{D}_{\dot{\a}}\Phi = 0$)
is {\em not} a chiral superfield, both for associative and nonassociative
products.
However, in the particular case where the only nonvanishing components of
the symplectic supermatrix $P^{AB}$ are $P^{\a\dot{\b}}$ and
$P^{\underline{a}\underline{b}}$, chiral superfields
are closed both under
the deformed product defined in (\ref{deformprod}) and under Kontsevich
star product (\ref{konts}) (for the latter up to terms of order
${\cal O}(\hbar^3)$).
Clearly for $P^{\a\dot{\b}} \neq 0$ the above star products
are no more associative. Because of this, it could be problematic to generalize for instance the Wess--Zumino model to non(anti)commutative superspace, since $(\Phi\star\Phi)\star\Phi\neq \Phi\star(\Phi\star\Phi)$.
However, one may notice that for chiral superfields the above star products become commutative\footnote{Generalized
star products that are commutative but nonassociative have been considered
in a different context in \cite{dastrivedi}.}.
This commutativity implies that there is no ambiguity in putting
the parenthesis in the cubic interaction term of a deformed
Wess--Zumino model, since, when $\Phi$ is a chiral superfield, $(\Phi\star\Phi)\star\Phi= \Phi\star(\Phi\star\Phi)$ holds. Therefore, the action for the deformed Wess--Zumino model
\begin{equation}
S = \int d^4x d^2\theta d^2\bar{\theta}\,\Phi\star\bar{\Phi}
+ \int d^4x \left[\int d^2\theta \left(\frac m2 \Phi\star\Phi + \frac g3
\Phi\star\Phi\star\Phi\right) + {\mbox{c.~c.}} \right].
\label{WZ}
\end{equation}
is well defined and can be studied.
Notice that in this case the $\star$--product in the kinetic term cannot
be simply substituted with the standard product as it happens in superspace
geometries where grassmannian coordinates anticommute \cite{ferrara, zanon}.
\subsubsection{Non(anti)commutative $N=2$ Euclidean superspace}
From the discussion of $d=4$ $N=1$ superspace it is clear that the superspace conjugation relations relating $\th^\a$ and $\thb^\ad$ are strong constraints on the non(anti)commutative algebra.  
The main difference in the description of euclidean superspace
with respect to Minkowski relies on the reality conditions
satisfied by the spinorial variables. So it is reasonable to hope that in an euclidean signature structures with a nonvanishing anticommutator between two fermions could appear.

As it is well known \cite{PVP}, in euclidean signature
a reality condition on spinors is applicable only in the presence of
extended supersymmetry. In \cite{mypaper1} we concentrated on
the simplest case, the $N=2$ euclidean superspace, even if our analysis
can be easily extended to more general cases. 

In a chiral description
the two--component Weyl spinors satisfy a symplectic Majorana condition
\beq
(\th^\a_i )^{\ast} ~=~ \th_\a^i ~\equiv~ C^{ij} \, \th^\b_j \, C_{\b \a}
\quad , \quad
(\thb^{\ad, i} )^{\ast} ~=~ \thb_{\ad, i} ~\equiv~
\thb^{\bd, j} \, C_{\bd \ad} \, C_{ji}
\label{simplmajorana}
\eeq
where $C^{12} = -C_{12} = i$.
This implies that the most general non(anti)commutative
algebra can be written as an obvious generalization of (\ref{coord2})
with the functions on the rhs now being in suitable representations
of the R--symmetry group. 
When imposing covariance under (super)translations we obtain
that the most general non(anti)commutative geometry in euclidean
superspace is
\bea
\left\{\th^{\a}_i,\th^{\b}_j\right\} &&=~ {A_1}^{\a\b,}_{~~~ij} \quad , \quad
\left\{ \thb^{\ad,i} , \thb^{\bd,j} \right\} ~=~ A_2^{\ad \bd,ij}
\quad , \quad
\left\{ \th^{\a}_i,\thb^{\dot{\a},j} \right\} ~=~ B^{\a\dot{\a},~~j}_{~~~~i}
\non\\
\left[ x^{\a\dot{\a}},\th^{\b}_i \right] &&=~
i{{\cal C}_1}^{\underline{a}\b,}_{~~~ i} (\th, \thb)
\non\\
\left[x^{\a\dot{\a}},\thb^{\dot{\b},i}\right] &&=~
i {\cal C}_2^{\a\dot{\a}\dot{\b},i}(\th, \thb)
\non\\
\left[x^{\a\dot{\a}},x^{\b\dot{\b}}\right] &&=~
i {\cal D}^{\a\dot{\a}\b\dot{\b}}(\th,\thb)
\label{nonanti2}
\eea
where 
\bea
&& {{\cal C}_1}^{\a\dot{\a}\b,}_{~~~~~i} (\th, \thb) ~\equiv~
{C_1}^{\a\dot{\a}\b,}_{~~~~~i}
~+~\frac{i}{2} \th^{\a}_j B^{\b \dot{\a},~~j}_{~~~~i}
~+~ \frac{i}{2} \thb^{\dot{\a}, j} {A_1}^{\a\b,}_{~~~ji}
\non\\
&& {\cal C}_2^{\a\dot{\a}\dot{\b},i}(\th, \thb) ~\equiv~
C_2^{\a\dot{\a}\dot{\b},i}
~+~\frac{i}{2} \th^{\a}_j A_2^{\ad \bd,ji}
~+~\frac{i}{2} \thb^{\ad, j} B^{\a\dot{\b},~~i}_{~~~~j}
\non\\
&& {\cal D}^{\a\dot{\a}\b\dot{\b}}(\th,\thb) ~\equiv~
D^{\a\dot{\a}\b\dot{\b}}
\non\\
&&~~~~~~~
+~\frac{1}{2} \left( \th^{\a}_i C_2^{\b\dot{\b}\dot{\a},i}
~-~ \th^\b_i C_2^{\a\dot{\a}\dot{\b},i}
~+~ \thb^{\dot{\a}, i} {C_1}^{\b\bd\a,}_{~~~~~ i}
~-~ \thb^{\bd,i}  {C_1}^{\a\ad\b,}_{~~~~~ i} \right)
\non\\
&&~~~~~~~
+~\frac{i}{4}\left( \th^{\a}_i \, A_2^{\dot{\a}\dot{\b},ij} \, \th^{\b}_j
~+~ \th^{\a}_i \, B^{\b \dot{\a},~~i}_{~~~~j} \, \thb^{\dot{\b},j}
~+~\thb^{\dot{\a},i} \, B^{\a \dot{\b},~~j}_{~~~~i} \, \th^{\b}_j
~+~\thb^{\dot{\a},i} \, {A_1}^{\a\b,}_{~~~ij} \, \thb^{\dot{\b},j} \right)
\non\\
&&~~~~~~~~~
\label{cal2}
\eea
with $A_1$, $A_2$, $B$, $C_1$, $C_2$ and $D$ constant.

Following the same steps as in the Minkowski case, we can look
for the most general associative algebra. The results we obtain for
{\em associative} non(anti)commuting geometries in euclidean superspace are
\bea
\left\{\th^{\a}_i ,\th^{\b}_j \right\} &&=~ {A_1}^{\a\b,}_{~~~ij} \quad ,\quad
\left\{ \thb^{\ad,i} , \thb^{\bd,j} \right\} ~=~ 0 \quad , \quad
\left\{ \th^{\a}_i,\thb^{\dot{\b},j} \right\} ~=~ 0
\non\\
\left[ x^{\a\dot{\a}},\th^{\b}_i \right] &&=~ i{C_1}^{\a\dot{\a}\b,}_{~~~~~i}
~-~ \frac12 \thb^{\ad,j} {A_1}^{\a \b,}_{~~~ji}
\non\\
\left[x^{\a\dot{\a}},\thb^{\dot{\b},i}\right] &&=~ 0
\\
\left[x^{\a\dot{\a}},x^{\b\dot{\b}}\right] &&=~
iD^{\a\dot{\a}\b\dot{\b}}
~+~ \frac{i}{2} \left(\thb^{\dot{\a},i}  {C_1}^{\b\dot{\b}\a,}_{~~~~~i}
~-~ \thb^{\bd,i} {C_1}^{\a\dot{\a}\b,}_{~~~~~i}  \right) ~-~
\frac14 \thb^{\ad,i} \, {A_1}^{\a \b,}_{~~~ij} \, \thb^{\bd,j}
\non
\label{assnonantie1}
\eea
or
\bea
\left\{\th^{\a}_i,\th^{\b}_j\right\} &&=~ 0 \quad , \quad
\left\{ \thb^{\ad,i} , \thb^{\bd,j} \right\}
~=~ A_2^{\ad \bd,ij} \quad ,\quad
\left\{ \th^{\a}_i,\thb^{\dot{\b},j} \right\} ~=~ 0
\non\\
\left[ x^{\a\dot{\a}},\th^{\b}_i \right] &&=~ 0
\non\\
\left[x^{\a\dot{\a}},\thb^{\dot{\b},i}\right] &&=~ iC_2^{\a\dot{\a}\dot{\b},i}
~-~ \frac12 \th^\a_j A_2^{\ad \bd , ji}
\\
\left[x^{\a\dot{\a}},x^{\b\dot{\b}}\right] &&=~
iD^{\a\dot{\a}\b\dot{\b}}
~+~\frac{i}{2} \left( \th^{\a}_i C_2^{\b\dot{\b}\dot{\a},i}
~-~ \th^\b_i C_2^{\a\dot{\a}\dot{\b},i}  \right)
~-~ \frac14 \th^\a_i \, A_2^{\ad \bd , ij} \, \th^\b_j
\non
\label{assnonantie2}
\eea
We notice that in this case associativity imposes less restrictive constraints
because of the absence of conjugation relations between $A_1$ and $A_2$. 
As a consequence, nontrivial anticommutation relations among $\theta$'s 
(or $\bar{\theta}$'s) are allowed.
Moreover the R--symmetry group of the $N=2$ euclidean superalgebra is
broken only by the constant terms $C_1$ and $C_2$. Setting these terms equal
to zero leads to nontrivial (anti)commutation relations preserving 
$R$--symmetry.

Again, explicit expressions for the corresponding graded brackets
can be obtained as an obvious generalization of (\ref{poisson1},
\ref{supermatrix}). In this case
they define a super Poisson structure on euclidean superspace.
A simple example of a super Poisson structure is
\beq
\{ \Phi , \Psi \}_P ~=~ - ~\Phi \overleftarrow{D}_{\a}^i \,
{A_1}^{\a \b,}_{~~~ij} \, \overrightarrow{D}_{\b}^j \Psi
\label{example2}
\eeq
We notice that this extension is allowed {\em only} in euclidean superspace,
where it is consistent with the reality conditions on the spinorial
variables.
\subsection{$N=\frac{1}{2}$ supersymmetry}
\subsubsection{Seiberg $N=1/2$ superspace}
In this section I will review nonanticommutative $N=\frac{1}{2}$ superspace introduced by Seiberg in \cite{seiberg} and I will explain its relation with my results in \cite{mypaper1}.
As I have discussed in detail in the last section, in \cite{mypaper1} I obtained the most general nonanticommutative deformation of $d=4$ $N=1$ superspace compatible with supersymmetry and associativity. This involves nontrivial $[x,\th]$, $[x,\thb]$ and $[x,x]$, while $\{\th,\th\}$, $\{\thb,\thb\}$ and $\{\th,\thb\}$ cannot be turned on.  The situation improves in euclidean signature, where it is possible to turn on the fermionic anticommutators because of the different spinor conjugation relations.
Rigorously, a superspace with euclidean signature can be defined only with extended supersymmetry, because of the impossibility to assign consistent reality conditions to $\th^\a$ and $\thb^\ad$ in $N=1$ \cite{PVP}.
This is why I have studied the $N=2$ euclidean superspace in \cite{mypaper1}.
However, one can formally define an $N=1$ euclidean superspace by temporarily doubling the fermionic degrees of freedom, as it is done by Seiberg in \cite{seiberg}. 

To show how this works, I will first redefine the $N=2$ euclidean spinor variables (\ref{simplmajorana}) as follows
\bea
&&\th^\a \equiv \th_1^\a - \th_2^\a ~~~~~~;~~~~~\thb^\a \equiv \th_1^\a + \th_2^\a\cr
&&\th^\ad \equiv \thb^{1\ad} - \thb^{2\ad} ~~~~;~~~~~\thb^\ad \equiv \thb^{1\ad} - \thb^{2\ad}
\ena
These satisfy the reality conditions
\bea
&&(\th^\a)^\ast = i\thb_\a~~~~~~;~~~~~(\thb^\a)^\ast = -i\th_\a\cr
&&(\th_\a)^\ast = -i \thb^\a~~~~;~~~~~(\thb_\a)^\ast=i\th^\a
\ena
and analogous for the dotted variables. Dotted and undotted variables are unrelated. I will refer to $\th^\a$ and $\th^\ad$ as the left-moving sector of the theory and to $\thb^\a$ and $\thb^\ad$ as the right-moving sector. This terminology will be clarified in section 1.2.4, where the string theory origin of deformed superspaces will be discussed.  
We will see in section 1.2.4 that open string boundary conditions relate left- and right-moving fermionic variables on the D-brane ($\th=\thb$ on the boundary). As a result, the effective field theory on the brane is described by an $N=1$ euclidean superspace with fermionic variables $\th^\a$ and $\th^\ad$. For this reason from now on I will only consider spinor coordinates in the left-moving sector.

In \cite{mypaper1} we have used in both Minkowski and euclidean signature a nonchiral representation for superspace covariant derivatives. As a result consistency with supersymmetry transformations implied that if we turned on a nonvanishing anticommutators between two fermionic variables, also nonzero boson-boson and boson-fermion commutators appeared.  This made the algebra coordinate dependent  and forced us to build a complicated Kontevich-like product. 

A way to obtain a simpler, constant algebra for the supercoordinates is the following \cite{silvia}. In four dimensions it is possible to use a chiral representation of supersymmetry, where, for each supersymmetry, one of the two covariant derivatives coincides with the ordinary one while the other gets dressed.
If we make this chiral choice in $N=2$ euclidean superspace, by defining
\bea
&&Q_\a=i(\pa_\a -i\th^\ad \pa_{\a\ad}) ~~~~~;~~~~~~ Q_\ad=i\pa_\ad\cr
&&D_\a=\pa_\a ~~~~~~~~~~~~~~~~~~~~~;~~~~~~D_\ad=\pa_\ad +i\th^\a\pa_{\a\ad}
\label{chiralD}
\ena
corresponding to the supersymmetry transformations
\beq
\d x^{\a\ad}=-i\e^\a \th^\ad~~~; ~~~\d\th^\a=\e^\a~~~~;~~~~\d\th^\ad=\e^\ad 
\label{chiralsusy}
\eeq
we immediately realize that the nonanticommutative algebra 
\beq
\{\th^\a,\th^\b\}=2P^{\a\b}\qquad {\rm the~rest}=0
\label{easyalgebra}
\eeq
with $P^{\a\b}$ a constant symmetric matrix, is consistent with (\ref{chiralsusy}) and trivially associative .
One can easily check that, comparing to (\ref{modalg}, \ref{rtensor2}, \ref{rtensor}), the supersymmetry algebra is not modified by the deformation but the algebra of covariant derivatives becomes
\bea
&&\{D_\a,D_\b\}=0 ;~~~~~~~~~~~~~~~~~~~\{D_\a,D_\ad\}=i\pa_{\a\ad}\cr
&&\{D_\ad,D_\bd\}=-2P^{\a\b}\pa_{\a\ad}\pa_{\b\bd}
\ena
This is a problem when one tries to construct chiral superfields. So, even if this algebra is very simple, it is not suitable for constructing deformations of ordinary theories in superspace.

Seiberg in \cite{seiberg} chose the opposite chiral representation with respect to (\ref{chiralD})
\bea
&&Q_\a=i\pa_\a ~~~~~~~~~~~~~~~~~~~~;~~~~~ Q_\ad=i(\pa_\ad-i\th^\a\pa_{\a\ad})\cr
&&D_\a=\pa_\a+i\th^\ad\pa_{\a\ad} ~~~~~~~~;~~~~~ D_\ad=\pa_\ad 
\label{antichiralD}
\ena
corresponding to the supersymmetry transformations
\beq
\d x^{\a\ad}=-i\e^\ad \th^\a~~~; ~~~\d\th^\a=\e^\a~~~~;~~~~\d\th^\ad=\e^\ad 
\eeq
When we turn on a nontrivial anticommutator between the $\th$'s, we obtain
\bea
&&\{\th^\a,\th^\b\}=2P^{\a\b}\qquad\qquad \{\th^\a,\th^\bd\}=\{\th^\ad,\th^\bd\}=0\cr
&&[x^{\a\ad},\th^\b]=-2iP^{\a\b}\th^\ad\cr
&&[x^{\a\ad},x^{\b\bd}]=2\th^\ad P^{\a\b} \th^\bd
\ena
This is consistent with supersymmetry and associativity (it is just the ``$N=1$" version of the algebra (\ref{assnonantie1}) I gave in \cite{mypaper1} and discussed in the previous section).
This is coordinate-dependent and would require the construction of a Kontsevich-like product for the superfield algebra. 
However, Seiberg observed that it is possible to make the change of variables
\beq
y^{\a\ad}=x^{\a\ad}-i\th^\a\th^\ad
\label{changevar}
\eeq 
In terms of $(y^{\a\ad},\th^\a,\th^\ad)$ the superspace algebra takes the form (\ref{easyalgebra}) and this makes it possible to define a Moyal-like associative product acting on superfields $\Phi=\Phi(y^{\a\ad},\th^\a,\th^\ad)$ as follows
\bea
\Phi\ast\Psi&=&\Phi \exp(-\overleftarrow{\pa}_\a P^{\a\b} \overrightarrow{\pa}_\b)\Psi\cr
&=&\Phi\Psi -\Phi\overleftarrow{\pa}_\a P^{\a\b} \overrightarrow{\pa}_\b\Psi -\frac{1}{2}P^2 \pa^2 \Phi\pa^2 \Psi
\ena
where $P^2\equiv P^{\a\b}P_{\a\b}$.
In contradistinction to the bosonic case, this Moyal-like product has a finite derivative expansion, because of the grassmannian nature of the variables involved in the deformation.

With the antichiral representation (\ref{antichiralD}) the covariant derivative algebra is not deformed, while the supersymmetry algebra is deformed by curvature terms analogous to the ones we found in (\ref{modalg}, \ref{rtensor2}, \ref{rtensor})
\bea
&&\{Q_\a,Q_\b\}=0;\qquad\qquad\qquad\{Q_\a,Q_\ad\}=i\pa_{\a\ad}\cr
&&\{Q_\ad,Q_\bd\}=-2P^{\a\b}\pa_{\a\ad}\pa_{\b\bd}
\ena
Since only the dotted sector is modified, in \cite{seiberg} its is argued that only $N=\frac{1}{2}$ is preserved.
Exactly as in the $N=2$ case discussed in the previous section, these susy-breaking terms do not affect the supercoordinate algebra, that is consistent with supersymmetry.

In \cite{seiberg} chiral and vector superfields in $N=\frac{1}{2}$ have been extensively studied. 
In particular, since the covariant derivatives are not modified by the deformation, it is possible to define (anti)chiral superfields whose class is closed under $\ast$ and to write down the action for a deformed Wess-Zumino model
\bea
S&=&\int d^8 z \bar\Phi\Phi - \frac{m}{2} \int d^6 z \Phi^2 - \frac{\bar m}{2} \int d^6 \bar z {\bar\Phi}^2\cr
&-& \frac{g}{3}\int d^6z \Phi\ast\Phi\ast\Phi - \frac{\bar g}{3}\int d^6 \bar z \bar\Phi\ast\bar\Phi\ast\bar\Phi\cr
&=&S(P=0)+\frac{g}{6}\int d^4 x P^2 F^3
\ena
where $F$ is the auxiliary field in the chiral multiplet and total superspace derivatives have been neglected to obtain the last equality. 
Similarly, one can see that the $N=\frac{1}{2}$ deformation of super Yang-Mills is characterized by explicitly susy-breaking P-dependent component terms \cite{seiberg}.
Moreover in \cite{seiberg} the antichiral ring defined by the operator relation $[Q_\a,\bar {\cal O}]=0$ has been studied and it has been shown that all its properties are preserved by the deformation, while the chiral ring cannot be defined since $Q_\ad$ is not a symmetry of the theory anymore.  
\subsubsection{Results in non(anti)commutative field theories}
After Seiberg's paper \cite{seiberg} appeared, deformed superspaces have attracted much attention and a lot of efforts have been done to elucidate the properties of nonanticommutative field theories in superspace. This interest is mostly due to the fact that nonanticommutative superspaces have been shown to naturally emerge in superstring theory in the presence of R-R backgrounds \cite{vafaooguri2,seiberg, antonio,seibergberkovits}. I will review the string theory side of the story in section 1.2.4.
Here I'm going to browse the huge bibliography and discuss the main results obtained, without giving any detail. I will first discuss the progress in understanding $N=\frac{1}{2}$ theories.

In \cite{seiberg} the $N=\frac{1}{2}$ deformation of WZ and super Yang-Mills theories have been proposed. The deformation of WZ model is easy to describe, since in component formulation it corresponds to adding to the ordinary action a cubic term in the auxiliary field $F$.  
In \cite{britto,terashima} some features of the deformed WZ model, such as non validity of standard nonrenormalization theorems, stability of the vacuum energy and existence of the antichiral ring have been discussed through some examples, in both component and superspace formulations.
In particular, since supersymmetry plays an important role to guarantee renormalization through partial cancellation of UV divergences associated to bosonic and fermionic degrees of freedom, it is compelling to study renormalizability properties of the $N=\frac{1}{2}$ theory where part of the supersymmetry is explicitly broken.

A systematical analysis of perturbative renormalizability of $N=\frac{1}{2}$ WZ model has been performed in \cite{silvia} by explicit calculations up to two loops. It has been shown that, even if new divergences appear, the model can be rendered renormalizable by adding ab initio $F$ and $F^2$ terms to the ordinary lagrangian. It is somehow expected that these terms may accompany the $F^3$ deformation, since they are allowed by the symmetry of the theory. 
These two-loop results have been extended to all orders in perturbation theory in both component and superspace formulations \cite{renorm}. The proof of renormalizability  has been given on the base of dimensional arguments and global symmetries.

The study of deformations of gauge theories is more interesting than the scalar case. $N=\frac{1}{2}$ $U(n)$ gauge theories have been proven to be renormalizable in \cite{renorm2} in WZ gauge. This result have been checked up to one loop in \cite{sadooghi} in component formulation, again in WZ gauge. The study of these gauge theories in a manifestly gauge independent superspace setup has been accomplished by generalizing the background field method to nonanticommutative case \cite{silvianew}.

In  \cite{imaanpur,antonio2,torinesi} $N=\frac{1}{2}$ super Yang-Mills instantons have been studied.

In \cite{wolfSB,mikulovic} the problem of constructing a Seiberg-Witten map analogous to the one discussed in section 1.2.1 for superfields in deformed superspaces has been considered.

The case with extended supersymmetry has also been considered \cite{sokatchev, ketov} and it has been shown that, while in general $N=(1,1)$ supersymmetry is broken to $N=(\frac{1}{2}, 0)$, there are particular cases where $N=(\frac{1}{2}, 1)$ survives. Moreover, super Yang-Mills theory on extended deformed superspaces has also been studied in \cite{wolfSB}. 

$d=2$ $N=2$ classical aspects of sigma models characterized by a general K\"ahler potential and arbitrary superpotential deformed by a nonanticommutative product have been studied in \cite{chandrasekhar}.

Finally, the connection between nonanticommutative geometry and supermatrix models have been studied in \cite{park, shibusa}.

\section{Non(anti)commutative field theory from the (super)string}
\subsection{Noncommutative Yang-Mills theory from the open string}
In this section I mostly refer to \cite{seibwit} and show that a constant Neveu-Schwarz Neveu-Schwarz (NS-NS) $B$-field background modifies string dynamics nontrivially when D-branes are present. The open string with extrema constrained to lay on Dp-branes sees a deformed target-space $G_{\m\n}$ metric and a noncommutative target space coordinate algebra, characterized by a matrix $\th^{\m\n}$.   Taking a zero slope limit $\a'\rightarrow 0$ in such a way to keep the above open-string parameters finite, one can obtain two different effective theory descriptions. Depending on the choice of the regularization prescription, one obtains a field theory in ordinary space where the background field appears explicitly, or a noncommutative field theory where the background field only appears implicitly in the noncommutativity matrix $\th^{\m\n}$.
We are not going to give a review of the basic string theory needed in this section. For this we suggest the textbooks \cite{pol, gsstring}.
\subsubsection{The open string effective metric and noncommutativity parameter}
Let us consider the bosonic sector of open string theory, in a 10-dimensional flat spacetime background with metric $g_{\m\n}$, in the presence of a constant NS-NS field $B_{\m\n}$ and Dp-branes. Let us assume $B_{0i}=0$, where $i$ is a generic spacelike direction and $0$ is timelike. This means that we are going to consider a magnetic $B$ field. At the end of this section I will briefly discuss the electric case, to show that a zero-slope limit giving a noncommutative effective field theory on the brane is not admitted in this case \cite{unitar}. 

It is well-known that a constant background $B$ field can be gauged away in the bulk, but not on the boundary, on the Dp-brane, where it acts as a constant magnetic (or electric) field . If  ${\rm rk}(B)=r$, we can assume $r\leq p+1$.
We will choose spacetime coordinates in such a way that $B_{ij}\not=0$ for $i,j=1,...,r$ only and $g_{ij}=0$ for $i=1,...,r, j\not=1,...,r$.
The worldsheet action is 
\bea
&&S=\frac{1}{4\p\a'}\int_{\S}\left(g_{\m\n}\pa_a x^{\m}\pa^a x^{\n}-2\p i\a' B_{ij}\e^{ab}\pa_a x^i \pa_b x^j\right)\cr
&&=\frac{1}{4\p\a'}\int_{\S}g_{\m\n}\pa_a x^{\m}\pa^a x^{\n}-\frac{i}{2}\int_{\pa \S}B_{ij}x^{i}\pa_t x^{j}
\label{string}
\ena
where the second equality shows that antisymmetry of $\e$ implies that the $B$-term is  a boundary term. $\S$ is the (euclidean) string worldsheet, $\pa \S$ is its boundary, $\pa_t$ is the derivative tangent to the boundary $\pa \S$. Because of the presence of Dp-branes, the boundary term cannot be eliminated. It modifies the boundary conditions for the open string in the directions $i$ along the brane 
\EQ
g_{ij}\pa_n x^{j}+2\p i\a' B_{ij}\pa_t x^{j}\vert_{\pa \S}=0
\label{BC}
\EN 
where $\pa_n$ is the derivative in the normal direction with respect to the boundary $\pa \S$.
In this equation both i and j are along the brane.
We observe that for $B=0$ we obtain Neumann boundary conditions, while for maximal $B$ rank on the brane and $B\rightarrow \infty$ we get Dirichlet ones. In the latter case the string extrema are constrained to a single point on the Dp-brane, since the coordinates along the brane that describe them do not move. We can thus think that string extrema are attached to a 0-brane on the p-brane. 

From now on we will consider the classical approximation to string theory. $\S$ is a disk, that can be mapped to the upper half plane, described by complex coordinates $z$, $\bar{z}$, because of conformal invariance. The boundary conditions (\ref{BC}) can then be rewritten as
\EQ
g_{ij}(\pa-\bar{\pa})x^{j}+2\p\a'B_{ij}(\pa+\bar{\pa})x^{j}\vert_{z=\bar{z}}=0
\label{BC2}
\EN
where $\pa=\frac{\pa}{\pa z}$, $\bar{\pa}=\frac{\pa}{\pa \bar{z}}$, $Im(z)\geq 0$.
The propagator $\langle x^{i}(z) x^{j}(z')\rangle$ with the boundary conditions (\ref{BC2}) is given by \cite{prop}
\bea
&&\langle x^{i}(z) x^{j}(z')\rangle=-\a'\left[g^{ij}\log\vert z-z'\vert- g^{ij}\log\vert z-\bar{z}'\vert\right.\cr
&&\left.+G^{ij} \log\vert z-\bar{z}'\vert^2+\frac{1}{2\p\a'}\th^{ij}\log\frac{z-\bar{z}'}{\bar{z}-z'}+D^{ij}\right]
\label{propag}
\ena
where
\bea
&& G^{ij}=\left(\frac{1}{g+2\p\a' B}\right)^{ij}_S=\left(\frac{1}{g+2\p\a' B}g\frac{1}{g-2\p\a' B}\right)^{ij}\cr
&& G_{ij}=g_{ij}-\left(2\p\a'\right)^2\left(Bg^{-1}B\right)_{ij}\cr
&&\th^{ij}=2\p\a'\left(\frac{1}{g+2\p\a' B}\right)^{ij}_A=-\left(2\p\a'\right)^2\left(\frac{1}{g+2\p\a' B}B \frac{1}{g-2\p\a' B}\right)^{ij}
\label{param}
\ena
and $(~~)_S$, $(~~)_A$ denote the symmetric and antisymmetric part of the matrix in brackets, respectively. The constant quantities $D^{ij}$ depend on $B$, but not on $z$ and $z'$. They can be fixed to a certain value by making use of the fact that $B$ is arbitrary. 

We are interested in taking the limit $z\rightarrow \t\in R$, $z'\rightarrow \t'\in R$ in  (\ref{propag}).
This is because in the open string case vertex operators are to be inserted on the boundary of the worldsheet. Taking the limit one obtains
\EQ 
\langle x^{i}(\t) x^{j}(\t')\rangle=-\a'G^{ij}\log(\t-\t')^2+\frac{i}{2}\th^{ij}\e(\t-\t')
\label{open}
\EN
The function $\e(\t)$ is $1$  ($-1$) for positive (negative) $\t$.
The discontinuity in the propagator can be expressed in terms of the function $\e$ when convenient values for the constants $D^{ij}$ are chosen.

$G^{ij}$ is interpreted as the effective metric seen by open strings. This can be understood by comparing with the closed string case, where the propagator between two internal worldsheet points has a short distance behavior given by 
\EQ
\langle x^{i}(z) x^{j}(z')\rangle=-\a'g^{ij}\log\vert z-z'\vert
\EN
It is clear from (\ref{open}) that, in the commutative limit $\th\rightarrow 0$, the metric $G^{ij}$
is for the open string what $g^{ij}$ is for the closed string.

By considering the second term in (\ref{open}), we can see that the coefficient $\th^{ij}$  can be interpreted as noncommutativity parameter for the coordinates along the Dp-brane.
In conformal field theory there exists a correspondence between time ordering and operator ordering, that in this case gives 
\EQ
\left[x^i(\t),x^j(\t)\right]=T\left(x^i(\t)~x^j(\t^-)-x^i(\t)~x^j(\t^+)\right)=i\th^{ij}
\EN
The first equality says that the path integral of the time-ordered combination in  the right hand side corresponds to a matrix element  of the equal-time commutator between  $x^i$ and $x^j$. So we deduce  that the coordinates  $x^i$ are noncommuting with parameter $\th^{ij}$.
\subsubsection{Correlation functions, effective action and Moyal product}
Let us now consider the product of two tachyon vertex operators $e^{ip\cdot x}(\t)$, $e^{iq\cdot x}(\t')$, with $\t>\t'$.  By contracting with the two point function (\ref{open}) we obtain the short distance behavior
\bea
e^{ip\cdot x}(\t) \cdot e^{iq\cdot x}(\t')\sim (\t-\t')^{2\a'G^{ij}p_i q_j}e^{-\frac{1}{2}i\th^{ij}p_i q_j}e^{i(p+q)\cdot x}(\t')+...
\ena
We observe that in the limit $\a'\rightarrow 0$ the OPE formula reduces to $\ast$ product
\bea
e^{ip\cdot x}(\t) e^{iq\cdot x}(\t')\sim e^{ip\cdot x}\ast e^{iq\cdot x}(\t') 
\label{openc}
\ena
Therefore we expect that in the limit $\a'\rightarrow 0$ the theory should be easily described in terms of Moyal product. However, many interesting aspects emerge even without taking that limit.

We have shown that open strings in the presence of a constant $B$ field on a Dp-brane can be described not only in terms of the two parameters $g^{ij}$ and $B^{ij}$, but also in terms of two functions  $G^{ij}(g,B;\a')$ and $\th^{ij}(g,B;\a')$, representing the effective metric and noncommutativity parameter seen by the open string ending on the brane. 

Now we are going to show that, if we choose to work with the second couple of parameters, the dependence of the effective theory on $\th$ is very simple. The theory with a nonvanishing $\th$ can be obtained from the one with $\th=0$ by simply substituting Moyal $\ast$ products characterized by the noncommutativity parameter $\th$ to ordinary products. This is a general feature, no limit has to be taken.

It is well-known that perturbative string theory requires the evaluation of the path integral of vertex operators the generally take the form $P\left(\pa x,\pa^2 x,...\right)e^{ip\cdot x}$, where $P$ is a polynomial in the derivatives of $x$. Let us now consider the expectation value of the product of $k$ vertex operators, with momenta $p^1,...,p^k$ and $x$ along the Dp-brane. We are interested in determining the explicit dependence on the parameter $\th$, while we would like to keep the dependence on $G$ implicit. The two-point function is the sum of two terms, one contains $G$ only, the other contains $\th$ only. When  we contract terms containing derivatives of $x$, the part of the propagator involving $\th$ does not contribute, because $\e(\t-\t')$ is a constant function. So we can obtain the $\th$ dependence by just taking into consideration the contraction between exponentials
\bea
&&\langle \prod_{n=1}^k P_n\left(\pa x(\t_n),\pa^2 x(\t_n),...\right)e^{ip^n\cdot x(\t_n)}\rangle_{G,\th}\cr
&&=e^{-\frac{i}{2}\sum_{n>m}p_i^n\th^{ij}p_j^m\e(\t_n-\t_m)}\langle \prod_{n=1}^k P_n\left(\pa x(\t_n),\pa^2 x(\t_n),...\right)e^{ip^n\cdot x(\t_n)}\rangle_{G,\th=0}\cr
&&~~~~~~~
\ena
So the $\th$-dependence is simply given by the phase factor
\EQ
e^{-\frac{i}{2}\sum_{n>m}p_i^n\th^{ij}p_j^m\e(\t_n-\t_m)}
\label{fase}
\EN
Because of momentum conservation $\sum_n p^n=0$  and antisymmetry of $\th$, (\ref{fase}) only depends on the cyclic order of the points $\t_1,...,\t_n$ on the worldsheet boundary.

By knowing the $S$-matrix of low energy massless particles, one can deduce order by order in $\a'$ an effective action for the theory.  This will be expressed in terms of a certain number of functions $\Phi_i$ that in general may have values in the space of $n\times n$ matrices (think of the nonabelian gauge theory case). $\Phi_i$ represents the wave function of the i-th field. By looking at (\ref{param}) one notes that $B=0\Leftrightarrow \th=0$.  So the general form of the effective action for $B=0$ will be as follows 
\EQ
\int d^{p+1}x~ \sqrt{\det G}~Tr\left(\pa^{n_1}\Phi_1~\pa^{n_2}\Phi_2...\pa^{n_k}\Phi_k\right)
\EN
where $\pa^{n_i}$ stands for the product of  $n_i$ partial derivatives with respect to certain unspecified coordinates.

Now it is easy to move on to a theory with $B\not=0$.  If the effective action is written in momentum space, then it is sufficient to insert the phase factor (\ref{fase}). In configuration space this corresponds to replacing ordinary products with Moyal $\ast$ products (see (\ref{momenti})).
To conclude, the effective action with $B\not=0$ takes the form
\EQ
\int d^{p+1}x~ \sqrt{\det G}~Tr\left(\pa^{n_1}\Phi_1\ast \pa^{n_2}\Phi_2\ast...\ast\pa^{n_k}\Phi_k\right)
\EN
So we have found an easy way to describe the theory with $B\neq 0$ when knowing the one with $B=0$. However we must stress that both theories have an equally complicated $\a'$  
expansion.
\subsubsection{The description in the zero-slope limit}
The formalism of noncommutative geometry becomes much more powerful when the zero-slope limit $\a'\rightarrow 0$ is taken. This is somehow expected from (\ref{openc}).\\
We would like to take the limit in such a way to keep the open string parameters $G$ and $\th$ finite. 
This can be done by choosing 
\bea
&&\a'\sim\e^{\frac{1}{2}}\rightarrow0\cr
&&g_{ij}\sim\e\rightarrow 0~~~~~\rm{per}~~ i,j=1,...,r
\ena
when all the rest is kept fixed (also the two form $B$). Equation (\ref{param}) becomes
\bea
&&G^{ij}=\left\{
\begin{matrix}
-\frac{1}{(2\p\a')^2}\left(\frac{1}{B}g\frac{1}{B}\right)^{ij}&\rm{per}~ i,j=1,...,r\cr
g^{ij}&\rm{otherwise}
\end{matrix}\right.\cr
&&G_{ij}=\left\{
\begin{matrix}
-(2\p\a')^2(Bg^{-1}B)_{ij}&\rm{per}~i,j=1,...,r\cr
g_{ij}&\rm{otherwise}
\end{matrix}\right.\cr
&&\th^{ij}=\left\{
\begin{matrix}\left(\frac{1}{B}\right)^{ij}&\rm{per}~i,j=1,...,r\cr
0&\rm{otherwise}
\end{matrix}\right.
\label{paramalpha}
\ena
The propagator for two points on the boundary becomes
\EQ
\langle x^i(\t)x^j(0)\rangle=\frac{i}{2}\th^{ij}\e(\t)
\label{propalpha}
\EN
For two generic functions $f$ and $g$ we then obtain (see (\ref{openc})) \EQ
:f(x(\t)):~:g(x(0)):~=~:e^{\frac{i}{2}\e(\t)\th^{ij}\frac{\pa}{\pa x^i(\t)}\frac{\pa}{\pa x^j(0)}}f(x(\t))g(x(0)):
\EN
and thus
\EQ
\lim_{\t\rightarrow 0^+}:f(x(\t)):~:g(x(0)):~=~:f(x(0))\ast g(x(0)):
\label{funstar}
\EN
where $\ast$ is Moyal product (\ref{moyal}).\\
As a result, correlation functions of exponential operators on the disk boundary are given by
\EQ
\langle\prod_n e^{ip_i^nx^i(\t_n)}\rangle=e^{-\frac{i}{2}\sum_{n>m}p_i^n\th^{ij}p_j^m\e(\t_n-\t_m)}\d\left(\sum p^n\right)
\EN
In the general case with $n$ functions $f_1,...,f_n$
\EQ
\langle \prod_n f_n(x(\t_n))\rangle=\int dx f_1(x)\ast...\ast f_n(x)
\label{correlatori}
\EN
\subsubsection{Adding gauge fields}
Let us now add to the action (\ref{string}) a term representing the coupling of the string worldsheet to a gauge field $A_i(x)$. For simplicity we will take ${\rm rk}(A)=1$.
\EQ
-i\int d\t A_i(x)\pa_{\t}x^i
\label{actiongauge}
\EN
Comparing with (\ref{string}) we see that the constant field $B$ can be replaced by a gauge field  $A_i=-\frac{i}{2}B_{ij}x^j$, whose field strength is $F=B$. The bosonic string coupled to the $B$-field background is invariant under the gauge symmetry
\beq
\d B_{\m\n}=\pa_{[\m} \L_{\n]}
\eeq
modulo boundary terms. These terms can be compensated by the shift
\beq
\d A_\m = \L_\m
\eeq
Therefore, physics will be described by the gauge invariant combination $\o=F+B$.

The action (\ref{actiongauge}) is invariant under the transformation
\EQ
\d A_i=\pa_i \l
\label{trasf}
\EN
since the variation of the integrand is a total derivative.

When we consider a quantum field theory we must pay attention to the regularization procedure. Physics of course must not depend on the particular choice of regularization! If we choose a Pauli-Villars regularization, we obtain an ordinary gauge theory, symmetric under the usual gauge transformation (\ref{trasf}). We can make a different choice, though. We can use a point-splitting regularization, characterized by the fact that the product of two operators at the same point never appears. Actually, one first eliminates the region $\vert \t-\t'\vert<\d$ and then takes the limit $\d\rightarrow 0$. Now we are going to evaluate the variation of the path integral of the exponential of the action (\ref{actiongauge}) under the gauge transformation (\ref{trasf}), having first expanded the exponential in series with respect to $A$.
The first term in the expansion gives 
\EQ
-\int d\t A_i(x)\pa_{\t}x^i \cdot\int d\t' \pa_{\t'}\l
\EN
Even though the integrand of the second factor is a total derivative, one gets a boundary contribution for  $\t-\t'=\pm\d$. In the limit $\d\rightarrow 0$ this contribution takes the form
\bea
&&-\int d\t:A_i(x(\t))\pa_{\t}x^i(\t):~:\left(\l(x(\t^-))-\l(x(\t^+))\right):\cr
&&=-\int d\t:\left(A_i(x)\ast\l-\l\ast A_i(x)\right)\pa_{\t}x^i:
\label{termine}
\ena
To obtain this results one makes use of the fact that there are no contractions between $\pa_{\t}x$ and $x$ with the constant propagator (\ref{propalpha}). Moreover, the relation (\ref{funstar}) between operators and $\ast$ product has been taken into account.

To cancel out the term (\ref{termine}), we have to modify the transformation (\ref{trasf}). So we discover that the point splitting regularized theory is not invariant under (\ref{trasf}), but under the new transformation
\EQ
\d \hat{A}_i=\pa_i \hat{\l}+i\hat{\l}\ast \hat{A}_i-i\hat{A}_i\ast \hat{\l}
\EN 
We recognize the gauge invariance (\ref{gaugenc}) of noncommutative gauge theories with $n=1$. We have introduced the ``hatted" notation $\hat{A}$ for fields in a noncommutative algebra.

It is possible to show \cite{seibwit} that at a generic order $m$ in the $A$-expansion of the exponential of the action the correct gauge invariance is the noncommutative one, when point-splitting regularization is performed. 

Moreover in \cite{seibwit} the calculation of the expectation value of three gauge vertex operators is performed.  The result of this computation could also be obtained by considering the following effective action as a starting point
\EQ
S_{\rm{eff}}\propto \int \sqrt{\det G}~G^{ii'}G^{jj'}Tr\left( \hat{F}_{ij}\ast \hat{F}_{i'j'}\right)
\EN
where the field strength can be expressed in terms of $\hat{A}$ as in (\ref{gaugenc}).
\vskip 24pt
We have shown that the same string theory in the limit $\a' \rightarrow 0$ can be either described by an effective action corresponding to an ordinary gauge theory or by one associated to a noncommutative gauge theory, depending on the choice of the regularization procedure. Since physics cannot depend on the way this procedure is performed, two results obtained with different regularizations must be related by a redefinition of the coupling constants. In the worldsheet action the spacetime-valued fields play the role of coupling constants, so we expect that commutative and noncommutative effective descriptions should be related by a redefinition of these fields.

A natural guess is that a local map $\hat{A}=\hat{A}(A,\pa A,\pa^2 A,...;\th)$ among gauge fields and a corresponding map $\hat{\l}=\hat{\l}(\l,\pa \l,\pa^2 \l,...;\th)$ among the group parameters exists. In section 1.1.2 we observed that noncommutative $U(1)$ group is nonabelian. This tells us that such a correspondence cannot exist. Actually, the existence of such a map would imply an isomorphism between the ordinary gauge group and the corresponding noncommutative one. Since an abelian group cannot be isomorphic to a non abelian one, the first proposal for the map is ruled out. 

However, what is really needed for physics is that the gauge-transformed field $\d_{\l} A$ corresponds to the gauge-transformed  field $\hat{\d}_{\hat{\l}}\hat{A}$. Therefore it is sufficient that
\EQ
\hat{A}(A)+\hat{\d}_{\hat{\l}}\hat{A}(A)=\hat{A}(A+\d_{\l}A)
\label{regola}
\EN
and we can look for a correspondence 
\bea
&&\hat{A}=\hat{A}(A)\cr 
&&\hat{\l}=\hat{\l}(\l,A)
\label{corrisp}
\ena
satisfying (\ref{regola}).
The $A$-dependence of $\hat{\l}$ solves the problem of the isomorphism between the two gauge groups. A relation like (\ref{corrisp}) does not imply any correspondence between the two group structures. A correspondence of the required form was found in \cite{seibwit} (Seiberg-Witten map). It is given in terms of a set of differential equations describing how $A$ and $\l$ must vary with $\th$ for the physics to remain unchanged. 

The results discussed in this section concerning an open string ending on a single Dp-brane can be easily generalized to the case of a stack of $n$ coincident D-branes. In this case one obtains a $U(n)$ noncommutative Yang-Mills theory as an effective field theory on the brane worldvolume. In section 1.2.2 I have commented on the fact that $SU(n)$, $SO(n)$ and $Sp(n)$ subgroups are not closed under Moyal product, so in principle one expects that restrictions to this subgroups should not emerge from the string. Actually, the case of $SU(n)$ is ruled out because the $U(1)$ degree of freedom in $U(n)$ ceases to decouple. $SO(n)$ and $Sp(n)$ restrictions can instead emerge, in a very nontrivial way, by orientifold projection, as shown in \cite{tomasiello}.
\subsubsection{Open string in the presence of an electric NS-NS background and the breakdown of unitarity and causality in time-space noncommutative field theories}
In section 1.1.2 I have discussed unitarity and causality problems in noncommutative field theories with time-space noncommutativity, i.e. with $\th^{0i}\neq 0$. 

In principle we could expect these theories to arise in string theory in the presence of D-branes and a constant electric $B_{0i}$ background.
However, the case of an electric background field is very different with respect to the magnetic one. It can be shown that if the background electric field $E$ exceeds the critical upper value $E_c$, string pairs are produced that destabilize the vacuum.  So, if the electric field is along the $x_1$ direction and the metric is diagonal in the $(x_0,x_1)$ plane with components given by $g$, the bound is given by
\beq
E\leq E_c,~~~~~~~~~{\rm where}~~ E_c=\frac{g}{2\pi\a'}
\eeq
In \cite{electric} it has been shown that in this case the open string parameters $(G,\th)$ are related to the electric field on the brane by the formula
\beq
\a' G^{-1}= \frac{1}{2\pi}\frac{E}{E_c}\th
\eeq

As before, to obtain the effective field theory on the brane we have to consider a zero-slope limit $\a'\rightarrow 0$. From the previous formula, it is clear that if we want to keep the open string metric $G$ finite in the limit, when $\a'\rightarrow 0$ then also $\th\rightarrow 0$. Therefore, it is possible to obtain a field theory description involving massless open string modes only, but this will be an ordinary field theory and not a noncommutative one. On the other hand, we can keep the noncommutativity parameter $\th$ finite, but then $\a'$ must also be finite and we are considering a string theory and not a field theory. 

Indeed, in \cite{electric}, it has been shown that a limit can be taken where 
\beq
\frac{E}{E_c}\rightarrow 1 \qquad\qquad {\rm and} \qquad\qquad g \sim \frac{1}{1-\left(\frac{E}{E_c}\right)^2}
\eeq 
and all the other parameters are kept fixed, in particular the open string metric $G$. In this limit 
\beq
\th=2\pi\a' G^{-1}
\eeq
is finite, so time-space noncommutativity is present. The theory obtained describes open strings in noncommutive spacetime (NCOS). Open strings decouple from closed strings, therefore also from gravity. In \cite{GMMS} it has been shown that NCOS theory is the S-dual description of strongly coupled, spatially noncommutative $N=4$ Yang-Mills theory (other works concerning NCOS are listed in \cite{NCOS}).
 
We conclude that noncommutative field theory with time-space noncommutativity does not emerge as a consistent truncation of string theory. Moreover, string theory in the presence of an electric background on the brane is unitary and acausal effects are not present \cite{causal}. So $\a'$ corrections to noncommutative ``electric" field theory restore unitarity and causality. 

It is clear that ``electric" noncommutative field theories are missing some degrees of freedom, related to the undecoupled massive string modes, that are necessary for unitarity and causality of the theory.
In \cite{timenc} it has been shown that tachyonic particles are produced in scattering processes of noncommutative field theory with $\th^{0i}$ turned on. Form the string theory point of view, these particles may be viewed as a remnant of a continuous spectrum of these undecoupled closed-string modes. 
\subsection[Generalization to the superstring in RNS and GS\\ formalisms]{Generalization to the superstring in RNS and GS formalisms}
In this section I will generalize to the superstring the results obtained in the previous section for the bosonic string. In \cite{seibwit} the string with $N=1$ worldsheet supersymmetry (RNS) is considered. In \cite{chuzamora} instead the manifestly target-space supersymmetric superstring (GS) is discussed.
In both cases the open superstring is coupled to a constant NS-NS background in the presence of D-branes. 
I'm not going to give an introduction to these two formalisms for the superstring. For this I suggest the textbooks \cite{pol,gsstring}
\subsubsection{Open RNS string in the presence of a constant $B$-field and D-branes}
In \cite{seibwit} the following action for the RNS string coupled to a constant magnetic NS-NS background field $B^{ij}$ is considered
\bea
&&S=\frac{1}{4\p\a'}\int d^2 z\left\{\bar{\pa} x^{\m}\pa x_{\m}+i\psi^{\m}\bar{\pa}\psi_{\m}+i\bar{\psi}^{\m}\pa\bar{\psi}_{\m}-2\p i\a'B_{ij}\e^{ab}\pa_a x^i\pa_b x^j\right\}\cr
&&~~~~~
\label{actionRNS}
\ena
In the directions $i$, $j$ along the brane the following boundary conditions are imposed
\bea
&&g_{ij}(\pa-\bar{\pa})x^{j}+2\p\a'B_{ij}(\pa+\bar{\pa})x^{j}\vert_{z=\bar{z}}=0
\cr
&&g_{ij}(\psi^j-\bar{\psi}^j)+2\p\a'B_{ij}(\psi^j+\bar{\psi}^j)\vert_{z=\bar{z}}=0
\label{BCRNS}
\ena
The first condition can be naturally obtained by requiring that there are no boundary terms in the Euler-Lagrange equations of motion. The second one, involving fermions, is obtained by requiring consistency under supersymmetry of the boundary conditions, but cannot be obtained from the action (\ref{actionRNS}). Actually, one finds an inconsistency when requiring both the vanishing of boundary terms in the variation of the action and the compatibility of boundary conditions with supersymmetry. In \cite{ulf1} (see also \cite{summary} for a nice summary of the methods used) it has been shown that boundary terms can be added to the action (\ref{actionRNS}) so that the supersymmetric boundary conditions (\ref{BCRNS}) follow as boundary contributions to the field equations. 
Therefore, the theory described by the action $S+S_b$, with
\beq
S_b=-\frac{1}{2} \int d^2 z B_{\m\n} \bar\psi^\m \rho^\a \pa_\a \psi^\n
\eeq
and by the boundary conditions (\ref{BCRNS}) is invariant under the rigid $N=1$ worldsheet supersymmetry transformations
\bea
&&\d x^i=-i\eta(\psi^i+\bar{\psi}^i)\cr
&&\d\psi^i=\eta\pa x^i\cr
&&\d\bar{\psi}^i=\eta\bar{\pa}x^i
\label{superconf}
\ena
where the parameter $\eta$ is a worldsheet spinor and a spacetime scalar.

In \cite{seibwit} the open string is coupled to a gauge field $A$ by adding to (\ref{actionRNS}) the following boundary term
\EQ
L_A=-i\int d\t\left(A_i(x)\pa_{\t}x^i-iF_{ij}\Psi^i\Psi^j\right)
\label{accopp}
\EN
where $F_{ij}=\pa_iA_j-\pa_jA_i$ is the ordinary field strength (we are considering $U(1)$ case for simplicity) and
\EQ
\Psi^i=\frac{1}{2}\left(\psi^i+\bar{\psi}^i\right)
\EN
The variation of (\ref{accopp}) under the supersymmetry transformations (\ref{superconf}) is a total derivative
\EQ
\d\int d\t\left(A_i(x)\pa_{\t}x^i-iF_{ij}\Psi^i\Psi^j\right)=-2i\eta\int d\t\pa_{\t}(A_i\Psi^i)
\EN
Exactly as in bosonic case, the theory is regularized by making use of a ``point splitting'' technique and extra boundary terms are produced. Expanding the exponential of the action to first order in $A$ we can compute the variation of the path integral up to first order in $L_A$
\EQ
i\int d\t\int d \t'\left(A_i\pa_{\t}x^i(\t)-iF_{ij}\Psi^i\Psi^j(\t)\right)\left(-2i\eta\pa_{\t'}A_k\Psi^k(\t')\right)
\EN
Extra boundary terms appear when $\t'\rightarrow \t^+$ and $\t'\rightarrow \t^-$.
If the following interaction term is added to the action
\EQ
\int d\t A_i\ast A_j\Psi^i\Psi^j(\t)
\EN
the extra terms are cancelled by its variation under (\ref{superconf}).
So we deduce that $L_A$ must be changed into
\EQ
-i\int d\t\left(A_i(x)\pa_{\t}x^i-i\hat{F}_{ij}\Psi^i\Psi^j\right)
\EN
being $\hat{F}$ the noncommutative field strength.
If we had performed a Pauli-Villars regularization, instead, (\ref{accopp}) would have been invariant under worlsheet supersymmetry (\ref{superconf}) and we would have ended up with field strength and gauge symmetry of ordinary $U(1)$.

Summarizing, in \cite{seibwit} it has been shown that the RNS open string in the presence of a constant $B$-field and D-branes can be described, in the zero-slope limit  $\a'\rightarrow 0$, either by ordinary gauge theory or by noncommutative gauge theory on the brane, depending on the choice of the regularization prescription. The two different descriptions are related by a Seiberg-Witten map, as in bosonic case.

In \cite{ulf1} a different approach to the problem of coupling a gauge field to the open string in the presence of the $B$ field is presented. The coupling to the $A$ field is reconsidered in a way to preserve both shift symmetry and supersymmetry. A coupling term different with respect to (\ref{accopp}) is found that is not supersymmetric by itself, but only together with the rest of the action $S+S_b$ and after making use of the corresponding  boundary conditions. 

\subsubsection{Open GS superstring in the presence of a constant $B$-field and D-branes }
The Green-Schwarz superstring has manifest target space $N=2$ supersymmetry. The target space is a ten dimensional superspace described by the coordinates $(x^\m,\th^{\a i})$, with $i=1,2$. When the theory is coupled to a certain background and D-branes are present, in principle target space fermionic coordinates could be involved in noncommutativity. In \cite{chuzamora}, it has been shown that this is not true in the simple case of a constant NS-NS background. In this case only bosonic coordinates become noncommutative. In section 1.2.4 we will see that fermionic coordinates are indeed involved in noncommutativity when a constant R-R background is present.

The action for the GS superstring coupled to a constant NS-NS background in flat spacetime is given by
\bea
S_{GS}&&=-\frac{1}{2\p\a'}\int d^2\xi \left\{\P_i^{~\m}\P^{i\n}g_{\m\n}+2i\e^{ij}\pa_i x^{\m}(\bar{\th}^1\G_{\m}\pa_j \th^1-\bar{\th}^2\G_{\m}\pa_j\th^2)\right.\cr
&&\left.-2\e^{ij}(\bar{\th}^1\G^{\m}\pa_i\th^1)(\bar{\th}^2\G_{\m}\pa_j \th^2)+\e^{ij}\pa_i x^{\m}\pa_jx^{\n}B_{\m\n}\right\}
\label{piatta}
\ena
where 
\beq
\Pi_i^\m=\pa_i x^\m - i \thb\Gamma^\m \pa_i \th
\eeq
are the supersymmetry invariant one-forms.
It is clear from (\ref{piatta}) that only bosonic coordinates couple to $B_{\m\n}$. However, boundary conditions along the D-brane must also be considered before we can deduce that fermions are not affected by the presence of the background.\\
We find that bosonic coordinates must satisfy the same boundary conditions we found in the bosonic case (\ref{BC}). The two fermionic coordinates $\th^1$ and $\th^2$ must satisfy $\th^2=\G_B\th^1$ on the boundary, where $\G_B$ is a suitable $B$-dependent matrix satisfying $\G_B^2=1$. So the action (\ref{piatta}) with these boundary conditions is invariant under the supersymmetry
\bea
&&\d x^{\m}=i\bar{\e}\G^{\m}\th\cr
&&\d \th=\e
\ena
only if the supersymmetry parameters $\e^i$ also satisfy $\e^2=\G_B\e^1$. So the D-brane breaks half of the supersymmetry. 

In \cite{chuzamora} an explicit computation is performed to dermine the coordinate algebra on the brane. Following the method introduced by Chu and Ho \cite{chuho}, the authors deal with boundary conditions along the brane by treating them as constraints on phase-space. The presence of these constraints makes it necessary to consider Dirac brackets instead of Poisson brackets. Working in a light-cone gauge, in \cite{chuzamora} it was shown that the fermionic variables surviving the gauge fixing procedure satisfy a standard anticommutative algebra and only the bosonic sector is affected by the presence of the constant NS-NS background, exactly as in \cite{seibwit}.
So it is clear that if we want target-space fermionic variables to be deformed, we must consider a different background.
\subsection[Noncommutative selfdual Yang-Mills from the ${\cal N}=2$\\ string]{Noncommutative selfdual Yang-Mills from the ${\cal N}=2$ string}
In this section I will briefly introduce the $N=2$ string and its peculiar properties. Referring to \cite{olaf1}, I will apply the analysis outlined in the previous section to the open $N=2$ string in the presence of $n$ spacefilling D3-branes and a constant $B$ field  to show that it coincides at tree level with $U(n)$ noncommutative selfdual Yang-Mills.

The $N=2$ fermionic string is characterized by an $N=2$ worldsheet supersymmetry.
For the string to be critical, the target space must be four-dimensional with signature $(2,2)$.
Its propagating degrees of freedom are the embedding coordinates $x^\m$ and the RNS Majorana spinors $\psi^\m$, $\m=1,2,3,4$. These matter fields are coupled to the $N=2$ supergravity multiplet. Using symmetries of the action, one can show that all the gravitational degrees of freedom can be gauged away and in superconformal gauge the action can be written as follows
\beq
S=-\frac{1}{4\pi\a'}\int_\Sigma d^2 \sigma~\eta^{\a\b}\left(\pa_\a x^\m \pa_\b x^\n + i \bar \psi ^\m \rho_\a \pa_\b \psi^\n\right)g_{\m\n}
\eeq
where $g_{\m\n}=\zeta {\rm diag}(+1,+1,-1,-1)$ ($\zeta>0$ scaling parameter) is the metric in $R^{(2,2)}$. This action has a residual symmetry given by the $N=2$ superconformal group, that also contains rigid $N=2$ supersymmetry corresponding to the transformations
\bea
&&\d x^\m =\bar \e_1 \psi^\m + J^\m_\n \bar \e_2 \psi^\n \cr
&&\d\psi^\m = -i \rho^\a \pa_\a x^\m \e_1 + i J^\m_\n \rho^\a \pa_\a x^\n \e_2
\label{2susy}
\ena
where $J_\m^\n$ is a complex structure compatible with the metric $g_{\m\n}J^\n_\l + J_\m^\n g_{\l\n}=0$ (with our flat metric $J_2^1=-J_1^2=J_4^3=-J_3^4=1$).

It has been shown that the open $N=2$ fermionic string at tree level is identical to self-dual Yang-Mills in $2+2$ dimensions \cite{oogurivafa}. 
The absence of massive states in the physical spectrum is related to the vanishing of all tree-level amplitudes beyond three-point. The vanishing of amplitudes implies the existence of symmetries and vice-versa. Since an infinite number of tree-level amplitudes vanish, we expect an infinite number of symmetries to be present. These were described in \cite{olaf2,olaf3}.

The $N=2$ string seems to be a master theory quantizing integrable systems. It is well-known that most integrable models in $d=2$ and $d=3$ can be obtained by dimensional reduction from selfdual Yang-Mills. In \cite{oogurivafa} a possible definition for integrability in $d=(2,2)$ has been proposed, inspired by the peculiar properties of the tree-level S-matrix of the $N=2$ string. A system in $d=(2,2)$ would be classically integrable if the $n$-point tree-level amplitudes vanish beyond $n=3$. This definition is reminiscent of the one that can be given in $d=2$ concerning factorization of the S-matrix.

The $N=2$ open fermionic string can be coupled to a two-form NS-NS background $B$ field.
In \cite{olaf1} it has been shown that additional boundary terms must be added to the action for the boundary conditions obtained from the Euler-Lagrange procedure to be $N=2$ supersymmetric, exactly as in the case of the $N=1$ string \cite{ulf1}. Moreover, the presence of the second supersymmetry implies a nontrivial constraint on the $B$ field. This is the compatibility condition with respect to the complex structure $B_{\m\n}J^\m_\l - J^\n_\m B_{\l\n}=0$, i.e. $B$ must be K\"ahler.  The consistent $N=2$ gauge fixed action is then
\bea
S&=&-\frac{1}{4\pi\a'}\int_\Sigma d^2 \sigma\left[\left(\eta^{\a\b}g_{\m\n} + \e^{\a\b}2\pi\a' B_{\m\n}\right)\pa_\a x^\m \pa_\b x^\n \right.\cr
&&\left.+ \left(g_{\m\n}+ 2\pi\a' B_{\m\n}\right) i \bar \psi ^\m \rho_\a \pa_\a\psi^\n\right]
\ena
This action functional cannot be obtained from a superspace formulation.

In \cite{olaf1} it has been shown that the open $N=2$ string dynamics in the presence of $n$ coincident D3-branes filling the target space is modified by a magnetic $B$ field so that the open string sees an effective metric $G_{\m\n}$ and a noncommutative algebra on the brane characterized by a $\th^{\m\n}$ parameter. The starting point for the analysis is again the expression for the open string correlators (\ref{propag}). A particular choice of the $SO(2,2)$ generators allows us to write the matrices $J$ and $B$ in terms of the generators of the $U(1)\times U(1)$ subgroup of $SO(2,2)$, so that in complete generality the NS-NS field is expressed in terms of the two quantities  $B_1$ and $B_2$ as
\beq
B_{12}=-B_{21}\equiv B_1 \qquad\qquad\qquad B_{34} =-B_{43} \equiv B_2
\eeq
In this basis the open string effective metric $G^{\m\n}$, noncommutativity parameter $\th^{\m\n}$ and coupling $G_s$ can be obtained and have expressions similar to the corresponding ones for the ten-dimensional string (\ref{param}). 

Exactly as in the ten dimensional case a zero-slope limit can be taken that keeps the open string parameters finite. In this limit it has been shown that tree-level three-string amplitudes can be obtained from the noncommutative version of self-dual Yang-Mills in Leznov gauge (this depends on the choice of the Lorentz frame, one can as well obtain Yang gauge).
The corresponding lagrangian is given by
\bea
{\cal L}=\frac{1}{2} G^{\m\n}{\rm tr}\left(\pa_\m \Phi\ast\pa_\n \Phi\right) + \frac{1}{3}\e^{\ad\bd}{\rm tr}\left(\Phi\ast\hat\pa_{0\ad} \Phi \ast \hat\pa_{0\bd}\Phi\right)
\ena
where $G^{\m\n}\equiv e^\m_{\hat\sigma} e^\n_{\hat\lambda} \eta^{\hat\sigma \hat\lambda}$ is the open string effective metric and $\hat\pa$ is the corresponding derivative defined by $\hat\pa_{\hat\m}\equiv e^\n_{\hat\m}\pa_\n$.
This result is deduced as in the ten-dimensional case by noting that the effect of turning on the $B$-field is the multiplication of any open string amplitude by a phase factor (\ref{fase}). This corresponds to replacing ordinary products with Moyal products in the worldvolume effective field theory.

As a nontrivial check, it has also been shown that the tree-level four-point function for $U(n)$ noncommutative self-dual Yang-Mills in Leznov gauge vanishes. Therefore, the natural deformation of selfdual Yang-Mills in Leznov gauge seems to preserve the nice scattering properties of the original, commutative theory. We have seen that the vanishing of tree-amplitudes beyond three-point defines integrable systems in $d=(2,2)$ exactly as the factorization of the S-matrix is a definition of a an integrable system in $d=2$.  
So the result in \cite{olaf1} suggests that noncommutative selfdual Yang-Mills is integrable. 

This result will be important for the further developments considered in \cite{mypaper3}, where, in collaboration with O. Lechtenfeld, L. Mazzanti, S. Penati and A. Popov, I have constructed a noncommutative version of the sine-Gordon theory with a factorized S-matrix, that is obtained as a dimensional reduction of $(2,2)$ selfdual Yang Mills. 
\subsection[Non(anti)commutative field theories from the covariant\\ superstring]{Non(anti)commutative field theories from the covariant superstring}
In this section I would like to discuss the superstring origin of the non\-(anti)\-com\-mu\-ta\-ti\-ve superspaces introduced in sections 1.1.5, 1.1.6, 1.1.7.

Up to now I have discussed in detail the case of the open bosonic string in flat space in the presence of a constant NS-NS background field and Dp-branes and I have shown that the presence of the background induces a noncommutativity in the brane coordinate algebra. I have generalized this result to the RNS superstring, that is characterized by an $N=1$ worldsheet supersymmetry and happens to exhibit target space supersymmetry after a consistent truncation of the spectrum (GSO projection) is performed. This theory is not manifestly supersymmetric in target space. I have also shown how the bosonic results can be generalized to the manifestly target-space supersymmetric GS superstring. In this case we have seen that the presence of a constant NS-NS background does not modify the anticommutators between fermionic coordinates. Finally, I have generalized the bosonic string results to the $N=2$ string, characterized by an $N=2$ worldsheet supersymmetry.  In all these cases a constant NS-NS two-form background has been considered.

In this section we face the problem of finding a suitable superstring background that could induce, in the presence of D-branes, a nonanticommutative deformation of target space fermionic variables, exactly as the NS-NS B-field induces a deformation of bosonic target space coordinates. So target space must be a superspace and thus we need to consider a manifestly supersymmetric version of the superstring.
An example of this is the Green-Schwarz superstring \cite{gsstring} we considered in the previous section. It lives in a ten-dimensional superspace and its action is manifestly supersymmetric. Unfortunately, because of its complicated worldsheet symmetries, its action in a flat background is not quadratic, it cannot be quantized in a Lorentz-covariant way and this renders the formalism terribly difficult to handle.

Recently \cite{berkovits} a new proposal was made for an action describing the ten-di\-men\-sio\-nal superstring, that is manifestly target-space supersymmetric and is quadratic in a flat background\footnote{An introduction to this formalism will be given in chapter 3.}.
In this formalism the R-R field strengths are all contained in a bispinor $P^{\a\hat \a}$ where the two indices may have opposite or same chirality depending whether we are in IIA or IIB superstring theory. It is natural to think that, as much as the $B^{\m\n}$ background is related to a noncommutative deformation $[x^\m,x^\n]=i\th^{\m\n}$, the $P^{\a\hat\a}$ background may be related to a $\{\th^\a,\hat\th^{\hat\a}\}=C^{\a\hat\a}$ deformation.

In a series of papers \cite{vafaooguri2,seiberg,seibergberkovits, antonio} it has been shown that this is indeed the case. In the first three papers the four-dimensional theory obtained from type II ten-dimensional superstring compactified on a Calabi-Yau three-fold \cite{vafaberkovits, berkovitsgs} has been considered, while in the last paper the full ten-dimensional superstring theory \cite{berkovits,berkovitsRR} has been discussed.
I will consider the four-dimensional case first, since, compared to the covariant quantization of the superstring in ten dimension, the formalism is much simpler in this case, because of the smaller amount of manifest supersymmetry.

The relevant part of the lagrangian density is
\beq
{\cal L}=\frac{1}{2} \pa x^{\a\ad} \bar\pa x_{\a\ad} + p_\a \bar\pa \th^\a + p_\ad \bar\pa \th^\ad +\bar p_\a \pa \thb^\a + \bar p_\ad \pa \thb^\ad
\label{freeaction}
\eeq
where $p_\a$ $\bar p_\a$ $p_\ad$ and $\bar p_\ad$ are conjugate momenta to the superspace fermionic variables $\th^\a$  $\thb^\a$ $\th^\ad$ and $\thb^\ad$ respectively\footnote{Berkovits covariant formalism is first order for fermionic variables, whose conjugate momenta are introduced as independent fields. This was first done by Siegel in his approach to the GS superstring \cite{siegel}}. We indicate with dots target space Weyl spinor chirality and with bars worldsheet holomorphicity.  
The four dimensional action corresponding to (\ref{freeaction}) describes a free conformal field theory. The fields $x$, $\th$, $\thb$, $p$ and $\bar p$ satisfy free equations of motion, second order for $x$ and first order for fermionic variables. This theory exhibits an $N=2$ target-space supersymmetry\footnote{It is interesting to note that, if one considers the undotted fermions only, the model can be regarded as a topological B-model on euclidean four-dimensional space and the topological BRST symmetry is strictly connected to the susy transformations generated by the dotted charges $Q_\ad$ and $\bar Q_\ad$.}.

It is useful to apply to (\ref{freeaction}) the following change of variables
\bea
&&y^{\a\ad}= x^{\a\ad} +i\th^\a\th^\ad +i \thb^\a\thb^\ad\cr
&&q_\a=-p_\a -i \th^\ad\pa x_{\a\ad} +\frac{1}{2} \th^\bd\th_\bd\pa\th_\a -\frac{3}{2}\pa(\th_\a \th^\bd \th_\bd)\cr
&&d_\ad=p_\ad -i \th^\a \pa x_{\a\ad} -\th^\b\th_\b\pa\th_\ad +\frac{1}{2}\pa(\th^\b\th_\b)
\label{changevarII}
\ena
and analogous for $\bar q_\a$ and $\bar d_\ad$.
In the first line the reader will recognize the change of variables (\ref{changevar}) introduced in section 1.1.7 to obtain Seiberg's deformed superspace from my $N=2$ non(anti)commutative superspace. This transformation was first introduced by Vafa and Ooguri in \cite{vafaooguri2}.
In the second and third lines I have written the worldsheet versions of the chiral supersymmetry charges and superspace covariant derivatives,
\bea
&&Q_\a=\frac{\pa}{\pa \th^\a}~~~~;~~~~\bar Q_\a =\frac{\pa}{\pa \thb^\a}\cr
&&D_\ad =-\frac{\pa}{\pa\th^\ad}~~~~;~~~~\bar D_\ad= - \frac{\pa}{\pa\thb^\ad}
\ena
($Q_\ad$, $ \bar Q_\ad$, $D_\a$ and $\bar D_\a$ are dressed in this representation).
So, $q$ and $d$ represent the conjugate momenta to $\th$'s at fixed $y$ exactly as $p$'s represent the same conjugate momenta at fixed $x$. 

The lagrangian (\ref{freeaction}) can be rewritten in terms of the new variables as follows
\beq
{\cal L}=\frac{1}{2} \pa y^{\a\ad} \bar\pa y_{\a\ad} - q_\a \bar\pa \th^\a + d_\ad \bar\pa \th^\ad -\bar q_\a \pa \thb^\a + \bar d_\ad \pa \thb^\ad + {\rm total~ derivative}
\label{freeactionnew}
\eeq
In the presence of D-branes, one obtains the fermionic boundary conditions
\bea
&&\th^\a = \bar\th^\a~~~~;~~~~q_\a =\bar q_\a\cr
&&\th^\ad=\thb^\ad~~~~;~~~~d_\ad=\bar d_\ad
\ena
that only preserve half of the supersymmetry, generated by the charges $Q_\a+\bar Q_\a$ and $Q_\ad +\bar Q_\ad$.

It is possible to couple the action to the R-R background described by the selfdual graviphoton field strength 
\beq
F^{\a\b}\neq 0;\qquad\qquad\qquad F^{\ad\bd}=0
\label{selfdualback}
\eeq
Only in euclidean signature it is possible to turn on the selfdual part of the super-two-form field strength $F^{\a\b}$ while setting the antiselfdual part $F^{\ad\bd}$ to zero. This is the stringy counterpart of the discussion we presented in section 1.1.5  for nonanticommutative superspaces and works exactly the same way.

It is possible to show that the background (\ref{selfdualback}) is an exact solution of the full nonlinear string equations of motion and that there is no backreaction to the metric. This can be seen from the fact that a purely selfdual field strength does not contribute to the energy momentum tensor and does not involve the dilaton field in its kinetic term. 

In the action the graviphoton field strength couples to the worldsheet supersymmetry currents $q_\a$ as follows
\beq
\int F^{\a\b}q_\a \bar q_\b
\eeq
This is actually the graviphoton integrated vertex operator \footnote{In chapter 3 I will give a detailed derivation of the corresponding vertex operator in the ten-dimensional case.}, that in this case gives the whole dynamics since there is no backreaction. 

We are interested in the effect of the background on the dynamics, so we can concentrate on the $(q,\bar q)$ sector, with the lagrangian
\beq
{\cal L}=\frac{1}{\a'}\left(- q_\a \bar\pa\th^\a -\bar q_\a \pa \thb^\a +\a' F^{\a\b} q_\a \bar q_\b\right)
\label{Leff}
\eeq
We can integrate out the fields $q_\a$ and $\bar q_\a$ by using their equations of motion
\bea
&&\bar \pa \th^\a =\a' F^{\a\b}\bar q_\b\cr
&&\pa \thb^\a = -\a' F^{\a\b} q_\b
\label{eomq}
\ena
and obtain the effective lagrangian
\beq
{\cal L}_{\rm eff}= \left(\frac{1}{{\a'}^2 F}\right)_{\a\b} \pa\thb^\a \bar\pa\th^\b
\label{Leffective}
\eeq
We obtain boundary conditions for both fermionic variables and their derivatives
\bea
\th^\a &=& \thb^\a\cr
\pa \thb^\a &=& -\bar\pa \th^\a
\ena
The first condition breaks half of the supersymmetry on the boundary. The second one corresponds to the equality of supersymmetry charges $q_\a=\bar q_\a$ on the boundary (see the equations of motion \ref{eomq}). The boundary conditions for derivatives of $\th$ has first appeared in \cite{ulfmartinpeter} for the GS superstring. In that case they were additional, unexpected conditions, required by consistency of boundary conditions under kappa-symmetry.  They were shown not to overconstrain the system, because they arise as restrictions of the field equations to the boundary. In this case instead the effective action (\ref{Leff}), obtained by integrating out the fermionic conjugate momenta, is second order for the fermions. A boundary condition for derivatives of $\th$ naturally arises from requiring that there are no surface terms in the Euler-Lagrange equations of motion.

One can determine the fermionic propagators
\bea
&&\langle\th^\a (z)\th^\b (w)\rangle=\frac{{\a'}^2 F^{\a\b}}{2\pi i}\log\frac{\bar z-w}{z-\bar w}\cr
&&\langle\thb^\a (z)\thb^\b (w)\rangle=\frac{{\a'}^2 F^{\a\b}}{2\pi i}\log\frac{\bar z-w}{z-\bar w}\cr
&&\langle\th^\a (z)\thb^\b (w)\rangle=\frac{{\a'}^2 F^{\a\b}}{2\pi i}\log\frac{(z-w)(\bar z-\bar w)}{(z-\bar w)^2}
\ena
On the boundary we get
\beq
\langle\th^\a (\tau)\th^\b(\tau')\rangle=\frac{{\a'}^2 F^{\a\b}}{2} \e(\tau-\tau')
\eeq
corresponding to the algebra
\beq
\{\th^\a,\th^\b\}={\a'}^2 F^{\a\b}
\label{nacstring}
\eeq
Since the coordinates $y$ and $\thb$ are not affected by the background coupling, they remain commuting. This means that the algebra written in terms of the original coordinate $x$ involves nontrivial  terms in $[x,x]$ and $[x,\th]$, as required by consistency and first shown in my paper \cite{mypaper1}.

In \cite{vafaooguri2} the deformation (\ref{nacstring}) was not welcome. It has been shown that when the open superstring is also coupled to a constant gluino superfield on the boundary, by adding the following term to the action
\beq
\oint W^\a q_\a
\eeq
the superspace deformation (\ref{nacstring}) is undone and supersymmetry is restored if the gluino fields satisfy the deformed algebra
\beq
\{W_\a,W_\b\}=F^{\a\b}
\eeq
Instead, in \cite{seiberg} the supersymmetry-deforming algebra (\ref{nacstring}) was accepted and it was shown that in the zero-slope limit the $N=\frac{1}{2}$ theories we discussed in section 1.1.7 naturally emerge in string theory.

The analysis pursued in \cite{seiberg} was further developed in \cite{seibergberkovits}. In this paper it was shown that the constant selfdual background deforms the original $N=2$ superPoincar\'e algebra into another algebra that has still eight supercharges, four of which are unaffected by the background. In the presence of a D-brane, $N=\frac{1}{2}$ supersymmetry is realized linearly and the remaining $N=\frac{3}{2}$ is realized nonlinearly. This interpretation of the new terms arising in the supersymmetry algebra is similar to the original one we gave in \cite{mypaper1}. There we didn't consider the new terms arising in the supersymmetry algebra as symmetry-breaking terms. We considered them as symmetry-deforming terms and we recast them in the known form of a q-deformation.  

It is very interesting to note that if both selfdual and antiselfdual field strength are turned on, there is a backreaction that warps spacetime to euclidean $AdS_2\times S^2$. The string in this background has been studied in \cite{berkovitsmore} and the structure of the action closely resembles the pure spinor version of the superstring in $AdS_5\times S^5$ \cite{berkovitsRR}. The action becomes quadratic in the limit $F^{\ad\bd}\rightarrow 0$ resembling the Penrose limit in the ten-dimensional case. $N=\frac{1}{2}$ super Yang-Mills on euclidean $AdS_2\times S^2$ has been studied in \cite{immanpur2}.

A comparison to noncommutativity in the bosonic case is due. First of all, we notice that the superspace deformation (\ref{nacstring}) vanishes in the zero slope limit $\a'\rightarrow 0$ unless we also take the limit $F^{\a\b}\rightarrow \infty$. This in principle can be done since $F^{\a\b}$ is an exact solution to the string equations.
Moreover, it is interesting to note that, in contradistinction to the bosonic case, where the $B_{\m\n}$ term only affects the boundary conditions, the $F^{\a\b}$ term only affects the bulk equations of motion.
It would be nice to see if there is a duality transformation connecting the two cases \footnote{In \cite{cornalbaschiappa2} S-duality was considered in the context of noncommutative geometry in the presence of both NS-NS  B-field and R-R {\it potentials}.}. 

The four-dimensional results in \cite{vafaooguri2,seiberg} have been generalized to ten dimensions in \cite{antonio}. The main difference between the compactified four dimensional case and the full ten dimensional case is that a constant R-R field strength background, represented by a bispinor $P^{\a\hat\a}$, is not a solution of the ten-dimensional string equations of motion (full nonlinear type II SUGRA equations), but it is only a solution of the linearized equations. Therefore in this case there is a backreaction.  Nevertheless, it is possible to compute the corresponding vertex operator to be added to the action. The resulting theory is not the complete sigma model.
In \cite{antonio} it was considered the general case where a NS-NS constant $B^{\m\n}$, a constant R-R field strength $P^{\a\hat\a}$ and constant gravitinos $\Psi_m^\a$ $\Psi_m^{\hat\a}$ are turned on.
The resulting algebra is characterized by the usual nontrivial bosonic commutator induced by $B$, a nontrivial anticommutator between $\th^\a$ and $\th^{\hat\a}$ induced by the R-R field strength and nontrivial commutators between the bosonic coordinate and the two fermionic ones induced by the two gravitinos. 
It would be nice to perform an analogous analysis in the pure spinor version of the superstring in a p-p wave background \cite{berkovitspp}. In this case the whole sigma model action is known that involves a constant R-R field strength coupled to fermionic variables in a similar way with respect to (\ref{Leffective}).

After the superspace deformation of a theory is known, one can integrate out the fermionic variables to obtain the component formulation of the theory. It has been shown in \cite{torinesi} that the component formulation of $N=\frac{1}{2}$ super Yang-Mills theory can be obtained from standard RNS type IIB string theory compactified of a Calabi-Yau three-fold in the presence of a constant graviphoton background with a definite duality. A graviphoton background can be obtained in euclidean space by wrapping a R-R 5-form around a 3-cycle of internal Calabi-Yau space.
Even if in the general case the RNS is not suitable to deal with a R-R background, in the constant graviphoton field strength case there are simplifications that allow for the computation of tree-level scattering amplitudes on the disk with the insertion of R-R vertex operators. The method used in \cite{torinesi} is intrinsically perturbative, but the results are exact in the $\a'\rightarrow 0$ limit. So the nonanticommutative $N=\frac{1}{2}$ super Yang-Mills action in its component formulation is recovered from RNS string computations.

\section{Generalization to non-constant backgrounds}
In this section I would like to discuss some results generalizing the connection between string theory and noncommutative geometry to the case of a nonconstant $B$ field.
Let us introduce the notation $\o=F+B$ for the gauge invariant combination in terms of which the physics of an open string in the presence of Dp-branes, gauge field $A^{i}$ and background $B^{ij}$ field is described. The latter will not be constrained to be constant from now on.

We have to consider three possible situations, of growing complexity
\begin{itemize}
\item{constant $\o$}
\item{nonconstant $\o$, $d\o\equiv H=0$}
\item{nonconstant $\o$, $d\o\equiv H\not=0$}
\end{itemize}
We have discussed the first case, corresponding to flat D-brane and flat background, in sections 1.2.1, 1.2.2, 1.2.3 . Summarizing the results in \cite{seibwit}, in this case the Dp-brane worldvolume is described in terms of noncommuting coordinates. $\o$ defines an associative symplectic structure associated to the noncommutativity matrix $\th$. In the limit $\a'\rightarrow 0$ physics can be described by an effective theory which is a gauge theory deformed by the noncommutative associative Moyal product. In the limit $\a'\rightarrow 0$ the noncommutative parameter $\th$ is given by the inverse of $\o$ (see \ref{paramalpha}).

The second case has been studied in \cite{cattaneo}. It describes a physical situation with curved Dp-branes in a flat background. The worldvolume deformation is decribed by Kontsevich $\star$ product, where the noncommutativity parameter is given by the inverse of $\o$, as before. In this case $\o$ is not constant anymore, so neither is $\th$ and the correct product is Kontsevich $\star$. $\o^{-1} $ is still a Poisson structure on the manifold, that now exhibits a coordinate dependence such that associativity is satisfied. The formula (\ref{correlatori}), valid in the limit $\a'\rightarrow 0$, is generalized to
\EQ
\langle \prod_{i=1}^n f_i\left(x(\t_i)\right)\rangle=\int V(\o)dx~f_1\star...\star f_n
\label{corr}
\EN
being $\star$ Kontsevich associative product.

The last case, discussed in \cite{cornalba}, is a further generalization to the case where the background is also curved. $\o$ is not a Poisson structure anymore, since associativity is lost. This is related to the emergence of a second geometrical object playing a role in the physics of the system. This is the 3-form  $H=d\o$, that has been shown to be the parameter governing nonassociativity. However, it is always possible to give a description of the Dp-brane worldvolume in terms of a Kontsevich-like product. Noncommutativity is still governed by $\o^{-1}$ and nonassociativity by  $H$. 

In \cite{cornalba} a metric $g^{\m\n}$ is considered that is a small perturbation from the flat metric of section 1.2.1. The string action in a generic curved background is still of the form (\ref{string}), where the target space metric is $g^{\m\n}(x)$, describing a curved spacetime. It is possible to expand the action around the flat spacetime metric as follows
\EQ
S=S_0+S_1+...
\label{appross}
\EN
where $S_0$ represents the action in a flat spacetime (\ref{string}) and $S_1$ is the first correction due to the presence of a small curvature.

In situations where a curved background is present, it is convenient to choose special coordinates, known as Riemann normal coordinates, defined along geodesics in target space starting from $x^{\m}=0$  (see for instance \cite{hullnormal}, where the normal coordinate expansion of the metric is introduced in the context of sigma models and string theory). In terms of these coordinates the Taylor expansion of every tensor around $x^{\m}=0$ is expressed in terms of covariant tensors evaluated at the origin.
In particular, the expansion for the metric up to second order in $x$ is given by
\EQ
g_{\m\n}(x)=g_{\m\n}-\frac{1}{3}R_{\m\r\n\s}x^{\r}x^{\s}+\mathcal{O}(x^3)
\EN
where $R_{\m\n\r\s}$ is the curvature tensor.
The analogous expansion for the $B$ field, in radial gauge, is
\EQ
B_{ij}(x)=B_{ij}+\frac{1}{3}H_{ijk}x^k+\frac{1}{4}\nabla_{l}H_{ijk}x^k x^l+\mathcal{O}(x^3)
\label{svib}
\EN
with $H=dB$.

In \cite{cornalba} a first order approximation in $x$ is applied. The action considered is  
\EQ
S=S_0+S_1+S_B
\EN
where $S_0+S_B$ is the action (\ref{string}) for flat $g$ and constant $B$, while $S_1$ is the first order correction in $x$, deriving from the $B$ expansion (\ref{svib})
\EQ
S_1=-\frac{i}{6}\int_{\S}H_{ijk}x^k\e^{ab}\pa_a x^i \pa_b x^j
\EN
$S_1$ is treated as an interaction term with respect to the free theory, described by the action $S_0$.

The perturbative analysis with the interaction $S_1$ gives the following results.
In the limit $\a'\rightarrow 0$ correlators are expressed in a form analogous to (\ref{corr})
\EQ
\langle \prod_{i=1}^n f_i\left(x(\t_i)\right)\rangle=\int V(\o)~d^{p+1}x~(f_1\bullet...\bullet f_n)
\label{nacorr}
\EN
where $\bullet$ is Kontsevich-like nonassociative product,  whose definition is completely analogous to (\ref{kont}, \ref{kont2}), with $P(x)=\o^{-1}(x)$. The relation (\ref{assoc}) is not valid, though, for the noncommutativity parameter $P=\o^{-1}$ and the violation of associativity is proportional to the 3-form $H$ as follows
\EQ
(f \bullet g)\bullet h-f\bullet (g\bullet h)=\frac{1}{6}P^{im}P^{jn}P^{kl}H_{mnl}\pa_i f \pa_j g \pa_k h+...
\EN
Note that (\ref{nacorr}) is problematic, because of the nonassociativity of $\bullet$. It is necessary to specify the positions of the points $\t_i$ on the boundary of the disk, so invariance under cyclic permutations is lost.

The coordinate algebra is given by
\EQ
[x^i, x^j]_{\bullet}=iP^{ij}(x)
\EN
showing that the role played by the two-form $\o$ is unchanged in the general curved case.
The relations concerning the nonassociativity of the coordinate algebra are instead \EQ  
\left(x^i\bullet x^j\right)\bullet x^k-x^i\bullet \left(x^j\bullet x^k\right)=\frac{1}{6}P^{im}P^{jn}P^{kl}H_{mnl}
\EN
\EQ
[x^i,[x^j,x^k]_{\bullet}]_{\bullet}+[x^k,[x^i,x^j]_{\bullet}]_{\bullet}+[x^j,[x^k,x^i]_{\bullet}]_{\bullet}=-P^{im}P^{jn}P^{kl}H_{mnl}
\EN
These formulas further clarify the role played by the 3-form $H$ as the parameter governing nonassociativity.

The deep relationship between spacetime geometry and nonassociativity on the D-brane worldvolume discovered in \cite{cornalba} is very interesting, showing that nonassociative geometry \cite{nonassoc} also plays a role in the new developments regarding spacetime pioneered by string theory.

It would be nice to generalize the results discussed in this section to the manifestly target space supersymmetric string (in Green-Schwarz or Berkovits formalism) to see whether a Kontsevich-like product similar to the one I proposed in \cite{mypaper1}, in collaboration with D. Klemm and S. Penati, may emerge in the presence of a nontrivial super-three-form field strength background .

It would be also interesting to consider a generalization of the discussion presented in section 1.2.4, where the R-R field strength is not constant. In particular, when the R-R field strength has a linear dependence on the bosonic coordinates, it would be nice to investigate whether Lie algebraic deformed superspaces, characterized by the fermionic anticommutators
\beq
\{\th^\a,\th^\b\}=\g_\m^{\a\b}x^\m
\label{substr}
\eeq
can appear. These kind of superspace was first studied in \cite{Schwarz:pf}, in a different context. There, the authors considered the possibility that bosonic spacetime had a fermionic substructure, given by the relation (\ref{substr}). In \cite{mypaper4}, in collaboration with P. A. Grassi, I have started to consider this problem by computing the vertex operator for the Berkovits covariant superstring for a R-R field strength with a linear dependence on the bosonic coordinates. This will be discussed in detail in chapter 3.


\chapter{Noncommutative deformation of integrable field theories}
\section[A brief introduction to selected topics concerning two-dimensional\\ classical integrable systems]{A brief introduction to selected topics concerning two-dimensional classical integrable systems}
{\bf Disclaimer:} This section is not a thorough introduction to the vast field of integrable systems. I will only review topics needed in the two following sections, where my study of the noncommutative sine-Gordon system will be presented. In most cases I will not give details and I will just give references for the interested reader. Moreover, for any single topic in this section I will exhibit a single example, the ordinary sine-Gordon model.
\subsection{Infinite conserved currents and the bicomplex approach}
In classical mechanics a system described by $n$ degrees of freedom is completely integrable when it is endowed with $n$ conserved currents. In classical field theory, a system with an infinite number of local conserved currents is also said to be integrable. This is a property of the equations of motion. For some integrable system an action is also known that generates the equations by an Euler-Lagrange procedure, but this is not true in general. Indeed, it is true for the sine-Gordon model that I will consider in the rest of this chapter.

By making use of the bicomplex technique \cite{bicomplex} it is possible to construct second order differential equations that are integrable. Moreover, with this approach it is very easy to generate the corresponding conserved currents by an iterative procedure.

In this section we use euclidean signature and complex coordinates 
\beq
z =\frac{1}{\sqrt{2}} (x^0 +
ix^1)~~~~;~~~~
\bar{z} =\frac{1}{\sqrt{2}} (x^0 - ix^1)
\eeq
A bicomplex is a triple $({\cal M}, d, \d)$ where ${\cal M}=\otimes_{r\ge 0}{\cal M}^r$ is an $N_0$-graded associative (but not necessarily commutative) algebra, ${\cal M}^0$ is the algebra of functions on $\R^2$ and $d,\d:{\cal M}^r\rightarrow {\cal M}^{r+1}$ are two linear maps satisfying the conditions $d^2=\d^2=\{d,\d\}=0$. ${\cal M}^r$ is therefore a space of $r$-forms.
Let us consider the linear equation
\beq
\d \xi=l d\xi
\label{linsis}
\eeq 
where $l$ is a real parameter and $\xi \in {\cal M}^s$ for a given ``spin'' $s$. 
Suppose a nontrivial solution $\tilde \xi$ exists. Expanding it in powers of the given parameter $l$ as 
\beq
\tilde\xi=\sum_{i=0}^\infty l^i \xi^{(i)}
\eeq 
one obtains the following equations relating the components $\xi^{(i)}\in {\cal M}^s$
\beq
\d\xi^{(0)}=0~~~~;~~~~\d\xi^{(i)}=d \xi^{(i-1)}~~~~,~~~~i\ge1
\label{claws}
\eeq
Therefore we obtain the chain of $\d$-closed and $\d$-exact forms
\beq
\Xi^{(i+1)}\equiv d\xi^{(i)}=\d\xi^{(i+1)}~~~~,~~~~i\ge 0
\eeq
For the chain not to be trivial $\xi^{(0)}$ must not be $\d$-exact.
When the two differential maps $d$ and $\d$ are defined in terms of ordinary derivatives in $\R^2$, the conditions $d^2=\d^2=\{d,\d\}=0$ are trivially satisfied. Therefore, the possibly infinite set of conservation laws (\ref{claws}) is not associated to any second order differential equation and it is not useful for our purpose. 

However, it is possible to gauge the bicomplex by dressing the two differential maps $d$ and $\d$ with the connections $A$ and $B$ as follows
\beq
D_d=d+ A~~~~;~~~~D_\d=\d + B
\eeq 
The flatness conditions 
$D^2_d=D^2_\d=\{D_d,D_\d\}=0$
are now nontrivial and give the differential equations
\bea
&&{\cal F}(A)\equiv dA +A^2=0\cr
&& {\cal F}(B)\equiv\d B + B^2 =0\cr
&& {\cal G}(A,B)\equiv d B+\d A +\{A,B\}=0 
\label{nonlin}
\ena
Exactly as before we can consider the linear equation corresponding to (\ref{linsis})
\beq
{\cal D}\xi \equiv (D_\d - l D_d)\xi=0
\label{lin2}
\eeq
The nonlinear equations (\ref{nonlin}) are the compatibility conditions for (\ref{lin2})
\beq
0={\cal D}^2 \xi = \left[{\cal F}(B) + l^2 {\cal F}(A) - l {\cal G}(A,B)\right]\xi
\eeq
Supposing that the linear equation (\ref{lin2}) admits a solution $\tilde\xi\in {\cal M}^s$ and expanding it as $\tilde\xi=\sum_{i=0}^\infty l^i\xi^{(i)}$, one obtains the possibly infinite chain of identities 
\beq
D_\d \xi^{(0)}=0~~~~;~~~~D_\d \xi^{(i)}=D_d \xi^{(i-1)}~~~~,~~~~i\ge 1
\label{chain}
\eeq
from which $D_\d$-closed and $D_\d$-exact forms $\Xi^{(i)}$ can be constructed, when $\xi^{(0)}$ is not $\d$-exact.
Therefore, with a suitable choice of $A$ and $B$, it is possible to construct interesting second order integrable differential equations from (\ref{nonlin}) and their conserved currents from  (\ref{chain}).
In general the currents $\xi^{(i)}$are nonlocal functions of the coordinates, since they may be expressed in terms of integrals, but it is possible to extract local currents from them that have a physical meaning. 

As an example, we can derive the ordinary sine-Gordon equation from this formalism.
Let ${\cal M}={\cal M}_0 \otimes L$, where ${\cal M}_0$ is the space of $2\times 2$ matrices with entries in the algebra of smooth functions on ordinary $\R^2$ and $L=\otimes_{i=0}^{2} L^i$ is a two-dimensional graded vector space with the $L^1$ basis $(\tau,\s)$ satisfying $\tau^2=\s^2=\{\tau,\s\}=0$.
If we take the differential maps
\beq
\d f=\pab f\tau -Rf\s~~~~~;~~~~~df=-Sf\tau + \pa f \s
\eeq
with commuting constant matrices $R$ and $S$, then the conditions $d^2=\d^2=\{d,\d\}=0$ are trivially satisfied. 

To obtain nontrivial second order differential equations we can gauge the bicomplex by dressing $d$ as follows
\beq
D_d f\equiv G^{-1} d(Gf) 
\label{matrixsgo}
\eeq
with $G$ generic invertible matrix in ${\cal M}_0$. The condition $D_d ^2=0$ is trivially satisfied, while $\{\d,D_d\}=0$ yields the nontrivial second order differential equation 
\beq
\pab\left(G^{-1}\pa G\right)=\left[R,G^{-1} S G\right]
\eeq
With the choice
\beq
R \= S \= 2\sqrt \g\, \Bigl(\begin{matrix} 0 & 0 \\ 0 & 1 \end{matrix}\Bigr) 
\eeq
and taking $G\in SU(2)$ as follows
\beq
G \= e^{\frac{\im}{2} \s_2 \Phi} \= \biggl( \begin{matrix} 
\phantom{-}\cos{\frac{\Phi}{2}} & \ \sin{\frac{\Phi}{2}} 
\\[4pt]
-\sin{\frac{\Phi}{2}} & \ \cos{\frac{\Phi}{2}} 
\end{matrix} \biggr)
\eeq
we obtain the sine-Gordon equation from the off-diagonal part of the matrix equation (\ref{matrixsgo})
\beq
\pab\pa\Phi=4\g \sin\Phi=0
\label{sinegordon}
\eeq
The diagonal part instead gives an equation which is trivially satisfied. 

From this derivation it is clear that the bicomplex approach can be naturally extended to noncommutative space, by replacing ordinary products with Moyal products in the whole discussion. In particular $D_d=d+A\ast$ and $D_\d=\d +B\ast$ and equations (\ref{nonlin}) are generalized accordingly. This will be discussed for the sine-Gordon model in section 2.2.4. 

Since in the noncommutative case the group $SU(2)$ is not closed any more and must be extended to $U(2)$, as we have seen in section 1.1.2, it is natural to expect that the noncommutative generalization of this construction for  the sine-Gordon equation will be nontrivial. 
\subsection{Reductions from selfdual Yang-Mills}
The self-duality equations for Yang-Mills fields in $\R^4$ with signature $(++++)$ or $(++--)$ \cite{Yang} 
\bea
&&\frac{1}{2}\e_{\m\n\rho\s}F^{\rho\s}=F_{\m\n}\cr
&&F_{\m\n}=\pa_\m A_\n - \pa_\n A_\m +[A_\m,A_\n]
\label{sdeq}
\ena
are a famous example of nonlinear integrable equations in four dimensions.
For $SU(n)$ gauge theory the potentials $A_\m^a$ are real. It is possible to consider an analytic continuation of $A_\m^a$ into complex space parametrized by the
complex coordinates $y$, $\bar y$, $z$, $\bar z$. The selfduality equations (\ref{sdeq}) can be rewritten in the form
\beq
F_{yz}=F_{\bar y \bar z}=0~~~~;~~~~F_{y \bar y}\pm F_{z \bar z}=0
\label{complexsd}
\eeq
where the last sign is $+$ is the euclidean case and $-$ in the kleinian case.
The first equation in (\ref{complexsd}) implies that the potentials $A_y$, $A_z$ ($A_{\bar y}$, $A_{\bar z}$) are pure gauges for fixed $\bar y$, $\bar z$ ($y$, $z$), therefore two $n\times n$ complex matrices $B$ and $\bar B$ can be found such that 
\bea
&&A_y=B^{-1}\pa_y B~~~~;~~~~A_z=B^{-1}\pa_z B\cr
&&A_{\bar y}={\bar B}^{-1}\pa_{\bar y} {\bar B}~~~~;~~~~A_{\bar z}={\bar B}^{-1}\pa_{\bar z} {\bar B}
\ena
Defining $J=B\bar B^{-1}\in SL(n,\C)$, the last equation can be rewritten as
\beq
\pa_{\bar y}(J^{-1}\pa_y J)\pm\pa_{\bar z}(J^{-1}\pa_z J)=0
\label{sdymyang}
\eeq
describing selfdual Yang-Mills in Yang formulation.
This equation resembles the sum of two WZW model equations (see section 1.1.3) involving $(y,\bar y)$ and $(z,\bar z)$ variables, respectively.
Therefore Yang equation (\ref{sdymyang}) can be obtained from the following action
\bea
S&=&\int d^2 y d^2 z ~{\rm tr}(\pa_y J \pa_{\bar y} J^{-1}) - \int d^2y d^2 z\int_0^1 d\rho ~{\rm tr} \left(\hat J^{-1}\pa_\rho \hat J[\hat J^{-1}\pa_{\bar y} \hat J,\hat J^{-1}\pa_y \hat J]\right)\cr
&+&\int d^2 y d^2 z~ {\rm tr}(\pa_z J \pa_{\bar z} J^{-1}) - \int d^2y d^2 z\int_0^1 d\rho ~{\rm tr} \left(\hat J^{-1}\pa_\rho \hat J[\hat J^{-1}\pa_{\bar z} \hat J, \hat J^{-1}\pa_z \hat J]\right)\cr
&&~~~~~~~~~
\ena
where $\hat J(y,\bar y, z, \bar z, \rho)$ is a homotopy path satisfying $\hat J(\rho=0)=1$ and $\hat J(\rho=1)=J$.

Ward conjectured that all integrable equations in $d=2$ can be obtained as dimensional reductions of selfdual Yang-Mills equations \cite{ward}, so that the latter play the role of a universal integrable system. Since then the conjecture has been tested on many integrable systems \cite{SDYMred} whose Lax pair can also be obtained by reduction from the one associated to the selfduality equations (\ref{sdeq}). The kleinian case is particularly interesting because of its connections with the $N=2$ string, discussed in section 1.2.2.

Reductions are obtained by requiring invariance under any arbitrary subgroup $G$ of the group of conformal transformations of $\R^{(2,2)}$ (or $\R^{(4,0)}$). Afterwards algebraic constraints can be applied to the arbitrary matrices involved in the equations to obtain known integrable models.
In most cases invariance under translations in certain directions is required. 
Therefore it is clear that there are many more possibilities in reducing selfdual Yang-Mills in $\R^{(2,2)}$ with respect to the euclidean case, since, instead of requiring invariance under the usual complex coordinates
\bea
&&\sqrt 2 y=x^1+ix^2~~~~;~~~~\sqrt 2\bar y=x^1-ix^2\cr
&&\sqrt 2z=x^3+i x^4~~~~;~~~~\sqrt 2\bar z= x^3 -i x^4
\ena
(combining space with space and time with time),
in the kleinian case one can also require invariance with respect to light-cone coordinates
\bea
&&s=\frac{1}{2}\left(x^2- x^4\right)~~~~;~~~~t=\frac{1}{2}\left(x^2+x^4\right)\cr
&&u=\frac{1}{2}\left(x^1- x^3\right)~~~~;~~~~v=\frac{1}{2}\left(x^1+ x^3\right)
\ena
(combining space and time).

As an example I will show how ordinary euclidean sine-Gordon model can be obtained from selfdual Yang-Mills equations.

The euclidean version of Yang equation (\ref{sdymyang}) gives the sine-Gordon equation if one choses 
\beq
B=e^{\frac{z}{2}\s_1} e^{i\frac{\Phi}{2}\s_3}~~~~;~~~~\bar B=e^{\frac{\bar z}{2}\s_1}
\eeq
where $\Phi=\Phi(y,\bar y)$. In fact one finds that the field $\Phi$ satisfies (\ref{sinegordon}) with $4\g=-1$.

It will be also useful to know that the sine-Gordon equation can be obtained from kleinian selfdual Yang-Mills equations through a two-step reduction procedure. First of all one requires independence of Yang equation (\ref{sdymyang}) under one of the real coordinates $x^i$, reducing to the $2+1$ model
\beq
(\eta^{\m\n}+ V_\a \e^{\a\m\n})\pa_\m (J^{-1}\pa_\n J)=0
\eeq
where $V_\a$ is a constant vector in spacetime. A nonzero $V_\a$ breaks Lorentz invariance but restores integrability when it is spacelike and with unit length (nonlinear sigma models in $2+1$ dimensions can be Lorentz invariant or integrable but not both \cite{ward}). From this equation, in the case $V_\a=(0,1,0)$, we can make a further reduction \cite{leese} by choosing
\beq
J=\left(\begin{matrix}\cos \frac{\Phi}{2} & e^{-\frac{i}{2}x} \sin\frac{\Phi}{2}\\
 -e^{\frac{i}{2}x} \sin\frac{\Phi}{2} & \cos \frac{\Phi}{2} 
\end{matrix}\right)\in SU(2)
\eeq
with $\Phi$ depending on only two coordinates with different signature.
As a result, we obtain the sine Gordon equation in $1+1$ dimensions for the field $\Phi$. \subsection{Properties of the S-matrix}
In section 2.1.1 we have seen how to construct nonlinear differential equations in two-dimensions with the property of having an infinite number of conserved currents. We have also said that from that construction it is possible to extract conserved currents that are local and thus have a physical meaning. In this case the corresponding equations are called integrable.
Moreover, we have seen an example of two-dimensional integrable system, the ordinary sine-Gordon equation. 

The S-matrix of a two-dimensional theory with an infinite set of conserved currents that are local and yield conserved charges which are components of Lorentz tensors of increasing rank enjoys several nice properties.
\begin{itemize} 
\item{The general multiparticle S-matrix is elastic, i.e. the number and set of momenta of particles of any given mass remains the same before and after the collision.}
\item{Any multiparticle S-matrix factorizes into a product of two-particle S-matrices.}
\item{These two-particle S-matrices satisfy a cubic equation that in most cases is sufficient to obtain exact expressions for them (unitarity must be used, though!). The sine-Gordon model is one of these cases.} 
\end{itemize}
These are of course very nice properties. Indeed, the task of computing the general S-matrix considerably simplifies, since it reduces to determining only the two-body S-matrix.

I'm not going to give a proof of this theorem. The interested reader can for instance refer to \cite{rajaraman}. The key ingredient in the proof is the sufficient complexity of the conserved charges (i.e. the growing Lorentz rank). However, locality is strongly used and its is unclear whether the kind of nonlocality introduced by the $\ast$ product can be a problem. As we discussed in section 1.1.2, unitarity and causality problems arise in theories with noncommutating time, such as 1+1 theories, together with a general breakdown of the quantum mechanics framework. 

In my paper \cite{mypaper2}, an explicit example of a noncommutative two dimensional theory with an infinite number of conserved currents that does not have a factorized S-matrix was constructed. Therefore, it seems that the theorem cannot be trivially extended to noncommutative case. However, in my paper \cite{mypaper3}, a noncommutative two-dimensional system which is integrable and has a factorized S-matrix was constructed. It might be that a nontrivial interplay between integrability and causality drives a system to exhibit or not a factorized S-matrix.
These issues will be discussed in the rest of this chapter.

\subsection{Solitons}
The name soliton refers to solutions of nonlinear equations that represent a localized packet travelling without changing shape or velocity and preserving these properties after collision with other packets. 
In complicated equations containing nonlinear and dispersive terms, the existence of this kind of solution is a highly nontrivial property, due to a special balance between the effects of these two kinds of contributions.

A very famous example of a system displaying this kind of classical solutions is the sine-Gordon model. 
It can be shown that this system, described by the equation of motion (\ref{sinegordon}), admits the following static finite-energy solution 
\beq
\Phi\propto \tan^{-1}[\exp(x-x_0)]=\Phi_{\rm sol} (x-x_0)
\eeq
(soliton)
and the corresponding one (antisoliton) obtained by the discrete transformation $\Phi\rightarrow -\Phi$, which is a symmetry of the system.
Moving solutions can be obtained from static ones by Lorentz transformation.
A third kind of solution is present, called doublet or breather solution, which can be interpreted as a bound system made of a soliton-antisoliton pair.
For a detailed derivation of these solutions the reader should refer to \cite{rajaraman}.
It can be shown by studying exact time-dependent solutions representing scattering of solitons that the colliding solitons do not change shape or velocity after collision. From direct inspection of these scattering solutions, representing two (or more) (anti)solitons far apart and approaching with a relative velocity, it is clear that the only effect of the collision in the distant future is some time delay (see \cite{rajaraman}).

Solitons solutions are not present in any scalar field theory with a potential bounded from below in spatial dimension greater than two, as the energy of any field configuration can always be lowered by shrinking. This follows from a simple scaling argument by Derrick \cite{nosoli}.

Let us consider a theory for a single scalar field in D+1 dimensions for simplicity (the discussion can be easily extended to a set of $N$ interacting scalar fields), described by the standard relativistic lagrangian. The corresponding energy functional for static configurations is
\beq
E=\frac{1}{g^2}\int d^D x \left(\frac{1}{2}(\pa\Phi)^2 + V(\Phi)\right)
\label{energyord}
\eeq
Let $\Phi_0(x)$ be an extremum of (\ref{energyord}). Consider the energy of the configuration $\Phi_\l (x)=\Phi_0(\l x)$
\beq
E(\l)=\frac{1}{g^2}\int d^D x \left(\frac{1}{2}\l^{2-D}(\pa\Phi_0(x))^2 +\l^{-D} V(\Phi_0(x))\right)
\eeq
Since we assumed that $\Phi_0(x)$ is an extremum, we require $\frac{\pa E(\l)}{\pa\l}\vert_{\l=1}=0$. This gives the equation
\beq
\frac{1}{g^2}\int d^D x \left(\frac{1}{2}(D-2)(\pa\Phi_0(x))^2 + D V(\Phi_0(x))\right)=0
\eeq
If the potential $V$ is bounded from below by zero and $D\ge 3$, than kinetic and potential terms in this equation must vanish separately and thus no nontrivial space-dependent solutions are admitted. In the case $D=2$ one obtains that $V(\Phi_0(x))=0$. If $V$ has only discrete minima, then also in $D=2$ no time-dependent solutions are allowed. However, when $V$ has a continuos set of minima (in the case with more than one scalar field), possible space-dependent solutions are permitted.

Notice that this proof is not valid when higher derivative terms are present and when the scalar field is described by a nonrelativistic lagrangian. Moreover, only static solutions are excluded, while time-dependent ones are allowed. 

\section[Deforming an integrable field theory: The sine-Gordon model (a\\ first attempt)]{Deforming an integrable field theory: The sine-Gordon model (a first attempt)}
\subsection{Noncommutative solitons}
In this section I will mostly refer to the ICTP lectures by R. Gopakumar \cite{ictpgopak}. To the interested reader, I also suggest the lectures by N. Nekrasov \cite{nikitasol} and by J. Harvey \cite{harvey}. Since the literature concerning noncommutative solitons is vast, I only give a partial list of references in \cite{ncsolitons}. To this, one should add references \cite{olaf4,olaf5,wolf,bieling} where solitons of a specific 2+1 integrable model are studied, which are related the noncommutative sine-Gordon solitons in my work \cite{mypaper3} that will be discussed in section 2.3.5.
\vskip 11pt
A quite universal feature of noncommutative field theories is that they admit classical finite energy soliton  solutions that have no counterpart in local field theories. This novel soliton solutions are more or less insensitive to the details of the specific theory considered, so in this section I will consider the scalar theory in 2+1 dimensions for simplicity, with only spatial noncommutativity. 

In section 2.1.5 we have seen that ordinary scalar theory, with a standard relativistic lagrangian and a potential with a discrete set of minima, does not have any localized solution in spatial dimension greater than one \cite{nosoli} (see section 2.1.5).
In the following we will see that spatial nonlocality induced by noncommutativity allows for the presence of novel localized solutions that vanish in the commutative limit.

Consider the energy functional for static configurations
\beq
E=\frac{1}{g^2}\int d^2z \left(\pa\Phi\pab\Phi + V(\Phi)_\ast\right)
\eeq
where $z$, $\bar z$ are complex coordinates in the two dimensional noncommutative space and $\ast$ is the corresponding Moyal product. As we have seen in section 1.1.1, integrated quadratic terms are unaffected by Moyal product, so only the potential term is modified with respect to the ordinary theory. 
Since we know that for $\th=0$ there are no solitonic solutions, we will first consider the limit $\th\rightarrow \infty$. It is useful to rescale the complex coordinates $z\rightarrow z\sqrt \th$, $\bar z\rightarrow \sqrt \th$, so that the $\ast$ product does not depend explicitly on the deformation parameter and the energy functional becomes
\beq
E=\frac{1}{g^2}\int d^2 z \left(\pa\Phi\pab\Phi +\th V(\Phi)_\ast\right)
\eeq
where all the $\th$ dependence is in front of the potential term.
In the limit $\th\rightarrow \infty$ the kinetic term is negligible with respect to the potential term, at least for localized configurations varying over a size of order one in the rescaled coordinates.
Therefore, we will look for solutions of 
\beq
\left(\frac{\pa V}{\pa\Phi}\right)_\ast=0
\eeq
For instance, in the case of a cubic potential, one has to solve the equation
\beq
m^2 \Phi + b_3 \Phi\ast\Phi=0
\label{ncsoli}
\eeq
In the commutative case this equation would only admit the constant solutions
\beq
\Phi=\l_i
\eeq
where $\l_i$ are the extrema of the function $V(\Phi)$.

Nonlocality introduced by Moyal product allows for more interesting solutions.
Recalling the Weyl-Moyal correspondence we introduced in section 1.1.1, relating functions in a noncommutative algebra to operators in a suitable Hilbert space, we see that functions $\Phi(z,\bar z)$ satisfying $\Phi\ast\Phi=\Phi$ exist and correspond to projectors $P$, $P^2=P$, in the Hilbert space.
Therefore it is clear that $\hat\Phi=\l_i P$ is a solution of (\ref{ncsoli}) when $P$ is a projection operator on some subspace of the Hilbert space and $\l_i$ is an extremum of $V$.
Since integration over coordinates $z$, $\bar z$ corresponds through the Weyl-Moyal correspondence to trace over the Hilbert space, the energy functional in operator language is
\beq
E=\frac{2\pi\th}{g^2} V(\l_i) \rm{tr} P
\eeq
The most general solution to (\ref{ncsoli}) is 
\beq
\hat\Phi=\sum_k a_k P_k
\label{gensol}
\eeq
where the coefficients $a_k$ are chosen among the ordinary constant extrema $\l_i$ of $V(\Phi)$ and $P_k$ are mutually orthogonal projection operators.

To understand the physical meaning of the solutions we have found, we have to go back to configuration space. One finds that the solutions (\ref{gensol}) are radially symmetric in space and with an $r$-dependence given by
\beq
\sum_{n=0}^\infty a_n \phi_n(r^2)
\eeq
where 
\beq
\phi_n(r^2)=2(-1)^n e^{-r^2} L_n(2r^2)
\eeq
and $L_n(x)$ is the n-th Laguerre polynomial. The simplest solution $\phi_0(r)$ is a gaussian.
Moreover, non radially symmetric solutions can be generated by noting that the action in the limit $\th\rightarrow \infty$ has the $U(\infty)$ symmetry $\hat\Phi\rightarrow U\Phi U^\dagger$, where $U$ is a unitary operator acting on the Hilbert space.
It can be easily proven that radially symmetric solutions are stable against small fluctuations when they are constructed around a local minimum configuration $\l_i$ of the potential $V$ and that non radially symmetric are stable too, since $U(\infty)$ rotations do not change the energy. So stable solitons are present when the potential has at least two minima.

The $U(\infty)$ symmetry is broken when $\frac{1}{\th}$ corrections are taken into account (i.e. the kinetic term is not negligible anymore). Most of the infinite solutions we found in the $\th=\infty$ case disappear, but it was found that an interesting finite dimensional moduli space remains. 

Finally, I would like to discuss the connection between solitons in noncommutative field theory and D-branes. Actually, these solitons are the D-branes of string theory manifested in a field theory. Therefore, their study allows for probing stringy features in the more controlled context of field theory. 

An example of a stringy application of our discussion of solitons in scalar noncommutative field theory is in the context of tachyon condensation.
It is well-known that bosonic string theory is unstable because of the presence of a tachyonic scalar field $T$. The effective action for the tachyon field can be obtained by integrating out massive string fields
and is expected to take the form
\beq
S=\frac{C}{g_s}\int d^{26} x \sqrt g \left(\frac{1}{2}f(T) g^{\m\n} \pa_\m T\pa_\n T- V(T)+\dots\right)
\label{tachyon}
\eeq
where higher derivative terms and terms involving massless modes have been neglected.
$V(T)$ is a general potential with an unstable extremum at $T=T_0$ and a minimum chosen to be $V(0)=0$. As we have seen in section 1.2.1, turning on a B-field is equivalent to replacing the closed string metric $g^{\m\n}$ with the effective open string metric $G^{\m\n}$ and ordinary products with Moyal products in (\ref{tachyon}). In the zero-slope limit derivative terms can be neglected. The solitons of the theory obtained in this limit are the noncommutative solitons we studied before.
For instance the gaussian solution $T=T_0\phi_0(r)$ localized in two of the noncommutative directions is a candidate for the $D23$-brane. These solitons display the same instability of the corresponding D-branes, since they correspond to an extremum of $V(T)$ that is a maximum.  

From this brief discussion it should be clear that noncommutative field theories exhibit stringy features, such as D-branes, in the simpler context of a field theory. Therefore, their study can be helpful in the understanding of many string theory issues and tachyon condensation is just one example among these.
\subsection{Noncommutative deformation of integrable field theories}
As we have seen in chapter 1, bosonic noncommutative field theories display a variety of interesting properties but also problematic features, when time is involved in noncommutativity.
In this context, an interesting question is how noncommutativity could 
affect the dynamics of exactly solvable field theories, as for instance 
two--dimensional integrable theories. As we have seen in section 2.1.1, a common feature of these systems is 
that the existence of an infinite chain of {\em local} conserved currents
is guaranteed by the fact that the equations of motion can be 
written as zero curvature conditions for a suitable set of covariant 
derivatives \cite{curvature,curvature2}. In some cases, as for example the ordinary 
sine--Gordon or sigma models, an action is also known which generates
the integrable equations according to an action principle. 

Constructing a consistent noncommutative generalization of a two-di\-men\-sio\-nal theory is a particularly challenging problem though, since, when working with a Minkowski signature, time must be necessarily involved in noncommutativity.

Noncommutative versions of ordinary models are intuitively defined as models 
which reduce to the ordinary ones when the noncommutation parameter 
$\theta$ is removed.
As we discussed in detail for the specific example of the free massless scalar field theory in section 1.1.3, in general noncommutative generalizations are not unique as one can construct 
different noncommutative equations of motion which collapse to the same expression
when $\theta$ goes to zero.
For two dimensional integrable systems, a general criterion
to restrict the number of possible noncommutative versions is to require classical
integrability to survive in noncommutative geometry. This suggests that any noncommutative 
generalization should be performed at the level of equations of motion by
promoting the standard zero curvature techniques. This program has been 
worked out 
for a number of known integrable equations in Refs. \cite{NCEOM,NCbur}.

In chapter 1 we have discussed how noncommutative theories naturally arise in the context of string theory. In particular, in section 1.2.2, we have shown how the open $N=2$ string in the presence of a constant NS-NS background and a stack of $n$ D3-branes can be described, in the zero-slope limit, by $U(n)$ noncommutative selfdual Yang-Mills theory. Tree-level S-matrix computations show that the vanishing of amplitudes beyond three point, which is characteristic of ordinary selfdual Yang-Mills,  is preserved in the noncommutative theory, suggesting that noncommutative selfdual Yang-Mills, as its ordinary counterpart, is endowed with classical integrability. In section 2.1.3, we have seen that many ordinary integrable models in two and three dimensions can be obtained through a dimensional reduction procedure from selfdual Yang-Mills. From all this it is clear that dimensional reduction from noncommutative selfdual Yang-Mills could be another useful technique to generate possibly integrable noncommutative systems in 1+1 and 2+1 dimensions.

It is well known that in integrable commutative field theories there is no 
particle production and the S-matrix factorizes. A priori the same relation between the existence of infinite conserved charges and factorization properties of scattering processes might be lost in the noncommutative case. Nonlocality in time, responsible for acausal behavior of scattering processes and non unitarity, may interfere in a way to spoil these nice scattering properties. On the other hand, one may also hope that classical integrability would alleviate these pathologies arising when time-space noncommutativity is present. In any case it would be nice to construct a noncommutative generalization of a given ordinary integrable theory characterized by a well-defined and factorized S-matrix.

Finally, as we have seen in section 2.1.5,  two-dimensional integrable field theories admit soliton solutions. In the noncommutative case, as we have seen in section 2.2.5, a new kind of soliton appears that vanishes in the commutative limit.
The class of soliton solutions of the noncommutative version of an integrable field theory is expected to display both solitons that reduce to ordinary ones in the commutative limit and new solitons that vanish in the limit.
\subsection[The natural noncommutative generalization of the\\ sine-Gordon model]{The natural noncommutative generalization of the sine-Gordon model}
In my papers \cite{mypaper2, mypaper3}, in collaboration with M.T. Grisaru, O. Lechtenfeld, L. Mazzanti, S. Penati and A.D. Popov, continuing the program initiated by M.T. Grisaru and S. Penati in \cite{us}, I have addressed the problem of generalizing the sine-Gordon theory to noncommutative space. 

The main motivation for this work was the evidence that the natural deformation of this theory, described by the action
\beq
S = \frac{1}{\pi\l^2} \int d^2z \left[ \pa \Phi \pab \Phi -
2 \g ( \cos_\ast{\Phi} - 1) \right]
\label{oaction}
\eeq
with the corresponding equations of motion
\beq
\pa \pab \Phi = \g \sin_{\ast}{\Phi}
\eeq
is affected by some 
problems both at the classical and the quantum level.
 
At the classical level it does not seem to be integrable 
since the ordinary currents promoted to noncommutative currents by replacing the products
with $\ast$--products are not conserved \cite{us}. Moreover, 
we don't know how to find a systematic procedure to construct
conserved currents since the equations of motion cannot be obtained as
zero curvature conditions (a discussion about the lack of integrability for
this system is also given in Ref. \cite{CM}).

Scattering properties of the natural generalization of the sine-Gordon model have been investigated in \cite{CM}. It was found that particle production occurs. The tree level $2 \to 4$ amplitude does not 
vanish.   

At the quantum level the renormalizability properties of the ordinary
model (\ref{oaction}) defined for $\l^2 < 4$ seem to be destroyed by noncommutativity. 
The reason is quite simple 
and can be understood by analyzing the structure of the divergences 
of the NC model compared to the ordinary ones \cite{coleman, amit}.
In the $\l^2 < 4$ regime the only divergences come from multitadpole diagrams.
In the ordinary case the $n$--loop diagram gives a contribution 
$(\log{m^2a^2})^n$ where $a$ and $m$ are the UV and IR cut--offs respectively.
This result is independent of the number $k$ of external fields and of
the external momenta. As a consequence the
total contribution at this order can be resummed as
$\g (\log{m^2a^2})^n ( \cos{\Phi} - 1)$ and the divergence is cancelled
by renormalizing the coupling $\g$. This holds at any order $n$ and the 
model is renormalizable. 

In the noncommutative case the generic vertex from the expansion of $\cos_{\ast}{\Phi}$
brings nontrivial phase factors which depend on the momenta 
coming out of the vertex and on the noncommutativity parameter. 
The final configuration of phase factors associated
to a given diagram depends on the order we use to contract the fields 
in the vertex. Therefore, in the noncommutative case the ordinary $n$--loop diagram 
splits into a planar and a certain number of nonplanar configurations, where
the planar one has a trivial phase factor whereas the nonplanar diagrams
differ by the configuration of the phases (for a general discussion see 
Refs. \cite{filk, tadpole}).
The most general noncommutative multitadpole diagram is built up by combining planar 
parts with nonplanar ones where two or
more tadpoles are intertwined among themselves or with external legs. 
Since the nonplanar subdiagrams are convergent \cite{tadpole, micu, liuba} a 
generic  $n$--loop diagram contributes to the divergences of the theory 
only if it contains a nontrivial planar 
subdiagram. However, different $n$--loop diagrams with different 
configurations of planar and nonplanar parts give divergent contributions
whose coefficients depend on the number $k$ of external fields and
on the external momenta.  
A resummation of the divergences to produce a cosine potential is not 
possible anymore and the renormalization of the couplings of the model is 
not sufficient to make the theory finite at any order. Noncommutativity 
seems to deform the cosine potential at the quantum level and the theory 
loses the renormalizability properties of the corresponding commutative model.

Thus, the ``natural'' generalization of sine--Gordon is not satisfactory 
and one must
look for a different noncommutative generalization compatible with integrability 
and/or renormalizability.
\subsection[A noncommutative version of the sine-Gordon equation\\ with an infinite number of conserved currents]{A noncommutative version of the sine-Gordon equation with an infinite number of conserved currents}
In Ref. \cite{us} M.T. Grisaru and S. Penati constructed a classically integrable noncommutative generalization of the sine-Gordon model, by implementing the bicomplex approach described in section 2.1.1 (as in that section, here we use euclidean signature and complex coordinates 
$z =\frac{1}{\sqrt{2}} (x^0 +
ix^1)$, $\bar{z} =\frac{1}{\sqrt{2}} (x^0 - ix^1)$).

The bicomplex $({\cal M},d,\d)$ is considered, where in this case ${\cal M}={\cal M}_0 \otimes L$, ${\cal M}_0$ is the space of $2\times 2$ matrices with entries in the algebra of smooth functions on noncommutative $\R^2$ and $L=\otimes_{i=0}^{2} L^i$ is a two-dimensional graded vector space with the $L^1$ basis $(\tau,\s)$ satisfying $\tau^2=\s^2=\{\tau,\s\}=0$.
The differential maps are given by
\beq
\d f=\pab f\tau -Rf\s~~~~~;~~~~~df=-Sf\tau + \pa f \s
\eeq
with commuting constant matrices $R$ and $S$. The conditions $d^2=\d^2=\{d,\d\}=0$ are trivially satisfied. 

As we have seen in section 2.1.1, to obtain nontrivial second order differential equations the bicomplex must be gauged. In the derivation of the noncommutative sine-Gordon given in \cite{us} the $d$ operator is dressed as follows
\beq
D_d f\equiv G^{-1}\ast d(G\ast f) 
\label{matrixsg}
\eeq
with $G$ generic invertible matrix in ${\cal M}_0$ to be determined in a way to obtain a generalization of the sine-Gordon. While the condition $D_d ^2=0$ is trivially satisfied, $\{\d,D_d\}=0$ gives rise to nontrivial second order differential equations 
\beq
\pab\left(G^{-1}\ast \pa G\right)=\left[R,G^{-1}\ast \pa G\right]_\ast
\eeq
With the choice
\bea
&&R \= S \= 2\sqrt \g\, \Bigl(\begin{matrix} 0 & 0 \\ 0 & 1 \end{matrix}\Bigr) \cr
&&G \= e_{\ast}^{\frac{\im}{2} \s_2 \Phi} \= \biggl( \begin{matrix} 
\phantom{-}\cos_{\ast}{\frac{\Phi}{2}} & \ \sin_{\ast}{\frac{\Phi}{2}} 
\\[4pt]
-\sin_{\ast}{\frac{\Phi}{2}} & \ \cos_{\ast}{\frac{\Phi}{2}} 
\end{matrix} \biggr)
\ena
equation (\ref{matrixsg}) is a matrix equation in $U(2)$, corresponding to the system of two coupled equations of motion
\bea
&&
2i\pab b \equiv \pab \left( e_{\ast}^{-\frac{i}{2} \Phi}  \ast 
\pa e_{\ast}^{\frac{i}{2} \Phi} 
- e_{\ast}^{\frac{i}{2} \Phi}  \ast \pa e_{\ast}^{-\frac{i}{2} \Phi}\right) 
~=~ i \g \sin_{\ast}{\Phi}
\nonumber \\
&& 
2 \pab a \equiv \pab \left( e_{\ast}^{\frac{i}{2} \Phi}  \ast 
\pa e_{\ast}^{-\frac{i}{2} \Phi} 
+ e_{\ast}^{-\frac{i}{2} \Phi}  \ast \pa e_{\ast}^{\frac{i}{2} \Phi}\right) 
~=~ 0
\label{sg}
\ena

The first equation contains the potential term which is the ``natural''
generalization of the ordinary sine potential, whereas the other one has 
the structure
of a conservation law and can be seen as imposing an extra condition
on the system. In the commutative limit, the first equation reduces to
the ordinary sine--Gordon equation, whereas the second one becomes trivial. 
The equations are in general complex and possess the $Z_2$ symmetry 
of the ordinary sine--Gordon ( invariance under $\Phi \to -\Phi$). 

The reason why integrability seems to require two equations of motions 
can be traced back to the general structure of unitary groups in noncommutative geometry.
In the bicomplex approach the ordinary equations are obtained as 
zero curvature conditions for covariant derivatives defined in terms of
$SU(2)$ gauge connections. If the same procedure is to be implemented
in the noncommutative case, the group $SU(2)$, which is known to be not
closed in noncommutative geometry, has to be extended to a noncommutative
$U(2)$ group and a noncommutative $U(1)$ factor enters necessarily into the game. 
The appearance of the second equation in (\ref{sg}) for this noncommutative integrable 
version
of sine-Gordon is then a consequence of the fact that the fields develop
a nontrivial trace part. We note that the pattern of equations found in \cite{us} seems to be quite general
and unavoidable if integrability is of concern. In fact, the same 
has been found in Ref. \cite{CM} where a different but equivalent set of 
equations was proposed.  

In \cite{us} classical integrability of the system described by the set of equations (\ref{sg}) was proven by extracting from the bicomplex chain a set of conserved currents that are local, in the sense that they are functions of the field $\Phi$ and its derivatives, but not of integrals of $\Phi$. Since Moyal product has an infinite expansion in derivatives of fields, this introduces a kind of nonlocality in the theory that is intrinsic and unavoidable when working in a noncommutative space. 
In \cite{us} the expansion in the noncommutativity parameter $\th$ has been studied for the first currents to check their relation with ordinary currents and to explicitly verify their conservation up to second order in the deformation parameter.
\subsection{Solitons}
The presence of two equations of motion is in principle very restrictive and
one may wonder whether the class of solutions is empty. To show that this
is not the case, in Ref. \cite{us} solitonic solutions were constructed 
perturbatively which reduce to the ordinary solitons when we take the
commutative limit.  Since a classical action was not found in \cite{us}, these solitonic solutions found are said to be localized in the sense that at order zero in the deformation parameter they reduce to the well-known euclidean solitons of the sine-Gordon theory. Since the solution at order zero determines the solution to all orders in the deformation parameter, in \cite{us} these solitons are called localized at all orders. More generally, we observe that the second equation
in (\ref{sg}) is automatically satisfied by any chiral or antichiral 
function.
Therefore, we expect the class of solitonic solutions to be at least as
large as the ordinary one. In the general case, instead, 
we expect the class of dynamical
solutions to be smaller than the ordinary one because of the presence
of the nontrivial constraint. However, since the constraint equation is one 
order higher with respect to the dynamical equation, order by order in the 
$\theta$-expansion a solution always exists. This means that a 
Seiberg--Witten map between 
the NC and the ordinary model does not exist as a mapping between 
physical configurations, but it might be constructed as a mapping between 
equations of motion or conserved currents.  

The kind of noncommutative solitons discussed in section 2.2.1 has not been studied in \cite{us}. These solutions in principle should be present in this model. However, the model described in \cite{us} was shown to display bad scattering properties \cite{mypaper2}, as I will show in detail in section 2.2.9. For this reason it had to be discarded and replaced with a new model described in section 2.3 \cite{mypaper3}. Both kinds of soliton solutions were studied in detail for this model (see section 2.3.5).

\subsection{Reduction from noncommutative selfdual Yang-Mills}
The material presented in this section and the following ones, until the end of section 2.2, is mostly taken from the paper \cite{mypaper2}, written in collaboration with M.T. Grisaru, L. Mazzanti and S. Penati.

The (anti-)selfdual Yang--Mills equation is well-known to describe a
completely integrable classical system in four dimensions \cite{Yang}. 
In the ordinary case the equations of motion for many two dimensional 
integrable systems, including sine--Gordon, 
can be obtained through dimensional reduction of the (anti)selfdual Yang-Mills equations 
\cite{SDYMred}.

We have seen in section 2.1.3 that a convenient description of the (anti)selfdual Yang-Mills system is the so called 
$J$-for\-mu\-la\-tion, given in terms of a $SL(N,\C)$ 
matrix-valued $J$ field satisfying
\beq
\pa_{\bar{y}}\left(J^{-1} \pa_y J\right)+\pa_{\bar{z}}\left(J^{-1}\pa_z 
J\right)=0
\label{selfdual}
\eeq
where $y$, $\bar{y}$, $z$, $\bar{z}$ are complex variables 
treated as formally independent.

In the ordinary case, the sine-Gordon equation can be obtained from
(\ref{selfdual}) by taking $J$ in $SL(2, \C)$ to be \cite{SDYMsine}
\beq
J=J(u,z,\bar{z})=e^{\frac{z}{2}\s_i}e^{\frac{i}{2}\Phi\s_j}
e^{-\frac{\bar{z}}{2}\s_i}
\eeq
where $\Phi=\Phi(y,\bar y)$ depends on $y$ and $\bar y$ only and $\s_i$ are 
the Pauli matrices.

A noncommutative version of the (anti-)selfdual Yang--Mills 
system can be naturally obtained \cite{NCSDYM1}
by promoting the variables $y$, $\bar y$,
$z$ and $\bar z$ to be noncommutative thus extending the ordinary products in
(\ref{selfdual}) to $\ast$--products. In this case the $J$ field lives
in $GL(N,\C)$. 

As we outlined in section 1.2.3, it has been shown \cite{olaf1} that noncommutative selfdual Yang-Mills naturally emerges 
from open $N=2$ strings in a 
B-field background.
Moreover, in \cite{NCSDYM1, NCSDYM2, NCbur} examples of reductions to 
two-dimensional 
noncommutative systems were given. It was also argued that  the noncommutative deformation should 
preserve the integrability of the systems \cite{NCbur,NCSDYM4}.  

We now show that our noncommutative version of the sine-Gordon equations can be 
derived through dimensional reduction from the noncommutative selfdual Yang-Mills equations.
For this purpose we consider the noncommutative version of equations (\ref{selfdual})
and choose $J_\ast$ in $GL(2,\C)$ as
\beq
J_\ast=J_\ast(u,z,\bar{z})=e_\ast^{\frac{z}{2}\s_i}\ast 
e_\ast^{\frac{i}{2}\Phi\s_j}\ast e_\ast^{-\frac{\bar{z}}{2}\s_i}
\eeq
This leads to the matrix equation
\beq
\pa_{\bar{y}}a~ I ~+~ i\left(\pa_{\bar{y}}b+\frac{1}{2} 
\sin_\ast{\Phi} \right)\s_j=0
\label{sgmatrix}
\eeq
where $a$ and $b$ have been defined in (\ref{sg}).
Now, taking the trace we obtain $\pa_{\bar{y}} a = 0$  which is 
the constraint equation in (\ref{sg}). As a consequence, the term proportional
to $\s_j$ gives rise to the dynamical equation in (\ref{sg}) for the
particular choice $\g = -1$.
Therefore we have shown that the equations of motion of the noncommutative version
of sine--Gordon proposed in \cite{us} can be obtained from a suitable 
reduction of the noncommutative selfdual Yang-Mills system as in the ordinary case. From this derivation 
the origin of the constraint appears even more clearly: 
it arises from setting to zero the trace part which the 
matrices in $GL(2,\C)$ naturally develop under $\ast$--multiplication. 

Solving (\ref{sgmatrix}) for the particular choice $\s_j = \s_3$ we obtain the 
alternative set of equations
\bea
&& \pab \left( e_{\ast}^{-\frac{i}{2} \Phi}  \ast \pa 
e_{\ast}^{\frac{i}{2} \Phi}\right) ~=~\frac{i}{2} \g \sin_{\ast}{\Phi} \cr
&& \pab \left(e_{\ast}^{\frac{i}{2} \Phi}  \ast \pa 
e_{\ast}^{-\frac{i}{2} \Phi} \right) ~=~ -\frac{i}{2} \g \sin_{\ast}{\Phi}  
\label{sg2}
\eea
Order by order in the $\theta$-expansion the set of equations (\ref{sg}) 
and (\ref{sg2}) are equivalent. Therefore, the set (\ref{sg2}) is equally
suitable for the description of an integrable noncommutative generalization 
of sine-Gordon.

Since
our noncommutative generalization of sine--Gordon is integrable, the present result
gives support to the arguments in favor of the integrability of noncommutative selfdual Yang-Mills
system.
\subsection{The action}
We are now interested in the possibility 
of determining an action for the scalar field $\Phi$ satisfying the
system of eqs. (\ref{sg}). We are primarily motivated 
by the possibility to move on to a quantum description of the system. 

In general, it is not easy to find an action for the dynamical equation 
(the first eq. in (\ref{sg})) since $\Phi$ is constrained by the second
one. One possibility could be to implement the constraint by the use
of a Lagrange multiplier. Another quite natural possibility is to try to obtain the action by a dimensional reduction procedure from $(4,0)$ selfdual Yang-Mills action in Yang formulation. Unfortunately, this does not work, since WZW-like terms disappear from the reduced action because of cyclicity of Moyal product in an integral. As a result one obtains a reduced action generating nonchiral equations, different from the chiral ones in eqs. (\ref{sg}) and (\ref{sg2}).

We consider instead the equivalent set of equations (\ref{sg2}). 
We rewrite them in the form
\bea
&&\pab(g^{-1}\ast \pa g)=\frac{1}{4}\g\left(g^2-g^{-2}\right)\cr
&&\pab(g\ast\pa g^{-1})=-\frac{1}{4}\g\left(g^2-g^{-2}\right)
\label{sg3}
\ena
where we have defined $g\equiv e_\ast^{\frac{i}{2}\Phi}$. Since $\Phi$ is 
in general complex $g$ can be seen as an element of a noncommutative
complexified $U(1)$. The gauge group valued function 
$\bar g\equiv (g^\dagger)^{-1}= e_\ast^{\frac{i}{2}\Phi^\dagger}$ 
is subject to the equations
\bea
&&\pab(\bar g \ast\pa\bar g^{-1})=-\frac{1}{4}\g \left(\bar g^2-\bar 
g^{-2}\right)\cr
&&\pab(\bar g^{-1}\ast\pa \bar g)=\frac{1}{4}\g \left(\bar g^2-\bar 
g^{-2}\right)
\label{sg4}
\ena
obtained by taking the h.c. of (\ref{sg3}).

In order to determine the action it is convenient to concentrate on the
first equation in (\ref{sg3}) and the second one in (\ref{sg4}) as the two
independent complex equations of motion which describe the dynamics of our 
system. 

We first note that the left-hand sides of equations (\ref{sg3}) and
(\ref{sg4}) have the chiral structure which is well known to correspond to a
noncommutative version of the WZNW action \cite{wzwnc} (see section 1.1.3). Therefore we are led to consider 
the action 
\beq
S[g,\bar g]=S[g]+S[\bar g] 
\label{action}
\eeq
where, introducing the homotopy path $\hat{g}(t)$ such that $\hat{g}(0)=1$,
$\hat{g}(1) = g$ ($t$ is a commuting parameter) we have defined
\bea
&S[g]& =  \int d^2z ~\left[ \pa  g \ast 
\pab g^{-1} + \int_0^1 dt ~ \hat{g}^{-1} \ast \pa_{t} \hat{g} \ast
[\hat{g}^{-1} \ast  \pa  \hat{g}, 
\hat{g}^{-1} \ast \pab  \hat{g} ]_{\ast} \right.\cr
&-&\left.\frac{\g}{4}(g^2+ g^{-2}-2) \right] 
\label{actionWZNW}
\eea
and similarly for $S[\bar{g}]$. The first part of the action can be recognized 
as the noncommutative generalization of a complexified $U(1)$ WZNW action \cite{NS}.

To prove that this generates the correct equations, we should take 
the variation with
respect to the $\Phi$ field ($g = e_\ast^{\frac{i}{2}\Phi}$)
and deal with complications which follow from
the fact that in the noncommutative case the variation of an exponential is not
proportional to the exponential itself.
However, since the variation $\d \Phi$ is arbitrary, 
we can forget about its $\theta$ dependence and write $\frac{i}{2}\d \Phi 
= g^{-1} \d g $, trading the variation 
with respect to $\Phi$ with the variation
with respect to $g$. Analogously, the variation with respect to 
$\Phi^{\dagger}$ can be traded with the variation with respect to $\bar{g}$.

It is then a simple calculation to show that
\beq
\d S[g] = \int d^2z ~2g^{-1} \d g \left[ \pab \left( g^{-1} 
\ast \pa g \right) ~-~ \frac{i}{2} \g \sin_{\ast}{\Phi}
\right] 
\eeq
from which we obtain the first equation in (\ref{sg3}). Treating $\bar g$
as an independent variable an analogous derivation gives the second equation 
in (\ref{sg4}) from $S[\bar g]$.

We note that, when $\Phi$ is real, $g=\bar g$ and the action 
(\ref{action}) reduces to $S_{\rm real}[g]=2S_{WZW}[g]-\g(\cos_\ast\Phi -1)$. 
In general, since the two equations (\ref{sg}) are complex it
would be inconsistent to restrict ourselves to real solutions. However, it is
a matter of fact that the equations of motion become real when the field
is real.
Perturbatively in $\theta$ this can be proved order by order 
by direct inspection of the equations in \cite{us}. In particular, at
a given order one can show that the imaginary part of the equations
vanishes when the constraint and the equations of motion at
lower orders are satisfied. 

\subsection{The relation to the noncommutative Thirring model}
In the ordinary case the equivalence between the Thirring and  
sine--Gordon models \cite{coleman} can be proven at the level of functional
integrals by implementing the bosonization prescription 
\cite{wittenwzw, bosonization} on the fermions.
The same procedure has been worked out in noncommutative geometry \cite{MS,schiappa}.
Starting from the noncommutative version of Thirring described by 
\beq
S_T = \int d^2x \left[ \bar{\psi} i \g^\mu \pa_\mu \psi + m \bar{\psi} \psi
- \frac{\l}{2}(\bar{\psi} \ast \g^\mu \psi)  (\bar{\psi} \ast \g_\mu \psi) 
\right]
\label{Thirring}
\eeq
the bosonization prescription gives rise to the action for the bosonized 
noncommutative massive Thirring model which turns out to be a 
noncommutative WZNW action supplemented by a cosine potential term for the noncommutative U(1)
group valued field which enters the bosonization of the fermionic currents.
In particular, in the most recent paper in Ref. \cite{MS} it has been
shown that working in Euclidean space the massless Thirring action 
corresponds to the sum of two WZNW actions once a suitable choice for the 
regularization parameter is made. Moreover, in Ref. \cite{schiappa} it was proven 
that the bosonization of the mass term in (\ref{Thirring}) gives rise to
a cosine potential for the scalar field with coupling constant proportional
to $m$.

The main observation is that our action (\ref{action}) is the sum of two
noncommutative WZNW actions plus cosine potential terms for the pair of 
$U(1)_\C$ group valued fields $g$ and $\bar g$, considered as independent.
Therefore, our action can be interpreted as coming from the bosonization
of the massive noncommutative Thirring model, in agreement with the results 
in \cite{MS,schiappa}.

We have shown that even in the noncommutative case
the sine--Gordon field can be interpreted
as the scalar field which enters the bosonization of the Thirring model,
so proving that the equivalence between the Thirring and 
sine--Gordon models can be maintained in noncommutative generalizations of these models. 
Moreover, the classical integrability of our noncommutative version of 
sine--Gordon proven in  \cite{us} should automatically guarantee the 
integrability of the noncommutative Thirring model.  

In the particular case of zero coupling ($\g = 0$), the equations 
(\ref{action})
and (\ref{sg2}) correspond to the action and the equations of motion
for a noncommutative $U(1)$ WZNW model \cite{wzwnc}, respectively. Again, we can use 
the results
of \cite{us} to prove the classical integrability of the noncommutative $U(1)$ WZNW
model and construct explicitly its conserved currents.  

\subsection{(Bad) properties of the S-matrix}
In section 2.1.4 we showed that in integrable commutative field theories there is no 
particle production and the S-matrix factorizes. In the noncommutative case
properties of the S-matrix have been investigated for two
specific models: The $\l \Phi^4$ theory in two dimensions \cite{causal} and
the nonintegrable ``natural'' NC generalization the the sine--Gordon model
\cite{CM}. In the first reference a very pathological acasual behavior was 
observed due to the space and time noncommutativity (see section 1.1.2). For an incoming wave
packet the scattering produces an advanced wave which arrives at the origin
before the incoming wave. In the second model investigated it was found
that particle production occurs. The tree level $2 \to 4$ amplitude does not 
vanish.   

It might be hoped that classical integrability would alleviate these 
pathologies. In the NC integrable sine-Gordon case, since we have an action,
it is possible to investigate these issues. As described below we have 
computed the scattering amplitude for the $2 \to 2 $ process and found that
the acausality of Ref. \cite{causal} is not cured by integrability.  
We have also computed the production amplitudes for the processes $2 \to 3$ 
and $2 \to 4$ and found that they don't vanish.

We started from our action (\ref{actionWZNW}) rewritten in terms of Minkowski 
space coordinates $x^0,x^1$ and real fields 
($g = e_{\ast}^{\frac{i}{2} \Phi}$, $\hat{g}(t) = 
e_{\ast}^{\frac{i}{2} t \Phi}$ with $\Phi$ real)

\bea
&&S[g]=-\frac{1}{2}\int d^2 x~ g^{-1}\ast \pa^\mu g \ast
g^{-1}\ast\pa_\mu g  +\frac{\g}{4}\int d^2 x (g^2+g^{-2}-2) \cr
&&~~~~~~~~~~~~-\frac{1}{3}\int d^3 x ~\e^{\m\n\rho}\hat{g}^{-1}\ast \pa_\mu \hat{g} \ast 
\hat{g}^{-1}\ast\pa_\n \hat{g} \ast \hat{g}^{-1}\ast\pa_\rho \hat{g} 
\ena
where $f \ast g = f e^{\frac{i}{2} \th \e^{\mu\nu} \overleftarrow{\pa}_\mu
\overrightarrow{\pa}_\nu} g$, and we derived the following Feynman's rules

\begin{itemize}
\item The propagator
\beq
G(q)=\frac{4i}{q^2-2\g}
\eeq

\item
The vertices 
\bea
&& v_3(k_1,\dots,k_3)= \frac{2}{2^3 \cdot 3!}
\e^{\m\n}k_{1\m}k_{2\n}F(k_1,\dots,k_3) \nonumber \\
&&v_4(k_1,\dots,k_4)= i \left(-\frac{1}{2^4 \cdot 4!}
\left(k_1^2+3k_1\cdot k_3\right)+\frac{\g}{2 \cdot 4!}\right)
F(k_1,\dots,k_4) \nonumber \\
&& v_5(k_1,\dots,k_5)= -\frac{2\e^{\m\n}}{2^5 \cdot 5!}
\left(k_{1\m}k_{2\n}-k_{1\m}k_{3\n}+2k_{1\m}k_{4\n}\right)F(k_1,\dots,k_5)
\nonumber \\
&&  v_6(k_1,\dots,k_6)= i\left[\frac{1}{2^6 \cdot 6!}
\left(k_{1}^2+5k_{1}\cdot k_{3}-5k_{1} \cdot k_{4}+5k_{1} \cdot k_{5}\right)
- \frac{\g}{2 \cdot 6!}\right]\cdot\cr
&&~~~~~~~~~~~~~~~~~~~~~~~~~~~\cdot F(k_1,\dots,k_6)
\nonumber \\
&&~~~~~~~~~
\label{vertices}
\ena
where
\beq
F(k_1,\dots,k_n)=\exp\left(-\frac{i}{2}\sum_{i<j}k_i\times k_j\right)
\eeq
is the phase factor coming from the $\ast$-products in the action 
(we have indicated $a\times b= \theta \e^{\m\n} a_\m b_\n$), $k_i$ are
all incoming momenta and we used momentum conservation.
\end{itemize}

At tree level the $2 \to 2$ process is described by the diagrams with
the topologies in Fig. 1.

\vskip 18pt
\noindent
\begin{minipage}{\textwidth}
\begin{center}
\includegraphics[width=0.60\textwidth]{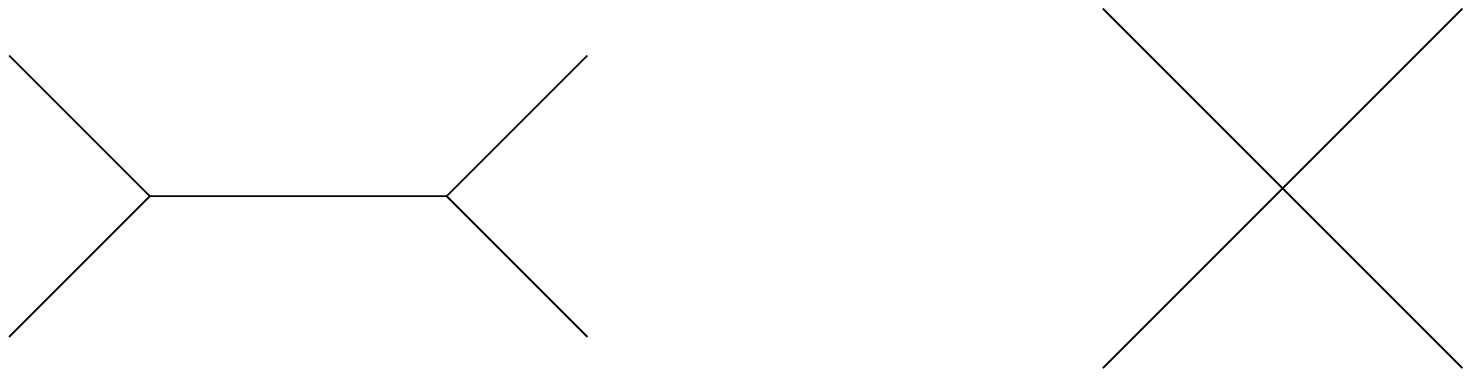}
\end{center}
\begin{center}
{\small{Figure 1:
Tree level $2 \to 2$ amplitude}}
\end{center}
\end{minipage}

\vskip 20pt

Including contributions from the various channels and using the three point
and four point vertices of eqs. (\ref{vertices}) we obtained for the
scattering amplitude the expression  
\beq
-\frac{i}{2} E^2 p^2 \left( \frac{1}{2E^2 - \g} - \frac{1}{2p^2 + \g}
\right) \sin^2{(pE \th)} ~+~ i \frac{\g}{2} \cos^2{(pE \th)}
\label{final}
\eeq
where $p$ is the center of mass momentum and $E = \sqrt{p^2 + 2\g}$.

For comparison with Ref. \cite{seiberg} this should be multiplied by an
incoming wave packet  
\beq
\phi_{in}(p) \sim \left( e^{-\frac{(p -p_0)^2}{\l}}+ 
e^{-\frac{(p +p_0)^2}{\l}}\right)
\eeq
and Fourier transformed with $e^{ipx}$.
We have not attempted to carry out the Fourier transform integration.
However, we note that for $p_0$ very large $E$ and $p$ are concentrated
around large values and the scattering amplitude assumes the form 
\beq
i\frac{\g}{4} \sin^2{(pE \th)} ~+~ i \frac{\g}{2} \cos^2{(pE \th)}
\eeq
which is equivalent to the result in Ref. \cite{causal}, leading to
the same acausal pathology \footnote{It is somewhat tantalizing that 
a change in the relative coefficient between the two terms 
would lead to a removal of the trigonometric factors which are responsible
for the acasual behavior.}. 

We describe now the computation of the production amplitudes $2 \to 3$ and 
$2 \to 4$. At tree level the contributions are drawn in Figures 2 and 3, 
respectively.

\vskip 18pt
\noindent
\begin{minipage}{\textwidth}
\begin{center}
\includegraphics[width=0.60\textwidth]{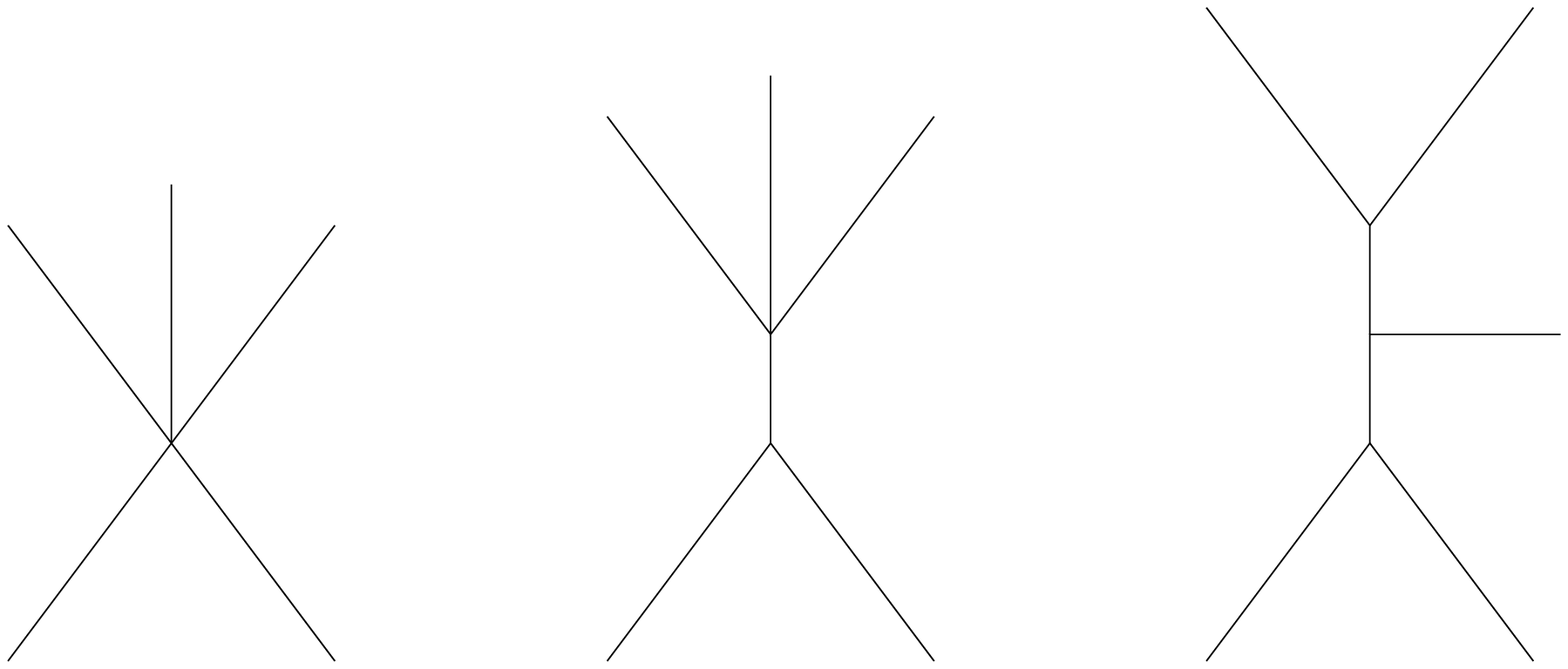}
\end{center}
\begin{center}
{\small{Figure 2:
Tree level $2 \to 3$ amplitude}}
\end{center}
\end{minipage}

\vskip 18pt
\noindent
\begin{minipage}{\textwidth}
\begin{center}
\includegraphics[width=0.60\textwidth]{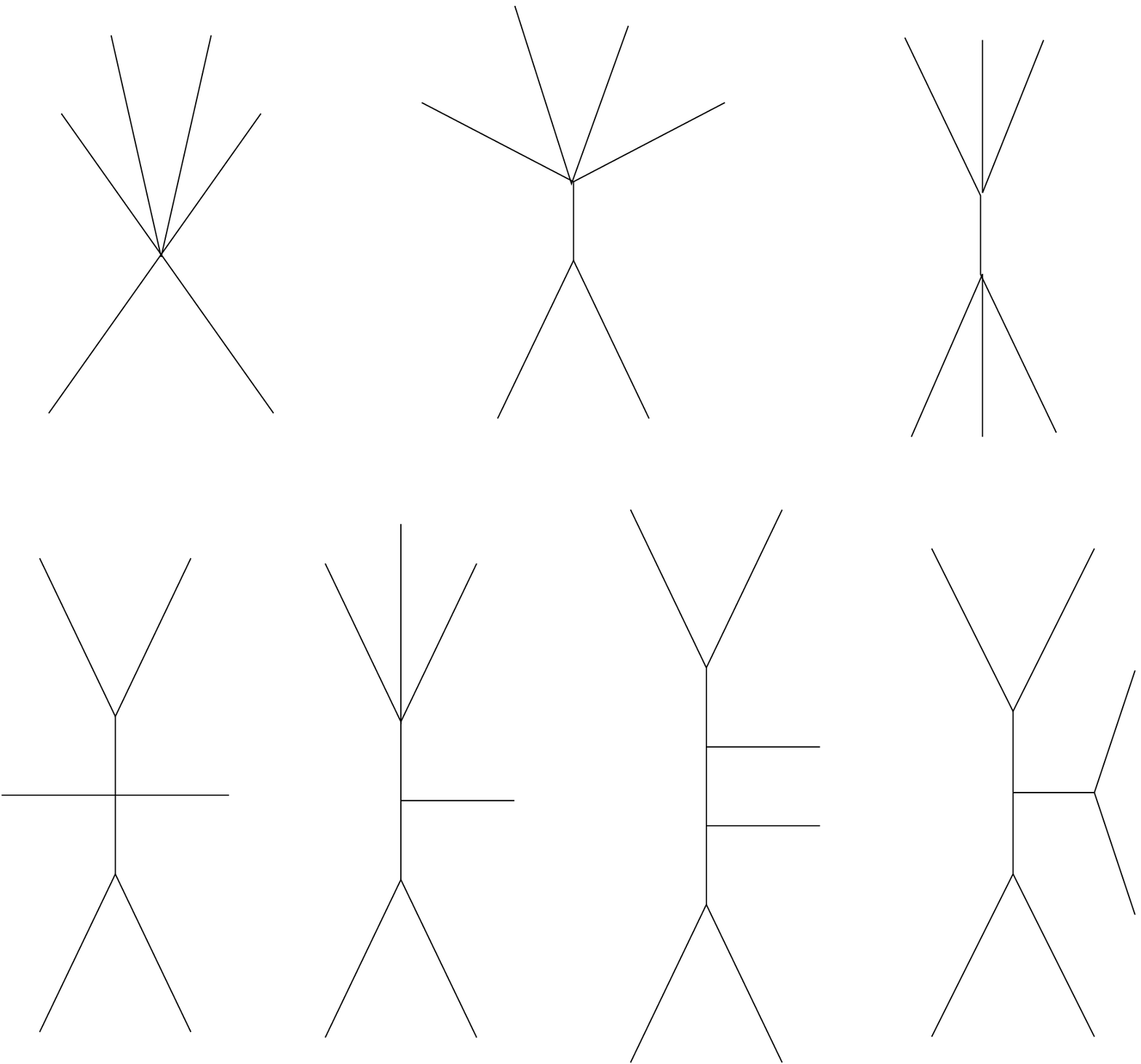}
\end{center}
\begin{center}
{\small{Figure 3:
Tree level $2 \to 4$ amplitude}}
\end{center}
\end{minipage}

\vskip 31pt

For any topology the different possible channels must be taken into account. 
This, as well as the complicated expressions for the vertices, has led us
to use an algebraic manipulation program computer.
We used {\em Mathematica}${}^{\copyright}$ to symmetrize completely 
the vertices 
(\ref{vertices}). This allows to take automatically into account the 
different diagrams obtained by exchanging momenta entering a given vertex. 
The contribution from each diagram was obtained as a product of 
the combinatorial factor, the relevant vertices and propagators.  
Due to the length of the program it was impossible to handle the calculation
in a completely analytic way. Instead, the program was run with assigned 
values of the momenta and arbitrary  $\theta$ and $\g$. For both the  
$2\rightarrow 3$ and $2\rightarrow 4$ processes the result is nonvanishing.
As a check of our calculation we mention that the production
amplitudes vanish  when we set $\th = 0$, for any value of the coupling and 
the momenta.

\subsection{Conclusions}
In \cite{mypaper2}, in collaboration with M.T. Grisaru, L. Mazzanti and S. Penati I have investigated some properties of the integrable noncommutative
sine--Gordon system proposed in \cite{us}. We succeeded in constructing
an action which turned out to be a WZNW action for a noncommutative, 
complexified $U(1)$ augmented by a cosine potential. We have shown that 
even in the noncomutative case there is a duality relation between our integrable
noncommutative sine--Gordon model and the noncommutative Thirring. 

Noncommutative WZNW models have been shown to be one--loop renormalizable \cite{LFI}. 
This suggests that the noncommutative sine--Gordon model proposed
in \cite{us} is not only integrable but it might lead to a well-defined
quantized model, giving support to the existence of a possible relation 
between integrability and renormalizability. 

Armed with our action we investigated some properties of the S--matrix
for elementary excitations.
However, in contradistinction to the commutative case, the S--matrix turned out
to be acasual and nonfactorizable \footnote{Other problems of the S-matrix have
been discussed in \cite{unitar,gmbk}.}. The reason for the acasual behavior has been
discussed in \cite{causal}~ where it was pointed out that 
noncommutativity induces a backward-in-time effect because of the presence
of certain phase factors (see section 1.1.2). It appears that in our case this effect is still
present in spite of integrability. 

It is not clear why the presence of an infinite number of local conserved
currents (local in the sense that they are not expressed as integrals of 
certain densities) does not guarantee factorization and absence of 
production in the
S-matrix as it does in the commutative case. The standard proofs of 
factorization use, among other assumptions, the mutual commutativity of the
charges - a property we have not been able to check so far because of the
complicated nature of the currents. But even if the charges were to commute
the possibility of defining them as powers of momenta, as required in the
proofs, could be spoiled by
acausal effects which prevent a clear distinction between incoming and
outgoing particles. 

In a series of papers \cite{doplicher} a different approach to quantum noncommutative field 
theories has 
been proposed when the time variable is not commuting. 
In those papers it has been argued that the problems associated to time-space noncommutativity are due to the fact that the time-ordering procedure does not commute with the star multiplication. Starting from the usual definition of the S-matrix in terms of the time-ordered exponential of the interaction term in the action and applying Wick theorem, one cannot combine the contraction functions of positive and negative frequency to obtain the causal Feynman propagator. Therefore, it has been suggested that, instead of the Feyman approach \cite{filk}, one should use the time ordered perturbation theory extended to the noncommutative case. Moreover, it has been shown that in this framework unitarity is preserved as long as the interaction lagrangian is explicitly hermitian.
It would be interesting to redo our calculations in that approach to see whether a 
well-defined factorized S-matrix for our model can be constructed. 
In this context it would be also interesting to investigate the scattering 
of solitons present in our model \cite{us}. 

In the next section a novel noncommutative sine-Gordon system, obtained by dimensional reduction from the $2+1$ model introduced in \cite{olaf5}, will be constructed and studied. We will see that it exhibits nice scattering properties, consistent with the usual relation between integrability and factorization of the S-matrix. 

\section{The noncommutative integrable sine-Gordon model}
\subsection{Introduction}
In this section I will present the results I obtained, in collaboration with O. Lechtenfeld, L. Mazzanti, S. Penati and A.D. Popov, in \cite{mypaper3}. The main goal of that work was to find a noncommutative generalization of the 
sine-Gordon system which, as a hallmark of integrability, possesses a 
well-defined {\em causal\/} and {\em factorized\/} S-matrix.
Furthermore, its equations of motion should admit noncommutative
multi-soliton solutions which represent deformations of the well known 
sine-Gordon solitons.

In sections 2.2.4 and 2.2.5 I have discussed the results obtained in \cite{us}, where a model was proposed which
describes the dynamics of a complex scalar field by a couple 
of equations of motion. These equations were obtained as flatness conditions
for a $U(2)$ bidifferential calculus~\cite{NCEOM} and automatically guarantee 
the existence of an infinite number of local conserved currents.
The same equations were also generated in~\cite{mypaper2} via a particular 
dimensional reduction of the noncommutative $U(2)$ selfdual Yang-Mills equations in euclidean
space. However, this reduction did {\em not\/} work at the level of 
the action, which turned out to be the sum of two WZW models augmented
by a cosine potential. 
Evaluating tree-level scattering amplitudes it was discovered, furthermore,
that this model suffers from acausal behavior and a non-factorized S-matrix,
meaning that particle production occurs.

At this point it is important to note that the noncommutative deformation
of an integrable equation is a priori not unique, because one may always add
terms which vanish in the commutative limit, as we have seen in section 1.1.3. For the case at hand, for example,
different inequivalent ans\"atze for the $U(2)$ matrices entering the bicomplex 
construction \cite{NCEOM} are possible as long as they all reproduce the ordinary
sine-Gordon equation in the commutative limit. It is therefore conceivable
that among these possibilities there exists an ansatz (different from the one 
in \cite{us,mypaper2}) which guarantees the classical integrability of the 
corresponding noncommutative model. What is already certain is the necessity
to introduce {\em two\/} real scalar fields instead of one, since in the
noncommutative realm the $U(1)$ subgroup of $U(2)$ fails to decouple. What has been
missing is a guiding principle towards the ``correct'' field parametrization.

Since the sine-Gordon model can be obtained by dimensional reduction from
2+2 dimensional selfdual Yang-Mills theory via a 2+1 dimensional integrable sigma model
\cite{ward}, and because the latter's noncommutative extension was shown to be
integrable in \cite{olaf4}, it seems a good idea to contruct an integrable 
generalization of the sine-Gordon equation by starting from the linear system
of this integrable sigma model endowed with a time-space noncommutativity. 
This is the key strategy of this paper.
The reduction is performed on the equations of motion first, but it also works
at the level of the action, so giving directly the 1+1 dimensional action 
we are looking for. This success is an indication that the new
field parametrization proposed in \cite{mypaper3} is the proper one.

To be more precise, in \cite{mypaper3} we proposed three different parametrizations,
by pairs of fields $(\phi_+,\phi_-)$, $(\rho,\vp)$ and $(h_1,h_2)$, all related
by nonlocal field redefinitions but all deriving from the compatibility
conditions of the underlying linear system \cite{olaf4}. The first two appear 
in Yang formulation~\cite{Yang} while the third one arises in Leznov formulation
\cite{L}. For either field pair in Yang formulation,
the nontrivial compatibility condition reduces to a pair of 
``noncommutative sine-Gordon equations'' which in the commutative
limit degenerates to the standard sine-Gordon equation for 
$\sfrac12(\phi_+{+}\phi_-)$ or $\vp$, respectively, while 
$\sfrac12(\phi_+{-}\phi_-)$ or $\rho$ decouple as free bosons. 
The alternative Leznov formulation has the advantage
of producing two polynomial (actually, quadratic) equations of motion for 
$(h_1,h_2)$ but retains their coupling even in the commutative limit.

With the linear system comes a well-developed technology for generating
solitonic solutions to the equations of motion. In \cite{mypaper3} the
dressing method \cite{dressing,curvature} was employed to explicitly outline the construction
of noncommutative sine-Gordon multi-solitons, directly in 1+1 dimensions as 
well as by reducing plane-wave solutions of the 2+1 dimensional integrable 
sigma model \cite{olaf5}.
The one-soliton sector was completely analyzed and it was found that the standard
soliton solution are recovered as undeformed. Noncommutativity becomes palpable only at the
multi-soliton level.

It was shown in \cite{olaf1} that the tree-level $n$-point amplitudes of 
noncommutative 2+2 dimensional SDYM vanish for $n>3$,
consistent with the vanishing theorems for the $N{=}2$ string.
Therefore, we were expecting nice properties of the S-matrix
to be inherited by this noncommutative sine-Gordon theory.
Indeed, a direct evaluation of tree-level amplitudes revealed that,
in the Yang as well as the Leznov formulation,
the S-matrix is {\em causal\/} and no particle production occurs.

\subsection[A noncommutative integrable sigma model in $2+1$\\ dimensions]{A noncommutative integrable sigma model in $2+1$ dimensions}
As has been known for some time, nonlinear sigma models in $2{+}1$ dimensions
may be Lorentz-invariant or integrable but not both~\cite{ward}.
Since the integrable variant, introduced in section 2.1.3, serves as our starting point for the derivation
of the sine-Gordon model and its soliton solutions, we shall present its 
noncommutative extension \cite{olaf4} in some detail in the present section.
\subsubsection{Conventions in noncommutative $\R^{2,1}$}
In $\R^{2,1}$ we shall use (real) coordinates $(x^a)=(t,x,y)$
in which the Minkowskian metric reads $(\eta_{ab})=\textrm{diag}(-1,+1,+1)$.
For later use we introduce the light-cone coordinates 
\begin{equation} \label{lightcone}
u\ :=\ \sfrac{1}{2}(t+y)\quad,\qquad
v\ :=\ \sfrac{1}{2}(t-y)\quad,\qquad
\pa_u\ =\ \pa_t+\pa_y\quad,\qquad
\pa_v\ =\ \pa_t-\pa_y \quad.
\end{equation} 
In view of the future reduction to $1{+}1$ dimensions, we choose the
coordinate~$x$ to remain commutative, so that the only non-vanishing
component of the noncommutativity tensor is 
\begin{equation}
\th^{ty}\ =\ -\th^{yt}\ =:\ \th\ >\ 0 \quad.
\end{equation}
\subsubsection{Linear system }
Consider on noncommutative $\R^{2,1}$ 
the following pair of linear differential equations~\cite{olaf4},
\begin{equation}\label{linsys}
(\z \pa_x -\pa_u)\Psi\ =\ A\ast\Psi \qquad\textrm{and}\qquad
(\z \pa_v -\pa_x)\Psi\ =\ B\ast\Psi \quad,
\end{equation}
where a spectral parameter~$\z\in\C P^1\cong S^2$ has been introduced.
The auxiliary field $\Psi$ takes values in U$(n)$ and depends on
$(t,x,y,\z)$ or, equivalently, on $(x,u,v,\z)$.
The $u(n)$ matrices $A$ and~$B$, in contrast, do not depend on~$\z$
but only on $(x,u,v)$.  
Given a solution~$\Psi$, they can be reconstructed via\footnote{
Inverses are understood with respect to the star product, 
i.e.~$\Psi^{-1}\ast\Psi=\mbf{1}$.} 
\begin{equation} \label{ABfromPsi}
A \= \Psi\ast(\pa_u-\z\pa_x)\Psi^{-1} \qquad\textrm{and}\qquad
B \= \Psi\ast(\pa_x-\z\pa_v)\Psi^{-1} \quad.
\end{equation}
It should be noted that the equations (\ref{linsys}) 
are not of first order but actually of infinite order in derivatives, 
due to the star products involved.
In addition, the matrix $\Psi$ is subject to the following reality
condition~\cite{ward}:
\begin{equation}\label{real}
\mbf{1} \= \Psi(t,x,y,\z)\,\ast\,[\Psi(t,x,y,\bar{\z})]^{\dagger} \quad,
\end{equation}
where `$\dagger$' is hermitian conjugation.
The compatibility conditions for the linear system~(\ref{linsys}) read
\begin{align}
\pa_x B -\pa_v A\ =\ 0 \quad ,\label{comp1} \\[4pt]
\pa_x A -\pa_u B -A\ast B +B\ast A\ =\ 0 \quad . \label{comp2}
\end{align}
By detailing the behavior of~$\Psi$ at small~$\z$ and at large~$\z$
we shall now ``solve'' these equations in two different ways, each one
leading to a single equation of motion for a particular field theory.
\subsubsection{Yang-type solution}
We require that $\Psi$ is regular at $\z{=}0$~\cite{olaf2}, 
\begin{equation} \label{asymp1}
\Psi(t,x,y,\z\to0)\= \Phi^{-1}(t,x,y)\ +\ O(\z) \quad,
\end{equation}
which defines a U$(n)$-valued field $\Phi(t,x,y)$,
i.e.~$\ \Phi^\dagger=\Phi^{-1}$.
Therewith, $A$ and~$B$ are quickly reconstructed via
\begin{equation} \label{ABfromPsi2}
A\=\Psi\ast\pa_u\Psi^{-1}\big|_{\z=0}
\=\Phi^{-1}\ast\pa_u\Phi \qquad\textrm{and}\qquad
B\=\Psi\ast\pa_x\Psi^{-1}\big|_{\z=0}
\=\Phi^{-1}\ast\pa_x\Phi 
\end{equation}
It is easy to see that compatibility equation (\ref{comp2}) is then automatic
while the remaining equation~(\ref{comp1}) turns into~\cite{olaf4}
\begin{equation} \label{yangtype}
\pa_x\,(\Phi^{-1}\ast\pa_x\Phi)-\pa_v\,(\Phi^{-1}\ast\pa_u\Phi)\ =\ 0 \quad.
\end{equation}
This Yang-type equation~\cite{Yang} can be rewritten as
\begin{equation} \label{yangtype2}
(\eta^{ab}+v_c\,\e^{cab})\,\pa_a (\Phi^{-1}\ast\pa_b \Phi)\ =\ 0\quad,
\end{equation}
where $\e^{abc}$ is the alternating tensor with $\e^{012}{=}1$ 
and $(v_c)=(0,1,0)$ is a fixed spacelike vector.
Clearly, this equation is not Lorentz-invariant but (deriving from a Lax pair)
it is integrable.

One can recognize (\ref{yangtype2}) as the field equation for
(a noncommutative generalization of) a WZW-like
modified U$(n)$ sigma model~\cite{ward,ioannidou} with the action\footnote{
which is obtainable by dimensional reduction from the Nair-Schiff
action~\cite{nair,moore} for SDYM in 2+2 dimensions}
\begin{equation} \label{Yaction}
\begin{aligned}
&S_{\textrm{Y}}\ =\ -\sfrac12\int\!\diff{t}\,\diff{x}\,\diff{y}\;\eta^{ab}\;
\tr\,\Bigl(\pa_a \Phi^{-1} \ast\, \pa_b \Phi \Bigr) \\
&~~~\quad\ -\sfrac13\int\!\diff{t}\,\diff{x}\,\diff{y} \int_0^1\diff{\l}\;
\widetilde{v}_{\r}\,\e^{\r\m\n\s}\;\tr\,\Bigl(
\Pht^{-1}\ast\,\pa_{\m}\Pht\,\ast\,
\Pht^{-1}\ast\,\pa_{\n}\Pht\,\ast\,
\Pht^{-1}\ast\,\pa_{\s}\Pht \Bigr) 
\end{aligned}
\end{equation}
where Greek indices include the extra coordinate~$\l$,
and $\e^{\r\m\n\s}$ denotes the totally antisymmetric tensor in~$\R^4$.
The field~$\Pht(t,x,y,\l)$ is an extension of~$\Phi(t,x,y)$, 
interpolating between
\begin{equation}
\Pht(t,x,y,0)\ =\ \textrm{const} \qquad\textrm{and}\qquad
\Pht(t,x,y,1)\ =\ \Phi(t,x,y) \quad,
\end{equation}
and `$\tr$' implies the trace over the U$(n)$ group space.
Finally, $(\widetilde{v}_{\r})=(v_c,0)$
is a constant vector in (extended) space-time.
\subsubsection{Leznov-type solution}
Finally, we also impose the asymptotic condition that 
$\ \lim_{\z\to\infty}\Psi=\Psi^0$ with some constant unitary 
(normalization) matrix~$\Psi^0$. The large~$\z$ behavior~\cite{olaf2}
\begin{equation} \label{asymp2}
\Psi(t,x,y,\z\to\infty)\= 
\bigl( \mbf{1}\ +\ \z^{-1}\Y(t,x,y)\ +\ O(\z^{-2}) \bigr)\,\Psi^0
\end{equation}
then defines a $u(n)$-valued field $\Y(t,x,y)$. 
Again this allows one to reconstruct $A$ and~$B$ through
\begin{equation} \label{ABfromPsi3}
A\=-\lim_{\z\to\infty} \bigl(\z\,\Psi\ast\pa_x\Psi^{-1}\bigr) \=\pa_x\Y
\quad\textrm{and}\quad
B\=-\lim_{\z\to\infty} \bigl(\z\,\Psi\ast\pa_v\Psi^{-1}\bigr) \=\pa_v\Y
\end{equation}
In this parametrization, compatibility equation (\ref{comp1}) becomes an
identity but the second equation~(\ref{comp2}) turns into~\cite{olaf4}
\begin{equation} \label{leznovtype}
\pa_x^2\Y -\pa_u\pa_v\Y -
\pa_x \Y \ast \pa_v \Y + \pa_v \Y \ast \pa_x \Y \ =\ 0 \quad.
\end{equation}

This Leznov-type equation~\cite{L} can also be obtained by extremizing
the action
\begin{equation} \label{Laction}
S_{\textrm{L}}\= \int\!\diff{t}\,\diff{x}\,\diff{y}\ \tr\,\Bigl\{
\sfrac12\,\eta^{ab}\,\pa_a \Y \,\ast\, \pa_b \Y \ +\ 
\sfrac13\,\Y \ast 
\bigl( \pa_x \Y\,\ast\,\pa_v \Y - \pa_v \Y\,\ast\,\pa_x \Y \bigr) 
\Bigr\} \quad,
\end{equation}
which is merely cubic.

Obviously, the Leznov field $\Y$ is related to the Yang field $\Phi$ 
through the non-local field redefinition
\begin{equation} \label{nonlocal}
\pa_x\Y\=\Phi^{-1}\ast\pa_u\Phi
\qquad\textrm{and}\qquad
\pa_v\Y\=\Phi^{-1}\ast\pa_x\Phi \quad.
\end{equation}
For each of the two fields $\Phi$ and~$\Y$, one equation from the pair 
(\ref{comp1}, \ref{comp2}) represents the equation of motion,
while the other one is a direct consequence of the parametrization
(\ref{ABfromPsi2}) or~(\ref{ABfromPsi3}).
\subsection{Reduction to noncommutative sine-Gordon}
\subsubsection{Algebraic reduction ansatz}
In section 2.1.3 we have seen that the (commutative) sine-Gordon equation can be obtained
from the self-duality equations for SU(2) Yang-Mills upon appropriate
reduction from $2{+}2$ to $1{+}1$ dimensions. In this process the integrable 
sigma model of the previous section appears as an intermediate step in
$2{+}1$ dimensions, and so we may take its noncommutative extension as our 
departure point, after enlarging the group to U(2). 
In order to avoid cluttering the formulae we suppress the `$\ast$' notation
for noncommutative multiplication from now on: all products are assumed to
be star products, and all functions are built on them, i.e.~$e^{f(x)}$
stands for $e_\ast^{f(x)}$ and so on.

The dimensional reduction proceeds in two steps, firstly, a factorization
of the coordinate dependence and, secondly, an algebraic restriction of
the form of the U(2) matrices involved.
In the language of the linear system~(\ref{linsys}) the adequate ansatz
for the auxiliary field~$\Psi$ reads
\begin{equation} \label{ansatzPsi}
\Psi(t,x,y,\z) \= V(x)\,\psi(u,v,\z)\,V^\+(x)
\qquad\textrm{with}\qquad 
V(x) \= \Ecal\,e^{\im\a\,x\,\s_1} \quad,
\end{equation}
where $\s_1=(\begin{smallmatrix} 0 & 1 \\ 1 & 0 \end{smallmatrix})$,
$\Ecal$ denotes some constant unitary matrix (to be specified later)
and $\a$ is a constant parameter. Under this factorization, 
the linear system~(\ref{linsys}) simplifies to\footnote{
The adjoint action means $\mathrm{ad}\s_1\,(\psi)=[\s_1,\psi]$.}
\begin{equation}
(\pa_u -\im\a\,\z\,\mathrm{ad}\s_1)\,\psi\=-a\,\psi \qquad\textrm{and}\qquad
(\z\pa_v -\im\a\,\mathrm{ad}\s_1)\,\psi\= b\,\psi 
\end{equation}
with $\ a=V^\+A\,V\ $ and $\ b=V^\+B\,V$.
Taking into account the asymptotic behavior (\ref{asymp1}, \ref{asymp2}), 
the ansatz~(\ref{ansatzPsi}) translates to the decompositions
\begin{align} \label{ansatzPhi}
\Phi(t,x,y) &\= V(x)\,g(u,v)\,V^\+(x)
\qquad\textrm{with}\qquad g(u,v) \in \textrm{U(2)}
\quad, \\[6pt] \label{ansatzUps}
\Y(t,x,y) &\= V(x)\,\chi(u,v)\,V^\+(x) 
\qquad\textrm{with}\qquad \chi(u,v) \in u(2) \quad.
\end{align}
To aim for the sine-Gordon equation, one imposes certain algebraic 
constraints on $a$ and $b$ (and therefore on~$\psi$).
Their precise form, however, is not needed, as we are ultimately interested
only in $g$ or~$\chi$. Therefore, we instead directly restrict $g(u,v)$
to the form
\begin{equation} \label{reducg}
g \= \Bigl(\begin{matrix} g_+ & 0 \\ 0 & g_- \end{matrix}\Bigr)
\= g_+ P_+ + g_- P_-
\qquad\textrm{with}\qquad g_+ \in \textrm{U(1)}_+ 
\quad\textrm{and}\quad    g_- \in \textrm{U(1)}_- 
\end{equation}
and with projectors $P_+=(\begin{smallmatrix}1&0\\0&0\end{smallmatrix})$
and $P_-=(\begin{smallmatrix}0&0\\0&1\end{smallmatrix})$.
This imbeds $g$ into a U(1)$\times$U(1) subgroup of~U(2).
Note that $g_+$ and $g_-$ do not commute, due to the implicit star product.
Invoking the field redefinition~(\ref{nonlocal}) we infer that the
corresponding reduction for~$\chi(u,v)$ should be\footnote{
Complex conjugates of scalar functions are denoted with a dagger
to remind the reader of their noncommutativity.}
\begin{equation} \label{reduch}
\chi \= \im \Bigl(\begin{matrix} 0 & h^\+ \\ h & 0 \end{matrix}\Bigr)
\qquad\textrm{with}\qquad h \in \C \quad,
\end{equation}
with the ``bridge relations''
\begin{equation} \label{nonlocal2}
\begin{aligned}
\a\,(h-h^\+) &\= - g_+^\+ \pa_u g_+ \= g_-^\+ \pa_u g_- \quad,\\[6pt]
\sfrac{1}{\a}\,\pa_v h &\= g_-^\+ g_+ - \mbf{1}
\qquad\textrm{and h.c.} \quad.
\end{aligned}
\end{equation}
In this way, the $u(2)$-matrix $\chi$ is restricted to be off-diagonal.

We now investigate in turn the consequences of the ans\"atze 
(\ref{ansatzPhi}, \ref{reducg}) and (\ref{ansatzUps}, \ref{reduch}) 
for the equations of motion (\ref{yangtype}) and (\ref{leznovtype}),
respectively.
\subsubsection{Reduction of Yang-type equation }
Let us insert the ansatz~(\ref{ansatzPhi}) into the
Yang-type equation of motion~(\ref{yangtype}).
After stripping off the $V$ factors one obtains
\begin{equation}
\pa_v (g^\+ \pa_u g) + \a^2 (\s_1 g^\+ \s_1 g - g^\+ \s_1 g \s_1) \= 0 \quad.
\end{equation}
Specializing with (\ref{reducg}) and employing the identities
$\ \s_1 P_\pm \s_1 = P_\mp\ $ we arrive at $\ Y_+P_+ + Y_-P_- =0$, 
with
\begin{equation} \label{Yg}
\begin{aligned}
Y_+ &\ \equiv\ 
\pa_v (g_+^\+ \pa_u g_+) + \a^2 (g_-^\+ g_+ - g_+^\+ g_-) \= 0 \quad, \\[6pt]
Y_- &\ \equiv\
\pa_v (g_-^\+ \pa_u g_-) + \a^2 (g_+^\+ g_- - g_-^\+ g_+) \= 0 \quad.
\end{aligned}
\end{equation}
Since the brackets multiplying~$\a^2$ are equal and opposite,
it is worthwhile to present the sum and the difference of the two equations:
\begin{equation} \label{Yg2}
\begin{aligned}
\pa_v\bigl( g_+^\+ \pa_u g_+ + g_-^\+ \pa_u g_- \bigr) &\= 0 \quad,\\[6pt]
\pa_v\bigl( g_+^\+ \pa_u g_+ - g_-^\+ \pa_u g_- \bigr) &\=
2\a^2 \bigl( g_+^\+ g_- - g_-^\+ g_+ \bigr) \quad.
\end{aligned}
\end{equation}

It is natural to introduce the angle fields $\phi_\pm(u,v)$ via
\begin{equation} \label{para1}
g \= e^{\frac{\im}{2}\phi_+P_+}\,e^{-\frac{\im}{2}\phi_-P_-}
\qquad\Leftrightarrow\qquad
g_+ \= e^{\frac{\im}{2}\phi_+} 
\qquad\textrm{and}\qquad 
g_- \= e^{-\frac{\im}{2}\phi_-} \quad.
\end{equation}
In terms of these, the equations~(\ref{Yg2}) read
\begin{equation} \label{Yphi}
\begin{aligned}
\pa_v\bigl( e^{-\frac{\im}{2}\phi_+}\,\pa_u e^{\frac{\im}{2}\phi_+} +
            e^{\frac{\im}{2}\phi_-}\,\pa_u e^{-\frac{\im}{2}\phi_-} \bigr) &\=0
\quad,\\[6pt]
\pa_v\bigl( e^{-\frac{\im}{2}\phi_+}\,\pa_u e^{\frac{\im}{2}\phi_+} -
            e^{\frac{\im}{2}\phi_-}\,\pa_u e^{-\frac{\im}{2}\phi_-} \bigr) &\=
2\a^2\bigl( e^{-\frac{\im}{2}\phi_+}e^{-\frac{\im}{2}\phi_-} -
            e^{\frac{\im}{2}\phi_-}e^{\frac{\im}{2}\phi_+} \bigr) \quad.
\end{aligned}
\end{equation}
We propose to call these two equations 
``the noncommutative sine-Gordon equations''.
Besides their integrability (see later sections for consequences)
their form is quite convenient for studying the commutative limit.
When $\th\to0$, (\ref{Yphi}) simplifies to
\begin{equation} \label{Ycomm}
\pa_u\pa_v (\phi_+{-}\phi_-) \= 0 \qquad\textrm{and}\qquad
\pa_u\pa_v (\phi_+{+}\phi_-) \= -8\a^2\,\sin\sfrac12(\phi_+{+}\phi_-) \quad.
\end{equation}
Because the equations have decoupled we may choose 
\begin{equation}
\phi_+ \= \phi_- \ =:\ \phi  \qquad\Leftrightarrow\qquad
g_+ \= g_-^\+ \qquad\Leftrightarrow\qquad
g \in \textrm{U(1)}_{\textrm{A}}
\end{equation}
and reproduce the familiar sine-Gordon equation
\begin{equation} \label{sG}
(\pa_t^2 -\pa_y^2)\,\phi \= -4\a^2\,\sin\phi \quad.
\end{equation}
One learns that in the commutative case the reduction is 
SU(2)$\to$U(1)$_{\textrm{A}}$ since the U(1)$_{\textrm{V}}$
degree of freedom $\phi_+{-}\phi_-$ is not needed.
The deformed situation, however, requires extending SU(2) to U(2),
and so it is imperative here to keep both U(1)s and work with 
{\it two\/} scalar fields.

Inspired by the commutative decoupling, one may choose another
distinguished parametrization of~$g$, namely
\begin{equation} \label{para2}
g_+ \= e^{\frac{\im}{2}\r}\,e^{\frac{\im}{2}\vp}
\qquad\textrm{and}\qquad
g_- \= e^{\frac{\im}{2}\r}\,e^{-\frac{\im}{2}\vp} \quad,
\end{equation}
which defines angles $\r(u,v)$ and $\vp(u,v)$ for the linear combinations 
$\textrm{U}(1)_{\textrm{V}}$ and $\textrm{U}(1)_{\textrm{A}}$, respectively.
Inserting this into (\ref{Yg}) one finds
\begin{equation} \label{Yrho}
\begin{aligned}
\pa_v\bigl( e^{-\frac{\im}{2}\vp}\,\pa_u e^{\frac{\im}{2}\vp} \bigr) + 
2\im\a^2\,\sin\vp &\= 
-\pa_v\bigl[e^{-\frac{\im}{2}\vp}e^{-\frac{\im}{2}\r}\,
(\pa_u e^{\frac{\im}{2}\r}) e^{\frac{\im}{2}\vp} \bigr] \quad,\\[6pt]
\pa_v\bigl( e^{\frac{\im}{2}\vp}\,\pa_u e^{-\frac{\im}{2}\vp} \bigr) -
2\im\a^2\,\sin\vp &\=
-\pa_v\bigl[e^{\frac{\im}{2}\vp}e^{-\frac{\im}{2}\r}\,
(\pa_u e^{\frac{\im}{2}\r}) e^{-\frac{\im}{2}\vp} \bigr] \quad.
\end{aligned}
\end{equation}
In the commutative limit, this system is easily decoupled to
\begin{equation} \label{sG2}
\pa_u \pa_v \r \=0 \qquad\textrm{and}\qquad
\pa_u \pa_v \vp + 4\a^2\,\sin\vp \= 0 \quad,
\end{equation}
revealing that $\ \r\to\frac12(\phi_+{-}\phi_-)\ $ and 
$\ \vp\to\frac12(\phi_+{+}\phi_-)=\phi\ $ in this limit.

It is not difficult to write down an action for (\ref{Yg})
(and hence for (\ref{Yphi}) or (\ref{Yrho})). The relevant action may be
computed by reducing (\ref{Yaction}) with the help of (\ref{ansatzPhi})
and~(\ref{reducg}). The result takes the form
\begin{equation} \label{gaction}
S[g_+,g_-] \= S_{W}[g_+]\,+\,S_{W}[g_-]\,+\,\alpha^2\int\!\diff{t}\,\diff{y}\;
\bigl( g_+^{\dag} g_- + g_-^\dag g_+ - 2 \bigr) \quad,
\end{equation}
where $S_W$ is the abelian WZW action
\begin{equation} \label{WZWaction}
S_{W}[f]\ \equiv\ - \sfrac12\int\!\intd t\,\intd y\; \pa_v f^{-1}\; \pa_u f 
\,-\,\sfrac13\int\!\intd t\,\intd y\int_0^1\!\intd \l\;\e^{\m\n\s}\,
\hat f^{-1}\pa_\m\hat f\;\hat f^{-1}\pa_\n\hat f\;\hat f^{-1}\pa_\s\hat f
\end{equation}
Here $\hat{f}(\l)$ is a homotopy path satisfying the conditions
$\hat{f}(0) = 1$ and $\hat{f}(1) = f$.
Parametrizing $g_{\pm}$ as in (\ref{para2}) and using the Polyakov-Wiegmann
identity, the action for $\rho$ and $\varphi$ reads
\begin{equation} \label{rhophiaction}
\begin{aligned}
S[\rho,\varphi] &\= 2 S_{PC}\bigl[ e^{\frac{\im}{2}\varphi}\bigr] \,+\,
2\a^2\int\!\intd t\,\intd y\;\bigl( \cos{\varphi} -1 \bigr)\,+\,
2 S_{W}\bigl[ e^{\frac{\im}{2}\rho}\bigr] \\ 
&\qquad - \int\!\intd t\,\intd y\; 
e^{-\frac{\im}{2}\rho}\,\pa_v e^{\frac{\im}{2}\rho} 
\bigl( e^{-\frac{\im}{2}\varphi}\,\pa_u e^{\frac{\im}{2}\varphi}
     + e^{\frac{\im}{2}\varphi}\,\pa_u e^{-\frac{\im}{2}\varphi}\bigr)\quad,
\end{aligned}
\end{equation}
where
\begin{equation}
S_{PC}[f]\ \equiv\ -\sfrac12\int\!\intd t\,\intd y\;\pa_v f^{-1}\;\pa_u f\quad.
\end{equation}
In this parametrization the WZ term has apparently been shifted entirely to
the $\rho$ field while the cosine-type self-interaction remains for the
$\varphi$ field only. This fact has important consequences for the scattering
amplitudes.

It is well known \cite{wittenwzw, bosonization, bosonization2}
that in ordinary commutative geometry the bosonization of $N$ free massless 
fermions in the fundamental representation of SU($N$) gives rise to a WZW model
for a scalar field in SU($N$) plus a free scalar field associated with 
the U(1) invariance of the fermionic system.
In the noncommutative case the bosonization of a single massless Dirac fermion 
produces a noncommutative U(1) WZW model~\cite{MS}, which becomes free only 
in the commutative limit. Moreover, the U(1) subgroup of U($N$) does no 
longer decouple~\cite{closure}, so that $N$ noncommuting free massless 
fermions are related to a noncommutative WZW model for a scalar in U($N$).
On the other hand, giving a mass to the single Dirac fermion leads to a 
noncommutative cosine potential on the bosonized side~\cite{quevedo,schiappa}.

In contrast, the noncommutative sine-Gordon model we propose in this paper 
is of a more general form.
The action~(\ref{gaction}) describes the propagation of a scalar field $g$ 
taking its value in U(1)$\times$U(1) $\subset$ U(2). Therefore, we expect it 
to be a bosonized version of two fermions in some representation of 
U(1)$\times$U(1). The absence of a WZ term for $\varphi$ and the lack
of a cosine-type self-interaction for~$\rho$ as well as the non-standard
interaction term make the precise identification non-trivial however.
\subsubsection{Reduction of Leznov-type equation }
Alternatively, if we insert the ansatz~(\ref{ansatzUps}) into the 
Leznov-type equation of motion~(\ref{leznovtype}) we get
\begin{equation}
\pa_u\pa_v\chi+2\a^2(\chi-\s_1\chi\s_1)+\im\a\bigl[[\s_1,\chi],\pa_v\chi\bigr]\=0\quad.
\end{equation}
Specializing with (\ref{reduch}) this takes the form $\ Z\s_-+Z^\+\s_+=0\ $
with $\ \s_-=(\begin{smallmatrix}0&0\\1&0\end{smallmatrix})\ $ and
$\ \s_+=(\begin{smallmatrix}0&1\\0&0\end{smallmatrix})$, where
\begin{equation} \label{Lh}
Z\ \equiv\ \pa_u\pa_v h + 2\a^2\,(h-h^\+)
+ \a\,\bigl\{ \pa_v h\,,\,h-h^\+ \bigr\} \=0 \quad.
\end{equation}
The decomposition
\begin{equation} \label{para3}
\chi \= \im(h_1\s_1 + h_2\s_2) \qquad\Leftrightarrow\qquad h \= h_1 + \im h_2
\end{equation}
then yields
\begin{equation}
\begin{aligned} \label{Lh12}
\pa_u\pa_v h_1 -2\a\,\bigl\{ \pa_v h_2\,,\,h_2 \bigr\} &\=0 \quad, \\[6pt]
\pa_u\pa_v h_2 +4\a^2 h_2 +2\a\,\bigl\{ \pa_v h_1\,,\,h_2 \bigr\} &\=0 \quad.
\end{aligned}
\end{equation}
These two equations constitute an alternative description of the
noncommutative sine-Gordon model; they are classically equivalent to
the pair of~(\ref{Yg2}) or, to be more specific, to the pair of~(\ref{Yrho}). 
For the real fields the ``bridge relations''~(\ref{nonlocal2}) read
\begin{equation} \label{nonlocal3}
\begin{aligned}
& 2\im\a\,h_2 \= -e^{-\frac{\im}{2}\vp}e^{-\frac{\im}{2}\r}\,
\pa_u ( e^{\frac{\im}{2}\r}e^{\frac{\im}{2}\vp} )
\= e^{\frac{\im}{2}\vp}e^{-\frac{\im}{2}\r}\,
\pa_u ( e^{\frac{\im}{2}\r}e^{-\frac{\im}{2}\vp} ) \quad, \\[6pt]
& \qquad\qquad \sfrac{1}{\a}\pa_v h_1 \= \cos\vp-1 \qquad\textrm{and}\qquad
\sfrac{1}{\a}\pa_v h_2 \= \sin\vp \quad.
\end{aligned}
\end{equation}
One may ``solve'' one equation of~(\ref{Yrho}) by an appropriate field 
redefinition from~(\ref{nonlocal3}), which implies already one member 
of~(\ref{Lh12}). The second equation from~(\ref{Yrho}) then yields 
the remaining ``bridge relations'' in~(\ref{nonlocal3}) as well as 
the other member of~(\ref{Lh12}). This procedure works as well in the 
opposite direction, from~(\ref{Lh12}) to (\ref{Yrho}). 
The nonlocal duality between $(\vp,\r)$ and $(h_1,h_2)$ is simply a 
consequence of the equivalence between (\ref{yangtype}) and (\ref{leznovtype})
which in turn follows from our linear system~(\ref{linsys}).

The ``$h$~description'' has the advantage of being polynomial.
It is instructive to expose the action for the system~(\ref{Lh12}).
Either by inspection or by reducing the Leznov action~(\ref{Laction})
one obtains
\begin{equation}
S[h_1,h_2] \= \int\!\diff{t}\,\diff{y}\;
\Bigl\{ \pa_u h_1 \pa_v h_1 + \pa_u h_2 \pa_v h_2 
-4\a^2 h_2^2 -4\a\,h_2^2\,\pa_v h_1 \Bigr\} \quad.
\label{haction}
\end{equation}
\subsection[Relation with the previous noncommutative\\ generalization of the sine-Gordon model]{Relation with the previous noncommutative generalization of the sine-Gordon model}
The noncommutative generalizations of the sine-Gordon model presented above
are expected to possess an infinite number of conservation laws, as they
originate from the reduction of an integrable model \cite{olaf4}. It is 
worthwhile to point out their relation to the noncommutative
sine-Gordon model I discussed in section 2.2, which also features an infinite number of local conserved 
currents.  

In \cite{us} an alternative noncommutative version of the sine-Gordon model 
was proposed. Using the bicomplex approach the equations of motion were 
obtained as flatness conditions of a bidifferential calculus,\footnote{
This subsection switches to Euclidean space $\R^2$, where $\pa$ and
$\bar\pa$ are derivatives with respect to complex coordinates.}
\beq
\bar{\pa} ( G^{-1} \ast \pa G) \= [ R\,,\, G^{-1} \ast R\,G ]_{\ast} \quad,
\label{eq}
\eeq
where  
\beq
R \= 2\alpha\, \Bigl(\begin{matrix} 0 & 0 \\ 0 & 1 \end{matrix}\Bigr) 
\eeq
and $G$ is a suitable matrix in U(2) or, more generally, in complexified
U(2). In \cite{us} the $G$ matrix was chosen as
\beq \label{sgdif}
G \= e_{\ast}^{\frac{\im}{2} \s_2 \Phi} \= \biggl( \begin{matrix} 
\phantom{-}\cos_{\ast}{\frac{\Phi}{2}} & \ \sin_{\ast}{\frac{\Phi}{2}} 
\\[4pt]
-\sin_{\ast}{\frac{\Phi}{2}} & \ \cos_{\ast}{\frac{\Phi}{2}} 
\end{matrix} \biggr)
\eeq
with $\Phi$ being a complex scalar field. This choice produces the 
noncommutative equations (all the products are $\ast$-products)
\bea
&& 
\bar{\pa} \bigl( e^{\frac{\im}{2} \Phi}  \pa e^{-\frac{\im}{2} \Phi} 
+ e^{-\frac{\im}{2} \Phi}   \pa e^{\frac{\im}{2} \Phi} \bigr) 
~=~ 0 \quad,
\nonumber \\
&&
\bar{\pa} \bigl( e^{-\frac{\im}{2} \Phi}  \pa e^{\frac{\im}{2} \Phi} 
- e^{\frac{\im}{2} \Phi}   \pa e^{-\frac{\im}{2} \Phi} \bigr) 
~=~ 4\im\alpha^2 \sin{\Phi} \quad.
\label{sgnew}
\eea
As shown in \cite{mypaper2} these equations (or a linear combination of them) 
can be obtained as a dimensional reduction of the equations of motion
for noncommutative U(2) SDYM in 2+2 dimensions.

The equations (\ref{sgnew}) can also be derived from an action which consists 
of the sum of two WZW actions augmented by a cosine potential,
\beq
S[f,\bar f]\=S[f]+S[\bar f]  \qquad\text{with}\qquad 
S[f] \ \equiv\ S_W[f] -
\alpha^2 \int\!\diff{t}\,\diff{y}\; \bigl( f^2+ f^{-2} -2 \bigr) \quad,
\label{sgaction}
\eeq
with $S_W[f]$ given in (\ref{WZWaction}) for $f\equiv e^{\frac{\im}{2}\Phi}$
in complexified U(1). 
However, this action cannot be obtained from the SDYM action in 2+2 dimensions
by performing the same field parametrization which led to (\ref{sgnew}). 

Comparing the actions (\ref{gaction}) and (\ref{sgaction}) 
and considering $f$ and $\bar{f}$ as independent U(1) group valued fields
we are tempted to formally identify $f \equiv g_+$ and $\bar{f} \equiv g_-$. 
Doing this, we immediately realize that the two models differ in
their interaction term which generalizes the cosine potential. 
While in (\ref{sgaction}) the fields $f$ and $\bar{f}$ show only self-interaction,
the fields $g_+$ and $g_-$ in (\ref{gaction}) interact with each other. 
As we will see in section 2.3.6 this makes a big difference when evaluating the
S-matrix elements.

We close this section by observing that the equations of motion (\ref{Yphi})
can also be obtained directly in two dimensions by using the bicomplex
approach described in \cite{us}. In fact, if instead of (\ref{sgdif}) we choose
\beq
G \= 
\biggl( \begin{matrix} 
e^{\frac{\im}{2}\phi_+} + e^{-\frac{\im}{2}\phi_-}
& \ -\im e^{\frac{\im}{2}\phi_+} +\im e^{-\frac{\im}{2}\phi_-} \\[4pt]
\im e^{\frac{\im}{2}\phi_+} -\im e^{-\frac{\im}{2}\phi_-}
& \phantom{-}\ e^{\frac{\im}{2}\phi_+} + e^{-\frac{\im}{2}\phi_-} 
\end{matrix} \biggr)
\eeq
it is easy to prove that (\ref{eq}) yields exactly the set of 
equations (\ref{Yphi}). Therefore, by exploiting the results in \cite{us}
it should be straightforward to construct the first nontrivial conserved 
currents for the present model. 
\subsection{Solitons}
\subsubsection{Dressing approach in 2+1 dimensions. }
The existence of the linear system allows for powerful methods to
systematically construct explicit solutions for $\Psi$ and hence
for $\Phi^\+=\Psi|_{\z=0}$ or $\Y$. 
For our purposes the so-called dressing method
\cite{dressing,curvature}
proves to be most practical, and so we shall first present it here for
our linear system~(\ref{linsys}), before reducing the results to
solitonic solutions of the noncommutative sine-Gordon equations.

The central idea is to demand analyticity in the spectral parameter~$\z$
for the linear system~(\ref{linsys}), which strongly restricts the possible
form of~$\Psi$. The most elegant way to exploit this constraint starts from
the observation that the left hand sides of the differential relations
(D):=(\ref{ABfromPsi}) as well as the reality condition (R):=(\ref{real})
do not depend on~$\z$ while their right hand sides
are expected to be nontrivial functions of~$\z$ (except for the
trivial case $\Psi=\Psi^0$). More specifically, $\C P^1$ being compact,
the matrix function~$\Psi(\z)$ cannot be holomorphic everywhere but must
possess some poles, and hence the right hand sides of (D) and
(R) should display these (and complex conjugate) poles as well.
The resolution of this conundrum demands that the residues of the 
right hand sides at any would-be pole in~$\z$ have to vanish. 
We are now going to evaluate these conditions.

The dressing method builds a solution $\Psi_N(t,x,y,\z)$ featuring
$N$~simple poles at positions $\m_1,\m_2,\ldots,\m_N$ by left-multiplying 
an $(N{-}1)$-pole solution $\Psi_{N-1}(t,x,y,\z)$ with a single-pole factor 
of the form $\ \bigl(1+\frac{\m_N{-}\mb_N}{\z{-}\m_N}P_N(t,x,y)\bigr)$, 
where the $n{\times}n$ matrix function $P_N$ is yet to be determined. 
In addition, we are free to right-multiply $\Psi_{N-1}(t,x,y,\z)$ with some
constant unitary matrix~$\Psh^0_N$.  Starting from $\Psi_0=\mbf{1}$, 
the iteration $\ \Psi_0\mapsto\Psi_1\mapsto\ldots\mapsto\Psi_N\ $
yields a multiplicative ansatz for $\Psi_N$ which, via partial fraction 
decomposition, may be rewritten in an additive form (as a sum of simple pole
terms). Let us trace this iterative procedure constructively.

In accord with the outline above, the one-pole ansatz must read
($\Psh^0_1=:\Psi^0_1$)
\begin{equation} \label{Psione}
\Psi_1 \= \Bigl(\mbf{1}\,+\,\frac{\m_1-\mb_1}{\z-\m_1}\,P_1\Bigr)\,\Psi^0_1
\= \Bigl(\mbf{1} \,+\,\frac{\Lambda_{11}S_1^\+}{\z-\m_1}\Bigr)\,\Psi^0_1
\end{equation}
with some $n{\times}r_1$ matrix functions $\Lambda_{11}$ and~$S_1$
for some $1{\le}r_1{<}n$. The normalization matrix~$\Psi^0_1$ is 
constant and unitary. It is quickly checked that
\begin{equation} \label{R1}
\res_{\z=\mb_1} (R) =0 \quad\Rightarrow\quad
P_1^\+ \= P_1 \= P_1^2 \quad\Rightarrow\quad
P_1 \= T_1\,(T_1^\+ T_1)^{-1} T_1^\+ \quad,
\end{equation}
meaning that $P_1$ is a rank~$r_1$ projector built from an $n{\times}r_1$ 
matrix function~$T_1$. The columns of~$T_1$ span the image of~$P_1$ and 
obey $P_1T_1=T_1$. When using the second parametrization of~$\Psi_1$
in~(\ref{Psione}) one finds that
\begin{equation} \label{R2}
\res_{\z=\mb_1} (R) =0 \quad\Rightarrow\quad
(\mbf{1}-P_1)\,S_1\Lambda_{11}^\+ \=0 \quad\Rightarrow\quad
T_1 \= S_1 \qquad\qquad\qquad{}
\end{equation}
modulo a freedom of normalization.
Finally, the differential relations yield
\begin{equation} \label{D1}
\res_{\z=\mb_1} (D) =0 \quad\Rightarrow\quad
(\mbf{1}-P_1)\,\Lb_1\,(S_1\Lambda_{11}^\+) \=0 \quad\Rightarrow\quad
\Lb_1\,S_1 \= S_1\,\G_1^{A,B}
\end{equation}
for some $r_1{\times}r_1$ matrices~$\G_1^A$ and $\G_1^B$, 
after having defined
\begin{equation}
\bar{L}_i^A\ :=\ \pa_u-\mb_i\pa_x \quad\textrm{and}\quad 
\bar{L}_i^B\ :=\ \m_i(\pa_x-\mb_i\pa_v) \quad\textrm{for}\quad
i=1,2,\ldots,N \quad.
\end{equation}
Because the $\Lb_i$ are linear differential operators it is easy to
write down the general solution for~(\ref{D1}): 
Introduce ``co-moving coordinates''
\begin{equation} \label{comoving}
w_i \ :=\ x + \mb_i u + \mb_i^{-1} v \quad\Rightarrow\quad
\bar{w}_i \= x + \m_i u + \m_i^{-1} v \qquad\textrm{for}\quad
i=1,2,\ldots,N 
\end{equation}
so that on functions of $(w_i,\bar{w}_i)$ alone the $\Lb_i$ act as
\begin{equation}
\bar{L}^A_i \= \bar{L}^B_i \= (\m_i{-}\mb_i)\frac{\pa}{\pa\bar{w}_i} \quad.
\end{equation}
Hence, (\ref{D1}) is solved by
\bea
&&S_1(t,x,y) \= \Sh_1(w_1)\,e^{\bar{w}_1 \G_1 /(\m_1-\mb_1)} \cr
&&~~~\textrm{ 
for any $w_1$-holomorphic $n{\times}r_1$ matrix function $\Sh_1$}
\ena
and $\G_1^A=\G_1^B=:\G_1$.
Appearing to the right of~$\Sh_1$, the exponential factor is seen to drop out
in the formation of~$P_1$ via (\ref{R1}) and~(\ref{R2}). Thus, no generality
is lost by taking $\G_1=0$. We learn that any $w_1$-holomorphic 
$n{\times}r_1$ matrix $T_1$ is admissible to build a projector~$P_1$
which then yields a solution $\Psi_1$ (and thus~$\Phi$) via~(\ref{Psione}).
Note that $\Lambda_{11}$ need not be determined seperately but follows from 
our above result.
It is not necessary to also consider the residues at $\z{=}\m_1$ since
their vanishing leads merely to the hermitian conjugated conditions.

Let us proceed to the two-pole situation. The dressing ansatz takes the form
($\Psi^0_1\Psh^0_2=:\Psi^0_2$)
\begin{equation} \label{Psitwo}
\Psi_2 \= \Bigl(\mbf{1} \,+\,\frac{\m_2-\mb_2}{\z-\m_2}\,P_2\Bigr)
\Bigl(\mbf{1} \,+\,\frac{\m_1-\mb_1}{\z-\m_1}\,P_1\Bigr) \,\Psi^0_2
\= \Bigl(\mbf{1} \,+\,\frac{\Lambda_{21}S_1^\+}{\z-\m_1}
\,+\,\frac{\Lambda_{22}S_2^\+}{\z-\m_2}\Bigr) \,\Psi^0_2 \quad,
\end{equation}
where $P_2$ and $S_2$ are to be determined but $P_1$ and $S_1$ can be copied 
from above. Indeed,
inspecting the residues of (R) and (D) at $\z=\mb_1$ simply confirms that
\begin{equation}
P_1 \= T_1\,(T_1^\+ T_1)^{-1} T_1^\+ \qquad\textrm{and}\qquad 
T_1\=S_1 \qquad\textrm{with}\qquad 
S_1 \= \Sh_1(w_1)
\end{equation}
is just carried over from the one-pole solution.
Relations for $P_2$ and $S_2$ arise from 
\begin{align}
\res_{\z=\mb_2} (R) =0 &\quad\Rightarrow\quad
(\mbf{1}{-}P_2)\,P_2\=0\quad\Rightarrow\quad
P_2 \= T_2\,(T_2^\+ T_2)^{-1} T_2^\+ \ ,\\[6pt]
\res_{\z=\mb_2} (R) =0 &\quad\Rightarrow\quad
\Psi_2(\mb_2)\,S_2\Lambda_{22}^\+ \= 
(\mbf{1}{-}P_2)(1-\sfrac{\m_1-\mb_1}{\m_1-\mb_2}P_1)\,S_2\Lambda_{22}^\+\=0\ ,
\label{TnotS}
\end{align}
where the first equation makes use of the multiplicative form of the
ansatz~(\ref{Psitwo}) while the second one exploits the additive version.
We conclude that $P_2$ is again a hermitian projector (of some rank~$r_2$)
and thus built from an $n{\times}r_2$ matrix function~$T_2$. Furthermore,
(\ref{TnotS}) reveals that $T_2$ cannot be identified with $S_2$ this time,
but we rather have
\begin{equation} \label{T2fromS2}
T_2 \= \Bigl(1-\frac{\m_1{-}\mb_1}{\m_1{-}\mb_2}\,P_1\Bigr)\,S_2
\end{equation}
instead. Finally, we consider
\begin{equation}
\res_{\z=\mb_2} (D) =0 \quad\Rightarrow\quad
\Psi_2(\mb_2)\,\Lb_2\,(S_2\Lambda_{22}^\+) \=0 \quad\Rightarrow\quad
\Lb_2\,S_2 \= S_2\,\G_2^{A,B}
\end{equation}
which is solved by
\beq
S_2(t,x,y) \= \Sh_2(w_2)\,e^{\bar{w}_2 \G_2 /(\m_2-\mb_2)} 
\eeq
for any $w_2$-holomorphic $n{\times}r_2$ matrix function $\Sh_2$
and $\G_2^A=\G_2^B=:\G_2$. Once more, we are entitled to put $\G_2=0$.
Hence, the second pole factor in (\ref{Psitwo})
is constructed in the same way as the first one, except for the small
complication~(\ref{T2fromS2}). Again, $\Lambda_{21}$ and $\Lambda_{22}$ can
be read off the result if needed.

It is now clear how the iteration continues. 
After $N$ steps the final result reads
\begin{equation}
\Psi_N \= \biggl\{ \prod_{\ell=0}^{N-1} \Bigl(\mbf{1} \,+\, 
\frac{\m_{N-\ell}-\mb_{N-\ell}}{\z-\m_{N-\ell}}\,P_{N-\ell} \Bigr)\biggr\}\,
\Psi^0_N
\=\biggl\{\mbf{1}\,+\,\sum_{i=1}^N\frac{\Lambda_{Ni}S_i^\+}{\z-\m_i}\biggr\}
\,\Psi^0_N \quad,
\end{equation}
featuring hermitian rank $r_i$ projectors~$P_i$ at $i=1,2,\ldots,N$, via
\begin{equation}
P_i \= T_i\,(T_i^\+ T_i)^{-1} T_i^\+ \qquad\textrm{with}\qquad
T_i \= \biggl\{ \prod_{\ell=1}^{i-1} \Bigl(\mbf{1} \,-\,
\frac{\m_{i-\ell}-\mb_{i-\ell}}{\m_{i-\ell}-\mb_i}\,P_{i-\ell}\Bigr)\biggl\}
\,S_i
\quad,
\end{equation}
where
\begin{equation}
S_i(t,x,y) \= \Sh_i(w_i)
\end{equation}
for arbitrary $w_i$-holomorphic $n{\times}r_i$ matrix functions $\Sh_i(w_i)$.
The corresponding classical Yang and Leznov fields are
\begin{align} \label{PhiN}
\Phi_N &\= \Psi_N^\+(\z{=}0) \= 
{\Psi^0_N}^\+\,\prod_{i=1}^N \bigl( \mbf{1}-\r_i\,P_i \bigr)
\qquad\textrm{with}\qquad \r_i \= 1-\frac{\m_i}{\mb_i} \quad, \\[6pt]
\Y_N &\=\lim_{\z\to\infty}\z\,\bigl(\Psi_N(\z)\,{\Psi^0_N}^\+-\mbf{1}\bigr)
\= \sum_{i=1}^N (\m_i{-}\mb_i)\,P_i \quad.
\end{align}
The solution space constructed here is parametrized (slightly redundantly) 
by the set $\{\Sh_i\}_1^N$ of matrix-valued holomorphic functions and the 
pole positions~$\m_i$.
The so-constructed classical configurations have solitonic character
(meaning finite energy) when all these functions are algebraic.

The dressing technique as presented above is well known in the commutative
theory; novel is only the realization that it carries over verbatim to the
noncommutative situation by simply understanding all products as star
products (and likewise inverses, exponentials, etc.). Of course, it may be
technically difficult to $\ast$-invert some matrix, but one may always
fall back on an expansion in powers of~$\th$.
\subsubsection{Solitons of the noncommutative sine-Gordon theory}
We should now be able to generate $N$-soliton solutions to the noncommutative
sine-Gordon equations, say~(\ref{Yrho}), by applying the reduction from 
$2{+}1$ to $1{+}1$ dimensions (see previous section) to the above strategy
for the group~U(2), i.e.~putting $n{=}2$. 
In order to find nontrivial solutions, we specify the constant matrix~$\Ecal$
in the ansatz~(\ref{ansatzPsi}) for~$\Psi$ as
\begin{equation}
\Ecal \= e^{-\im\frac{\pi}{4}\s_2} \= \sfrac{1}{\sqrt{2}}
\Bigl(\begin{matrix} 1 & -1 \\ 1 & \phantom{-} 1 \end{matrix}\Bigr)
\end{equation}
which obeys the relations
$\ \Ecal\s_3=\s_1\,\Ecal\ $ and $\ \Ecal\s_1=-\s_3\,\Ecal$.
Pushing $\Ecal$ beyond~$V$ we can write
\begin{equation} \label{Wdef}
\Phi(t,x,y) \= W(x)\,\gt(u,v)\,W^\+(x)
\qquad\textrm{with}\qquad W(x) \= e^{-\im\a\,x\,\s_3}
\end{equation}
and
\begin{equation} \label{grot}
\gt(u,v) \= \Ecal\,g(u,v)\,\Ecal^\+ \= \Ecal\,\biggl( \begin{matrix} 
g_+ & \ 0 \\[4pt] 0 & \ g_- \end{matrix}\biggr)\,\Ecal^\+
\= \sfrac12 \, \biggl(\begin{matrix}
g_+{+}g_- & \ \ g_+{-}g_- \\[4pt] g_+{-}g_- & \ \ g_+{+}g_- 
\end{matrix}\biggr) \quad.
\end{equation}

With hindsight from the commutative case~\cite{curvature} we choose
\begin{equation}
\Psh^0_i \= \s_3 \quad\forall i \qquad\Longleftrightarrow\qquad 
\Psi^0_N \=\s_3^N
\end{equation}
(which commutes with $W$)
and restrict the poles of~$\Psi$ to the imaginary axis, 
$\m_i=\im p_i\ $ with $\ p_i\in\R$.
Therewith, the co-moving coordinates~(\ref{comoving}) become
\begin{equation}
w_i \= x - \im (p_i\,u - p_i^{-1} v) \ =:\ x - \im\h_i(u,v) \quad,
\end{equation}
defining $\h_i$ as real linear functions of the light-cone coordinates.
Consequentially, from~(\ref{PhiN}) we get $\ \r_i=2\ $ and find that
\begin{equation}
\gt_N(u,v) \= \s_3^N\,\prod_{i=1}^N \bigl( \mbf{1}-2\,\Pt_i(u,v) \bigr)
\qquad\textrm{with}\qquad P_i \= W\,\Pt_i\,W^\+ \quad.
\end{equation}
Repeating the analysis of the previous subsection, 
one is again led to construct hermitian projectors
\begin{equation}
\Pt_i \= \Tt_i\,(\Tt_i^\+ \Tt_i)^{-1} \Tt_i^\+ \qquad\textrm{with}\qquad
\Tt_i \= \prod_{\ell=1}^{i-1} \Bigl(\mbf{1} \,-\,
\frac{2\,p_{i-\ell}}{p_{i-\ell}+p_i}\,\Pt_{i-\ell}\Bigr)\,\St_i \quad,
\end{equation}
where $2{\times}1$ matrix functions $\St_i(u,v)$ are subject to
\begin{equation} \label{Dred}
\Lbt_i\,\St_i \= \St_i\,\widetilde{\G}_i
\qquad\textrm{for}\quad  i=1,2,\ldots,N
\end{equation}
and some numbers~$\widetilde\G_i$ (note that now rank $r_i{=}1$) 
which again we can put to zero.
On functions of the reduced co-moving coordinates~$\h_i$ alone, 
\begin{equation}
\Lbt_i \= W^\+ \Lb_i W \=
(\m_i{-}\mb_i)\, W^\+ \frac{\pa}{\pa\bar{w}_i} W \= 
p_i\,\Bigl( \frac{\pa}{\pa\h_i} + \a \,\s_3 \Bigr)
\end{equation}
so that (\ref{Dred}) is solved by
\begin{equation}
\St_i(u,v) \= \widehat{\St}_i(\h_i) \= 
\biggl(\begin{matrix}\g_{i1}e^{-\a\,\h_i}\\[4pt] 
                     \im\g_{i2}e^{+\a\,\h_i}\end{matrix}\biggr)
\=e^{-\a\,\h_i\s_3}\,
\biggl(\begin{matrix} \g_{i1} \\[4pt] \im\g_{i2} \end{matrix}\biggr)
\qquad\textrm{with}\quad \g_{i1}, \g_{i2} \in\C \quad.
\end{equation}
Furthermore, it is useful to rewrite
\begin{equation}
\g_{i1}\g_{i2} =: \l_i^2 \quad\textrm{and}\quad \g_{i2}/\g_{i1} =: \g_i^2
\qquad\Longleftrightarrow\qquad
\biggl(\begin{matrix} \g_{i1} \\[4pt] \im\g_{i2} \end{matrix}\biggr)\=\l_i\,
\biggl(\begin{matrix} \g_i^{-1} \\[4pt] \im\g_i \end{matrix}\biggr)
\end{equation}
because then $|\g_i|$ may be absorbed into $\h_i$ by shifting 
$\a\h_i\mapsto\a\h_i+\ln|\g_i|$. The multipliers~$\l_i$ drop out in the
computation of~$\Pt_i$.
Finally, to make contact with the form~(\ref{grot}) we restrict the
constants $\g_i$ to be real.

Let us check the one-soliton solution, i.e.~put $N{=}1$.
Suppressing the indices momentarily, 
absorbing $\g$ into~$\h$ and dropping~$\l$, we infer that
\bea
\label{onesol}
&&\Tt\=\biggl(\begin{matrix} e^{-\a\h} \\[4pt] \im e^{\a\h} \end{matrix}\biggr)
\quad\Longrightarrow\quad
\Pt\=\frac{1}{2\,\ch 2\a\h} \biggl(\begin{matrix}
e^{-2\a\h} & -\im \\[4pt] \im & e^{+2\a\h} \end{matrix}\biggr)\cr
&&\Longrightarrow\quad
\gt\= \Biggl(\begin{matrix}
\rm{th} 2\a\h & \ \frac{\im}{\ch 2\a\h}\\[6pt] \frac{\im}{\ch 2\a\h} & \ \rm{th} 2\a\h 
\end{matrix}\Biggr)
\ena
which has $\det\,\gt=1$. 
Since here the entire coordinate dependence comes in the single 
combination~$\h(u,v)$, all star products trivialize and the one-soliton 
configuration coincides with the commutative one. Hence, the field~$\r$ 
drops out, $\gt\in$ SU(2), and we find, comparing (\ref{onesol}) with 
(\ref{grot}), that
\begin{equation}
\sfrac12(g_+{+}g_-) \= \cos\sfrac{\vp}{2} \= \rm{th} 2\a\h
\qquad\textrm{and}\qquad
\sfrac{1}{2\im}(g_+{-}g_-) \= \sin\sfrac{\vp}{2} \= \sfrac{1}{\ch 2\a\h}
\end{equation}
which implies
\begin{equation}
\tan\sfrac{\vp}{4} \= e^{-2\a\h} \qquad\Longrightarrow\qquad
\vp \= 4\,\arctan e^{-2\a\h} \= -2\,\arcsin (\rm{th} 2\a\h) \quad,
\end{equation}
reproducing the well known sine-Gordon soliton with mass $\ m=2\a$. 
Its moduli parameters are the velocity $\ \n=\frac{1-p^2}{1+p^2}\ $ 
and the center of inertia $\ y_0=\frac{1}{\a}\sqrt{1{-}\n^2}\ln|\g|\ $ 
at zero time~\cite{curvature}. In passing we note that in the 
``$h$~description'' the soliton solution takes the form
\begin{equation}
h_1 \= p\,\rm{th} 2\a\h \qquad\textrm{and}\qquad 
h_2 \= \sfrac{p}{\ch 2\a\h} \qquad\Longrightarrow\qquad
h \= p\,\rm{th}(\a\h{+}\sfrac{\im\pi}{4}) \= p\,e^{\frac{\im}{2}\vp} \quad.
\end{equation}

Noncommutativity becomes relevant for multi-solitons. 
At $N{=}2$, for instance, one has
\begin{equation}
\begin{aligned}
& \gt_2 \= (1-2\Pt_1)\,(1-2\Pt_2) \qquad\textrm{with}\quad
\Pt_1 = \Pt \quad\textrm{from (\ref{onesol})}\cr
&\textrm{and}\quad
\Pt_2 \= \Tt_2\,(\Tt_2^\+ \Tt_2)^{-1} \Tt_2^\+ \\[6pt]
& \textrm{where}\qquad
\Tt_2 \= \bigl(\mbf{1} - \sfrac{2 p_1}{p_1+p_2} \Pt_1\bigr)\,\widehat{\St}_2
\qquad\textrm{and}\qquad \widehat{\St}_2 \= e^{-\a\,\h_2\s_3}\,
\bigl(\begin{smallmatrix} \g_2^{-1} \\ \im\g_2 \end{smallmatrix}\bigr)\cr 
&\textrm{with}\quad \g_2\in\R \quad.
\end{aligned}
\end{equation}
We refrain from writing down the lengthy explicit expression for~$\gt_2$
in terms of the noncommuting coordinates $\h_1$ and~$\h_2$, 
but one cannot expect to find a unit (star-)determinant for~$\gt_2$
except in the commutative limit. This underscores the necessity of extending
the matrices to~U(2) and the inclusion of a nontrivial $\r$ 
at the multi-soliton level.

It is not surprising that the just-constructed noncommutative sine-Gordon 
solitons themselves descend directly from BPS solutions of the $2{+}1$
dimensional integrable sigma model. Indeed, putting back the $x$~dependence
via~(\ref{Wdef}), the $2{+}1$ dimensional projectors~$P_i$ are built from
$2{\times}1$ matrices
\begin{equation}
S_i \= W(x)\,\Sh_i(\h_i) 
\= \e^{-\im\a\,w_i\s_3}\, 
\biggl(\begin{matrix} \g_i^{-1} \\[4pt] \im\g_i \end{matrix}\biggr) \= 
\biggl(\begin{matrix} 1 \\[4pt] \im\g_i^2\e^{2\im\a\,w_i}\end{matrix}\biggr)\,
\g_i^{-1}e^{-\im\a\,w_i} \quad.
\end{equation}
In the last expression the right factor drops out on the computation of
projectors; the remaining column vector agrees with the standard conventions
\cite{ward,olaf4,curvature,bieling}. Reassuringly, the coordinate dependence has 
combined into~$w_i$. The ensueing $2{+}1$ dimensional configurations~$\Phi_N$
are nothing but noncommutative multi-plane-waves the simplest examples of 
which were already investigated in~\cite{bieling}.
\subsection{(Nice) properties of the S-matrix}
In this section we compute tree-level amplitudes for the noncommutative 
generalization of the sine-Gordon model proposed in section 2.3.3, 
both in the Yang and the Leznov formulation.
In commutative geometry the sine-Gordon S-matrix factorizes in two-particle
processes and no particle production occurs, as a consequence of 
the existence of an infinite number of conservation laws. 
In the noncommutative case it is interesting to investigate whether 
the presence of an infinite number of conserved currents is still
sufficient to guarantee the integrability of the system in the sense 
of having a factorized S-matrix.

The previous noncommutative version of the sine-Gordon model we introduced and studied in section 2.2 is endowed with an infinite 
set of conserved currents. 
In section 2.2.9 we have seen that, despite the existence of an infinite chain of conservation laws, 
particle production occurs in that model and that
the S-matrix is neither factorized nor causal.\footnote{
Acausal behaviour in noncommutative field theory was first observed in 
\cite{causal} and shown to be related to time-space noncommutativity.}
As already stressed in section 2.3.4, the noncommutative generalization of the 
sine-Gordon model we proposed in \cite{mypaper3} and discussed in the present section 2.3 differs from the one studied 
in \cite{us} in the generalization of the cosine potential. Therefore, both 
theories describe the dynamics of two real scalar fields, but 
the structure of the interaction terms between the two fields is different. 
We then expect the scattering amplitudes of the present
theory to behave differently from those of the previous one. To this end 
we will compute the amplitudes corresponding to $2\to 2$ processes for 
the fields $\rho$ and $\vp$ in the $g$-model (Yang formulation) as well as 
for the fields $h_1$ and $h_2$ in the $h$-model (Leznov formulation). 
In the $g$-model we will also compute $2\to 4$ and $3\to 3$ amplitudes 
for the massive field~$\vp$. In both models the S-matrix will turn out to be 
{\em factorized\/} and {\em causal\/} in spite of their time-space 
noncommutativity.
\subsubsection{Amplitudes in the ``$g$-model''. Feynman rules}
We parametrize the $g$-model with $(\rho, \varphi)$ as in (\ref{rhophiaction})
since in this parametrization the mass matrix turns out to be diagonal,
with zero mass for $\rho$ and $m{=}2\alpha$ for $\varphi$.
Expanding the action (\ref{rhophiaction}) up to the fourth order in the
fields, we read off the following Feynman rules: 
\begin{itemize}
\item{The propagators 
\bea
\parbox{2cm}{\includegraphics[width=1.9cm]{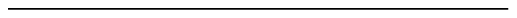}}
&\ \equiv\ &\langle \varphi\varphi\rangle\=\frac{2\im}{k^2-4\a^2}\quad,\\[4pt]
\parbox{2cm}{\includegraphics[width=1.9cm]{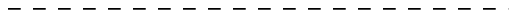}}
&\ \equiv\ &\langle \rho\,\rho \rangle\=\frac{2\im}{k^2}\quad.
\ena}
\item{The vertices 
(including a factor of ``i'' from the expansion of $\e^{\im S}$)
\bea
\parbox{2cm}{\includegraphics[width=1.9cm]{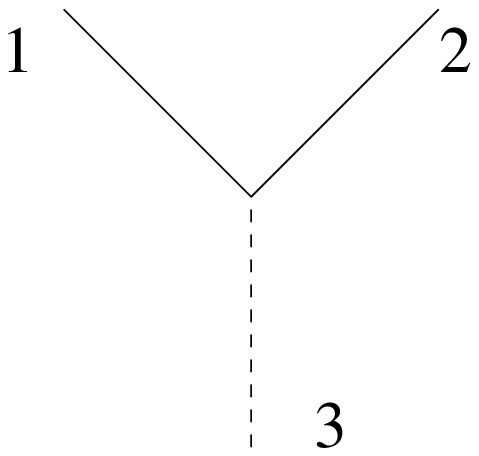}}
&\=&-\frac{1}{2^3}(k_{2}^2-k_1^2-2k_1\wedge k_2) F(k_1,k_2,k_3)\\
\parbox{2cm}{\includegraphics[width=1.9cm]{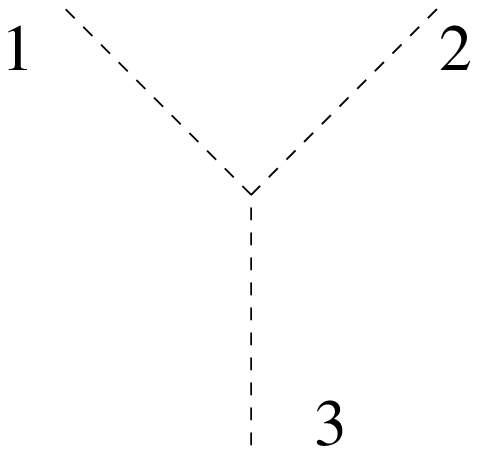}}
&\=&\frac{1}{2\cdot 3!}\;k_{1}\wedge k_2\; F(k_1,k_2,k_3)
\ena}
\bea
\parbox{2cm}{\includegraphics[width=1.9cm]{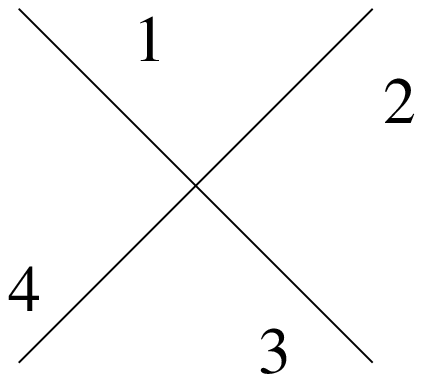}}
&\=&\Bigl[-\frac{\im}{2^3\cdot4!}\,(k_1^2+3k_1\cdot k_3)+
\frac{2\im\a^2}{4!}\Bigr]F(k_1,k_2,k_3,k_4)\cr
&&~~~~\\
\parbox{2cm}{\includegraphics[width=1.9cm]{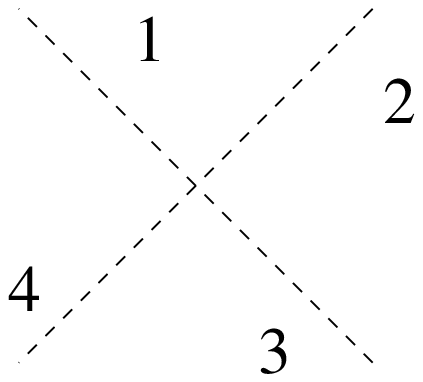}}
&\=&-\frac{\im}{2^3\cdot 4!}\,(k_1^2+3k_1\cdot k_3)\,F(k_1,k_2,k_3,k_4)\cr
&&~~~~\\
\parbox{2cm}{\includegraphics[width=1.9cm]{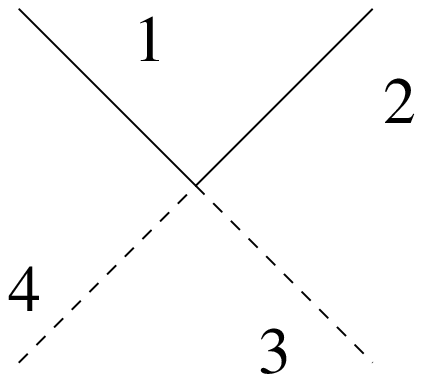}}
&\=&-\frac{\im}{2^5}\,(k_1^2-k_2^2+2 k_1\cdot k_3-2 k_2\cdot k_3+\,2k_1\wedge k_2+2k_1\wedge k_3 \non\\
&&+2k_3\wedge k_2)\,F(k_1,k_2,k_3,k_4)
\ena
\end{itemize}
where we used the conventions of section 2 with the definitions
\beq
u\cdot v \= - \eta^{ab}\,u_a v_b   
\= u_tv_t - u_yv_y \qquad\text{and}\qquad
u\wedge v\=   u_tv_y-u_yv_t \quad.
\eeq
Moreover, we have defined
\beq
F(k_1,\dots, k_n) \= \exp \bigl\{ -\sfrac{\im}{2} \theta
\textstyle{\sum_{i<j}^n} k_i \wedge k_j \bigr\} \quad.
\eeq
and use the convention that all momentum lines are entering the
vertex and energy-momentum conservation has been taken into account. 

We now compute the scattering amplitudes 
$\varphi\varphi \to \varphi\varphi$, $\rho\rho \to \rho\rho$ and 
$\varphi\rho\to \varphi\rho$ 
and the production amplitude $\varphi\varphi \to \rho\rho$. 
We perform the calculations in the center-of-mass frame. We assign the
convention that particles 
with momenta $k_1$ and $k_2$ are incoming, while those with momenta 
$k_3$ and $k_4$ are outgoing.
\subsubsection{Amplitude $\varphi\varphi\to \varphi\varphi$}
The four momenta are explicitly written as
\bea\label{momenta aa-aa}
k_1=(E,p)\ ,\quad k_2=(E,-p)\ ,\quad k_3=(-E,p)\ ,\quad k_4=(-E,-p)\ ,
\ena 
with the on-shell condition $E^2-p^2=4\a^2$.
There are two topologies of diagrams contributing to this process. 
Taking into account the leg permutations corresponding to the same 
particle at a single vertex, the contributions read\\
\begin{center}\begin{tabular}{r@{}clr@{}cl}
\parbox{1.9cm}{\includegraphics[width=1.9cm]{vertice1111.eps}}
&=&$2\im\a^2 \cos^2 (\theta Ep)\quad,$ & 
\parbox{2.4cm}{\includegraphics[width=2.4cm]{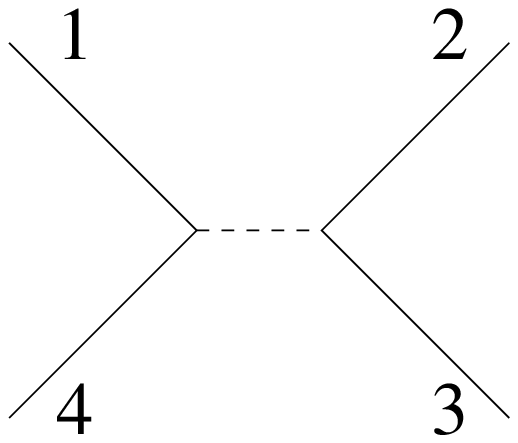}}
&=&$0\quad,$\\
\parbox{1.9cm}{\includegraphics[width=1.9cm]{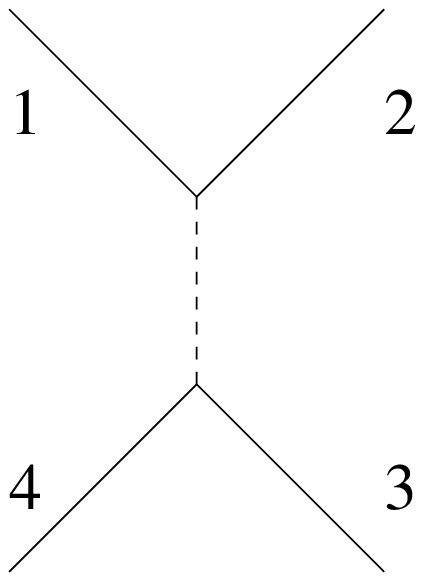}}
&=&$-\frac{\im}{2}p^2\sin^2 (\theta Ep)\quad,$ &
\parbox{2.4cm}{\includegraphics[width=2.4cm]{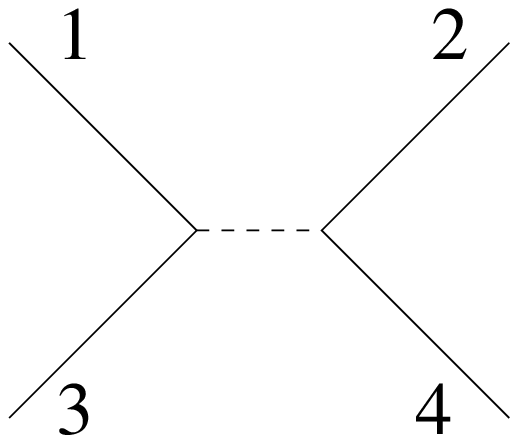}}
&=&$ \frac{\im}{2}E^2\sin^2 (\theta Ep)\quad.$ 
\end{tabular}\\ \end{center}
\noindent
The second diagram is actually affected by a collinear divergence
since the total momentum $k_1+k_4$ for the internal massless particle
is on-shell vanishing. 
We regularize this divergence by temporarily giving a small mass to
the $\rho$ particle. It is easy to see that the amplitude is 
zero for any value of the small mass since the wedge products $k_1 \wedge 
k_4$ and $k_2 \wedge k_3$ from the two vertices always vanish. 
As an alternative procedure we can put one of the external particles
slightly off-shell, so obtaining a finite result which vanishes in the 
on-shell limit. 

Summing all the contributions, for the $\varphi\varphi \to \varphi\varphi$ 
amplitude we arrive at 
\beq
A_{\varphi\varphi \to \varphi\varphi}\=2\im\a^2 \quad,
\eeq
which perfectly describes a {\em causal\/} amplitude.

A nonvanishing $\varphi\varphi \to \varphi\varphi$ amplitude appears also in 
the noncommutative sine-Gordon proposal of \cite{us,mypaper2}. However, there 
the amplitude has a nontrivial $\theta$-dependence which is responsible for 
acausal behavior. 
Comparing the present result with the result in \cite{mypaper2}, we observe 
that the same kind of diagrams contribute. The main difference is that 
the exchanged particle is now massless instead of massive. 
This crucial difference leads to the cancellation of the 
$\theta$-dependent trigonometric behaviour which in the previous case
gave rise to acausality.
\subsubsection{Amplitude $\rho\rho \to \rho\rho$}
In this case the center-of-mass momenta are given by
\beq
k_1=(E,E)\ ,\quad k_2=(E,-E)\ ,\quad k_3=(-E,E)\ ,\quad k_4=(-E,-E)\ ,
\label{rhomomenta}
\eeq  
where the on-shell condition $E^2-p^2=0$ has already been taken into account. 
For this amplitude we have the following contributions\\
\begin{center}\begin{tabular}{r@{}clr@{}cl}
\parbox{2cm}{\includegraphics[width=1.9cm]{vertice2222.eps}}
&=& $0\quad,$ &
\parbox{2.5cm}{\includegraphics[width=2.4cm]{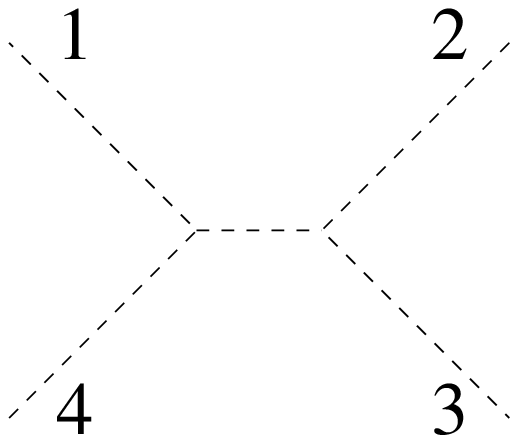}}
&=& $0\quad,$ \\
\parbox{2cm}{\includegraphics[width=1.9cm]{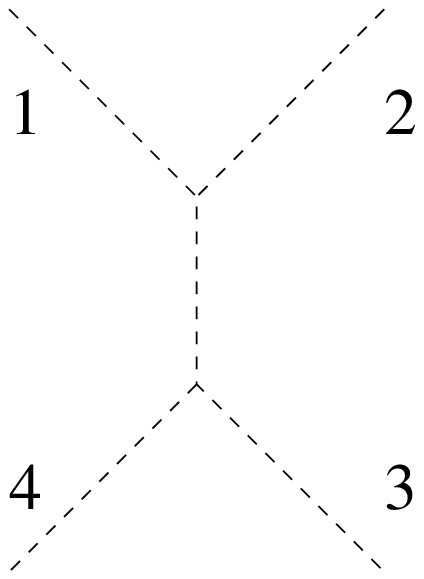}}
&=& $-\frac{\im}{2}E^2\sin^2 (\theta E^2)\quad,$ &
\parbox{2.5cm}{\includegraphics[width=2.4cm]{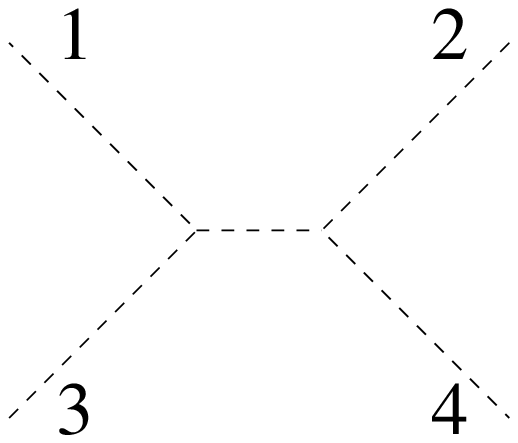}}
&=& $\frac{\im}{2}E^2\sin^2 (\theta E^2)\quad.$
\end{tabular}\\ \end{center}
Again, a collinear divergence appears in the second diagram.
In order to regularize the divergence we can proceed as before by
assigning a small mass to the $\rho$ particle. The main difference with respect
to the previous case is that now the $\rho$ particle also appears as an 
external particle, with the consequence that the on-shell momenta in 
(\ref{rhomomenta}) will get modified by the introduction of a regulator mass.
A careful calculation shows that the amplitude is zero for any value
of the regulator mass, due to the vanishing of the factors 
$k_1 \wedge k_4$ and $k_2 \wedge k_3$ from the vertices.

Therefore, the two nonvanishing contributions add to
\beq
A_{\rho\rho \to \rho\rho}\=0 \quad.
\eeq
\subsubsection{Amplitude $\varphi\rho \to \varphi\rho$}
There are two possible 
configurations of momenta in the center-of-mass frame, describing the
scattering of the massive particle with either a left-moving or a right-moving
massless one. In the left-moving case the momenta are
\bea\label{momenta ab-ab 1} 
k_1=(E,p)\ ,\quad  k_2=(p,-p)\ ,\quad  k_3=(-E,p)\ , \quad  k_4=(-p,-p)\ ,
\ena
while in the right-moving case we have
\bea\label{momenta ab-ab 2} 
k_1=(E,-p)\ ,\quad k_2=(p,p)\ ,\quad k_3=(-E,p)\ , \quad  k_4=(-p,-p)\ .
\ena
For the left-moving case (\ref{momenta ab-ab 1}) the results are\\
\begin{center}
\begin{tabular}{r@{}cl@{\hspace{1cm}}r@{}cl}
\parbox{2cm}{\includegraphics[width=1.9cm]{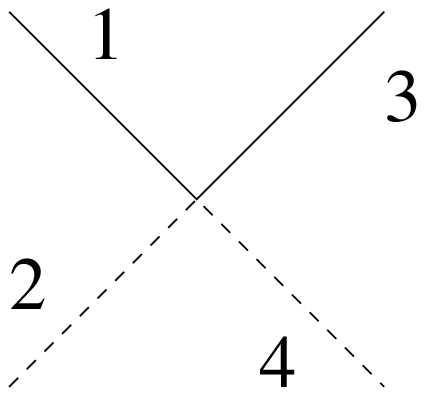}}
&=&\multicolumn{4}{l}{
$-\frac{\im}{2}Ep\,\sin(\theta Ep)\,\sin(\theta p^2)\quad,$}\\
\parbox{2cm}{\includegraphics[width=1.9cm]{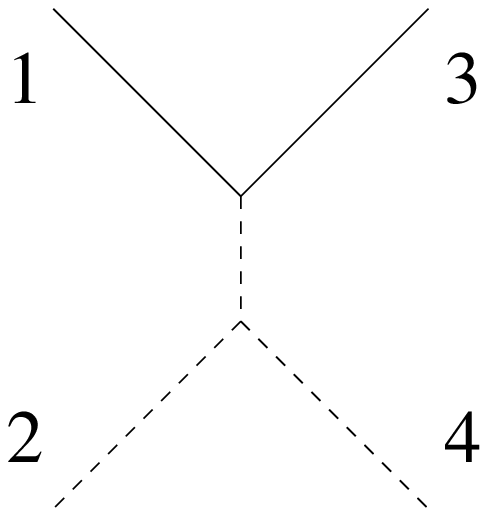}}
&=&\multicolumn{4}{l}{
$\frac{\im}{2}Ep\,\sin(\theta Ep)\,\sin(\theta p^2)\quad,$}\\
\parbox{2.5cm}{\includegraphics[width=2.4cm]{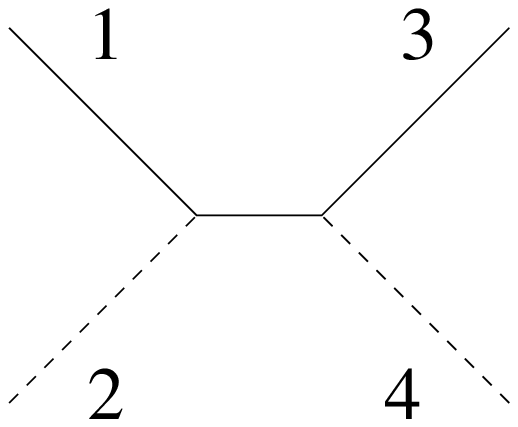}}
&=&$0\quad,$ &
\parbox{2.5cm}{\includegraphics[width=2.4cm]{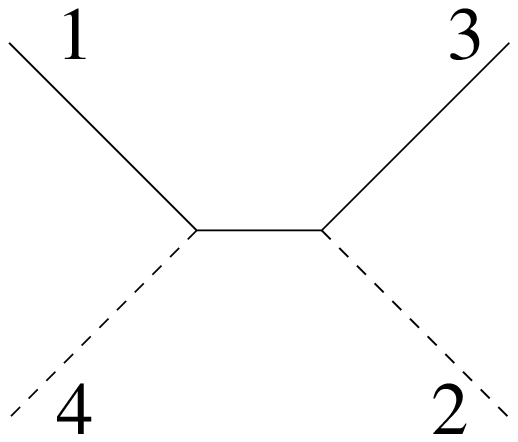}}
&=&$0\quad.$
\end{tabular}\\
\end{center}
For the right-moving choice (\ref{momenta ab-ab 2}), we obtain instead\\
\begin{center}
\begin{tabular}{r@{}cl@{\hspace{1cm}}r@{}cl}
\parbox{2cm}{\includegraphics[width=1.9cm]{varabab.eps}}
&=&$0\quad,$ &
\parbox{2cm}{\includegraphics[width=1.9cm]{varaapbb.eps}}
&=&$0\quad,$ \\
\parbox{2.5cm}{\includegraphics[width=2.4cm]{varabpab.eps}}
&=&$0\quad,$ &
\parbox{2.5cm}{\includegraphics[width=2.4cm]{abpab3.eps}}
&=&$0\quad.$
\end{tabular}\\ 
\end{center}
In this second case an infrared divergence is present due to the massless 
propagator, but again it can be cured as described before.
In both cases the scattering amplitude vanishes,
\beq
A_{\varphi\rho \to \varphi\rho}\=0 \quad.
\eeq
\subsubsection{Amplitude $\varphi\varphi \to \rho\rho$ }
The momenta in the center-of-mass frame are given by
\bea\label{momenta aa-bb}
k_1=(E,p)\ ,\quad k_2=(E,-p)\ ,\quad k_3=(-E,E)\ ,\quad k_4=(-E,-E)\ .
\ena 
In this case we have three kinds of diagrams contributing. 
The corresponding results are\\
\begin{center}
\begin{tabular}{r@{}cl@{\hspace{1cm}}r@{}cl}
\parbox{2cm}{\includegraphics[width=1.9cm]{vertice1122.eps}}
&=& \multicolumn{4}{l}{
$\frac{\im}{2}Ep\,\sin(\theta Ep)\,\sin(\theta E^2)\quad,$} \\
\parbox{2cm}{\includegraphics[width=1.9cm]{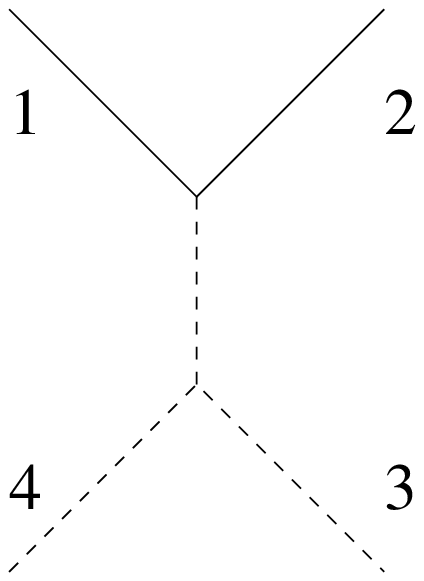}}
&=& \multicolumn{4}{l}{
$-\frac{\im}{2}Ep\,\sin(\theta Ep)\,\sin(\theta E^2)\quad,$} \\
\parbox{2.5cm}{\includegraphics[width=2.4cm]{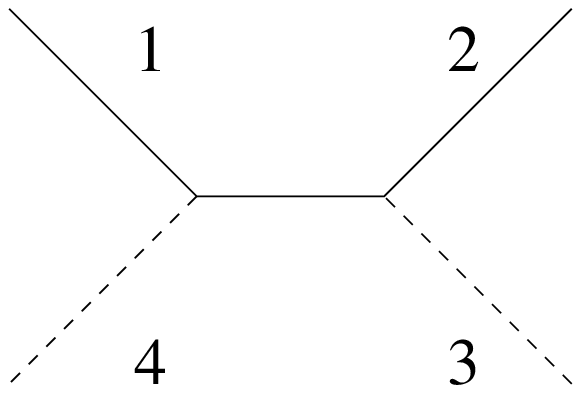}}
&=&$0\quad,$ &
\parbox{2.5cm}{\includegraphics[width=2.4cm]{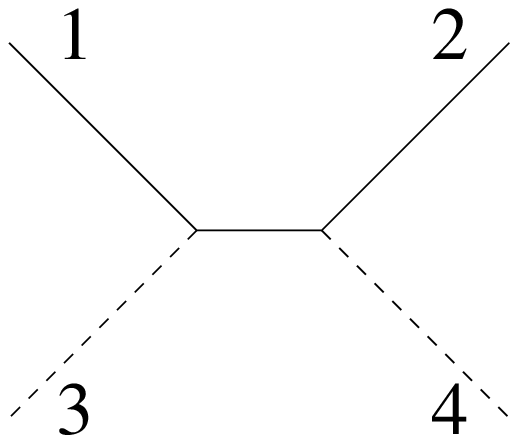}}
&=&$0\quad.$
\end{tabular}\\
\end{center}
\noindent
Summing the four contributions, we obtain
\beq
A_{\varphi\varphi \to \rho\rho}\=0
\eeq
as it should be expected for a production amplitude in an integrable model.
The same is true for the time-reversed production,
\beq
A_{\rho\rho \to \varphi\varphi}\=0 \quad.
\eeq

Summarizing, we have found 
that the only nonzero amplitude for tree-level $2 \to 2$ processes 
is the one describing the scattering among two of the massive excitations. 
The result is constant, independent of the momenta and so describes a 
perfectly {\em causal\/} process.
Since the result is independent of the noncommutation parameter $\theta$ 
it agrees with the four-point amplitude for the ordinary sine-Gordon model.
Finally, we have found that the production amplitudes 
$\varphi \varphi \to \rho \rho$ and $\rho\rho \to \varphi\varphi$ vanish, 
as required for ordinary integrable theories.

As a further check of our calculation and an additional test of our model  
we have computed the production amplitude 
$\varphi \varphi \to \varphi \varphi \varphi \varphi$ 
and the scattering amplitude 
$\varphi \varphi \varphi \to \varphi \varphi \varphi$. 
In both cases the topologies we have to consider are
\vspace{0.5cm}
\begin{center}
\begin{tabular}{c@{\hspace{0.8cm}}c@{\hspace{0.8cm}}c}
\parbox{2.5cm}{\includegraphics[width=2.4cm]{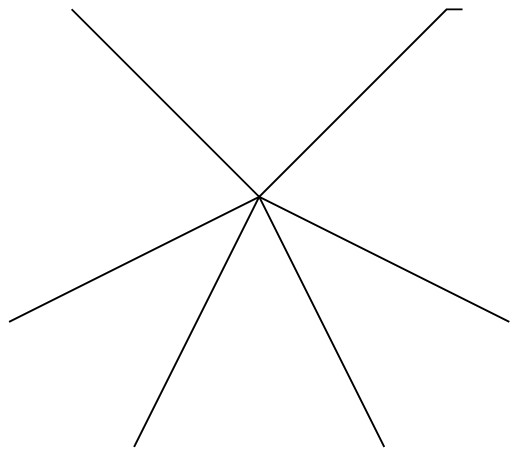}}&
\parbox{2.3cm}{\includegraphics[width=2.2cm]{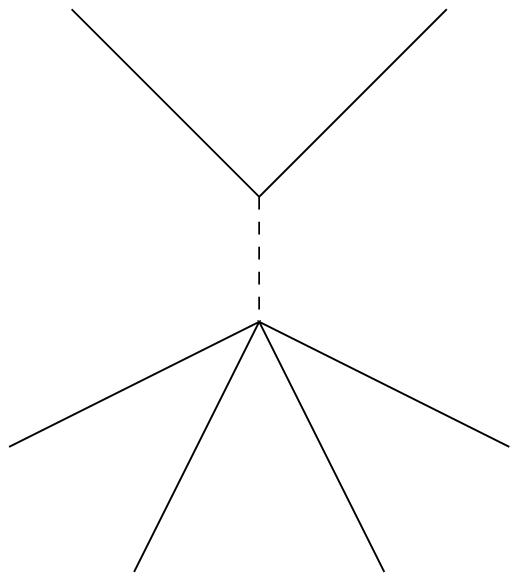}}& 
\parbox{2.0cm}{\includegraphics[width=1.9cm]{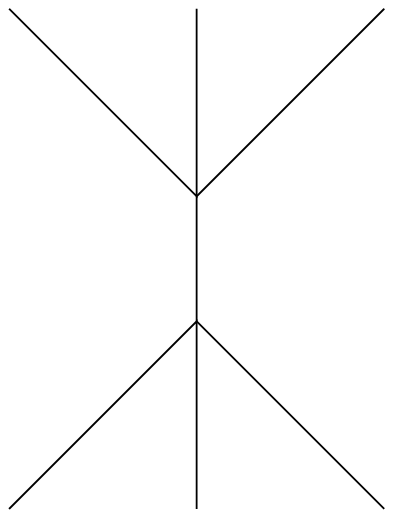}} \end{tabular} \\
\begin{tabular}{c@{\hspace{0.8cm}}c@{\hspace{0.8cm}}c@{\hspace{0.8cm}}c}
\parbox{2.0cm}{\includegraphics[width=1.9cm]{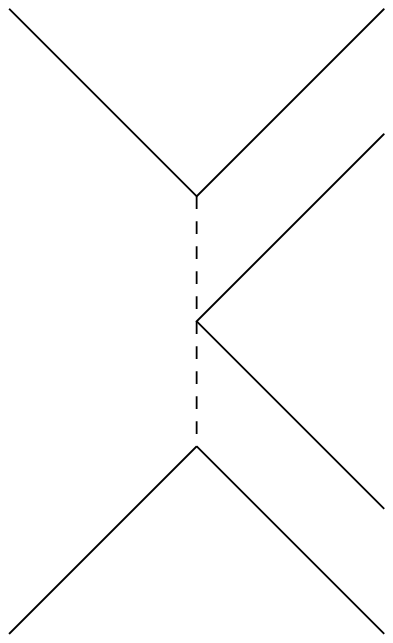}}&
\parbox{2.0cm}{\includegraphics[width=1.9cm]{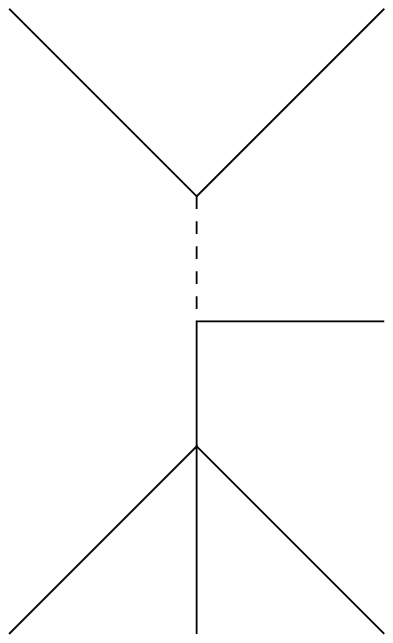}}&
\parbox{2.0cm}{\includegraphics[width=1.9cm]{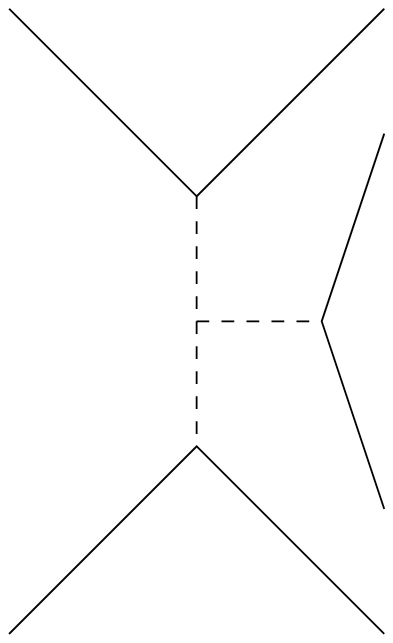}}&
\parbox{1.8cm}{\includegraphics[width=1.7cm]{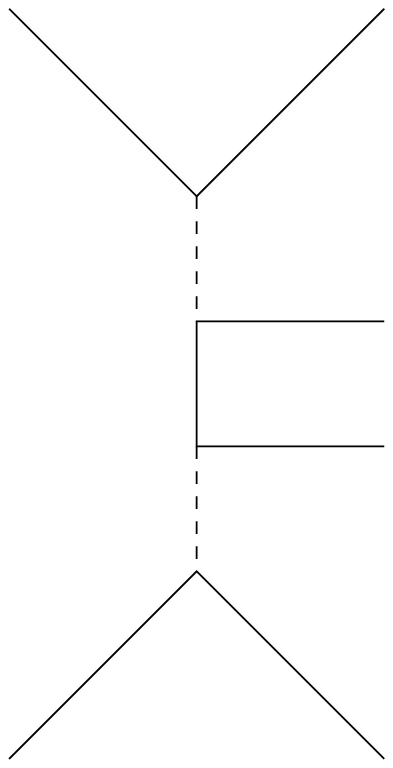}}\quad.
\end{tabular}\\ \end{center}
\vspace{0.5cm}
\noindent
Due to the growing number of channels and ordering of vertices, it is
no longer practical to perform the calculations by hand.
We have used {\it Mathematica}$^{\copyright}$ 
to symmetrize the vertices and take automatically into account the 
different diagrams obtained by exchanging momenta entering a given vertex.
The computation has been performed with assigned values of the external
momenta but arbitrary values for $\alpha^2$ and $\theta$. 
We have found a vanishing result for both the 
scattering and the production amplitude. 
This is in agreement with the commutative sine-Gordon model results.
\subsubsection{Amplitudes in the ``$h$-model'' }
We now discuss the $2\to 2$ amplitudes in the Leznov formulation. 
The theory is again described by two interacting fields, $h_1$ 
massless and $h_2$ massive. Referring to the action (\ref{haction})
we extract the following Feynman rules,
\begin{itemize}
\item{The propagators
\bea
\parbox{2cm}{\includegraphics[width=1.9cm]{prop2.eps}}
&\ \equiv\ &\langle h_1 h_1 \rangle\=\frac{\im}{2k^2}\quad,\\
\parbox{2cm}{\includegraphics[width=1.9cm]{prop1.eps}}
&\ \equiv\ &\langle h_2 h_2 \rangle\=\frac{\im/2}{k^2-4\a^2}\quad.
\ena}
\item{The vertex
\bea
\parbox{2cm}{\includegraphics[width=1.9cm]{vertice112.eps}}
&\=&- 4\a\,(k_{3t}-k_{3y})\, F(k_1,k_2,k_3)\quad.
\ena}
\end{itemize}
Again, we compute scattering amplitudes in the center-of-mass frame.
Given the particular structure of the vertex, at tree level there
is no $h_1 h_1 \to h_1 h_1$ scattering.
To find the $h_2 h_2 \to h_2 h_2$ amplitude we assign the momenta
(\ref{momenta aa-aa}) to the external particles. 
The contributions are \\
\begin{center}\begin{tabular}{r@{}clr@{}cl}
\parbox{2.3cm}{\includegraphics[width=2.2cm]{aapaa2.eps}}
&=& $-16\im\a^2\cos^2(\theta Ep)\quad,$ &
\parbox{2cm}{\includegraphics[width=1.9cm]{11p11fig.eps}}
&=& $16\im\a^2\cos^2(\theta Ep)~,$ \\
\parbox{2.3cm}{\includegraphics[width=2.2cm]{aapaa3.eps}}
&=&$0\quad.$ 
\end{tabular}\\ 
\end{center}
\noindent
We note that a collinear divergence appears in the last diagram which
can be regularized as described before.
Summing the two nonvanishing contributions we obtain complete cancellation.

For the $h_2 h_2 \to h_1 h_1$ amplitude
the center-of-mass-momenta are given in (\ref{momenta aa-bb}).
The only topology contributing to this production amplitude has 
two channels, yielding \\
\begin{center}\begin{tabular}{r@{}clr@{}cl}
\parbox{2.5cm}{\includegraphics[width=2.4cm]{12p12fig.eps}}
&=&$0\quad,$ &
\parbox{2.3cm}{\includegraphics[width=2.2cm]{abpab2.eps}}
&=& $0\quad,$
\end{tabular}\\ 
\end{center}
\noindent 
which are both zero, so giving a vanishing result once more.
The same is true for the $h_1 h_1 \to h_2 h_2$ production process.

Finally, for the $h_1 h_2 \to h_1 h_2$ amplitude, we refer to the 
center-of-mass momenta defined in (\ref{momenta ab-ab 1}) 
and (\ref{momenta ab-ab 2}). In both cases the contributions are\\
\begin{center}\begin{tabular}{r@{}clr@{}cl}
\parbox{2.3cm}{\includegraphics[width=2.2cm]{varabpab.eps}}
&=&$0\quad,$ &
\parbox{2.3cm}{\includegraphics[width=2.2cm]{abpab3.eps}}
&=&$0\quad,$
\end{tabular}\\ \end{center}
\noindent 
and so we find that the sum of the two channels is always equal to zero.

Since all the $2 \to 2$ amplitudes vanish, the S-matrix is trivially causal
and factorized.

Both in the ordinary and noncommutative cases the ``$h$-model''  is dual to 
the ``$g$-model''. In the commutative limit the ``$g$-model'' gives rise to a 
sine-Gordon model plus a free field which can be set to zero. 
In this limit our amplitudes exactly reproduce the  sine-Gordon amplitudes. 
On the other hand, the amplitudes for the ``$h$-model'' all vanish. Therefore,
in the commutative limit they do not reproduce anything immediately 
recognizable as an ordinary sine-Gordon amplitude. This can be understood by 
observing that, both in the ordinary and in the noncommutative case, the
Leznov formulation is an alternative description of the sine-Gordon dynamics
and obtained from the standard Yang formulation by the {\em nonlocal field 
redefinition\/} given in~(\ref{nonlocal3}). Therefore, it is expected
that the scattering amplitudes for the elementary exitations, which are
different in the two formulations, do not resemble each other.
\subsection{Conclusions}
In this section 2.3 I have introduced and discussed a novel noncommutative sine-Gordon system \cite{mypaper3} based on
{\em two\/} scalar fields, which seems to retain all advantages of
$1{+}1$ dimensional integrable models known from the commutative limit.
The rationale for introducing a second scalar field was provided by
deriving the sine-Gordon equations and action through dimensional and 
algebraic reduction of an integrable $2{+}1$ dimensional sigma model: 
In the noncommutative extension of this scheme it is natural to generalize 
the algebraic reduction of SU(2)$\to$U(1) to one of U(2)$\to$U(1)$\times$U(1).
We gave two Yang-type and one Leznov-type parametrizations of the coupled 
system in (\ref{Yphi}), (\ref{Yrho}) and (\ref{Lh12}) and provided the actions
for them, including a comparison with previous proposals.
It was then outlined how to explicitly construct noncommutative sine-Gordon
multi-solitons via the dressing method based on the underlying linear system.
We found that the one-soliton configuration agrees with the commutative one
but already the two-soliton solutions gets Moyal deformed.

What is the gain of doubling the field content as compared to the standard
sine-Gordon system or its straightforward star deformation?
Usually, time-space noncommutativity adversely affects the causality and
unitarity of the S-matrix (see, e.g.~\cite{CM, us, mypaper2}), even in the
presence of an infinite number of local conservation laws. In contrast,
the model described in \cite{mypaper3} seems to possess an S-matrix which is {\em causal\/} 
and {\em factorized\/}, as we checked for all tree-level $2\to 2$ processes
both in the Yang and Leznov formulations. Furthermore, we verified the
vanishing of some $3\to 3$ scattering amplitudes and $2\to 4$ production
amplitudes thus proving the absence of particle production.

It would be nice to understand what actually drives a system to be 
integrable in the noncommutative case. A hint in this direction might
be that the model proposed in~\cite{us} has been constructed directly in 
two dimensions even if its equations of motion (but {\em not\/} the
action) can be obtained by a suitable reduction of a four dimensional
system (noncommutative self-dual Yang-Mills). The model proposed in this 
paper, instead, originates directly, already at the level of the action,
from the reduction of noncommutative self-dual Yang-Mills theory 
which is known to be integrable and related to the $N{=}2$ string~\cite{olaf1}. 

Several directions of future research are suggested by our results.
First, one might hope that our noncommutative two-field sine-Gordon model
is equivalent to some two-fermion model via noncommutative bosonization.
Second, it would be illuminating to derive the exact two-soliton solution
and extract its scattering properties, either directly in our model or by
reducing wave-like solutions of the 2+1 dimensional sigma model 
\cite{bieling,wolf}.
Third, there is no obstruction against applying the ideas and techniques 
of this paper to other 1+1 dimensional noncommutative integrable systems 
in order to cure their pathologies as well.

\chapter[Covariant superstring vertices and a possible nonconstant\\ superspace deformation]{Covariant superstring vertices and a possible nonconstant superspace deformation}
\section{An introduction to the pure spinor superstring}
In this section I will mostly refer to the review paper \cite{berkICTP}.
\subsection{Motivation: Problems with RNS and GS formalisms}
In section 1.2.2 I have briefly outlined how the discussion of the stringy origin of bosonic noncommutative geometry presented in section 1.2.1 can be generalized to the superstring, in both RNS and GS formalisms. As anticipated, both of them display some awkward features, due to their target-space or worldsheet symmetry structure, respectively.

The RNS formalism is characterized by an $N=1$ worldsheet supersymmetry. The field content is the set of bosonic coordinate fields $x^m$ (worldsheet scalars and spacetime vectors) together with the worldsheet spinors (and spacetime vectors) $\psi^m$. The worldsheet action for the string in a flat background is quadratic, therefore the quantization in this formalism is straightforward. After a suitable consistent truncation of the spectrum (GSO projection), the theory also enjoys target space supersymmetry, but clearly this symmetry is not manifest. As a result, a series of problems arises, for instance in the computation of amplitudes with more than four external fermions and in dealing with general R-R backgrounds. 

The GS formalism instead is manifestly target-space supersymmetric, but the worldsheet symmetry structure is quite complicated. Target space is a ten-dimensional superspace described by the bosonic coordinates $x^m$ and their superpartners $\th^\a$, $\hat\th^{\hat\b}$ (in type II case). For the number of physical fermionic degrees of freedom to be related to the bosonic ones as required by target-space supersymmetry, a worldsheet fermionic local symmetry must be present ($\kappa$-symmetry, \cite{kappa}). Therefore, the natural supersymmetric generalization of the bosonic string action
\beq
S_1=\int d^2 \sigma \sqrt h h^{ij}\Pi_i\cdot \Pi_j
\label{gsnowz}
\eeq
where $h^{ij}$ is the worldsheet metric and $\Pi_i^m$ is the supersymmetrized bosonic momentum, does not work, not being $\kappa$-symmetric. When $N\le 2$ a WZ term $S_2$ can be added, so that the resulting action $S=S_1+S_2$ is $\kappa$-symmetric (for this discussion the reader can refer to \cite{gsstring} and references therein). $S$ in conformal gauge and in a flat background is given by (\ref{piatta}) with $B=0$. It is nonquadratic and describes a complicated, interacting worldsheet field theory. This fact prevents quantization except in light-cone, where the action reduces to a quadratic form. Since light-cone gauge is not manifestly Lorentz covariant, problems in the computation of amplitudes emerge also in this formalism and only four-point tree and one-loop amplitudes have been computed. Moreover, backgrounds that do not allow for a light-cone choice cannot be dealt with at the quantum level.

An alternative approach to the GS formalism was introduced by Siegel \cite{siegel}.
The main problem of the GS superstring is that a set of phase-space constraints arise at the classical level whose structure do not allow for a Dirac quantization procedure. In particular, since the conjugate momenta $p_\a$ to the fermionic variables $\th^\a$ do not appear in the action, one has phase space constraints $d_\a=0$ together with the Virasoro constraint $T=-\half \Pi\cdot\Pi=0$ related to the conformal gauge choice. The anticommutator of the $d$'s is proportional to the $\Pi$'s. As a result, half of the fermionic constraints are first class and half are second class \cite{goteborg}. The separation of the two different kinds of constraint cannot be achieved in a manifestly Lorentz covariant way. This explains why quantization of the model only works in light-cone gauge. 

In \cite{siegel} Siegel proposed to rewrite the GS action in a first order formalism for the fermionic variables, hoping that a set of phase space constraints that are all first class could be found. These contraints were to be constructed out of the supersymmetric objects $\Pi^m$, $\pa\th^\a$ and the GS constraint $d_\a$, no longer constrained to vanish. The explicit form of the left-moving contribution to the GS action in conformal gauge in Siegel formalism is 
\beq
S=\int d^2 z \left(\half\pa x_m\bar\pa x^m +p_\a\bar\pa\th^\a\right)
\label{siegelgs}
\eeq
where the fermionic conjugate momenta $p_\a$ are independent variables.
This approach was shown to work for quantizing the superparticle, but not the superstring, since its correct physical spectrum was never obtained. 
We should note that the action (\ref{siegelgs}) is quadratic and therefore it is immediate to determine the OPE's between the free fields. Quantization of this theory is as simple as the in RNS formalism.

In the next section we will see how Siegel action (\ref{siegelgs}) is the starting point for the construction of a formalism for the superstring that allows for a covariant quantization and describes the same physics as the RNS and GS strings.
\subsection{Pure spinor superstring basics}
From now on I will use the Weyl representation of the $32\times 32$ ten-dimensional Dirac matrices, where they are off-diagonal and $\g^{m}_{\a\b}$ and $\g^{m}_{\hat\a\hat\b}$ are the real symmetric
$16\times 16$ off-diagonal blocks, satisfing the Fierz identities 
$\g_{m\a(\b} \g^m_{\g\d)} =0$.

Useful properties to keep in mind are the following. Every symmetric bispinor can be decomposed in terms of a vector and a five form as
\beq
f_{\a\b}=\g^m_{\a\b}f_m + \g_{\a\b}^{mnpqr}f_{mnpqr}
\label{symm}
\eeq
while every antisymmetric bispinor can be decomposed in terms of a three form as
\beq
\tilde f_{\a\b}=\g^{mnp}_{\a\b}\tilde f_{mnp}
\label{antisymm}
\eeq

Our conventions for $d=10$ $N=2$ superspace covariant derivatives and 
supersymmetry charges are
\bea
\label{susuI}
&&D_{\a} = \pa_{\a} + {1\over 2}(\g^{m}\th)_{\a} \pa_{m}\,, ~~~~~~
Q_{\a} = \pa_{\a} - {1\over 2}(\g^{m}\th)_{\a} \pa_{m}\,, \cr
&&\hat D_{\hat\a} = \pa_{\hat\a} + {1\over 2}(\g^{m}\hat\th)_{\hat \a} \pa_{m}\,, ~~~~~~
\hat Q_{\hat\a} = \pa_{\hat\a} - {1\over 2}(\g^{m}\th)_{\hat\a} \pa_{m}\,,
\ena
which satisfy
\bea
\label{derivI}
&&\left\{D_\a,D_\b\right\}=\g^m_{\a\b}\pa_m\,, \quad\quad
\left\{\hat D_{\hat\a},\hat D_{\hat\b}\right\}=\g^m_{\hat\a\hat\b}\pa_m\,, 
\quad\quad
\left\{D_\a,\hat D_{\hat\b}\right\}=0\cr
&&
\{D_{\a}, Q_{\b} \} =0\,, ~~~~~
\{\hat D_{\hat\a},\hat Q_{\hat \b} \} =0\,.
\ena
\vskip 11pt

Berkovits completed Siegel action (\ref{siegelgs}) by adding some missing worldsheet ghost degrees of freedom.
The evaluation of what's missing in Siegel approach can be achieved by  ``counting". 
Siegel action (\ref{siegelgs}) gives the free-field OPE's
\bea
&&x^m(y)x^n(w)\sim -2\eta^{mn}\log |y-w|\cr
&&p_\a(y) \th^\b(w)\sim \d_\a^\b (y-w)^{-1}
\label{freeope}
\ena
From this, we can determine the contributions of the different fields to the conformal anomaly. Since fermionic fields contribute -32 and bosonic ones +10, the missing fields should contribute +22.
Moreover, one can consider the contribution to the Lorentz current coming from fermionic degrees of freedom $M_{mn}=\half p\g_{mn}\th$ and compare to the analogous term in RNS formalism $M_{mn}=\psi_m\psi_n$. The current-current OPE's are similar, except for the coefficient of the double pole term, which is +4 in Siegel case and +1 in RNS case. Therefore, the missing ghost variables should contribute to the Lorentz current in a way to produce a -3 in the double pole.

Indeed, Berkovits found that an irreducible representation of $SO(9,1)$ with these characteristics exists.
This is a bosonic pure-spinor satisfying 
\beq
\l\g^m\l=0
\label{purespin}
\eeq  
To solve this constraint and find the free ghost fields, one has to break the manifest Lorentz covariance to a $U(5)$ subgroup of (Wick rotated) $SO(10)$. In terms of this parametrization one can write down the ghost-field action, check that the OPE of the ghost contribution to the Lorentz current has a -3 coefficient in the double pole and that the stress tensor has central charge +22, as required.
Apparently, one goes back to the old problem of the lack of manifest Lorentz covariance. However, one can formally write down an action in the form
\beq
S=\int d^2 z \left(\half\pa x_m\bar\pa x^m +p_\a\bar\pa\th^\a+w_\a\bar\pa\l^\a\right) + {\rm right~ moving}
\label{formalA}
\eeq 
where the independent conjugate momenta $w_\a$ of the ghost field $\l^\a$ have been introduced, and then ``remember" that the $\l$ fields are constrained by equation (\ref{purespin}). Both the action and the pure spinor constraint are manifestly Lorentz covariant. The problem is how to deal with constrained fields in a path integral approach.

When there are first class constraints in a theory, a BRST quantization procedure can be applied and the BRST charge is constructed out of the constraints themselves multiplied by ghost fields. 
When the constraints are second class, this does not work because the BRST charge one obtains is not nilpotent. In Berkovits approach to the superstring, the (left-moving) BRST-like charge is defined as 
\beq
Q=\int dz \l^\a d_\a
\label{BRSTcharge}
\eeq
where $d_\a$ is the constraint of the GS superstring that in this formalism plays the role of the supersymmetric version of the fermionic conjugate momentum $p_\a$
\beq
d_{\a} = p_{\a} - {1\over 2} \pa x^{m} (\g_{m} \th)_{\a} - {1\over 8} (\g^{m} \th)_{\a} (\th \g_{m} \pa\th)
\eeq
Therefore, this construction of $Q$ could be reminiscent of some sort of BRST quantization of the GS superstring. 

From the free OPE's (\ref{freeope}) one can compute the OPE's between the composite variables $d_\a$ and find
\beq
d_\a(y)d_\b(w)\sim -(y-w)^{-1}\g^m_{\a\b}\Pi_m(w)
\eeq
Therefore the BRST charge (\ref{BRSTcharge}) is nilpotent because of the pure spinor condition (\ref{purespin}). 
Also because of the pure spinor condition, one observes that the ghost conjugate momentum $w_\a$ can only appear in combinations that are invariant under the gauge transformation
\beq
\label{conD}
\d w_{\a} = \L_{m} (\g^{m} \l)_{\a}
\eeq
with arbitrary $\L_m$.
These gauge-invariant combinations are the pure spinor contribution to the Lorentz current $N_{mn}=\frac{1}{2}:(w\g_{mn}\l):$ and the ghost number current $J=:w_\a \l^\a:$.

In the superparticle case, the BRST charge (\ref{BRSTcharge}) and pure spinor condition (\ref{purespin}) can be obtained by an honest, although unusual, gauge fixing procedure starting from the Brink-Schwarz action in semi-light-cone gauge rewritten in the Siegel formalism. 

Unfortunately, no analogous procedure works for the superstring.
Following the usual prescription of the BRST 
quantization rules, we could start from the GS superstring and define the quantum action as follows
\beq
\label{conDA}
S_{0} =  S_{GS} + 
Q\int d^{2}z w_{\a} \bar\pa \th^{\a}
\eeq
where $S_{GS}$ is the Green-Schwarz action in conformal gauge \cite{gsstring}. By moving on to a Siegel description for fermionic fields and by explicitly writing down all the contributions to (\ref{conDA}), one obtains (\ref{formalA}).
Even if this looks like the usual BRST procedure, we have to notice that 
the BRST-like operator $Q$ is nilpotent up to gauge 
transformations (\ref{conD}). This compensates the fact that the Green-Schwarz 
action is not invariant under BRST transformations. In addition, we can always add BRST invariant terms to the 
action. However, there is no procedure 
to get (\ref{conDA}) from an honest gauge fixing of the Green-Schwarz action 
(a suggestion is given in \cite{tonin}). 
\vskip 11pt
Now I'm going to discuss pure spinor superstring vertex operators.
I will first derive the open superstring vertices for simplicity. Closed superstring vertices will be studied in much detail in section 3.1.3. 

Since in the open string case vertices are to be inserted on the boundary of the worldsheet, where the boundary condition $\th=\hat\th\vert_{z=\bar z}$ holds, they can be expressed in terms of the left-moving fermions only (or, more correctly, in terms of the linear combination $\th_+=\frac{1}{\sqrt 2}(\th+\hat\th)$ and the corresponding one for the fermionic momenta).  

Physical states for the open superstring are defined as ghost-number one states in the cohomology of $Q$, defined in (\ref{BRSTcharge}), with $\l$ satisfying the pure spinor condition (\ref{purespin}).
Open superstring vertex operators with $({\rm mass})^2=\frac{n}{2}$ are constructed out of the fields $x^m$, $\th^\a$, $d_\a$, $\l^\a$ and the gauge invariant objects $N_{mn}$ and $J$ containing the ghost momenta. They are obtained as the generic combinations with ghost number one and conformal weight $n$ at zero momentum. Since the composite objects $d_\a$, $N_{mn}$ and $J$ carry conformal weight one and $\l^\a$ carries ghost number one, it is clear that for instance the most general vertex operator at $({\rm mass})^2=0$ is
\beq
{\cal V}^{(1)}=\l^\a A_\a(x,\th)
\eeq
By requiring $Q{\cal V}^{(1)}=0$, one obtains the equations of motion for the spinor superfield $A_\a(x,\th)$.
By making use of the OPE
\beq
d_\a(y)A_\b(x,\th)(w)\sim D_\a A_\b(w)
\eeq
one finds that the superfield $A_\a(x,\th)$ must satisfy the equation $\l^\a\l^\b D_\a A_\b=0$. Because of the property (\ref{symm}) and the pure spinor condition (\ref{purespin}), this is equivalent to 
\beq
\g_{mnpqr}^{\a\b}D_\a A_\b=0
\eeq
It can be shown that these are the superMaxwell equations written in terms of a spinor superfield. Equation $QU=0$ is invariant under the gauge transformation $\d U=Q\Omega$ where $\Omega$ is a generic scalar superfield. Indeed, this implies the gauge transformation for the spinor superfield $\d A_\a=D_\a\Omega$, which is the expected gauge transformation of superMaxwell theory.

Going on to the next mass level, $({\rm mass)}^2=\half$, one finds that the most general vertex operator is
\bea
{\cal V}^{(1)}_1&=&\pa\l^\a A_\a(x,\th) + :\pa\th^\b \l^\a B_{\a\b}(x,\th):+:d_\b \l^\a C^\b_{~~\a}(x,\th): \cr
&+&:\Pi^m \l^\a H_{m\a}(x,\th): + :J\l^\a E_\a(x,\th): + :N^{mn} \l^\a F_{\a mn}(x,\th):\cr
&&~~~
\label{massiveU}
\ena
Cohomology equations and gauge transformations imply that the superfields appearing in the vertex describe a spin two multiplet.

The integrated massless open superstring vertex operator $\int dz {\cal V}^{(0)}$ can be obtained by making use of the cohomology descent equation
\beq
[Q,{\cal V}^{(0)}]=\pa {\cal V}^{(1)}
\label{opendescent}
\eeq
${\cal V}^{(0)}$ is expanded in terms of the 1-forms ${\bf X}=\left(\pa\th^\a, \Pi^m, d_\a,\half N_{mn}\right)$ as follows
\beq
{\cal V}^{(0)}=\pa\th^\a A_\a(x,\th) + \Pi^m A_m(x,\th) +d_\a W^\a(x,\th) +\half N_{mn}F^{mn}
\label{openV}
\eeq
The descent equation (\ref{opendescent}) is satisfied when the superfields $A_\a$, $A_m$, $W^\a$ and $F_{mn}$ are governed by the superMaxwell equations
\bea
&&D_\a A_\b +D_\b A_\a -\g^m_{\a\b}A_m=0\cr
&&D_\a A_m -\pa_m A_\a -\g_{m\a\b}W^\b=0\cr
&&D_\a W^\b -\frac{1}{4}(\g^{mn})_\a^{~~\b}F_{mn}\cr
&&\l^\a\l^\b (\g_{mn})_\b^{~~\g}D_\a F^{mn}=0
\label{openEOM}
\ena
The last equation is redundant, since it is implied by the previous one and by the pure spinor condition (\ref{purespin}). The vertex (\ref{openV}) was first found by Siegel in \cite{siegel}, except for the pure spinor term, by making use of superspace arguments.
\vskip 22pt
All this discussion can be easily generalized to type II closed superstrings. The field content is $x^{m}$ where $m=0,\dots,9$, two Majorana-Weyl spinors $\th^{\a}$, $\hat \th^{\hat \a}$ 
with $\a=\hat \a =1, \dots, 16$ (with opposite or same chirality depending whether one is in IIA or IIB case), their conjugate momenta $p_{\a}$, $\hat p_{\hat \a}$, two ghosts $\l^\a$, $\hat\l^{\hat\a}$ satisfying the pure spinor conditions 
$$\l
 \g^{m} \l =0\,, ~~~~~
  \hat\l \g^{m} \hat\l=0\,,$$ 
  and the corresponding conjugate momenta $w_{\a}$, $\hat w_{\hat\a}$.
  Again, supersymmetric versions of the fermionic conjugate momenta can be introduced as follows
 \bea
\label{ccA}
&&d_{\a} = p_{\a} - {1\over 2} \pa x^{m} (\g_{m} \th)_{\a} - {1\over 8} (\g^{m} \th)_{\a} (\th \g_{m} \pa\th)
\,, \cr
&&\hat d_{\hat\a} = \hat p_{\bar\a} - {1\over 2} \bar\pa x^{m} (\g_{m} \hat\th)_{\hat\a} - 
{1\over 8} (\g^{m} \hat\th)_{\hat \a} (\hat\th \g_{m} \bar\pa\hat\th) 
\,, 
\ena
The BRST operators are defined by 
\beq
\label{conC}
Q_{L} = \oint dz \l^{\a} d_{\a}\,, ~~~~~~~~ Q_{R} = \oint d\bar z \hat\l^{\hat\a} \hat d_{\hat\a}\,.
\eeq
which satisfy
\beq
\label{conCA}
Q_{L}^{2} = -\oint dz\, \l\g^{m}\l \Pi_{m}\,, ~~~~~
[Q_{L}, Q_{R}]=0\,, ~~~~
Q_{R}^{2} =- \oint d\bar z\, \hat\l\g^{m}\hat\l \hat\Pi_{m}\,, ~~~~~
\eeq
where $\Pi_z^m$ and 
$\hat\Pi_{\bar z}^m$ are the left- and right-moving supersymmetrized bosonic momenta
\beq
\Pi_z^m = \pa x^m + {1\over 2}\th \g^m \pa \th \qquad;\qquad\hat\Pi_{\bar z}^m = \bar\pa x^m +{1\over 2} \hat\th \g^m \bar\pa \hat\th
\label{bosomom}
\eeq
Due to pure spinor constraints, the BRST charges
are nilpotent up to gauge transformations of $w_{\a}$, $\hat w_{\hat\a}$, 
given by 
\beq
\label{conDclosed}
\d_{L} w_{\a} = \L_{m} (\g^{m} \l)_{\a}\,, ~~~~~~~~
\d_{R} \hat w_{\a} = \hat \L_{m} (\g^{m} \hat\l)_{\a}\,.
\eeq
with arbitrary local parameters $\L_{m}$ and $\hat \L_{m}$. Gauge invariant operators are
 \vskip 11pt 
\begin{center}
\begin{tabular}{l l}
$J_{L} = : w_{\a} \l^{\a} :\,,$ & $J_{R} = : \hat w_{\a} \hat\l^{\a} :\,,$\\
${N}_{L} =  {1\over 2}:w \g^{mn} \l :\,,$ & ${N}_{R} =  {1\over 2}:\hat w \g^{mn} \hat\l :\,, $
\end{tabular}
\end{center}
\beq
\label{gi}
\eeq
By formally following the usual prescription of the BRST 
quantization rules, we can define the quantum action starting from the GS superstring in Siegel formalism as follows
\beq
\label{conDAI}
S_{0} =  S_{GS} + 
Q_{L} \int d^{2}z w_{\a} \bar\pa \th^{\a} + Q_{R} \int d^{2}z \hat w_{\hat\a} \pa \hat\th^{\hat\a}\,.
\eeq 

By exploiting the different contributions in (\ref{conDAI}), we obtain 
\beq
\label{conE}
S_{0} = \int d^2 z 
\Big({1\over 2}\pa x^m \bar \pa x_m + p_\a \bar \pa\th^\a +\hat p_{\hat\a}\pa \hat\th^{\hat\a}+ w_\a \bar\pa \l^\a + \hat w_{\hat\a}\pa \hat\l^{\hat\a}  \Big)\,,
 \eeq
which is BRST invariant and invariant under the gauge transformation (\ref{conDclosed}) if the spinors 
$\l^{\a}, \hat\l^{\hat\a}$ are pure. The action is also invariant under the $N=2$ supersymmetry 
transformations generated by $Q_{\e} = \e^{\a} \oint dz q_{\a} + 
\hat \e^{\hat \a} \, \oint d\bar z \hat q_{\hat \a}$
where the explicit expressions for the supersymmetry currents are
\bea
\label{conF}
&&q_{\a} = p_{\a} + {1\over 2} \pa x^{m} (\g_{m} \th)_{\a} + {1\over 24}(\th \g^{m} \pa\th) (\g_{m} \th)_{\a} 
\,, \cr
&&
\hat q_{\hat\a} = \hat p_{\bar\a} +{1\over 2} \bar\pa x^{m} (\g_{m} \hat\th)_{\hat\a} + 
{1\over 24}(\hat\th \g^{m} \bar\pa\hat\th)  (\g_{m} \hat\th)_{\hat \a} 
\,. 
\ena
It is interesting to note that these do not anticommute with the BRST operators $Q_L$ and $Q_R$, since
\beq
\label{susyQ}
[Q_L,q_\a]=\pa \chi_\a\,,~~~~~~~~~~[Q_R,\hat q_{\hat\b}]=\bar\pa \hat\chi_{\hat\b}
\eeq
where $\chi_\a$ and $\hat\chi_{\hat\b}$ are the BRST-invariant quantities 
\beq
\label{susyQII}\chi_\a\equiv{1\over 3}(\l\g^m\th)(\g_m\th)_\a\,,~~~~~~~~~~~\hat\chi_{\hat\b}={1\over 3}(\hat\l\g^p\hat\th)(\g_p\hat\th)_{\hat\b}
\eeq
We also introduce the Lorentz currents
\bea
\label{lorentz}
&&L^{mn}=
 {1\over 2}: \pa x^{[m} x^{n]} : +{1\over 2}:(p\g^{mn}\th): + : N^{mn}:\,,\cr
&&\hat L^{pq}={1\over 2}:\bar\pa x^{[p} x^{q]}:+{1\over 2}: (\hat p\g^{pq}\hat\th): + :\hat N^{pq}:\,,
\ena
which satisfy the following commutation relations with the BRST charges
\beq
\label{lorentzQ}
[Q_L,L^{mn}]=\pa {\cal G}^{mn};~~~~~[Q_R,\hat L^{pq}]=\bar\pa \hat {\cal G}^{pq}
\eeq
where 
\beq
\label{lorentzQII}{\cal G}^{mn}={1\over 4}(\th \g^r \l)(\d_r^{[m} x^{n]}+{1\over 4}(\th\g_r\g^{mn}\th));~~~\hat {\cal G}^{pq}={1\over 4}(\hat\th \g^r \hat\l)(\d_r^{[p} x^{q]}+{1\over 4}(\hat\th\g_r\g^{pq}\hat\th))
\eeq
are BRST invariant.
By using the equations of motion from (\ref{conE}) it is easy to show that 
the currents $q_{\a}$, $\hat q_{\hat \b}$, $\l^{\a} d_{\a}$, $\hat \l^{\hat\b} \hat d_{\b}$, $L^{mn}$ and
$\hat L^{pq}$ are holomophic and anti-holomorphic, respectively. 

In the following section I will describe in detail type II vertex operators, their descent equations and the corresponding superfield equations of motion and gauge transformations. These closed string vertices are as usual obtained by taking the left-right product of the open superstring vertices I described in the present section. 

\subsection{Type II superstring vertex operators}
In this section I will describe in detail the construction of the closed superstring ghost number 
  $(1,1)$ local vertex  operator ${\cal V}^{(1,1)}$ and of the integrated vertex operators $\int dz~ {\cal
  V}_z^{(0,1)}$, $\int d\bar z~ {\cal V}_{\bar z}^{(1,0)}$, and $\int dz \wedge d\bar z~ {\cal V}_{z\bar z}^{(0,0)}$, related to it by the closed string descent equations.  
  
  Introducing the notation 
  ${\cal O}^{(a,b)}_{c,d}$ for local vertex operators with ghost number $a (b)$ 
  in the left (right) sector and (anti)holomorphic indices $c (d)$, 
  we identify 
  \bea
  \label{deA}
  {\cal O}^{(1,1)}_{0,0} &=& {\cal V}^{(1,1)}\,, \cr
  {\cal O}^{(0,1)}_{1,0} = {\cal V}^{(0,1)}_z dz\,&,&\,
  {\cal O}^{(1,0)}_{0,1} = {\cal V}^{(1,0)}_{\bar z} d \bar z\,,\cr
  {\cal O}^{(0,0)}_{1,1} &=& {\cal V}^{(0,0)}_{z\bar z} dz\wedge d\bar z\,.
  \ena
  The descent equations read\footnote{Here we use the square brackets to denote both 
  commutation and anti-commutation relations. The difference is established by 
  the nature of the operators involved in the relations.} 
  \beq
  \label{deB}
  [Q_L, {\cal O}^{(a,b)}_{c,d}] = \pa \, {\cal O}^{(a+1,b)}_{c-1,d}\,, 
  ~~~~~
  [Q_R, {\cal O}^{(a,b)}_{c,d}] =\bar \pa \, {\cal O}^{(a,b+1)}_{c,d-1}\,,
  \eeq
  where $\pa = dz \pa_z$ and $\bar \pa = d\bar z \pa_{\bar z}$ 
  are the holomorphic and antiholomorphic differentials. $Q_L$ and 
  $Q_R$ are the BRST charges for holomorphic and antiholomorphic 
  sectors we introduced in (\ref{conC}). 
  More explicitly, at the first level we have 
  \beq
  \label{deC}
  [Q_L, {\cal V}^{(1,1)}] = 0 \,,
  ~~~~~
  [Q_R, {\cal V}^{(1,1)}] = 0\,,
  \eeq
  while at the next level we get 
  \beq
  \label{deCC}
  [Q_L, {\cal V}^{(0,1)}_z] = \pa_z {\cal V}^{(1,1)}\,, 
  ~~~~~
  [Q_R, {\cal V}^{(0,1)}_z] = 0\,,
  \eeq
  \beq
  [Q_R, {\cal V}^{(1,0)}_{\bar z}] = \pa_{\bar z} {\cal V}^{(1,1)}\,, 
  ~~~~~
  [Q_L, {\cal V}^{(1,0)}_{\bar z}] = 0\,,
  \eeq
  and, finally, 
  \beq
  \label{deD}
  [Q_L, {\cal V}^{(0,0)}_{z \bar z}] = \pa_z  {\cal V}^{(1,0)}_{\bar z} \,, 
  ~~~~~
  [Q_R, {\cal V}^{(0,0)}_{z \bar z}] = - \pa_{\bar z} {\cal V}^{(0,1)}_{z} \,.
  \eeq
The vertex operators ${\cal O}^{(a,b)}_{c,d}$ are to be expanded 
in powers of ghost fields $\l^\a$ and $\hat\l^{\hat \a}$ or in powers 
of the supersymmetric holomorphic and antiholomorphic 1-forms
 \beq
 \label{vectors}
 {\bf X}_z =  \left( \pa_z\theta^\a,~ \Pi^m_z,~ d_{z\a},~ 
\half N^{mn}_z \right)\,, ~~~~~~~ 
{\bf \hat X}_{\bar z} = \left( \pa_{\bar z}\hat\theta^{\hat\b}~,~
\hat\Pi^p_{\bar z}~,~ \hat d_{\bar z\hat \b}~,~ \half \hat N^{pq}_{\bar z}\right)\,.
\eeq
The explicit expressions of these 1-form operators in terms of sigma model fields are given in (\ref{ccA}, \ref{bosomom}, \ref{gi}). 
The coefficients are superfields of the coordinates $x^m$, $\th^\a$ and 
$\hat\th^{\hat \a}$. A further relation is obtained by acting from the left on the first equation of (\ref{deD}) with $Q_R$ or on the second with $Q_L$. Using eqs. 
(\ref{deCC}), 
one obtains 
\beq
\label{deE}
[Q_R,[Q_L, {\cal V}^{(0,0)}_{z \bar z}]] = \pa_z  \pa_{\bar z} {\cal V}^{(1,1)}\,,
\eeq
which is the closed string analogue of (\ref{opendescent}) and relates the integrated vertex ${\cal V}^{(0,0)}_{z \bar z}$ to the unintegrated one ${\cal V}^{(1,1)}$ . 

Equations (\ref{deC}, \ref{deCC}, \ref{deD}) 
are invariant under the gauge transformations given by 
\beq
\d {\cal V}^{(1,1)}= [Q_L, \L^{(0,1)}]+ [Q_R, \L^{(1,0)}] 
\eeq
\beq
\label{gaugegen}
\d {\cal V}_{\bar z}^{(1,0)} = [Q_L,  \tau _{\bar z}^{(0,0)}] + \pa_{\bar z} \L^{(1,0)}\,, 
~~~~~~~~~
\d {\cal V}_z^{(0,1)}= [Q_R, \tau_z ^{(0,0)}] + \pa_z \L^{(0,1)}
\eeq
\beq
\d {\cal V}_{z \bar z}^{(0,0)}=\pa_z \tau ^{(0,0)}_{\bar z} - 
\pa_{\bar z} \tau_z^{(0,0)}
\eeq
where the zero forms $\L^{(0,1)}$ and $ \L^{(1,0)}$ have ghost number $(1,0)$ and 
$(0,1)$ and are proportional to  $\l^\a$ and $\hat\l^{\hat \a}$, 
and the coefficients are superfields. 
The holomorphic and antiholomorphic 
1-forms $\tau_z ^{(0,0)}$ and $\tau ^{(0,0)}_{\bar z}$ are 
to be expanded in terms of the 1-forms ${\bf X}_z$ and ${\bf \hat X}_{\bar z}$ given in (\ref{vectors}) and coefficients are again superfields. 

In addition, the gauge parameters $\L^{(0,1)}$, $\L^{(1,0)}$, $\tau_z ^{(0,0)}$ and $\tau ^{(0,0)}_{\bar z}$ must satisfy the following consistency conditions
\beq
\label{consI} 
[Q_L, \L^{(1,0)}]=0 \quad\quad\quad 
[Q_R, \L^{(0,1)}]=0\,,
\eeq
and
\beq
\label{consII}
[Q_L, \tau_z^{(0,0)}] + \pa_z \L^{(1,0)}=0 \quad\quad\quad
[Q_R, \tau_{\bar z}^{(0,0)}] + \pa_{\bar z} \L^{(0,1)}=0\,.
\eeq
These equations resemble the descent equations 
for the  open string vertex operator ${\cal V}^{(1)}= \l^\a A_\a$, 
but in that case there are boundary conditions 
for the fermionic fields: $\th^\a(z) = \hat\th^{\hat \a}(z)$ at $z = \bar z$.  

Equations (\ref{consI}) and (\ref{consII}) are further invariant under the gauge transformations 
\beq
\label{constrans}
\delta \Lambda^{(1,0)} = [Q_{L}, \Upsilon^{(0,0)}]\,, ~~~~~
\delta \Lambda^{(0,1)} = [Q_{R}, \hat\Upsilon^{(0,0)}]\,,
\eeq
\beq
\delta \tau^{(0,0)}_{z} =- \pa_{z} \Upsilon^{(0,0)}\,, ~~~~~
\delta \tau^{(0,0)}_{\bar z} =- \pa_{\bar z} \hat\Upsilon^{(0,0)}\,. 
\eeq
where $\Upsilon^{(0,0)}$ and $\hat\Upsilon^{(0,0)}$ are generic superfields. However, 
consistency with (\ref{gaugegen}) imposes $\Upsilon^{(0,0)}= \hat\Upsilon^{(0,0)}$. 
The superfield $\Upsilon^{(0,0)}$ will be useful to define a suitable gauge fixing  procedure 
and to take into account the reducible gauge symmetry of the 
NS-NS two form of 10-dimensional supergravity. 

To derive equations (\ref{deCC}) we can view the vertex operators 
${\cal V}^{(0,1)}_z$ and 
${\cal V}^{(1,0)}_{\bar z}$ 
as deformations of the BRST charges
\beq
\label{defBRST}
Q_L \rightarrow Q_L + \oint d\bar z \, {\cal V}^{(1,0)}_{\bar z}\,, 
\quad\quad\quad
Q_R \rightarrow Q_R + \oint d z \, {\cal V}^{(0,1)}_{z}\,, 
\eeq
and the vertex operator ${\cal V}^{(0,0)}_{z \bar z}$ as
the deformation of the action
\beq
\label{defAction}
S \rightarrow S + \int dz d\bar z \, {\cal V}^{(0,0)}_{z \bar z}\,. 
\eeq
Eqs. (\ref{deC}) are derived by requiring the nilpotency of the new charges and the vanishing of their anticommutation relation. 
\subsection{Amplitudes}
Since the worldsheet ghost variables $\l^\a$ are constrained by the pure spinor condition (\ref{purespin}), it is not obvious how the define a path integral in this variables and therefore how to compute superstring amplitudes. For this reason in \cite{grassi} a different formulation was proposed where the pure spinor constraint is relaxed by adding more fields to the theory. Clearly this should be done without modifying the BRST cohomology, that was shown to reproduce the correct superstring physical spectrum \cite{BerkovitsNN, BerkovitsMX}.
However, Berkovits recently showed \cite{multiloop} that multiloop superstring amplitudes can be computed in the pure spinor formalism, by introducing an analogue of the RNS ``picture changing" operators.

When only tree-level amplitudes are under concern, a prescription can be given to compute them relying on properties of BRST cohomology, as shown in \cite{BerkovitsPH}. The prescription given there was shown to coincide with the standard RNS one in \cite{BerkovitsUS}.
In terms of the vertex operators ${\cal O}^{(a,b)}_{c,d}$, the amplitudes on the 
sphere are defined as
\beq
\label{ampl}
{\cal A}_{n+3} = 
\Big\langle  
{\cal V}^{(1,1)}(z_{1},\bar z_{1}) {\cal V}^{(1,1)}(z_{2},\bar z_{2}) {\cal V}^{(1,1)}(z_{3},\bar z_{3})
\prod_{n} \int dz d\bar z {\cal V}^{(0,0)}
\Big\rangle
\eeq
 where the three unintegrated vertex operators are needed to fix the 
 $SL(2,{\bf C})$ invariance on the sphere. An unintegrated vertex 
 ${\cal V}^{(1,1)}(z_{1},\bar z_{1})$ can be replaced by a product of $(1,0)$ and 
 $(0,1)$ vertices $\oint dz {\cal V}_{z}^{(0,1)} \oint d\bar z {\cal V}_{\bar z}^{(1,0)}$ 
 which has the same total ghost number and the same total 
 conformal spin as the original vertex ${\cal V}^{(1,1)}$. In \cite{BerkovitsPH} 
 supersymmetry and gauge invariance were proven 
 under the assumption that the prescription for the zero modes is the following
 \beq
 \label{zeromodi}
 \langle {\cal V}^{(3,3)} \rangle = 1
\eeq
where 
\beq
 {\cal V}^{(3,3)} =  
 (\l_{0}\g^{m}\th_{0} 
\l_{0}\g^{n}\th_{0} \l_{0}\g^{p}\th_{0} \th_{0} \g_{mnp}\th_{0}) 
(\hat\l_{0}\g^{m} \hat\th_{0} 
\hat\l_{0}\g^{n}\hat\th_{0} \hat\l_{0}\g^{p}\hat\th_{0} \hat\th_{0} \g_{mnp}\hat\th_{0}) \,. 
\eeq
As anticipated, this zero-mode prescription is justified by cohomological arguments. In fact, by analogy with the RNS case, one deduces that the the expectation value for the +3 ghost-number vertex operator for the Yang-Mills antighost must be fixed to one and there is a unique state of such ghost number in the pure-spinor BRST cohomology, given by ${\cal V}^{(3,3)}$.

In \cite{multiloop} Berkovits has given a general prescription for the computation of multiloop amplitudes in the pure spinor formalism for the superstring. Since in my work I didn't compute amplitudes, I'm not going to give technical details about this. However, I will discuss the various difficulties that were overcome in \cite{multiloop}.

As outlined in section 3.1.1, the ghost variables can appear only as $\l^\a$ (with conformal weight 0) or in the two gauge-invariant combinations $N_{mn}$ and $J$ (with conformal weight 1).
Because of the pure spinor constraint (\ref{purespin}), $\l^\a$ has only eleven free components.
As a result, on a genus $g$ surface $\l^\a$ has eleven independent zero-modes and $N_{mn}$ and $J$ have $11g$ ones. It is not obvious how to determine a Lorentz covariant prescription to integrate over these ghost zero-modes. A Lorentz-invariant measure for $\l$ $[{\cal D} \l]$ was constructed in \cite{multiloop}. Moreover, it has been noted that zero modes of $N$ and $J$ are related by a constraint following from the pure spinor relation (\ref{purespin}), such that all these zero-modes can be expressed in terms of ten free $N$ zero-modes. The measure for these free modes, $[{\cal D}N]$, is also given in \cite{multiloop}. 

On a genus $g$ surface, one integrates out all the non-ghost fields and the ghost non-zero-mode fields by making use of the given OPE's. One is then left with an expression like
\beq
{\cal A}=\langle f(\l, N_1, J_1,\dots, N_g, J_g)\rangle
\eeq
that only depends on the ghost zero-modes. Then one uses the Lorentz invariant measures to define the integration over these zero-modes
\beq
{\cal A}=\int [{\cal D} \l][{\cal D}N_1]\dots[{\cal D}N_g]f(\l,N_1,J_1,\dots,N_g,J_g)
\eeq
To compute the integral, as in RNS formalism \cite{fms}, it is necessary to insert picture changing operators involving the delta-functions $\d(C_\a\l^\a)$, $\d(B_{mn}N^{mn})$ and $\d(J)$, where the constant spinor and antisymmetric tensor $C_\a$ and $B_{mn}$ should not affect the amplitudes (that otherwise could not be Lorentz invariant!). These picture changing operators will be called $Y_C$, $Z_B$ and $Z_J$ respectively. Their insertion in loop amplitudes is necessary to absorb the ghost zero-modes. Exactly as in the RNS formulation, picture changing operators must be BRST-invariant with a BRST-trivial worldsheet derivative, this second requirement needed for the amplitudes to be independent of PCO's positions on the worldsheet. In \cite{multiloop} operators with these properties were constructed and it was also shown that even if they are not supersymmetric, their variation is BRST-trivial. As a result, supersymmetry is preserved (up to surface terms). 
In \cite{multiloop} it was also shown that the computation of tree amplitudes with the Lorentz covariant measure for pure spinor zero-modes agrees with the previous prescription obtained by cohomology arguments. 

To compute loop amplitudes a last ingredient is missing. In fact, in RNS formalism, in the computation of a $g$-loop amplitude the insertion of $(3g-3)$ $b$-ghosts is necessary. The $b$-ghost has $-1$ ghost number and satisfies the relation $\{Q,b(u)\}=T(u)$ where $T$ is the stress tensor.
The problem is that in the pure spinor formalism the ghost conjugate momentum $w_\a$ only appears in the gauge invariant combinations $N_{mn}$ and $J$ that have both ghost-number zero. So apparently there is no candidate for an analogue of the $b$ ghost in this formalism.
However, in \cite{multiloop} it was shown that the picture raising operator $Z_B$, that will be present in the expression of a general multiloop amplitude, can be used to construct a nonlocal operator $\tilde b_B$, carrying ghost-number zero, such that $\{Q,\tilde b_B(u,z)\}=T(u)Z_B(z)$. From $\tilde b_B(u,z)$ it is possible to define a local $b_B(u)$, however for the computation of the amplitudes it will be sufficient to know the nonlocal operator (about this topic, see also \cite{odatonin}).

With all these ingredients one can give the following super-Poincar\'e covariant prescription for the computation of a generic $N$-point $g$-loop closed string amplitude 
\bea
{\cal A}&&=\int d^2 \tau_1 \dots d^2 \tau_{(3g-3)}\langle \vert \prod_{P=1}^{3g-3}\int d^2 u_P \m_P(u_P)\tilde b_{B_P}(u_P,z_P)\cr
&&\prod_{P=3g-2}^{10g}Z_{B_P}(z_P)\prod _{R=1}^g Z_J(v_R)\prod_{I=1}^{11}Y_{C_I}(y_I)\vert^2 \prod_{T=1}^N \int d^2 t_T {\cal V}_T(t_T)\rangle\cr
&&~~~~
\label{prescr}
\ena
where $\vert ~\vert^2$ means left-right product, $\tau_P$ are the Teichmuller parameters associated to the Beltrami differentials $\m_P(u_P)$ and ${\cal V}_T(t_T)$ are the dimension $(1,1)$ closed superstring vertex operators for the $N$ external states.
When $g=1$, the general prescription (\ref{prescr}) must be modified, as usual, by changing one integrated vertex operator with an unintegrated one.
\vskip 22pt
Even though this formalism is quite cumbersome for computing superstring amplitudes, it makes it easy to prove some general vanishing theorems, as it is somehow expected because of the great amount of manifest symmetries. Actually, the general vanishing of certain amplitudes can be deduced by just counting zero-modes. For instance, S-duality of type IIB superstrings imply that $R^4$ terms in the low energy effective action do not receive perturbative corrections above one loop \cite{Sth}. In RNS formalism, this was checked only up to two loops in \cite{Spert}. In the pure spinor formalism for the superstring, instead, this can be easily proven for general $g$ \cite{multiloop}.
Furthermore, it is well-known that massless N-point superstring $g$-loop amplitudes vanish for $N<4$. This is equivalent to perturbative finiteness of the theory, when unphysical divergences are not present in the interior of moduli space \cite{finiteness}. This was proven in \cite{mandelstam} by an argument that made use of both GS and RNS formalisms. In the pure spinor formalism this can also be proven by counting zero-modes \cite{multiloop}.
\subsection{What we can (cannot) do with this formalism, up to now}
In this section I would like to summarize the results obtained with the pure spinor formalism for the superstring.

First of all, I should say that in \cite{BerkovitsNN, BerkovitsMX} it was proven that the pure spinor BRST cohomology reproduces the correct superstring spectrum. Therefore, the pure spinor formalism describes the same physics as the RNS and the GS formalisms. In the following I will describe how the pure spinor superstring proved to be superior to analyze many aspects of string theory.

As discussed in the previous sections, the main motivation to introduce a superPoincar\'e covariant formulation for the superstring is the number of difficulties one encounters when trying to compute a general superstring amplitude in the RNS and GS formalisms. As we have seen in the previous section, the pure spinor formalism for the superstring in principle allows to compute arbitrary $N$-point multiloop amplitudes (even though, in practice, only tree-level \cite{tree} and four-point one-loop \cite{multiloop} amplitudes (that can be computed also in RNS or GS formalisms) have been explicitly evaluated up to now). 
Furthermore, we have already stressed that a big advantage of the pure spinor formulation is the possibility to prove vanishing theorems at arbitrary order in the perturbative expansion.

When I wrote my paper \cite{mypaper4}, in collaboration with P.A. Grassi, the prescription for computing a general multiloop amplitude was not known. However it was already clear that the computation of the amplitudes in the pure spinor formalism is rather involved, even in the tree-level case. One of the biggest problems is the complexity of the expression for unintegrated and integrated vertex operators, due to the manifest symmetries that render the formulation redundant. As we have seen in section 3.1.3, vertices are written in terms of superfields satisfying a set of linearized equations of motion, where the physical fields, such as the graviton, dilaton, R-R field strength and so on, do not appear explicitly.  To write down the component expansion of the vertex, one has to solve the equations of motion, after having chosen a specific gauge. This procedure is quite complicated and determining the vertex for a given physical field is not an easy task. In \cite{mypaper4}, P.A. Grassi and I described an iterative procedure to compute type II vertices that eliminates auxiliary fields from the vertices and allows to determine the whole $\th$ and $\hat\th$ vertex expansion given the physical fields.

Another serious problem of RNS formalism is the difficulty in dealing with general R-R background.
We have seen that vertices for closed superstring can be written and that the equations of motion and gauge trasformations for the superfields appearing in the vertices are the linearized supergravity equations and gauge transformations.
Moreover, the pure spinor superstring can be naturally coupled to a general supergravity background. It has been shown in \cite{BerkovitsUE} that nilpotency and holomorphicity of the pure spinor BRST charge imply the on-shell superspace constraints of the supergravity background. 
Aspects of the superstring in specific R-R backgrounds, such as $AdS_5\times S^5$ and the pp-wave were considered in \cite{berkovitspp, BerkovitsYR,AdS}.
Furthermore, as we have seen in section 1.2.4, non(anti)commutative superspaces were shown to emerge when open superstrings in the presence of D-branes and R-R backgrounds are considered. The natural setting for this discussion was the ten-dimensional pure spinor superstring \cite{antonio} and its compactification on a CY three-fold \cite{vafaooguri2,seiberg}. 

In the open string case, requiring BRST invariance also implies the correct equations of motion for the background fields. Indeed, in \cite{borninf} it was shown that classical BRST invariance of the open pure spinor superstring implies the supersymmetric Born-Infeld equations, that were first determined by making use of superembedding techniques \cite{superembedding} in \cite{kerstan}\quad (for the application of the superembedding formalism to determine higher-order corrections to the effective dynamics of string/M theory branes, see also \cite{kerstan2}).  $\kappa$-symmetry in the GS string and superembedding formalism is replaced by BRST symmetry in the pure spinor formalism.To obtain the supersymmetric Born-Infeld equations from the pure spinor formalism for the superstring, one requires that the left and right pure spinor BRST currents are equal on the worldsheet boundary in the presence of the background.

I would also like to say that the pure spinor formalism has been successfully used to quantize the $d=10$ superparticle \cite{superparticle}. Moreover, by replacing ten-dimensional pure spinors with eleven-dimensional pure spinors, the formalism has been extended to the $d=11$ superparticle and supermembrane \cite{membrane}. The covariant prescription to compute loop amplitudes in the covariant formalism for the superstring I briefly discussed in section 3.1.4 has been generalized to the eleven-dimensional superparticle in \cite{grassi11}.

The supermembrane is a quite problematic object in string theory, since the impossibility of performing a covariant quantization has made it difficult to study its properties. 
However, the supermembrane is worth studying, since it is expected to be related to M-theory, which is the underlying eleven dimensional theory from which the nonpertubative symmetries of string theory are believed to come. In \cite{membrane} Berkovits ``replaced" $\kappa$-symmetry with BRST symmetry as he did for the superstring. However, not all the problems of the supermembrane are solved by moving on to a pure spinor description, since the pure spinor supermembrane action is not quadratic in a flat background.  In \cite{membrane} a conjecture is made that supermembrane amplitudes can however be computed and that they are M-theory scattering amplitudes which, after a suitable compactification that reduces to ten dimensions, reproduce Type IIA superstring scattering amplitudes. These amplitudes would contain non-perturbative information about the Type IIA superstring which might be useful for studying M-theory. 

Finally, in my paper \cite{mypaper4}, P. A. Grassi and I discussed an application of pure spinor techniques to the construction of off-shell vertex operators in the asymmetric picture, that could be useful to study the coupling of R-R potentials to D-branes. We also proposed an application to closed string field theory, by studying antifields in this formalism and constructing a kinetic term for the closed string field theory action that seems to respect the correct symmetries and generate the right equations. 

To conclude, I would like to stress that, even if the construction of the pure spinor formalism might seem ``ad hoc" and not justified by an underlying general principle, it is the first successful covariant method to quantize the superstring. It has already proven to be useful to deal with many aspect of string theory that were unreachable by the other formalisms. It might be that in the end a more natural and elegant construction of the covariant superstring will be found. However, it is now very clear that the pure spinor approach at least goes in the right direction and is very effective and useful in string theory.
\section{An iterative procedure to compute the vertex operators}
In this section, I will present the general procedure to compute pure spinor closed string vertex operators I introduced in \cite{mypaper4}, in collaboration with P.A. Grassi.
\subsection{Warm up: The open superstring case}
Motivated by the increasing interest in the covariant 
techniques for computation of the amplitudes 
in string theory, in \cite{mypaper4} P.A. Grassi and I provided a calculation scheme for superstring vertex operators 
in pure spinor approach  
\cite{berkovits,grassi}. 
Since the amount of symmetries that are manifest in the covariant formulation increases, also the number of auxiliary fields 
increases and a useful technique to compute 
the basic ingredients is needed.  In \cite{mypaper4} we provided such a 
procedure and some applications (that will be discussed in section 3.3). 
First of all I will briefly review the open superstring case, 
to explain the main idea that will be applied in the next sections to the more complicated closed string case.  

In the case of the open superstring, we have seen in section 3.1.2 that the massless sector is described by a 
vertex operator ${\cal V}^{(1)} = 
\l^{\a} A_{\a}$ at ghost number one, where $\l^{\a}$ is 
a pure spinor satisfying (\ref{purespin}) and $A_{\a}(x,\th)$ 
is the spinorial component of the superconnection. 
The superfield $A_{\a}$ can be completely 
expressed in terms of the 
gauge field $a_{m}(x)$ and the gluino $\psi^{\a}(x)$, for example as
\beq
\label{I}
A_{\a}(x,\th) = \half(\g^m\th)_\a a_m (x) +\frac{1}{3} (\g^m\th)_\a (\g_m\th)_\g \psi^\g (x) + {\cal O}(\th^3). 
\eeq
The vertex operator ${\cal V}^{(1)}$ belongs to the cohomology of the 
BRST charge $Q = \int d\sigma \l^{\a} d_{\sigma\a}$, where $d_{\sigma\a}$ 
is defined in section 3.1.2, if and only if 
the components of $A_{\a}$ satisfy the linear Maxwell and Dirac equations
\beq
\label{IA}
\pa^{m} (\pa_{m} a_{n} - \pa_{n} a_{m}) =0 \,, ~~~~~
\g^{m}_{\a\b} \pa_{m} \psi^{\b} = 0\,.
\eeq
The contributions ${\cal O}(\th^{3})$ are given in terms of the derivatives 
of $a_{m}$ and $\psi$ and are completely fixed by the equations 
of motion (\ref{openEOM}) given in \cite{howe}, where 
$A_{m}$ is the vectorial part of the superconnection and  
$D_{\a} = \pa_\a+ \half(\g^m\th)_\a \pa_m$ is the superderivative.  
The lowest components of $A_{\a}$ in (\ref{I}) are eliminated by  a gauge fixing condition. 

Even though the computation of all terms in the expansion of $A_{\a}$ seems a 
straightforward procedure, technically it is rather involved. However, 
there exists a powerful technique which simplifies the task. 
The main idea is to choose a suitable gauge fixing such as for instance
\beq\label{II}
\th^{\a} A_{\a}(x, \th) = 0\,,
\eeq
which reduces the independent components in the superfield $A_{\a}$.
This choice\footnote{The following 
gauge condition has a counterpart in bosonic string theory: $x^{m} A_{m}(x) =0$.
This fixes the gauge invariance under $\delta A_{m} = \pa_{m} \omega(x)$ and 
it coincides with the Lorentz gauge in momentum space $\pa_{p_{\mu}} \tilde A_{m} =0$. 
The gauge fixing yields the equation $( 1 + x^{n} \pa_{n})  A_{m} = x^{n} F_{mn}$
which can be solved directly by inverting $(1 + x^{n} \pa_{n})$ and obtaining 
$A_{m} = \int^{x} d^{26}y [(1 + y^{p} \pa_{p})^{-1} (y^{n} F_{mn}(y) ]$.
} fixes part of the super-gauge transformation 
$\delta A_{\a} = D_{\a} \Omega$, where $\Omega$ is a scalar 
superfield with ghost number zero. To reach the gauge (\ref{II}), we have 
to impose $\th^{\a}(A_{\a} + \delta A_{\a}) = 0$, which implies 
that $\th^{\a} D_{\a} \Omega = - \th^{\a}A_{\a}$. Expanding 
$\Omega$ as $\Omega = \sum_{n \geq 0} 
\Omega_{[\a_{1} \dots \a_{n}]} \th^{\a_{1}} \dots \th^{\a_{n}}$, all components with 
$n\geq 1$ are fixed except the lowest component $\Omega_{0}$, which corresponds to the usual bosonic gauge transformation 
of Maxwell theory. 
 
Acting with $D_{\a}$ on (\ref{II}) and 
using the equations of motion (\ref{openEOM}), one gets the recursive relations 
\bea
\label{III}
&&(1 + {\bf D}) A_{\a} =  (\g^{m} \th)_\a A_{m}\cr
&&{\bf D}A_m=(\g_m\th)_\g W^\g \cr
&&{\bf D}W^\a=-\frac{1}{ 4}(\g^{mn}\th)^\a F_{mn}\cr
&&{\bf D}F_{mn}=-(\g_{[m}\th)_\g \pa_{n]} W^\g
\ena
where ${\bf D} \equiv \th^{\a} \pa_{\a}$. So, given the zero-order component of $A_{m}$, we can compute the order-$\th$ component of $A_{\a}$. 
The same can be done for $A_{m}$, the spinorial field strength $W^{\a}$ and 
the bosonic curvature $F_{mn} = \pa_{[m} A_{n]}$ making use of the other three equations. 
This renders the  
task of computing all components of $A_{\a}$ in terms of initial data 
$A_{m}(x) = a_{m}(x) + {\cal O}(\th)$ and 
$W^{\a}(x)=\psi^{\a}(x) + {\cal O}(\th)$ a purely algebraic problem
(\cite{har} and \cite{OoguriPS}). 
Moreover, 
one is able to compute all components of the superfields 
appearing in the (descent) ghost-number-zero vertex operator ${\cal V}^{(0)}_{\sigma}$ 
\beq
\label{IV}
{\cal V}^{(0)}_{\sigma} = \pa_{\sigma} \th^{\a} A_{\a} + \Pi^{m}_{\sigma} A_{m} 
+ d_{\sigma\a} W^{\a} + \half\, N^{mn}_{\sigma} F_{mn}\,, 
\eeq
which satisfies the descent equation 
$[Q, {\cal V}^{(0)}_{\sigma}] = \pa_{\sigma} {\cal V}^{(1)}$. 
Here $\sigma$ is the boundary worldsheet coordinate and 
$N^{mn}_{\sigma} = \half w_{\sigma} \g^{mn} \l$ is the pure spinor part of the Lorentz current. As we have seen in section 3.1.2, the operators 
$\Pi^{m}_{\sigma}$ and $d_{\sigma\a}$ are the supersymmetric line 
element and the fermionic constraint of the Green-Schwarz superstring \cite{gsstring}, 
respectively. 	

In my paper \cite{mypaper4}, we applied the same technique to 
IIA/IIB supergravity. Starting from the vertex operators for closed 
superstrings, we derived the complete set of equations 
from the BRST cohomology and we defined all curvatures and 
gauge transformations. Then, we imposed a set of gauge fixing 
conditions to remove the lowest components of the superfields and 
we derived an iterative procedure to compute all components. We showed
that a further gauge fixing is needed to fix the reducible gauge symmetries 
and we showed that all chosen gauges can indeed be reached. 

The procedure for closed strings is original by itself, but, more importantly, 
our analysis leads to a generalization of (\ref{II}) to all 
vertex operators, associated to both massless and massive states. Indeed, in \cite{mypaper4} we showed that the 
gauge fixing (\ref{II}) can be written in terms of a new nilpotent 
charge ${\cal K}$ (with negative ghost number) as follows
\beq
\label{IVA}
\{ {\cal K}, {\cal V}^{(1)} \}= 0 \,.
\eeq
This imitates the Siegel gauge in string field theory. When restricted to massless states, this generalized gauge fixing condition reduces to the gauge fixing (\ref{II}) for open superstrings and to the corresponding gauge fixing for closed strings. 
When applied to massive states, (\ref{IVA}) also leads to a suitable gauge fixing. In our paper, we explicitly derived the gauge conditions for the first massive state for the open superstring. 
Again, (\ref{IVA}) fixes all auxiliary fields in terms of the physical on-shell data and 
eliminates the lowest components. 

I would like to stress that in \cite{mypaper4} we only considered deformations (vertex operators) 
at first order in the coupling constant, neglecting the backreaction 
of background fields. 
\subsection{Linearized IIA/IIB supergravity equations}
In the present section we derive the equations of motion for the 
massless background 
fields in superspace from the BRST cohomology of the superstring. 
Let us start from the simplest equations (\ref{deC}) for the 
vertex ${\cal V}^{(1,1)}$ whose general expression is 
  \beq
  \label{verB} 
  {\cal V}^{(1,1)} = \l^\a A_{\a\hat\b}\hat\l^{\hat\b} \,. 
  \eeq
 The superfield $A_{\a \hat\b}(x,\th,\hat\th)$ 
  satisfies the equations of motion  \cite{howe}
  \beq
  \label{verC}
  \g^{\a\b}_{mnopq} D_\a A_{\b \hat\b} =0\,, ~~~~~~ 
  \g^{\hat\a\hat\b}_{mnopq} D_{\hat\a} A_{\a \hat\b} =0\,, ~~~~~~ 
  \eeq
  where $\g^{\a\b}_{mnopq}$ is the antisymmetrized product of 
  five gamma matrices. The pure spinor conditions imply that only the 
 5-form parts of the $D_\a A_{\b \hat\b}$ and $D_{\hat\a} A_{\a \hat\b}$ 
 are indeed constrained \cite{berkovits,berkICTP,BerkovitsWM}. 
 By using Bianchi identities, one can show that 
 they yield the type IIA/IIB supergravity equations of motion at the linearized level. 
 All auxiliary fields present in the superfield $A_{\a\hat \b}$ are fixed by 
 eqs. (\ref{verC}). 
 
 As outlined before, one can use  
 different types of vertices to simplify the computations. 
 Integrated vertices are written 
 in terms of a huge number of different superfields, whose components 
 are completely fixed by the equations of 
 motion. As a result, these vertices are quite complicated espressions.

The set of superfields needed to compute 
 ${\cal V}^{(0,0)}, \dots, {\cal V}^{(1,1)}$
 can be grouped into the following matrix
 \beq
 \label{vertmatrix} 
{\bf A } = 
 \left[ \begin{matrix}
 A_{\a\hat\b} & A_{\a p} & E_{\a}^{~~\hat\b}& 
 \Omega_{\a, pq}\cr
A_{m\hat\b} & A_{mp} & E_m^{~~\hat\b} & 
\Omega_{m, pq}\cr
 E^{\a}_{~~\hat\b} & E^\a_{~~p} & P^{\a\hat\b} & 
 {C}^\a_{~~pq}\cr
 \Omega_{mn, \hat\b} & \Omega_{mn,p} & 
 C_{mn}^{~~~~\hat\b} & S_{mn,pq}
 \end{matrix}\right]
 \eeq
The first components of $A_{mp}$, 
$E_m^{~~\hat\b}$, $E^\a_{~~p}$ and $P^{\a\hat\b}$ 
are identified with the  supergravity fields as follows
\beq
\label{sugra}
A_{mp}= g_{mp} + b_{mp} + \eta_{mp} \phi + {\cal O}(\th, \hat\th)\,, 
\eeq
\beq
E_m^{~~\hat\b} = \psi_m^{~~\hat \b} +  {\cal O}(\th, \hat\th)\,,  ~~~~~~~~~E^\a_{~~p}  =  \psi_{~~p}^\a + {\cal O}(\th, \hat\th)\,, 
\eeq
\beq 
P^{\a\hat\b} = f^{\a\hat\b} + {\cal O}(\th, \hat\th)\,.
\eeq
The fields $g_{mn}$, $b_{mn}$, $\phi$, 
 $\psi_{~~p}^\a$, $\psi_m^{~~\hat \b}$ and $f^{\a \hat \b}$ are 
 the graviton, the NS-NS two-form, the dilaton, the two gravitinos (the gamma-traceless 
 part of $\psi_{~~p}^\a$, $\psi_m^{~~\hat \b}$), the two dilatinos (the 
 gamma-trace part of $\psi_{~~p}^\a$, $\psi_m^{~~\hat \b}$) and the
 RR field strengths. IIA and IIB differ in the chirality 
 of the two spinorial indices $\a$ and $\hat \a$. This 
 changes the type of RR fields present in the spectrum. The first 
 components of the superfields $\Omega_{m,pq}$ ($\Omega_{mn,p}$), $C_{mn}^{~~~~\b}$  ($C^\a_{~~ pq}$) and $S_{mn,pq}$ are identified with the linearized 
 gravitational connection $\Gamma_{rs}^t$, the curvature of 
 the gravitinos and the linearized Riemann tensor,  respectively.
 The remaining superfields are the spinorial partners of the above superfields. 
 Those constraints are 
 given in terms of the spinorial components $A_{\a\hat\b}$, $A_{\a p}$, $E_{\a}^{~~\hat\b}$ 
 and $\Omega_{\a, pq}$. The structure of superspace formulation 
 of type IIA and IIB supergravity in the present framework is also 
 discussed in \cite{BerkovitsUE}. 

Given the vectors ${\bf X}_z$ and ${\bf \hat X}_{\bar z}$ (see (\ref{vectors}))
we can explicitly write the vertex operator ${\cal V}^{(0,0)}_{z \bar z} ={\bf X}^T_z {\bf A} {\bf \hat X}_{\bar z} $ 
as
\bea
\label{vert}
{\cal V}^{(0,0)}_{z \bar z} 
&=& \pa_z\theta^\a~ A_{\a\hat\b}~ \pa_{\bar z}\hat\theta^{\hat\b} 
+ \pa_z \theta^\a~ A_{\a p}~ \hat\Pi^p_{\bar z} 
+ \Pi^m_z~ A_{m \hat\b}~ \pa_{\bar z}\hat\theta^{\hat\b}
+ \Pi^m_z~ A_{mp}~ \hat\Pi^p_{\bar z}  \cr
&+& d_{z\a}~ E^\a_{~~\hat\b}~ \pa_{\bar z}\hat\theta^{\hat\b}~ 
+ d_{z\a}~ E^\a_{~~p}~ \hat\Pi^p_{\bar z} 
+ \pa_z\theta^\a~ E_\a^{~~\hat\b}~\hat d_{\bar z\hat\b}
+ \Pi^m_z~ E_m^{~~\hat\b}~\hat d_{\bar z\hat\b} \cr
&+& d_{z\a}~ P^{\a\hat\b}~ \hat d_{\bar z\hat\b}
+\half~ N^{mn}_z~ \Omega_{mn,\hat\b}~ \pa_{\bar z}\hat\theta^{\hat\b} 
+\half~ N^{mn}_z~ \Omega_{mn,p}~ \hat\Pi^p_{\bar z} \cr
&+&\half~ \pa_z\theta^\a~ \Omega_{\a,pq} \hat N^{pq}_{\bar z} 
+\half~ \Pi^m_z~  \Omega_{m,pq} \hat N^{pq}_{\bar z} 
+\half~ N^{mn}_z~ C_{mn}^{\quad\hat\b}~\hat d_{\bar z\hat\b} \cr
&+&\half~ d_{z\a}~ C^{\a}_{\quad pq}~ \hat N^{pq}_{\bar z} 
+\frac{1}{4}~ N^{mn}_z~ S_{mn,pq}~ \hat N^{pq}_{\bar z}
\ena
From equations (\ref{deC}), (\ref{deCC}), (\ref{deD}) and (\ref{deE}) 
in the previous section we derive the complete set of 
equations for the background fields 
\vskip 11pt
\begin{center}
\begin{scriptsize}
\begin{tabular}{c l l}
$\left(\half,\half,\half\right)$ & 
$D_\a A_{\b\hat\g} + D_\b A_{\a\hat\g} - \g^m_{\a\b} A_{m\hat\g}=0$ & $\hat D_{\hat\a} A_{\b\hat\g} + \hat D_{\hat\g} A_{\b\hat\a} - \g^m_{\hat\a\hat\g} A_{\b m}=0$\\
$\left(\half,\half, 1\right)$ & $D_\a A_{m\hat\b} -\pa_m A_{\a\hat\b}-\g_{m\a\g}E^{\g}_{~~\hat\b}=0$ & $\hat D_{\hat\a} A_{\b p} - \pa_p A_{\b\hat\a} - \g_{p\hat\a \hat\g} E_\b^{~~\hat\g}=0$\\
$\left(\half,\half, 1\right)$ & $D_\a A_{\b p} + D_\b A_{\a p} -\g^m_{\a\b} A_{mp}=0$ & $\hat D_{\hat\a}A_{m\hat\b}+ \hat D_{\hat\b} A_{m\hat\a} + \g^p_{\hat\a\hat\b}A_{mp}=0$\\
$\left(\half,\half, \frac{3}{2}\right)$ & $D_\a E^{\b}_{~~\hat\g} - \frac{1}{ 4} (\g^{mn})_\a^{~~\b} \Omega_{mn,\hat\g}=0$ & $\hat D_{\hat\a} E_\b^{~~\hat\g} -\frac{1}{4} (\g^{pq})_{\hat\a}^{~~\hat\g} \Omega_{\b, pq}=0$\\
$\left(\half,\half, \frac{3}{ 2}\right)$ & $D_\a E_\b^{~~\hat\g} + D_\b E_\a^{~~\hat\g} - \g^m_{~~\a\b} E_m^{~~\hat\g}=0$ & $\hat D_{\hat\a} E^\b_{~~\hat\g} + \hat D_{\hat\g} E^\b_{~~\hat\a} - \g^p_{~~\hat\a\hat\g}E^\b_{~~p}=0$\\
$\left(\half,1, 1\right)$ & $D_\a A_{mp} -\pa_m A_{\a p} - \g_{m\a\g} E^\g_{~~p}=0$ & $\hat D_{\hat\a}A_{mp}+ \pa_p A_{m\hat\a} + \g_{p\hat\a\hat\b}E_m^{~~\hat\b}=0$\\
$\left(\half,\frac{3}{2}, 1\right)$ & $D_\a E^\b_{~~p} - \frac{1}{ 4} (\g^{mn})_\a^{~~\b}\Omega_{mn,p}=0$ & $\hat D_{\hat\a} E_m^{~~\hat\b}+\frac{1}{ 4} \Omega_{m,pq}(\g^{pq})_{\hat\a}^{~~\hat\b}=0$\cr
$\left(\half,\frac{3}{2}, 1\right)$& $D_\a E_m^{~~\hat\b} -\pa_m E_\a^{~~\hat\b} - \g_{m\a\g}P^{\g\hat\b}=0$ & $\hat D_{\hat\a} E^\b_{~~p} - \pa_p E^\b_{~~\hat\a} - \g_{p\hat\a\hat\g}P^{\b\hat\g}=0$\\
$\left(\half,\half, 2\right)$ & $D_\a \Omega_{\b, pq} + D_\b \Omega_{\a, pq} - \g^m_{~~\a\b}\Omega_{m,pq}=0$ & $\hat D_{\hat\a} \Omega_{mn,\hat\b} + \hat D_{\hat\b} \Omega_{mn,\hat\a}+ \g^p_{\hat\a\hat\b}\Omega_{mn,p}=0$\\
$\left(\half,\frac{3}{2}, \frac{3}{2}\right)$ & $D_\a P^{\b\hat\g} -\frac{1}{4} (\g^{mn})_\a^{~~\b}C_{mn}^{~~~~\hat\g}=0$ & $\hat D_{\hat\a} P^{\b\hat\g} - \frac{1}{4} (\g^{pq})_{\hat\a}^{~~\hat\g}C^{\b}_{~~pq} =0$\\
$\left(\half,1, 2\right)$& $D_\a \Omega_{m,pq} -\pa_m \Omega_{\a, pq}- \g_{m\a\g}C^{\g}_{~~ pq}=0$ & $ \hat D_{\hat\a}\Omega_{mn,p} + \pa_p \Omega_{mn,\hat\a} + \g_{p\hat\a\hat\b}C_{mn}^{~~~~\hat\b}=0$\\
$\left(\half,\frac{3}{2}, 2\right)$ & $D_\a C^{\b}_{~~pq} \frac{1}{4} (\g^{mn})_\a^{~~\b}S_{mn,pq}=0$ & $ \hat D_{\hat\a} C_{mn}^{~~~~\hat\b} + \frac{1}{4} (\g^{pq})_{\hat\a}^{~~\hat\b}S_{mnpq}=0 $
\end{tabular}
\end{scriptsize}
\end{center}
\beq
\label{eom}
\eeq
where the labels $(a,b,c)$ denote the scaling dimensions of the generators of  
the extended super-Poincar\'e algebra \cite{siegel,GreenNN} 
which the equations belong to. 

Moreover, one obtains the following eight equations, which do not provide further information, since they are implied by (\ref{eom}) and pure spinor conditions 
\bea
\label{trivialeq}
&&N^{mn}\l^\g D_\g \Omega_{mn,\hat\b}=0~~~~~~~~~~ \hat\l^{\hat\g}\hat D_{\hat\g}\Omega_{\a, mn}\hat N^{mn}=0\cr
&&N^{mn} \l^\g D_\g \Omega_{mn,p}=0~~~~~~~~~~ \hat\l^{\hat\g}\hat D_{\hat\g}\Omega_{m,pq}\hat N^{pq}=0\cr
&&N^{mn} \l^\g D_\g C_{mn}^{~~~~\hat\b}=0~~~~~~~~~~\hat\l^{\hat\g}\hat D_{\hat\g}C^{\a}_{~~ mn}\hat N^{mn}=0\cr
&&N^{mn} \l^\g D_\g S_{mn,pq}\hat N^{pq}=0~~~~N^{mn}\hat\l^{\hat\g}\hat D_{\hat\g} S_{mn,pq}\hat N^{pq}=0
\ena
Since we assumed that the superfields 
$\Omega_{mn,p}, \Omega_{m,pq}, C_{mn}^{~~~~\hat\b}, C^\a_{~~pq}$ and $S_{mn,pq}$
correspond to the linearized curvatures of the connections, we can derive new equations 
needed for the iterative procedure outlined in the introduction. 
By contracting equations (\ref{eom}) with respect to the bosonic derivative and antisymmetrizing 
the bosonic indices, one obtains
\bea
\label{neweom}
&& D_\a \Omega_{mn,\hat\b}=\pa_{[m}\g_{n]\a\g}E^\g_{~~\hat\b}~~~~~~~~
\hat D_{\hat\b}\Omega_{\a, pq}=\pa_{[p}\g_{q]\hat\b\hat\g}E_\a^{~~\hat\g}\cr
&& D_\a \Omega_{mn,p}=\pa_{[m}\g_{n]\a\g}E^\g_{~~p}~~~~~~~~~\hat D_{\hat\b}\Omega_{m,pq}=-\pa_{[p}\g_{q]\hat\b\hat\g}E_m^{~~\hat\g}\cr
&& D_\a C_{mn}^{~~~~\hat\b}=\pa_{[m}\g_{n]\a\g}P^{\g\hat\b}~~~~~~~~~\hat D_{\hat\b}C^\a_{~~pq}=\pa_{[p}\g_{q]\hat\b\hat\g}P^{\a\hat\g}\cr
&& D_\a S_{mn,pq} = \pa_{[m}\g_{n]\a\g}C^\g_{~~pq}~~~~~~~\hat D_{\hat\b} S_{mn,pq} =- \pa_{[p}\g_{q]\hat\b\hat\g}C_{mn}^{~~~~\hat\g}
\ena
(we define $a_{[m}b_{n]}=a_m b_n-a_n b_m$).  
The identification of the superfields $\Omega_{mn,p}$, $\Omega_{m,pq}$, $C_{mn}^{~~~~\hat\b}$, $C^\a_{~~pq}$ and 
$S_{mn,pq}$ with the linearized curvatures is automatically derived in the formalism \cite{grassi}, and 
equations (\ref{neweom}) are the usual Bianchi identities.

In order to show that the above equations imply the supergravity
equations of motion we proceed as follows.
We first consider the third line of (\ref{neweom}) and 
the $(\half, \frac{3}{ 2}, \frac{3}{2})$ line of (\ref{eom}), that we recall for the reader convenience
\bea
\label{neweomP}
&&D_\a P^{\b\hat\g} -\frac{1}{4} (\g^{mn})_\a^{~~\b}C_{mn}^{~~~~\hat\g}=0 ~~~~~~~~~~~~ \hat D_{\hat\a} P^{\b\hat\g} - \frac{1}{ 4} (\g^{pq})_{\hat\a}^{~~\hat\g}C^{\b}_{~~pq} =0\cr
&& D_\a C_{mn}^{~~~~\hat\b}=\pa_{[m}\g_{n]\a\g}P^{\g\hat\b} 
~~~~~~~~~~~~~~~~~~~
\hat D_{\hat\b}C^\a_{~~pq}=\pa_{[p}\g_{q]\hat\b\hat\g}P^{\a\hat\g}\cr
&&
\ena
Acting with $\g^{m}_{\a\sigma} \pa_{m}$ on $P^{\sigma \hat \b}$ and 
using the commutation relations of the $D$'s, one gets
\bea
\label{newI}
&&\g^{m}_{\a\sigma} \pa_{m} P^{\sigma \hat \b} = 
(D_{\a} D_{\sigma} + D_{\sigma} D_{\a}) P^{\sigma \hat \b} 
= \frac{1}{4}  (\g^{mn})_\a^{~~\sigma} D_{\sigma} C_{mn}^{~~~~\hat\b}\,,\cr
&&= - \half  (\g^{mn})_\a^{~~\sigma} \g_{m \sigma\g} \pa_{n} P^{\g\hat\b} = 
 \frac{9}{ 2}  \g^{m}_{\a\sigma} \pa_{m} P^{\sigma \hat \b}
\ena
Here we also used the first equation of (\ref{neweomP}) and 
$D_{\a} P^{\a\hat\b} =0$ (which follows from (\ref{neweomP})). In the 
second line we used the first equation in the second line on (\ref{neweomP}) and  
the identity $(\g^{mn}\g_{m})_{\a\b} = -9 \g^{n}_{\a\b}$. By performing the 
same manipulations on the hatted quantities we derive the equations
\beq
\label{newII}
\g^{m}_{\a\sigma} \pa_{m} P^{\sigma \hat \b} = 0\,, ~~~~~~~~~~
\g^{m}_{\hat \a \hat\sigma} \pa_{m} P^{\a \hat\sigma} = 0\,. 
\eeq
Decomposing $P^{\a\hat \b}$ in terms of Dirac matrices, it is 
straightforward to show that (\ref{newII}) implies the equations of motion 
for the RR fields. 

Acting again with $\g^{\a\g}_{n} D_{\a}$ on (\ref{newII}) and using equations 
(\ref{neweomP}) one gets 
\beq
\label{newIII}
0 = \g_{n}^{\a\g} \g^{m}_{\a\b} \pa_{m} D_{\g} P^{\b\hat\b} = 
(\g_{n}\g^{m})^{\g}_{\b} (\g^{pq})_{\g}^{\b} \pa_{m} C_{pq}^{~~~\hat\b} = 
\pa^{m} C_{mn}^{~~~\hat\b}\,,
\eeq
and analogously for $C^{\a}_{~~pq}$. These equations are 
the Maxwell equations for the curvature of the gravitinos. They are not 
enough to describe the dynamic of gravitinos and we have to invoke new 
equations coming from the second line of (\ref{neweom}) and the 
$(\half, \frac{3}{ 2}, 1)$ line of (\ref{neweom}). 

Applying $\g^{m}_{\a\sigma} \pa_{m}$ on $E^{\sigma}_{~p}$ and 
with  $\g^{p}_{\hat\a\hat\sigma} \pa_{p}$ on $E^{~\hat \sigma}_{m}$, the 
same algebraic manipulations yield
\beq
\label{newIV}
\g^{m}_{\a\sigma} \pa_{m} E^{\sigma}_{~p} = 0\,, ~~~~~~~~~~
\g^{p}_{\hat \a \hat\sigma} \pa_{p} E_{m}^{~\hat\sigma} = 0\,. 
\eeq
which are the Dirac equations for the gravitinos. These 
equations are gauge invariant under the gauge transformations 
discussed in the next section since the gauge parameters have 
to satisfy a field equation. In addition, 
as above, we find the equations
\beq
\label{newV}
\pa^{m} \Omega_{mn,p} =0\,, ~~~~~~~~~~
\pa^{p}  \Omega_{m,pq} =0\,, ~~~~~~~~
\eeq
which are, at the lowest component 
of the superfield $\Omega_{mn,p}$ and $\Omega_{m,pq}$,  
the equations of motion of the graviton, the dilaton and the NS-NS form
\bea
\label{newVI}
&&\pa^{m} ( \pa_{[m} g_{n]p} + \pa_{[m} b_{n]p} + \eta_{p[n} \pa_{m]}\phi ) = 0\,, 
\cr
&&\pa^{p} ( \pa_{[p} g_{|m|q]} + \pa_{[p} b_{|m|q]} + \eta_{nm[q} \pa_{p]}\phi) = 0\,. 
\ena

Pursuing this line of reasoning, one can derive similar
equations for $E^{\b}_{~~\hat \a}$, $E_{\a}^{~~\hat \b}$, $\Omega_{mn,\hat \g}$ 
and $\Omega^{\a}_{~~pq}$, which guarantee that the 
fields are either pure gauge or auxiliary fields. Finally, 
by studying the last line of (\ref{neweom}) and the line 
$(\half, \frac{3}{2}, 2)$ of (\ref{eom}), one derives new equations for 
$C^{\b}_{~~pq}, C_{mn}^{~~~~\hat\b}$ and $S_{mn,pq}$, which do not give further information since they are
implied by the previous ones. 
\subsection{Gauge transformations and gauge fixing}
In order to solve the equations of motion (\ref{eom}) and (\ref{neweom}) it is convenient to choose 
a suitable gauge. Indeed, for supersymmetric theories, the large amount of 
auxiliary fields can be reduced by choosing the Wess-Zumino gauge. 
We first discuss the general structure of the gauge transformations 
(\ref{gaugegen}), 
we then provide a gauge fixing and we finally check that this gauge can be reached. In the present framework, the gauge parameters 
$\L^{(0,1)}$, $\L^{(1,0)}$, $\tau_z ^{(0,0)}$ and 
$\tau ^{(0,0)}_{\bar z}$ 
satisfy equations (\ref{consI}) and (\ref{consII}) 
and they are defined up to the gauge transformation (\ref{constrans}). 
This additional gauge invariance is fixed by a further 
gauge fixing. 

The general structure of the gauge parameters $\L^{(0,1)}$, $\L^{(1,0)}$, $\tau_z ^{(0,0)}$ and 
$\tau ^{(0,0)}_{\bar z}$ is given by
\bea
\label{gaugeparI}
\L^{(1,0)} = \l^\a \Theta_\a  \quad\quad\quad 
\L^{(0,1)} = \hat \Theta_{\hat\a} \hat\l^{\hat\a}\,,
\ena
and 
\bea
\label{gaugeparII}
&& \tau_z^{(0,0)} = \pa_z\theta^\a \Xi_\a + \Pi^m_z \Sigma_m + d_{z\a} \Phi^\a + \half N^{mn}_z \Psi_{mn}\cr
&& \tau_{\bar z}^{(0,0)} = \hat  \Xi_{\hat\a}\pa_{\bar z}\hat\theta^{\hat\a} + \hat \Sigma_p \hat\Pi^p_{\bar z} + 
\hat \Phi^{\hat\a}\hat d_{\bar z\hat\a} + \half \hat \Psi_{pq}\hat N^{pq}_{\bar z}\,.
\ena
where $ \Theta_\a, \dots, \hat \Psi_{mn}$ are superfields in the variables $x^m,\th^\a$, and $\hat \th^{\hat \a}$. 
In terms of these superfields, eq. (\ref{consI}) gives
\beq
\label{consIII}
(\g^{mnpqr})^{\a\b}D_\b \Theta_\a=0~~~~~~~~~~
(\g^{mnpqr})^{\hat\a\hat\b}\hat D_{\hat\b} \hat \Theta_{\hat\a}=0\,,
\eeq
while eq. (\ref{consII}) gives 
\bea
\label{consIV}
&& \Theta_\a + \Xi_\a =0 ~~~~~~~~~~~~~~~~~~~~~~~~~~~~~~ \hat\Theta_{\hat\a}- \hat\Xi_{\hat\a}=0\cr
&& D_\a \Theta_\b - D_\b \Xi_\a +\g^m_{\a\b}\Sigma_m=0 ~~~~~~~~~\hat D_{\hat\a}\hat\Theta_{\hat\b} + \hat D_{\hat\b}\hat\Xi_{\hat\a} + \g^p_{\hat\a\hat\b}\hat\Sigma_p=0\cr
&& D_\a \Sigma_m +\pa_m \Theta_\a -\g_{m\a\b}\Phi^\b=0 ~~~~~~~ \hat D_{\hat\a}\hat\Sigma_p +\pa_p \hat\Theta_{\hat\a}+ \g_{p\hat\a\hat\b}\hat\Phi^{\hat\b}=0\cr
&& D_\a \Phi^\b -\frac{1}{4} (\g^{mn})_\a^{~~\b}\Psi_{mn}=0~~~~~~~~~~\hat D_{\hat\a}\hat\Phi^{\hat\b} + \frac{1}{ 4} \left(\g^{pq}\right)_{\hat\a}^{~~\hat\b}\hat\Psi_{pq}\cr
&& N^{mn}\l^\g D_\g \Psi_{mn}=0 ~~~~~~~~~~~~~~~~~~~~~~ \hat\l^{\hat\g}\hat D_{\hat\g}\hat\Psi_{pq}\hat N^{pq}=0\,.
\ena
These equations look like the superspace field equations for superMaxwell theory (\ref{openEOM}), however 
the superfields $\Theta_\a$, $\Sigma_m$, $\Phi^\a$ and $\Psi_{mn}$ and the corresponding hatted quantities depend 
on $x^m$, $\th^\a$ and $\hat\th^{\hat\a}$. Therefore, the eqs. (\ref{consIV}) are not sufficient to determine completely the 
components of those superfields. The free independent components are indeed the gauge parameters.
We also note that the last pair of equations is trivial when the previous equations and 
the pure spinor conditions are imposed. 
Finally, because of the similarity with SYM case, it is quite natural to impose the condition that $\Psi_{mn}$ and $\hat\Psi_{pq}$ are the linearized curvatures of $\Sigma_m$ and $\hat\Sigma_p$. Again, this assumption is automatic in \cite{grassi}. 
The gauge transformations of the superfields in ${\cal V}^{(0,0)}_{z \bar z}$ are given by
\vskip 11pt
\begin{center}
\begin{tabular}{c l l}
$\left(\half,\half\right)$ & $\d A_{\a\hat\b}=D_\a \hat\Xi_{\hat\b}+\hat D_{\hat\b}\Theta_\a$ & \\
$\left(\half,1\right)$ & $\d A_{\a p}=\pa_p \Theta_\a +D_\a \hat\Sigma_p$ &
$\d A_{m \hat\b}=\pa_m \hat\Theta_{\hat \b} + \hat D_{\hat\b} \Sigma_m$\\
$\left(1,1\right)$ & $\d A_{mp}=\pa_m \hat\Sigma_p -\pa_p \Sigma_m$ & \\
$\left(\frac{3}{2},\half\right)$ & $\d E^\a_{~~\hat\b}= - \hat D_{\hat\b} \Phi^\a$ &
$\d E_\a^{~~\hat\b}=D_\a \hat\Phi^{\hat\b}$\\
$\left(\frac{3}{2},1\right)$ &$\d E^\a_{~~p}= -\pa_p \Phi^\a$ &
$\d E_m^{~~\hat\b}= \pa_m \hat\Phi^{\hat\b}$\\
$\left(\half,2\right)$ &$\d \Omega_{\a,pq}=D_\a \hat\Psi_{pq}$ &
$\d \Omega_{mn,\hat\b}=\hat D_{\hat\b} \Psi_{mn}$\\
$\left(1,2\right)$ &
$\d \Omega_{m,pq}=\pa_m \hat\Psi_{pq}$ &
$\d \Omega_{mn,p}=-\pa_p \Psi_{mn}$\\
$\left(\frac{3}{2},\frac{3}{2}\right)$ &
$\d P^{\a\hat\b}=0$ & \\
$\left(\frac{3}{2},2\right)$ & 
$\d C^{\a}_{~~ pq}=0$ & 
$\d C_{mn}^{~~~~\hat\b}=0$\\
$\left(2,2\right)$ &$ \d S_{mn,pq}=0$\,.
\end{tabular}
\end{center}
\beq
\label{gaugetrans}
\eeq
From these equations, we easily see that the superfields $P^{\a \hat \b}$, $C^\a_{~~pq}$, $C_{mn}^{~~~~\hat\b}$ and 
$S_{mn,pq}$ are indeed gauge invariant, as expected, being linearized 
field strengths. At zero order in $\th$ and $\hat\th$ eq. (\ref{gaugetrans}) gives the gauge transformations of 
supergravity fields. For example, the 
first components of $\hat\Sigma_p = \zeta_p + \xi_p + {\cal O}(\th, \hat\th)$ and 
$\Sigma_m = \zeta_m - \xi_m + {\cal O}(\th, \hat\th)$ are to  be identified with the parameters of 
diffeomorphisms $\delta g_{mp} = \pa_m \xi_p + \pa_p \xi_m$ and with the gauge transformations 
of the NS-NS form $\delta b_{mp} = \pa_m \zeta_p - \pa_p \zeta_m$. 
So, the zero-order terms of the gauge parameter superfields $\Theta_\a$, $\Sigma_m$, $\Phi^\a$ 
and of the corresponding hatted quantities are
\bea
\label{gaugezero}
&&\Theta_\a={\cal O}(\th, \hat\th);~~~~~~~~~~~~~~~~~~~~~~\hat\Theta_{\hat\b}={\cal O}(\th, \hat\th)\cr
&&\Sigma_m = \zeta_m - \xi_m + {\cal O}(\th, \hat\th);~~~~~~~
\hat\Sigma_p = \zeta_p + \xi_p + {\cal O}(\th, \hat\th)\cr
&&\Phi^\a=\varphi^\a+ {\cal O}(\th, \hat\th); ~~~~~~~~~~~~~~~\hat\Phi^{\hat\b}=\hat\varphi^{\hat\b}+ {\cal O}(\th, \hat\th) 
\ena
Furthermore, the large amount of gauge parameters allows us to choose the gauge
\bea
\label{gaugefix}
&&\th^\a A_{\a\hat\b}=0~~~~~~~~A_{\a\hat\b}~\hat\th^{\hat\b}=0\cr
&&\th^\a A_{\a p}=0~~~~~~~A_{m\hat\b}~\hat\th^{\hat\b}=0\cr
&& \th^\a E_\a^{~~\hat\b}=0~~~~~~~E^\a_{~~\hat\b}~\hat\th^{\hat\b}=0\cr
&&\th^\a~ \Omega_{\a,pq}=0~~~~~\Omega_{mn,\hat\b}~\hat\th^{\hat\b}=0\,.
\ena
Indeed, we have at our disposal the parameters $\Theta_\a$, $\Sigma_m$, $\Phi^\a$ and $\Psi_{mn}$ and 
the corresponding hatted quantities to impose the gauge (\ref{gaugefix}). Before showing that the gauge can be reached 
we have to notice that the transformations (\ref{gaugetrans}) and the 
equations (\ref{consIII}) are invariant under the residual gauge transformations (\ref{constrans})
\bea
&&\d \Theta_\a=D_\a \Omega~~~~~~ \d \hat\Xi_{\hat\b}=\hat D_{\hat\b}\Omega\cr
&&\d \Sigma_m = - \pa_m \Omega~~~~~\d \hat\Sigma_p= - \pa_p \Omega\cr
&&\d \Phi^\a=0~~~~~~~~~~~~ \d \hat\Phi^{\hat\b}=0\cr
&&\d  \Psi_{mn}=0~~~~~~~~~  \d \hat\Psi_{pq}=0\,,
\label{resid}
\eea
depending on the scalar superfield $\Upsilon^{(0,0)}=\hat\Upsilon^{(0,0)}\equiv\Omega$.
This requires an additional gauge fixing
\beq
\label{gaugefixII}
\theta^\a \Theta_\a + \hat\theta^{\hat\b}\hat\Xi_{\hat\b}=0\,.
\eeq

To show that the gauge choice (\ref{gaugefix}) can be reached by the gauge transformations 
(\ref{gaugetrans}), we have to solve, for instance, the equations
\beq
\label{reacI}
\th^\a (A_{\a\hat\b} + \delta A_{\a\hat\b}) = 0\,, ~~~~~~~~
(A_{\a\hat\b} + \delta A_{\a\hat\b}) \hat\th^{\hat\b} = 0\,,
\eeq
and analogously for all other gauge conditions (\ref{gaugefix}). By using the properties 
of the superderivative, gauge fixing (\ref{gaugefixII}), consistency conditions (\ref{consIV}), and by defining the operators
\beq
\label{defD}
{\bf D}\equiv \th^\a D_\a=\th^\a{\frac{\pa}{\pa\th^\a}}\,,~~~~~~~~{\bf \hat D}\equiv \hat\th^{\hat\b} \hat D_{\hat\b}=\hat\th^{\hat\b}{\frac{\pa}{\pa\hat\th^{\hat\b}}}\,,
\eeq
we get the following recursive equations 
\vskip 11pt
\begin{footnotesize}
\begin{tabular}{l l}
$(1 + {\bf D} + {\bf \hat D}) \Theta_{\a} = - A_{\a \hat \b} \hat\th^{\hat \b}  
- (\g^m \th)_\a \Sigma_m$ &
$(1 + {\bf D} + {\bf \hat D}) \hat\Theta_{\hat \b} = - \th^\a A_{\a \hat \b} - (\g^p\hat\th)_{\hat\b} \hat\Sigma_p$\\
$({\bf D}+{\bf \hat D})\Sigma_m = A_{m\hat\b}\hat\th^{\hat\b} + (\g_m\th)_\b\Phi^\b$ &
$({\bf D}+{\bf \hat D})\hat\Sigma_p= -\th^\a A_{\a p} - (\g_p\hat\th)_{\hat\g}\hat\Phi^{\hat\g}$\\
$({\bf D}+{\bf \hat D})\Phi^\a = E^\a_{~~\hat\b}\hat\th^{\hat\b} - \frac{1}{4}(\g^{mn}\th)^\a \Psi_{mn}$ &
$({\bf D}+{\bf \hat D})\hat\Phi^{\hat\b} =- \th^\a E_\a^{~~\hat\b} + \frac{1}{4}(\g^{pq}\hat\th)^\b \hat\Psi_{pq}$\\
$({\bf D}+{\bf \hat D})\Psi_{mn}=\Omega_{mn,\hat\b}\hat\th^{\hat\b} - (\g_{[m}\th)_\g\pa_{n]}\Phi^\g$ & $({\bf D}+{\bf \hat D})\hat\Psi_{pq}=-\th^\a\Omega_{\a,pq}+(\g_{[p}\hat\th)_{\hat\g}\pa_{q]}\hat\Phi^{\hat\g}$
\end{tabular}
\end{footnotesize}
\beq
\label{reacII}
\eeq
The operator $({\bf D}+{\bf \hat D})$ acts on homogeneous polynomials 
in $\th^\a$ and $\hat\th^{\hat \a}$ by multiplication by the degree of homogeneity 
and  it does not change its degree. Therefore, the relations (\ref{reacII}) are recursive in powers of $\th$ and 
$\hat\th$. 
They can be solved algebraically given $A_{\a \hat \b}, \dots, \Omega_{\a,pq}$ order by order in $\th$ and $\hat\th$ and this proves that the gauge 
can indeed be imposed.
Of course, to reconstruct the gauge-parameter superfields by means of the recursive equations (\ref{reacII}), we also need lowest order data for them. These are the zero order supergravity gauge parameters (\ref{gaugezero}).  
 To obtain the last couple of equations we used the additional condition that $\Psi_{mn}$ and $\hat\Psi_{pq}$ are the linearized curvatures of $\Sigma_m$ and $\hat\Sigma_p$.
\subsection[Recursive equations and explicit solution (up to order\\ $\th^2\hat\th^2$)]{Recursive equations and explicit solution (up to order $\th^2\hat\th^2$)}
The next step is the derivation of the recursion equations for supergravity superfields. Acting 
with $D_\a$ and $\hat D_{\hat \b}$ on the gauge fixing conditions (\ref{gaugefix}), and using the 
definition (\ref{defD}), it is straightforward to derive the recursion relations from eq. (\ref{eom})
\vskip 11pt
\begin{center}
\begin{tabular}{l l}
$(1+{\bf D})A_{\a\hat\b}=(\g^m \th)_\a A_{m\hat\b}$ & $(1+{\bf \hat D})A_{\a\hat\b}=(\g^p \hat\th)_{\hat\b}A_{\a p}$\\
${\bf D}A_{m\hat\b}=(\g_m \th)_\g E^\g_{~~\hat\b}$ & ${\bf \hat D}A_{\a p}=(\g_p\hat\th)_{\hat\g}E_\a^{~~\hat\g}$\\
${\bf D} E^\a_{~~\hat\b}=-\frac{1}{4} (\g^{mn}\th)^\a \Omega_{mn,\hat\b}$ & ${\bf \hat D} E_\a^{~~\hat\b}=- \frac{1}{4}(\g^{pq}\hat\th)^{\hat\b}\Omega_{\a, pq}$\\
${\bf D} \Omega_{mn,\hat\b}=-(\g_{[m}\th)_\g \pa_{n]} E^\g_{~~\hat\b}$ & ${\bf \hat D} \Omega_{\a, pq}=-(\g_{[p}\hat\th)_{\hat\g} \pa_{q]} E_\a^{~~\hat\g}$
\end{tabular}
\end{center}
\beq
\label{recursionDI}
\eeq
\begin{center}
\begin{tabular}{l l}
$(1+{\bf D})A_{\a p}= (\g^m \th)_\a A_{mp}$ & $(1+{\bf \hat D})A_{m\hat\b}=- (\g^p \hat\th)_{\hat\b}A_{mp}$ \\
${\bf D}A_{mp}=(\g_m \th)_\b E^\b_{~~p}$ & ${\bf \hat D}A_{mp}=-(\g_p\hat\th)_{\hat\b}E_m^{~~\hat\b}$\\
${\bf D}E^{\a}_{~~p}=-\frac{1}{ 4}(\g^{mn}\th)^\a\Omega_{mn,p} $ & ${\bf \hat D}E_m^{~~\hat\b}=\frac{1}{4}(\g^{pq}\hat\th)^{\hat\b}\Omega_{m,pq}$\\
${\bf D} \Omega_{mn,p}=-(\g_{[m}\th)_\g \pa_{n]}E^\g_{~~p} $ & ${\bf \hat D} \Omega_{m,pq}=(\g_{[p}\hat\th)_{\hat\g} \pa_{q]}E_m^{~~\hat\g}$
\end{tabular}
\end{center}
\beq
\label{recursionDII}
\eeq
\begin{center}
\begin{tabular}{l l}
$(1+{\bf D})E_\a^{~~\hat\b}=(\g^m \th)_\a E_m^{~~\hat\b}$ & $(1+{\bf \hat D})E^{\a}_{~~\hat\b}=(\g^p \hat\th)_{\hat\b}E^\a_{~~p}$\\
${\bf D}E_m^{~~\hat\b}=(\g_m \th)_\g P^{\g\hat\b}$ & ${\bf \hat D}E^\a_{~~p}=(\g_p\hat\th)_{\hat\g}P^{\a\hat\g}$\\
${\bf D}P^{\a\hat\b}= -\frac{1}{4} (\g^{mn}\th)^\a C_{mn}^{~~~~\hat\b}$ & ${\bf \hat D}P^{\a\hat\b}= -\frac{1}{4} (\g^{pq}\hat\th)^{\hat\b}C^\a_{~~pq}$\\
${\bf D} C_{mn}^{~~~~\hat\b}=-(\g_{[m}\th)_\g \pa_{n]}P^{\g\hat\b}$ & ${\bf \hat D} C^\a_{~~pq}=-(\g_{[p}\hat\th)_{\hat\g} \pa_{q]}P^{\a\hat\g}$
\end{tabular}
\end{center}
\beq
\label{recursionDIII}
\eeq
\begin{center}
\begin{tabular}{l l}
$(1+{\bf D})\Omega_{\a, pq}=(\g^m \th)_\a\Omega_{m,pq}$ & $(1+{\bf \hat D})\Omega_{mn,\hat\b}=-(\g^p\hat\th)_{\hat\b}\Omega_{mn,p}$\\
${\bf D}\Omega_{m,pq}=(\g_m\th)_\b C^{\b}_{~~pq}$ & ${\bf \hat D}\Omega_{mn,p}=-(\g_p\hat\th)_{\hat\b}C_{mn}^{~~~~\hat\b}$\\
${\bf D}C^{\a}_{~~pq}=-\frac{1}{ 4}(\g^{mn}\th)^\a S_{mn,pq}$ & ${\bf \hat D}C_{mn}^{~~~~\hat\b}=\frac{1}{4}(\g^{pq}\hat\th)^{\hat\b}S_{mn,pq}$\\
${\bf D}S_{mn,pq}=-(\g_{[m}\th)_\g \pa_{n]}C^\g_{~~pq},$ & ${\bf \hat D}S_{mn,pq}=(\g_{[p}\hat\th)_{\hat\g}\pa_{q]}C_{mn}^{~~~~\hat\g}$
\end{tabular}
\end{center}
\beq
\label{recursionDIV}
\eeq
A given superfield appears in two groups of equations in order that both its $\th$ and $\hat\th$ components are fixed. 
Inside each group there is an iterative structure (see \cite{har} and \cite{OoguriPS}) which allows us to solve 
those equations recursively given the initial conditions and there is a hierarchical structure among 
the different groups of equations which allows us to solve them subsequently.  
To provide the initial data, we identify the lowest-components of 
the matrix superfield ${\bf A}$ in (\ref{vertmatrix}) with supergravity fields 
 \beq
 \label{vertmatrixbc} 
{\bf A } = 
 \left[ \begin{matrix}
  0 & 0 & 0 & 0 \cr
0 & g_{mp} + b_{mp} + \eta_{mp} \phi 
& \psi_m^{~~\hat\b} & 
\omega_{m, pq} \cr
0 & \psi^\a_{~~p} & f^{\a\hat\b} & 
 {c}^\a_{~~pq}\cr
0 & \omega_{mn,p} & 
 c_{mn}^{~~~~\hat\b} & s_{mn,pq}\end{matrix}\right] + {\cal O}(\th, \hat\th)\,,
 \eeq
where the linearized gravitational connection and curvatures are given by 
\bea
\label{curva}
&&\omega_{m, pq} = (\pa_p g_{mq} - \pa_q g_{mp}) + (\pa_p b_{mq} - \pa_q b_{mp}) + (\eta_{mq} \pa_p  -  \eta_{mp} \pa_q) \phi\,, \cr
&&c_{mn}^{~~~~\hat \b} = (\pa_m \psi_{n}^{~~\hat \b} - \pa_n \psi_{m}^{~~\hat \b})\,,  \cr
&&s_{mn,pq} = (\pa_m \omega_{n,pq} - \pa_n \omega_{m, pq}) \,,
\ena
and, analogously, for $\omega_{mn,p}$ and $c^\a_{~~pq}$. 

In the following we give the component-expansion for the physical superfields $A_{mp}$, $E_{m}^{~~\hat\b}$, $E^{\a}_{~~p}$ and $P^{\a\hat\b}$, up to second order in both $\th$ and $\hat\th$. The corresponding curvatures can be easily computed from the defining equations (\ref{curva}).
\bea
A_{mp} &&= (g+b+\eta\phi)_{mp}+ (\g_m\th)_\b\psi^\b_{~~p}-(\g_p\hat\th)_{\hat\b}\psi_m^{~~\hat\b}+(\g_m\th)_\b(\g_p\hat\th)_{\hat\g}f^{\b\hat\g} \cr
&&-\frac{1}{ 8}(\g_m\th)_\b (\g^{nr}\th)^\b\omega_{nr,p} -\frac{1}{8}(\g_p\hat\th)_{\hat\b}(\g^{qr}\hat\th)^{\hat\b}\omega_{m,qr}\cr
&&+\frac{1}{8}(\g_m\th)_\b(\g^{nr}\th)^\b(\g_p\hat\th)_{\hat\g}c_{nr}^{~~~\hat\g}-\frac{1}{8}(\g_m\th)_\g(\g_p\hat\th)_{\hat\b}(\g^{qr}\hat\th)^{\hat\b}c^\g_{~~qr}\cr
&&+\frac{1}{64}(\g_m\th)_\b(\g^{nr}\th)^\b (\g_p\hat\th)_{\hat\g}(\g^{qs}\hat\th)^{\hat\g}s_{nr,qs}+\dots 
\ena
\bea
\label{compo}
E_{m}^{~~\hat\b} &&= \psi_m^{~~\hat\b} + (\g_m\th)_\g f^{\g\hat\b}+\frac{1}{4}(\g^{pq}\hat\th)^{\hat\b}\omega_{m,pq}-\frac{1}{4}(\g_m\th)_\g (\g^{pq}\hat\th)^{\hat\b} c^\g_{~~pq}\cr
&&-\frac{1}{8}(\g_m\th)_\g (\g^{nr}\th)^\g c_{nr}^{~~~\hat\b} + \frac{1}{4}(\g^{pq}\hat\th)^{\hat\b}(\g_p\hat\th)_{\hat\g}\pa_q\psi_m^{~~\hat\g}\cr
&&-\frac{1}{32}(\g_m\th)_\g(\g^{nr}\th)^\g(\g^{pq}\hat\th)^{\hat\b}s_{nr,pq}+\frac{1}{4}(\g_m\th)_\g(\g^{pq}\hat\th)^{\hat\b}(\g_p\hat\th)_{\hat\g}\pa_q f^{\g\hat\g}\cr
&&-\frac{1}{32}(\g_m\th)_\g (\g^{nr}\th)^\g (\g^{pq}\hat\th)^{\hat\b}(\g_p\hat\th)_{\hat\g} \pa_q c_{nr}^{~~~\hat\g}+\dots
\ena
\bea
E^{\a}_{~~p} &&= \psi^\a_{~~p} - \frac{1}{4}(\g^{mn}\th)^\a\omega_{mn,p} + (\g_p\hat\th)_{\hat\g}f^{\a\hat\g}+ \frac{1}{4}(\g^{mn}\th)^\a(\g_p\hat\th)_{\hat\b} c_{mn}^{~~~\hat\b}\cr
&&+\frac{1}{ 4}(\g^{mn}\th)^\a (\g_m\th)_\g\pa_n\psi^\g_{~~p} - \frac{1}{ 8}(\g_p\hat\th)_{\hat\g}(\g^{qr}\hat\th)^{\hat\g} c^\a_{~~qr}\cr
&&+\frac{1}{4}(\g^{mn}\th)^\a(\g_m\th)_\g (\g_p\hat\th)_{\hat\b}\pa_n f^{\g\hat\b}+\frac{1}{ 32}(\g^{mn}\th)^\a(\g_p\hat\th)_{\hat\g}(\g^{qr}\hat\th)^{\hat\g}s_{mn,qr}\cr
&&+\frac{1}{32}(\g^{mn}\th)^\a (\g_m\th)_\g(\g_p\hat\th)_{\hat\b}(\g^{qr}\hat\th)^{\hat\b} \pa_n c^\g_{~~qr} +\dots
\ena
\bea
P^{\a\hat \b} &&= f^{\a\hat\b} - \frac{1}{ 4}(\g^{mn}\th)^\a c_{mn}^{~~~~\hat\b}-\frac{1}{4}(\g^{pq}\hat\th)^{\hat\b}c^\a_{~~pq}- \frac{1}{16}(\g^{mn}\th)^\a (\g^{pq}\hat\th)^{\hat\b}s_{mn,pq}\cr
&&+\frac{1}{ 4}(\g^{mn}\th)^\a (\g_m\th)_\g \pa_n f^{\g\hat\b} +\frac{1}{4}(\g^{pq}\hat\th)^{\hat\b}(\g_p\hat\th)_{\hat\g} \pa_q f^{\a\hat\g}\cr
&&+\frac{1}{16}(\g^{mn}\th)^\a(\g_m\th)_\g (\g^{pq}\hat\th)^{\hat\b}\pa_n c^\g_{~~pq}-\frac{1}{16}(\g^{mn}\th)^\a(\g^{pq}\hat\th)^{\hat\b}(\g_p\hat\th)_{\hat\g}\pa_q c_{mn}^{~~~\hat\g}\cr
&& +\frac{1}{16}(\g^{mn}\th)^\a (\g_m\th)_\g(\g^{pq}\hat\th)^{\hat\b}(\g_p\hat\th)_{\hat\g}\pa_n \pa_q f^{\g\hat\g}+\dots
\ena
In the following we list the solution up to second order in both $\th^{\a}$ and $\hat\th^{\hat \a}$ 
for the auxiliary superfields $A_{\a\hat\b}, A_{\a p}, A_{m \hat \b}, E_{\a}^{~~\hat\b}$ and $E^{\a}_{~~\hat \b}$. 
\bea
A_{\a\hat \b}&& = -\frac{1}{ 4}(\g^m\th)_\a(\g^p\hat\th)_{\hat\b}(g+b+\eta\phi)_{mp}\cr
&&+\frac{1}{ 6}(\g^m\th)_\a(\g_m\th)_\g(\g^p\hat\th)_{\hat\b}\psi^\g_{~~p}+\frac{1}{ 6}(\g^m\th)_\a(\g^p\hat\th)_{\hat\b}(\g_p\hat\th)_{\hat\g}\psi_m^{~~\hat\g}\cr
&&+\frac{1}{9}(\g^m\th)_\a (\g_m\th)_\b (\g^p\hat\th)_{\hat\b}(\g_p\hat\th)_{\hat\g}f^{\b\hat\g}+\dots \cr
&& \cr
A_{\a p}&& = \half\th^\b\g^m_{\b\a}(g+b+\eta\phi)_{mp} - \half(\g^m\th)_\a(\g_p\hat\th)_{\hat\g}\psi_m^{~~\hat\g}+\frac{1}{3}(\g^m\th)_\a(\g_m\th)_\b\psi^\b_{~~p}\cr
&&+\frac{1}{3}(\g^m\th)_\a(\g_m\th)_\b(\g_p\hat\th)_{\hat\g}f^{\b\hat\g}-\frac{1}{16}(\g^m\th)_\a(\g_p\hat\th)_{\hat\g}(\g^{qr}\hat\th)^{\hat\g}\omega_{m,qr}\cr
&&-\frac{1}{24}(\g^m\th)_\a(\g_m\th)_\g(\g_p\hat\th)_{\hat\b}(\g^{qr}\hat\th)^{\hat\b}c^\g_{~~qr}+\dots 
\ena
\bea
A_{m \hat \b}&& = -\half \hat\th^{\hat\g}\g^p_{\hat\g\hat\b}(g+b+\eta\phi)_{mp} +\half(\g_m\th)_\g(\g^p\hat\th)_{\hat\b}\psi^\g_{~~p}+ \frac{1}{3}(\g^p\hat\th)_{\hat\b}(\g_p\hat\th)_{\hat\g}\psi_m^{~~\hat\g}\cr
&&+\frac{1}{16}(\g_m\th)_\g(\g^{nr}\th)^\g(\g^p\hat\th)_{\hat\b}\omega_{nr,p}+\frac{1}{3}(\g_m\th)_\b(\g^p\hat\th)_{\hat\b}(\g_p\hat\th)_{\hat\g}f^{\b\hat\g}\cr
&&-\frac{1}{ 24}(\g_m\th)_\g(\g^{nr}\th)^\g (\g^p\hat\th)_{\hat\b}(\g_p\hat\th)_{\hat\g}c_{nr}^{~~~\hat\g}+\dots 
\ena
\bea
E_{\a}^{~~\hat\b}&& = \half\th^\g \g^m_{\g\a}\psi_m^{~~\hat\b}+\frac{1}{8} (\g^m\th)_\a (\g^{pq}\hat\th)^{\hat\b}\omega_{m,pq}+ \frac{1}{ 3}(\g^m\th)_\a(\g_m\th)_\g f^{\g\hat\b}\cr
&&-\frac{1}{12}(\g^m\th)_\a(\g_m\th)_\g(\g^{pq}\hat\th)^{\hat\b}c^\g_{~~pq}+\frac{1}{8}(\g^m\th)_\a(\g^{pq}\hat\th)^{\hat\b}(\g_p\hat\th)_{\hat\g}\pa_q\psi_m^{~~\hat\g}\cr
&& +\frac{1}{12}(\g^m\th)_\a (\g_m\th)_\g (\g^{pq}\hat\th)^{\hat\b}(\g_p\hat\th)_{\hat\g} \pa_q f^{\g\hat\g}+\dots
\ena
\bea
E^{\a}_{~~\hat\b}&& = \half\hat\th^{\hat\g}\g^p_{\hat\g\hat\b}\psi^\a_{~~p}+ \frac{1}{8}(\g^{mn}\th)^\a(\g^p\hat\th)_{\hat\b}\omega_{mn,p}+ \frac{1}{3}(\g^p\hat\th)_{\hat\b}(\g_p\hat\th)_{\hat\g}f^{\a\hat\g}\cr
&&-\frac{1}{8}(\g^{mn}\th)^\a(\g_m\th)_\g(\g^p\hat\th)_{\hat\b}\pa_n\psi^\g_{~~p}-\frac{1}{12}(\g^{mn}\th)^\a(\g^p\hat\th)_{\hat\b}(\g_p\hat\th)_{\hat\g}c_{mn}^{~~~\hat\g}\cr
&&+\frac{1}{12}(\g^{mn}\th)^\a(\g_m\th)_\g (\g^p\hat\th)_{\hat\b}(\g_p\hat\th)_{\hat\g} \pa_n f^{\g\hat\g}+\dots
\label{compoII}
\ena

It is easy to verify that this expansion satisfies the gauge conditions (\ref{gaugefix})  and that all 
auxiliary fields have been eliminated and reexpressed in terms of derivatives of physical supergravity fields.

The next step is to insert the expansion (\ref{compo}) into the definition of the vertex operator (\ref{vert}) and 
recombine the worldsheet one-forms ${\bf X}_z$ and ${\bf X}_{\bar z}$ in order to 
get a more manageable expression. However, it makes sense to provide such expression for an interesting example in section 3.3.  

We have to notice that the vertex operator ${\cal V}^{(1,1)}$ contains only the superfield 
$A_{\a \hat \b}$ which encodes all the needed information regarding the supergravity fields, which however
appear at higher orders in $\th$'s and $\hat\th$'s. This is sufficient for amplitudes computations, even 
though the measure factor on zero modes in the correlation functions has to 
soak up plenty of $\th$'s and $\hat\th$'s (\cite{berkovits, BerkovitsPH, GrassiNZ}). 
\subsection{Gauge fixing for massive states}
In the previous sections, we explored the gauge fixing 
for the massless sector of open and closed string theory. However, 
the spectrum of string theory contains infinitely many massive states 
defined, in the closed string case, by the equations 
\beq
\label{maA}
 \Big[Q_L, {\cal V}^{(1,1)}_{n}\Big] = 0 \,,
  ~~~~~
 \Big[Q_R, {\cal V}^{(1,1)}_{n}\Big] = 0\,,
  ~~~~~
\Big[L_{0,L} + L_{0,R} -n, {\cal V}^{(1,1)}_{n}\Big]= 0 \,,
\eeq
where $L_{0,L} = \oint dz~ z~ T_{zz}$ and $L_{0,R} = \oint d\bar z~ \bar z ~\bar T_{\bar z\bar z}$. 
The index $n$ denotes the mass of the state. Even if these equations can be solved by 
expanding the vertex operators ${\cal V}^{(1,1)}_{n}$ in terms of the building-blocks 
$\pa \th^{\a}$, $\bar\pa \hat\th^{\hat\a}$, $\Pi^{m}$, $\bar \Pi^{m}$,..., it is convenient to fix 
a gauge as in the massless case 
and then solve the equations by an iterative construction 
as shown in the previous section. However, since we cannot explore the complete set of vertices and provide a gauge fixing for each of them, we propose a definition of gauge fixing based on new anticommuting and 
nilpotent charges to be imposed 
on the physical states. This resembles the Siegel gauge (where 
the corresponding charges are $b_{L,0}  = \oint dz \,z \,b_{zz}$ and 
$b_{L,0}  = \oint d\bar z \,\bar z \, \hat b_{\bar z\bar z}$ where 
$b_{zz}$ and $\hat b_{\bar z\bar z}$ are the left- and right-moving 
antighosts) used in string field theory to eliminate all auxiliary fields and 
to define the propagator for the string field. 

We introduce the following charges ``dual'' to the BRST operators
\beq
\label{maB}
{\cal K}_{L} = \oint dz \, \th^{\a} w_{\a}\,, ~~~~~
{\cal K}_{R} = \oint d\bar z \, \hat\th^{\hat \b} \hat w_{\hat \b}\,.
\eeq
They are nilpotent and anti-commute. They are not supersymmetry invariant 
 as can be directly seen by the presence of $\th^{\a}$ and $\hat \th^{\hat \b}$. This in fact 
 implies that we are choosing a non symmetric gauge which can be viewed as a generalization 
 of the Wess-Zumino gauge condition in 10 dimensions. It eliminates the lowest non physical 
 component of the superfields and it fixes the auxiliary fields -- appearing at higher order 
 in the superspace expansion -- in terms of the physical fields and their derivatives. In addition, 
 ${\cal K}_{L/R}$ are not invariant under the gauge transformations (\ref{conDclosed}) , but 
 their gauge variations are BRST invariant because of the pure spinor conditions
 \beq
 \label{maBB}
 \{Q_{L}, \Delta_{L} {\cal K}_{L} \} = 0 \,, ~~~~~~
 \{Q_{R}, \Delta_{R} {\cal K}_{R} \} = 0 \,, 
 \eeq
 Moreover,  ${\cal K}_{L/R}$ have  
 the following commutation relations with the BRST operators
 \bea
 \label{maC}
 &&\{Q_{L}, {\cal K}_{L}\} =  {\cal D} + J_{L}\,, ~~~~~
 \{Q_{R}, {\cal K}_{L} \} = 0\,,\cr
 &&
\{Q_{R}, {\cal K}_{R}\} =  \hat{\cal D} + J_{R}\,, ~~~~~
 \{Q_{L}, {\cal K}_{R} \} = 0\,,
 \ena
where 
\bea
\label{maD}
&&{\cal D} = \oint dz : \th^{\a} d_{\a} :\,, ~~~~~~ J_{L} = \oint dz : \l^{\a} w_{\a} :\cr
&&\hat{\cal D} = \oint d\bar z :\hat \th^{\hat \a} \hat d_{\hat \a} :\,, ~~~~~~ 
J_{R} = \oint d\bar z : \l^{\hat \a} \hat w_{\hat \a} :
\ena
Acting on superfields $F(x,\th, \hat \th)$, 
we have that  $\{ {\cal D}, F\} = {\bf D} F$ and 
$\{ \hat{\cal D}, F\} = {\bf \hat D} F$. 
The ordering of fields in the operators ${\cal D}$, $\hat{\cal D}$, $J_{L}$ and 
$J_{R}$ is needed to define the corresponding currents. The operators are 
gauge invariant under (\ref{conDclosed}) because of (\ref{maBB}). The main difference 
with respect to Siegel gauge fixing in string field theory is that in that case  
$b_{zz}$ and $\hat b_{\hat z \hat z}$ are 
holomorphic and antiholomorphic anticommuting currents of 
spin 2. 

In the case of the open superstring, denoting by $Q$ and by ${\cal K}$ the BRST and gauge fixing operators, the gauge condition on the massless vertex operator ${\cal V}^{(1)} = \l^{\a} A_{\a}$ is given by 
\beq
\label{maE}
\{{\cal K}, {\cal V}^{(1)} \} = \oint dw \, \Big(\th^{\a} w_{\a} \Big)(w) \, \Big(\l^{\a} A_{\a}(x,\th)\Big)(z) = \th^{\a} A_{\a} = 0\,.
\eeq
We notice that the field $\th^{\a}$ in ${\cal K}$ is 
harmless for massless vertices, but it will give a nontrivial contribution in the massive case. In the latter 
case one has to add a compensating non-gauge invariant contribution on the 
r.h.s. of (\ref{maE}) in order to compensate the fact that ${\cal K}$ is not gauge invariant 
under (\ref{conDclosed}).

Applying $Q$ on the left hand side of (\ref{maE})
applying ${\cal K}$ on the equation 
$\{Q, {\cal V}^{(1)}\} = \l \g^{m} \l A_{m}(x,\th)=0$  
and using the commutation 
relations (\ref{maC}), we obtain
\beq
\label{maEE}
({\bf D} + 1) {\cal V}^{(1)} = \l \g^{m} \th A_{m}\,. 
\eeq
Eliminating the ghost $\l^{\a}$, we end up with 
equation (\ref{III}) for the superfields $A_{\a}$ and $A_{m}$. This 
procedure can be clearly generalized to massive states. First, we 
discuss the closed string case, then we show an example for the 
first massive state for open superstrings and, finally, we show that 
the zero momentum cohomology satisfies the same equations generalized 
to zero modes. 

For closed strings, we reproduce the gauge fixing (\ref{gaugefix}) by
the following conditions 
\beq
\label{maF}
\{{\cal K}_{L}, {\cal V}^{(1,1)} \} = 0\,, ~~~~~~  \{{\cal K}_{R}, {\cal V}^{(1,1)} \} = 0
\eeq
and, for the gauge parameters $\Lambda^{(1,0)}$ and $\Lambda^{(0,1)}$ in eq. 
(\ref{gaugeparI}), by the gauge condition
\beq
\label{maG}
\{ {\cal K}_{L}, \L^{(1,0)}\} + \{ {\cal K}_{R}, \L^{(0,1)} \} = 0\,.
\eeq
which coincides with (\ref{gaugefixII}). Applying the BRST 
charge on the left hand sides of (\ref{maF}), acting 
with ${\cal K}_{L}$ and ${\cal K}_{R}$ on equations (\ref{deC}), 
and finally using the commutation relations (\ref{maC}), we derive the conditions 
for the iterative equations given in the previous section. 

Let us show that the gauge fixing (\ref{maE}) also fixes the gauge transformations in a suitable way for the first massive state of the open superstring ${\cal V}^{(1)}_{1}$, leading to a 
recursive procedure to compute the vertex operator in term of the initial data, a 
multiplet of on-shell fields containing a massive spin 2 field \cite{gsstring}. 

A general decomposition of ${\cal V}^{(1)}_{1}$ in terms of fundamental building-blocks is given in (\ref{massiveU}) and its gauge transformation is generated by 
\beq
\label{maI}
\delta {\cal V}^{(1)}_{1} =\Big[ Q, \Omega^{(0)}_{1} \Big]\,,
\eeq
with
\beq
\Omega^{(0)}_{1} =  \pa\th^{\b}  \Omega_{\b} 
+ :\Pi^{m} \Omega_{m}: + :d_{\b} \Omega^{\b}: 
+:N^{mn}: \Omega_{mn} + :w_{\b} \l^{\b}: \Omega \,. 
\eeq
The decompositions are based on the requirement that the 
vertex operator should be invariant under the gauge transformation $\Delta$ 
given in (\ref{conDclosed}). A further gauge transformation of $\Omega^{(0)}_{1}$ 
would be a variation of a negative ghost number field. The only one is the antighost 
$w_{\a}$, but there is no gauge invariant operator only with $w_{\a}$ without $\l^{\a}$. Notice 
that we have to add a (BRST-invariant) 
compensating term of the form $w \g^{mnpq} \l$ in order to reabsorb the non-invariance of 
${\cal K}$. 

Imposing (\ref{maE}), we get 
\bea
\label{maJ}
&&A_{\a} + \th^{\b} B_{\b\a} = 0\,, ~~~~~~~ \th^{\a} H_{\a m} = 0\,,~~~~~~~ \th^{\b} C_{\b}^{~\a} = 0 \,,
\cr
&&\th^{\b} F_{\b mn} + \frac{1}{1440}\left[(\g_{mn})^{\a}_{~\g} C^{\g}_{~~\a}-(\g_{mn})^{\a}_{~\g}\th^\g E_\a\right] = 0\,, \cr
&&\half(\g^{mn}\th)^\b F_{\b mn}-C^{\a}_{~~\a}+2\th^\a E_\a = 0\,. 
\ena
This gauge fixing can be reached by adjusting the parameters 
$\Omega_{\a}, \Omega_{m}, \Omega^{\a}, \Omega_{mn}$ and $\Omega$. Using 
 equations (\ref{maJ}) and applying the operator ${\bf D}$, we obtain 
the iterative relations to compute the vertex. 
The gauge fixing (\ref{maJ}) fixes only the supergauge part of the 
gauge transformation. This gauge 
does not fix the physical gauge transformation of the massive spin 2 system \cite{BerkovitsQX}. 

Finally, we show that the measure for zero modes satisfies the gauge fixing proposed 
above. In fact, by restricting the attention to zero momentum cohomology, 
we supersede ${\cal K}$ with the differential 
\beq
\label{maZ}
{\cal K}_{0} = \th^{\a}_{0} \frac{\partial}{\partial \l^{\a}_{0}} 
\eeq
which acting on ${\cal V}^{(3)} = ( \l_{0}\g^{m}\th_{0}) 
(\l_{0}\g^{n}\th_{0})( \l_{0}\g^{p}\th_{0})( \th_{0} \g_{mnp}\th_{0}) $, 
yields 
\beq
\label{maZZ}
{\cal K}_{0} {\cal V}^{(3)} = 0\,. 
\eeq
Similarly, for the closed superstring, the 
ghost number $(3,3)$ cohomology representative 
satisfies the corresponding gauge fixing. 

Even if the gauge fixing is not manifestly supersymmetric, 
the supersymmetry of the target space theory is still a symmetry. 
As usual, in the Wess-Zumino gauge, a supersymmetry 
transformation must be accompanied by a gauge transformation to 
bring the vertex to the original gauge. This means that 
\beq
\label{susA}
\delta_{\e} [{\cal K}, {\cal V}] + [{\cal K}, \delta {\cal V}] =0 
\eeq
where $\delta {\cal V} = [Q, \Omega_{\e}]$, $\delta_{\e} {\cal V} = 
[ \e^{\a} Q_{\a}, {\cal V}]$ and $Q_{\a} = \oint dz \, q_{\a}$ (the 
supersymmetry generator $q_{\a}$ is given in (\ref{conF})).  
As an example, we 
show that $\Omega_{\e}$ can be indeed found for 
the massless sector of the open superstring and the extension 
is similar for the other cases. Equation (\ref{susA}) reduces 
to 
\beq
\label{susB}
\e^{\a} A_{\a} + \th^{\a} \e^{\b} Q_{\b} A_{\a} + \th^{\a} D_{\a} \Omega_{\e} =0\,, 
\eeq
which yields
\beq
\label{susC}
{\bf D} \Omega_{\e} = 0\,.
\eeq
Again, this equation can be solved iteratively in powers of $\th$'s and 
it follows that $\Omega = \Omega_{0}(x)$. (\ref{susC}) can be checked explicitly 
on the solutions (\ref{compo}). 
\section[An example: Linearly $x$-dependent Ramond-Ramond field\\ strength and possible Lie-\-al\-ge\-bra\-ic superspace deformations]{An example: Linearly $x$-dependent Ramond-Ramond field strength and possible Lie-\-al\-ge\-bra\-ic superspace deformations}
In this section I will discuss in detail one application of the iterative procedure I presented in the previous section. The unintegrated and integrated vertices for a particular nonconstant R-R field-strength will be computed, that are expected to be related to a nonconstant deformation of ten-dimensional superspace.
At the end of the section, other two applications will be briefly described. No details will be given for these two, since they are not related to noncommutative geometry, which is the main subject of this thesis. However, the interested reader can refer to my work \cite{mypaper4}, where the three applications are discussed.
\subsection{Motivation: Nonconstant superspace deformations}
In section 1.2.4, we have seen how non(anti)commutative superspaces, first studied in \cite{ferrara} and in my paper \cite{mypaper1}, arise in the context of superstring theory, when open superstrings in the presence of R-R backgrounds and D-branes are considered. Up to now, only supergeometries with constant fermion-fermion anticommutators have been derived, associated to a constant R-R background, for the superstring compactified on a CY three-fold \cite{vafaooguri2,seiberg} and for the uncompactified ten-dimensional superstring \cite{antonio}. In both cases the covariant formulation for the superstring has been used.

In section 1.3 we have discussed how, in the bosonic case, nonconstant deformations of the coordinate algebra are related to the presence of a general curved NS-NS background. Therefore, it is natural to expect that more general, nonconstant R-R backgrounds can lead to nonconstant superspace deformations. In particular, in \cite{antonio} it was conjectured that from non-constant 
RR field strengths one can derive new equal-time commutation 
relations between coordinates $x^{m}$ and $\th^{\a}$ living on the 
boundaries such as 
\beq
\label{nnon}
\{\th^{\a}, \th^{\b} \} = \g^{\a\b}_{m} x^{m}\,, 
\eeq
 generalizing the construction of Lie-algebraic 
 non-commutative geometries to supermanifolds \cite{Schwarz:pf}~ (for a different example of a Lie-\-al\-ge\-bra\-ic geometry in superspace see \cite{kosinski}). 
 
 The vertex operator 
 for non-constant R-R fields strengths is the basic ingredient of this kind of analysis. 
\subsection{The ansatz for the RR field strength}
 Applying the iterative method I introduced in the previous section, I will show how to 
 compute the vertex for linearly $x$-dependent RR field strengths. This is the most simple $x$-dependent background one can consider. Moreover, it is interesting since it is supposed to be related to the nonconstant deformation (\ref{nnon}).
 
We will consider the following ansatz for the R-R field strength
\beq
\label{defapprox}
P^{\a\hat\b}=f^{\a\hat\b}+{\cal C}_m^{~~\a\hat\b}x^m
\eeq
where ${\cal C}_m^{~~\a\hat\b}$ is constant. $P^{\a\hat\b}$ must satisfy equations (\ref{newII}), which become $\g^{m}_{\a\b} {\cal C}_{m}^{\b\hat\g} = \g^{m}_{\hat\a\hat\b} {\cal C}^{\g \hat\a} =0$ for the specific ansatz (\ref{defapprox}).
Equations (\ref{newII}) can be rewritten in terms of forms by decomposing $P^{\a\hat\b}$ according to 
Dirac equations. For example, for type IIB we have the 1-form $P_{m}$, the 3-form $P_{[mnp]}$ and 
the 5-form $F_{[mnpqr]}$. Solving the Bianchi identities we get $P_{m} = \pa_{m} A$, $ P_{[mnp]} = \pa_{[m} A_{np]}$,... 
and the field equations are $\pa^{m} P_{m} = \pa^{2} A = 0$, $\pa^{m}  P_{[mnp]} = \pa^{m} \pa_{[m} A_{np]}$,... 
These can be solved in terms of quadratic polynomials 
$A(x) = (10 \, a_{(mn)} - a_{r}^{~r} \eta_{mn}) x^{m} x^{n}$, $A_{[mn]} = (10 \, a_{[mn],(rs)} - 
a_{[mn,t]}^{~~t} \eta_{rs})  x^{r} x^{s}$,... where $a_{(mn)}$, $a_{[mn], (rs)}$,... are 
constant background fields. 

In the constant field strength case, our iterative procedure can be applied to compute the integrated and unintegrated vertex operators. One obtains
\beq
{\cal V}_{z\bar z}^{(0,0)}=q_\a f^{\a\hat\b} \hat q_{\hat\b},
\eeq
where $q_\a$ and $\hat q_{\hat\b}$ are the supersymmetry currents given in (\ref{conF}).
So it is easy to see that equation (\ref{deE}) is verified with
\beq
{\cal V}^{(1,1)}=\chi_\a f^{\a\hat\b}\hat\chi_{\hat\b},
\eeq
which is
clearly BRST invariant (see (\ref{susyQ}) and (\ref{susyQII})). 

Since in the $\th$ and $\hat\th$ expansions of the physical and auxiliary superfields $A_{\a\hat\b}$,...,$P^{\a\hat\b}$ (see eqs. (\ref{compo} and (\ref{compoII}))
the number of bosonic derivatives acting on physical zero-order components grows with growing order in $\th$ and $\hat\th$, it is clear that the ansatz (\ref{defapprox}) will correspond to only a few non-zero terms in the expansion.
Actually, the highest-order contributions are $\th^4\hat\th^2$ and $\th^2\hat\th^4$ terms. Here 
we give the explicit expressions 
\bea
\label{nonconstRRII}
A_{\a\hat \b}& = &\frac{1}{ 9}(\g^m\th)_\a (\g_m\th)_\b (\g^p\hat\th)_{\hat\b}(\g_p\hat\th)_{\hat\g}(f^{\b\hat\g}+{\cal C}_n^{~~\b\hat\g}x^n)\cr
&+&\frac{1}{180}(\g^m\th)_\a(\g_m\th)_\g(\g^p\hat\th)_{\hat\b}(\g_p\hat\th)_{\hat\g}(\g^{qr}\hat\th)^{\hat\g}(\g_q\hat\th)_{\hat\d}{\cal C}_r^{~~\g\hat\d}\cr
&+&\frac{1}{180}(\g^m\th)_\a(\g_m\th)_\d(\g^{nr}\th)^\d(\g_n\th)_\g(\g_p\hat\th)_{\hat\b}(\g^p\hat\th)_{\hat\g}{\cal C}_r^{~~\g\hat\g} \,, 
\ena
\bea 
A_{\a p}& =&\frac{1}{ 3}(\g^m\th)_\a(\g_m\th)_\b(\g_p\hat\th)_{\hat\g}(f^{\b\hat\g}+{\cal C}_n^{~~\b\hat\g}x^n) \cr
&+&\frac{1}{36}(\g^m\th)_\a (\g_m\th)_\g(\g_p\hat\th)_{\hat\b}(\g^{qr}\hat\th)^{\hat\b}(\g_q \hat\th)_{\hat\g}{\cal C}_r^{~~\g\hat\g}\cr
&+&\frac{1}{60}(\g^m\th)_\a(\g_m\th)_\b(\g^{nr}\th)^\b(\g_n\th)_\g(\g_p\hat\th)_{\hat\b}{\cal C}_r^{~~\g\hat\b}
\ena
\bea 
A_{m \hat \b}& = &\frac{1}{3}(\g_m\th)_\b(\g^p\hat\th)_{\hat\b}(\g_p\hat\th)_{\hat\g}(f^{\b\hat\g}+{\cal C}_n^{~~\b\hat\g}x^n)\cr
&+&\frac{1}{ 36}(\g_m\th)_\a(\g^{nr}\th)^\a(\g_n\th)_\g(\g_p\hat\th)_{\hat\b}(\g^p\hat\th)_{\hat\g}{\cal C}_r^{~~\g\hat\g}\cr
&+&\frac{1}{60}(\g_m\th)_\g(\g^p\hat\th)_{\hat\b}(\g_p\hat\th)_{\hat\g}(\g^{rs}\hat\th)^{\hat\g}(\g_r\hat\th)_{\hat\d}{\cal C}_s^{~~\g\hat\d} 
\ena
\bea
E_{\a}^{~~\hat\b}& =& \frac{1}{3}(\g^m\th)_\a(\g_m\th)_\g (f^{\g\hat\b}+{\cal C}_n^{~~\g\hat\b}x^n)\cr
& +&\frac{1}{12}(\g^m\th)_\a (\g_m\th)_\g (\g^{pq}\hat\th)^{\hat\b}(\g_p\hat\th)_{\hat\g}{\cal C}_q^{~~\g\hat\g}\cr
&+&\frac{1}{ 60}(\g^m\th)_\a(\g_m\th)_\b(\g^{nr}\th)^\b(\g_n\th)_\g {\cal C}_r^{~~\g\hat\b}
\ena
\bea
E^{\a}_{~~\hat\b}& = &\frac{1}{ 3}(\g^p\hat\th)_{\hat\b}(\g_p\hat\th)_{\hat\g}(f^{\a\hat\g}+{\cal C}_m^{~~\a\hat\g}x^m)\cr
&+&\frac{1}{12}(\g^{mn}\th)^\a(\g_m\th)_\g (\g^p\hat\th)_{\hat\b}(\g_p\hat\th)_{\hat\g}{\cal C}_n^{~~\g\hat\g}\cr
&+&\frac{1}{60}(\g^p\hat\th)_{\hat\b}(\g_p\hat\th)_{\hat\g}(\g^{qr}\hat\th)^{\hat\g}(\g_q\hat\th)_{\hat\d}{\cal C}_r^{~~\a\hat\d}
\ena
\bea
A_{mp} &=& (\g_m\th)_\b(\g_p\hat\th)_{\hat\g}(f^{\b\hat\g}+{\cal C}_n^{~~\b\hat\g}x^n) \cr
&+&\frac{1}{ 12}(\g_m\th)_\b(\g^{nr}\th)^\b(\g_n\th)_\g(\g_p\hat\th)_{\hat\b}{\cal C}_r^{~~\g\hat\b}\cr
&+&\frac{1}{12}(\g_m\th)_\g(\g_p\hat\th)_{\hat\b}(\g^{rs}\hat\th)^{\hat\b}(\g_r\hat\th)_{\hat\g}{\cal C}_s^{~~\g\hat\g}
\ena
\bea
E_{m}^{~~\hat\b} &=& (\g_m\th)_\g (f^{\g\hat\b}+{\cal C}_n^{~~\g\hat\b}x^n)\cr
&+&\frac{1}{4}(\g_m\th)_\g(\g^{pq}\hat\th)^{\hat\b}(\g_p\hat\th)_{\hat\g}{\cal C}_q^{~~\g\hat\g}\cr
&+&\frac{1}{12}(\g_m\th)_\a(\g^{nr}\t)^\a(\g_n\th)_\g {\cal C}_r^{~~\g\hat\b}
\ena
\bea
E^{\a}_{~~p} &=& (\g_p\hat\th)_{\hat\g}(f^{\a\hat\g}+{\cal C}_m^{~~\a\hat\g}x^m)\cr
&+&\frac{1}{4}(\g^{mn}\th)^\a(\g_m\th)_\g (\g_p\hat\th)_{\hat\b}{\cal C}_n^{~~\g\hat\b}\cr
&+&\frac{1}{12}(\g_p\hat\th)_{\hat\b}(\g^{qr}\hat\th)^{\hat\b}(\g_q\hat\th)_{\hat\g}{\cal C}_r^{~~\a\hat\g}
\ena
\bea
P^{\a\hat \b} &=& (f^{\a\hat\b}+{\cal C}_m^{~~\a\hat\b}x^m)\cr
&+&\frac{1}{4}(\g^{mn}\th)^\a (\g_m\th)_\g {\cal C}_n^{~~\g\hat\b}\cr
&+&\frac{1}{4}(\g^{pq}\hat\th)^{\hat\b}(\g_p\hat\th)_{\hat\g} {\cal C}_q^{~~\a\hat\g}\,.
\label{nonconstRRIII}
\ena
\subsection{The vertex for linearly $x$-dependent RR field strength}
To obtain the vertices ${\cal V}^{(1,1)}$ and ${\cal V}^{(0,0)}_{z\bar z}$ for the linearly $x$-dependent RR field strength we have to insert 
(\ref{nonconstRRII}) and (\ref{nonconstRRIII}) back into (\ref{verB}) and (\ref{vert}).

For the unintegrated vertex operator we find 
\bea
\label{Unonconst}
&&{\cal V}^{(1,1)}=\chi_\a f^{\a\hat\b}\hat\chi_{\hat\b}\cr
&&~~~+\chi_\a\left[\left(x^m \d_\g^\a \d_{\hat\g}^{\hat\b}+\frac{1}{20}(\g^{qm}\hat\th)^{\hat\b}(\g_q \hat\th)_{\hat\g} \d_\g^\a + \frac{1}{20}(\g^{nm}\th)^\a (\g_n\th)_\g \d_{\hat\g}^{\hat\b}\right){\cal C}_m^{~~\g\hat\g}\right]\hat\chi_{\hat\b}\cr
&&
\ena
while for the integrated vertex operator ${\cal V}^{(0,0)}_{z\bar z}$ we obtain
\bea
\label{Vnonconst}
{\cal V}^{(0,0)}_{z\bar z}&=&q_\a f^{\a\hat\b} q_{\hat\b}\cr
&+&q_\a\left[x^s \d_\g^\a \d_{\hat\g}^{\hat\b} + \frac{1}{ 4}(\g^{rs}\th)^\a (\g_r\th)_\g \d_{\hat\g}^{\hat\b} + \frac{1}{ 4} (\g^{ps}\hat\th)^{\hat\b}(\g_p\hat\th)_{\hat\g}\d_\g^\a\right]{\cal C}_s^{~~\g\hat\g}\hat q_{\hat\b}\cr
&+& \left[-\frac{1}{6}(\pa x^m + \frac{1}{ 10}\th\g^m\pa\th)(\th\g_m\g^{rs}\th)-N^{rs}\right](\g_r\th)_\a {\cal C}_s^{~~\a\hat\b}\hat q_{\hat\b}\cr
&+&q_\a {\cal C}_s^{~~\a\hat\b}(\g_r\hat\th)_{\hat\b}\left[-\frac{1}{ 6}(\bar\pa x^p + \frac{1}{10}\hat\th \g^p\bar\pa\hat\th)(\hat\th\g_p\g^{rs}\hat\th)-\hat N^{rs}\right]
\ena

Unfortunately, the complicated structure of ${\cal V}^{(0,0)}_{z\bar z}$ prevents 
from a simple analysis of superspace deformations as in \cite{antonio}. In the future it would be nice to find a way to study superspace deformations deriving from nonconstant R-R backgrounds. The computation of this vertex is a first step, but it is clear that plenty of work is needed to understand how to use it to compute the way it deforms the supergeometry.
\section{Other applications}
\subsection{Vertex operators with R-R gauge potentials}
In the presence of D-branes, one can ask which states couple to them and 
which vertex operators describe such interaction. As it was discussed in  \cite{asymI} 
in the framework of RNS formalism, one has to construct the 
vertex operators for R-R fields in the asymmetric picture. In addition, 
a propagating closed string (i.e. with non vanishing momentum) 
emitted from a disk or a D-brane, has to be off-shell. Therefore, 
one needs to break the BRST invariance by allowing 
a non vanishing commutator with $Q_{L, 0} + Q_{R,0}$ where 
$Q_{L/R, 0}$ are the picture conserving parts of BRST charges in the 
RNS formalism. In particular in \cite{asymII} the authors construct a 
solution of $[ Q_{L, 1} + Q_{R,1}, W]=0$, where $W$ is the vertex operator 
in the asymmetric picture. The off-shell vertex operators directly couple the R-R potentials to the worldvolume of the
D-brane.

In \cite{mypaper4}, we constructed analogous vertices for 
closed superstrings which do not satisfy the classical supergravity 
equations of motion, but modified superfield constraints. 
They allow for a description of the 
R-R gauge potentials, in contradistinction to the on-shell formalism case, 
where only the field strengths $P^{\a\hat \b}$ 
appear. First of all, there are some important differences. The 
two BRST charges $Q_{L}$ and $Q_{R}$ contain a single 
term and therefore the decomposition used in \cite{asymII} is 
not viable. In addition, there are no different 
 pictures (in the usual sense) for a given vertex since there are no superghosts 
 associated to local worldsheet supersymmetry.  
 There is, however, the possibility of constructing two operators 
 which resemble the picture lowering and raising operator \cite{multiloop}, as we briefly discussed in section 3.1.4, but the implications 
 of this new idea in the present context have not been explored yet. 
 
Nevertheless we can construct an off-shell formalism 
by considering the 
following combination of vertices with different ghost numbers: 
\beq
\label{vI}
{\cal V}^{(2)} = {\cal V}^{(2,0)} + {\cal V}^{(1,1)} + {\cal V}^{(0,2)}\,,
\eeq
where the notation ${\cal V}^{(a,b)}$ stands for vertex operators 
with the left ghost number $a$ and with the right ghost number $b$. 
The ghost number of the l.h.s. is just the sum of the ghost numbers. 
By expanding this in terms of the pure spinor ghost and by applying the modified condition
\beq
[Q_L+Q_R,{\cal V}_2]=0
\eeq
instead of the usual two conditions for the left and right sectors, we showed that this leads to equations of motion that are deformations of the usual supergravity constraints. The way constraints are relaxed to go off-shell follows very closely the case of $N=1$ super-Yang-Mills presented in \cite{spincoho}. By following a procedure analogous to the one described in section 3.2.4, a suitable gauge-fixing can be applied to eliminate auxiliary fields. The superfields can then be expanded in $\th$ and $\hat\th$ and one finds that R-R gauge-potentials explicitly appear.
The construction of vertices with R-R potentials in covariant formulation 
has been also discussed in 
 \cite{cornalbaschiappa2}. There the authors considered only the constant case. 
\subsection[Antifields and the kinetic terms for closed superstring\\ field theory]{Antifields and the kinetic terms for closed superstring field theory}
The linearized form of supergravity equations written in terms of the BRST charges of the pure spinor 
sigma model gives us the framework to analyze some aspects of 
closed string field theory action. As it is well-known, the action for closed string field theory 
has to take into account the presence of selfdual forms (for example the 
five form in type IIB supergravity). This can be done either by breaking explicitly 
the Lorentz invariance, or by admitting an infinite number of fields 
in the action \cite{dualactions}. 

In \cite{mypaper4} we showed
that this action can be indeed constructed by mimicking the bosonic 
closed string field theory action 
discussed in \cite{zwie} (and in references therein). 

To this purpose, we derived the set of antifields for the massless sector of closed string theory, we discussed the coupling of the fields to the antifields for a closed string field theory action and we finally proposed a kinetic term which leads to the correct equations of motion. We showed that we could 
easily account for new fields which nevertheless do not propagate and 
we checked that the action had the correct symmetries leading to the complete BV 
action for type IIA/IIB supergravity.  

Since in \cite{mypaper4} we only dealt with linearized supergravity equations, we did not discuss generalizations of Witten string field $\star$-product for the open superstring. Similarly, it was outside the scope of \cite{mypaper4} to construct a full-fledged closed string field theory.

\chapter{Conclusions and outlook}
In this thesis I have presented my papers \cite{mypaper1,mypaper2,mypaper3,mypaper4}, where I investigated aspects of noncommutative geometry  and superstring theory. 

When I started my research activity, it was already known that field theories defined on a noncommutative space arise as a low energy description of D-brane dynamics in the presence of a constant NS-NS background. This had been proven for the bosonic string, the RNS superstring and the $N=2$ string. Moreover, it had been shown that extending this discussion to superstring theory in GS formalism, where spacetime fermions are present, the (anti)commutation relations involving the fermions are not modified by the constant NS-NS background.
These string theory results had already induced a growing interest in noncommutative field theory and many results had been obtained. For instance, it was already known that noncommutative field theories with time-space noncommutativity display awkward features, such as acausality and nonunitarity, and that these ill-defined theories cannot be obtained as a low energy limit of string theory. Moreover, the behavior of noncommutative field theory with respect to Poincar\'e symmetry was well-known. Moyal noncommutative deformation breaks Lorentz-invariance but preserves translation invariance.  
The generalization of the string results concerning D-branes in a constant NS-NS background had been generalized to nonconstant backgrounds afterwards, to show that a Kontsevich-like product replaces Moyal product in this case. The deformation is associative when the background is a closed two-form and nonassociative otherwise. 
All these results are reviewed in the first chapter of my thesis, so that my work can be put into context.
\vskip 11pt
In modern physics symmetries have such an importance that a theory is actually defined in terms of its symmetries and its field content. If a classical theory constructed out of the chosen fields does not contain all the interactions that are allowed by its symmetry structure, the missing terms will be generated at the quantum level when the theory is renormalized.

A unifying aspect of my contributions to the field of noncommutative geometry is the way symmetries can be implemented in noncommutative generalizations of known ordinary theories.
Since the noncommutative generalization of a given theory is not unique, preserving its symmetries (or at least some of them) in its noncommutative generalization is the first criterium one should take into account in the evaluation of the various possibilities, given the importance symmetries have in the physics of our world. 

In my paper \cite{mypaper1}, written in collaboration with D. Klemm and S. Penati, I have considered the very special symmetry, known as supersymmetry, relating bosonic and fermionic degrees of freedom in a theory. This symmetry has not been observed in nature yet, however there are hopes that it will make itself manifest at higher energies. For people who believe that the ultimate, fundamental theory of reality is string theory, supersymmetry is expected to be one of the characteristics of nature at sufficiently high energy, since string theory requires it for consistency.  

Supersymmetric theories are better studied in a formalism known as superspace. Superspace is an extension of spacetime where bosonic coordinates are accompanied by fermionic ones. There supersymmetry is realized in the form of generalized translations. The geometry of superspace is not flat,  since a nonvanishing torsion is present. 

In \cite{mypaper1} we studied the way supersymmetric theories can be deformed by implementing a non(anti)commutative geometry, without supersymmetry to be lost.  We first considered four-dimensional superspace with a Minkowski metric. We wrote down the most general algebra for bosonic and fermionic coordinates of superspace and then required covariance with respect to supertraslations and associativity. We also required the reality properties of the spinors in (anti)commutative superspace to be still valid in the deformed superspace. This was the crucial difference with respect to the previous work \cite{ferrara}. We found that nontrivial coordinate algebras are allowed that involve fermionic coordinates. We also noted that, for consistency with respect to supersymmetry, turning on the anticommutators between the fermions implies that also terms depending on the fermionic coordinates appear in the fermion-boson and boson-boson commutators. The ``trivial" superspace deformation where only bosonic coordinates are rendered noncommutative with a constant commutator is found as a special case.
However, in Minkowski signature deformations involving nonzero fermion-fermion anticommutators are ruled out because of spinor reality conditions together with associativity requirements.
Therefore, it is clear that more general deformations are allowed if spinor conjugation relations can be relaxed. In \cite{mypaper1} we noted that this is possible when moving to a four-dimensional superspace with Euclidean signature, that can only be defined when extended supersymmetry is present. By applying our procedure to Euclidean $N=2$ superspace, we found that deformations with constant fermion-fermion anticommutators are allowed.

In \cite{mypaper1} we also discussed how to construct a non(anti)commutative $\star$ product between superfields. Since our supergeometries involve coordinate-dependent terms in the coordinate algebra, a Moyal-like product is not associative, because of the superspace nontrivial torsion. However, we suggested that a Kontsevich-like product can be constructed, with the property of being associative if and only if the supercoordinate algebra is also associative. We gave the formula for the product up to second order in the deformation parameter and argued that there was no objection of principle to extending Kontsevich procedure to all orders. This was done later in \cite{chepelev}.

In \cite{mypaper1} we also noted that when superspace is rendered non(anti)commutative the supersymmetry algebra is deformed by terms quadratic in bosonic derivatives. These terms do not affect the coordinate algebra, but modify the supersymmetry transformation of a general superfield.
Finally, we made some speculations on how the deformed superspaces we found could emerge from string theory, formulated in a manifestly target-space supersymmetric way (i.e. in Green-Schwarz \cite{gsstring} or Berkovits formalisms \cite{berkovits}).

Indeed, it was found that deformed superspaces similar to the ones I studied in \cite{mypaper1} naturally arise when the open superstring in the manifestly superPoincar\'e covariant formalism introduced by Berkovits \cite{berkovits}  is compactified on a CY three-fold in the presence of D-branes and a constant R-R background \cite{vafaooguri2,seiberg}. The nonanticommutative superspace found there is related to the one I studied in the $N=2$ euclidean case by a change of variables and a reduction to $N=1$ by identification of the two fermionic coordinates on the boundary of the string worldsheet. Since in this superspace only one of the two Weyl spinor supersymmetry charges are deformed by quadratic terms in bosonic derivatives, Seiberg called it $N=\half$ superspace. The constant R-R background considered in \cite{vafaooguri2,seiberg} is allowed only in an euclidean signature. This is the stringy counterpart of the algebraic discussion in my paper \cite{mypaper1}, justifying by a geometrical argument why superspace deformations involving nonzero fermion-fermion anticommutators can only appear in an Euclidean signature. Deformed superspaces were also found to emerge in the uncompactified ten dimensional superstring when a constant R-R background and D-branes are present \cite{antonio}. However, since this background is not an exact  solutions to the string equations, but only to the linearized equations, it is not obvious that this deformation survives the zero-slope limit necessary to obtain the low energy D-brane dynamics. 

After the discovery that non(anti)commutative superspaces can be obtained from the superstring, a lot of efforts in the study of non(anti)commutative field theories have been done and many interesting properties of $N=\half$ Wess-Zumino model and super-Yang-Mills theory have been elucidated. Different deformations of theories with extended supersymmetry have been considered. The connection between deformed superspaces and supersymmetric matrix models has been elucidated. 

Up to now only constant R-R backgrounds have been considered. It would be interesting to study what happens in the more general, nonconstant case. A first step in this direction has been taken in my paper \cite{mypaper4}, written in collaboration with P.A. Grassi, where superstring vertex operators for linearly $x$-dependent R-R field strength have been computed. These objects are the main ingredient to generalize the string analysis presented in \cite{antonio} to the nonconstant case.  Linearly $x$-dependent R-R backgrounds are expected to be related to a special kind of superspace deformation, with a Lie algebraic structure where the anticommutator between two fermions gives the bosonic coordinate. This can be interpreted by saying that spacetime bosonic coordinates have a fermionic substructure. In these terms, this issue was investigated in \cite{Schwarz:pf}. 
\vskip 11pt
Now I would like to go back to the main theme unifying my work, the implementation and implications of symmetries in noncommutative theories. Up to know I have discussed how I approached the problem of deformation of supersymmetric theories and the developments that followed my work, in both field and string theory. 

In two papers of mine \cite{mypaper2,mypaper3}, I considered instead the problem of constructing the noncommutative generalization of field theories possessing an infinite number of conserved local symmetry charges, i.e classically integrable. As it is well-known, in the commutative case the underlying symmetry structure of these theories has strong implications on their dynamics. In particular, two dimensional integrable theories have a factorized S-matrix and particle production does not occur.  Most ordinary integrable theories in two and three dimensions have been shown to be related to a four dimensional theory, selfdual Yang-Mills, from which they can be obtained by dimensional reduction. The S-matrix of selfdual Yang-Mills also displays a peculiar property, all tree-level amplitudes beyond three-point being vanishing. This property has been used as a definition of integrability in four dimensions. Selfdual Yang-Mills is related to the $N=2$ string, characterized by an $N=2$ worldsheet supersymmetry. It has been proven that tree-level $N=2$ string dynamics coincides with selfdual Yang-Mills theory. 

The $N=2$ string can be coupled to a constant NS-NS background. In the presence of D-branes it has been shown that the low energy limit of the brane dynamics can be described by noncommutative selfdual Yang-Mills theory. 
Therefore, given an ordinary integrable two-dimensional field theory, one can first try to construct a noncommutative deformation that preserves classical integrability, in the sense of having an infinite number of conserved local charges (local in the sense that they are not written in terms of integrals). Then one can check if the usual S-matrix properties are still valid after the deformation. This is not obvious, since noncommutativity introduces nonlocality in the theory and, in the two-dimensional case, this necessarily affects the time coordinate, causing in general an acausal behavior and the breakdown of unitarity. Furthermore, one can investigate whether the theory can be obtained by reduction from noncommutative selfdual Yang-Mills, or even use this method to construct the two-dimensional theory in the first place.

Another interesting issue is the study of solitons solutions. Commutative integrable theories usually display this kind of solutions. Noncommutative theories have new soliton solutions disappearing in the commutative limit, that exist thanks to the nonlocality introduced by Moyal product. Therefore, noncommutative integrable theories are an interesting setting to study both kinds of solitons.
Noncommutative solitons are a field theory realization of the D-branes that are present in the string theory from which they have emerged in the low energy limit, thus the study of noncommutative solitons can give clues on D-brane dynamics. 

In \cite{mypaper2,mypaper3}, all these issues have been studied in the special case of the sine-Gordon model. This is an integrable two-dimensional field theory, describing the dynamics of a scalar field selfinteracting through an oscillating potential. Apart from the many general nice properties related to its integrability, the sine-Gordon model also exhibits very nice renormalization properties.
Moreover, the ordinary sine-Gordon model is related to the Thirring model through bosonization. The relation between these two theories is  a simple example of duality relating the weak coupling limit of a theory with the strong-coupling limit of the other.  Bosonization has been studied in noncommutative geometry and it has been shown that the ordinary abelian $U(1)$ case is modified in the noncommutative setting so that a free fermion is not related to a free scalar, but to a scalar governed by a $U(1)$ WZW model.

In \cite{us}, M.T. Grisaru and S. Penati have shown that an integrable noncommutative version of the sine-Gordon model can be constructed, where two equations govern the dynamics of a scalar (and generically complex) field. One equation reduces to the sine-Gordon equation in the commutative limit, the other vanishes in the commutative limit, has the form of a conservation law and in fact gives the first of the infinite conserved currents. This system does not seem to be overconstrained, since the class of its localized solutions is at least as large as the one in ordinary sine-Gordon. The doubling of the equations of motion is related to the fact that  the $U(1)$ factor in the noncommutative group $U(n)$ does not decouple, in contradistinction to the ordinary case. Since commutative sine-Gordon theory is obtained from zero curvature conditions for gauge connections in $SU(2)$ group, in the noncommutative case this has to be enlarged to $U(2)$, which causes an additional equation of motion to appear.

In my paper \cite{mypaper2}, written in collaboration with M.T. Grisaru, L. Mazzanti and S. Penati, we showed that the equations describing this noncommutative version of the sine-Gordon model can be obtained by dimensional reduction from noncommutative selfdual Yang-Mills in Yang formulation. Unfortunately, an action cannot be obtained by an analogous reduction procedure. However, in \cite{mypaper2}, we found an action that gives the correct equations of motion. We then studied tree-level scattering amplitudes and found that acausality is present and the S-matrix is not factorized. Therefore, it seems that in general the presence of an infinite number of conserved currents is not sufficient for the S-matrix to be factorized in noncommutative case. We finally discussed the relation of our model with the noncommutative $U(1)$ Thirring model.
The fact that dimensional reduction does not work at the level of action, but only at the level of the equations of motion, is a sign that the parametrization of the degrees of freedom we considered in \cite{mypaper2} is not the correct one. 

In \cite{mypaper3}, in collaboration with O. Lechtenfeld, L. Mazzanti, S. Penati and A.D. Popov, we considered a different noncommutative generalization of the sine-Gordon model, that is also obtained by dimensional reduction from noncommutative selfdual Yang-Mills, by considering the $U(1)\times U(1)$ subgroup of $U(2)$. While in our first attempt to construct a noncommutative integrable sine-Gordon model we had modified the kinetic term of the ordinary theory to a $U(1)$ WZW-like term, while maintaining the usual cosine structure of the interaction term, in this new noncommutative theory we modified also the interaction term structure, in a way that two real scalars parametrizing the $U(1)\times U(1)$ subgroup of $U(2)$ are coupled by a nontrivial term. Again, the theory is described by two equations of motion. One of the two becomes trivial in the commutative limit, while the other one gives the ordinary sine-Gordon  equation. 

Solitons solutions of this model have been studied by making use of the dressing method. Moreover, tree-level scattering amplitudes have been computed and proven to be factorized and causal, in spite of the presence of time-space noncommutativity. 
Therefore this second noncommutative version of the sine-Gordon model has inherited the nice classical properties of its ordinary counterpart. It would be interesting to move on to a quantum description of the model to study its renormalizability properties and to investigate whether integrability survives at the quantum level. 

A quite striking feature of the noncommutative sine-Gordon system we constructed is that its tree-level amplitudes are completely independent of the noncommutativity parameter. It would be interesting to understand why this happens, for instance if this is a general feature of noncommutative integrable systems or a peculiar property of the sine-Gordon model. Another aspect that still needs to be investigated is the relation of this $U(1)\times U(1)$ theory with noncommutative fermions models. Somehow it would be more natural to see what happens in the bosonization of $U(2)$ fermion models and then consider a reduction to $U(1)\times U(1)$. 
Another possible development would be the study of other kinds of duality in this system, for instance T-duality.
\vskip 11pt
As I anticipated when discussing the superstring theory origin of the deformed superspaces I introduced in \cite{mypaper1}, the covariant formulation for the superstring has been proven to be superior in dealing with many string issues that were not treatable with the other known formalisms.
In particular this was true for deformed superspaces, that were shown to arise in the presence of R-R backgrounds. These cannot be dealt with when using the RNS formalism, while they can be treated with the GS formalism, that however is quite clumsy because of the lack of manifest Lorentz covariance.
Because of this connection with the work I did in the field of noncommutative geometry, I started to study this formalism. As I previously said, to generalize the results of \cite{vafaooguri2,seiberg,antonio} to a more general, nonconstant background one needs to determine the corresponding vertex operator first.
This is not an easy task in the covariant formulation, since the great amount of manifest symmetry makes the formalism redundant. The vertices are written in terms of superfields that have to satisfy the linearized supergravity equations of motion.

In my paper \cite{mypaper4}, written in collaboration with P.A. Grassi, we described an iterative procedure to compute the vertex operators in terms of the physical fields only. To do this a suitable gauge fixing must be imposed that removes the auxiliary fields from the vertices.
We used this technique to compute the vertices for linearly $x$-dependent R-R field strength that may be related to a Lie-algebraic deformation of superspace, as I already said, but we also discussed other two applications of our analysis that are a little bit out of the path, having no direct connection to noncommutative geometry. We showed how an off-shell formulation of the superstring vertices can be constructed, how the corresponding equations of motion are a deformation of the usual superfield constraints and how the gauge fixing necessary to implement our iterative procedure can be applied to this case. The motivation for this discussion relies in the fact that a propagating closed string emitted from a D-brane has to be off-shell. The off-shell vertex operators couple the R-R potential (and not the field strength) to the worldvolume of the brane. In our discussion, we showed that the vertices we obtain explicitly contain the R-R potential. Therefore our analysis could be useful in the study of D-brane dynamics in the covariant formalism.
Finally, we used our framework to determine the antifield equations of motion and to make a proposal for the kinetic term of closed superstring field theory. Since this kinetic term is written in superspace formulation, the first thing one should now do to make nontrivial checks about its properties is to rewrite it in component formulation.
\vskip 11pt
To conclude, in my work I have mostly investigated the way one can implement noncommutativity in theories with a special symmetry structure, such as supersymmetric and classically integrable theories.
Requiring the symmetries to be preserved in the noncommutative deformation is a strong constraint that allows to make a selection between the many possibile noncommutative versions of the same theory.

In the case of integrable theories, I explicitly discussed the case of the sine-Gordon theory and eventually found its noncommutative generalization that has infinite conserved currents and also displays all the nice features that in ordinary theories are implied by the symmetry structure, such as factorization of the S-matrix. However, many issues still have to be investigated, such as for instance the connection between the noncommutative sine-Gordon system and $U(2)$ fermion models and the reason for the complete absence of a dependence on the noncommutativity parameter in the tree-level amplitudes.

In the case of supersymmetric theories, I constructed superspace deformations that preserve supersymmetry, are associative and respect the usual spinor reality properties. The connection between the deformations I found and superstring theory in the presence of R-R backgrounds, described in the covariant formalism, has led me to deepen my knowledge of this formulation of superstring theory. Even if this might seem constructed ``ad hoc", without a strong underlying principle, it is the first superPoincar\'e covariant formulation of the superstring that works and it has already proven to very suitable to handle R-R backgrounds and to prove general theorems about superstring amplitudes. Since the computation of superstring vertex operators is not an easy task in the covariant formalism, I provided a recursive technique to compute the vertices in terms of physical fields only. This analysis of vertex operators has many possible applications. I discussed applications to the computation of superspace deformations in the presence of nonconstant R-R backgrounds, to the computation of vertices in the off-shell formulation, that are useful in the study of D-brane dynamics, and to the construction of kinetic terms for a closed superstring field theory. Since the covariant superstring in the presence of NS-NS backgrounds has not been studied in detail yet, it would be very interesting to perform an analysis of the string origin of noncommutative geometry in this formalism. Indeed, since NS-NS and R-R backgrounds are treated in a very ``symmetric" way in this formulation of the superstring, this setting should be very useful to study S-duality.

\appendix
\chapter{Conventions}
\section{Superspace conventions in $d=4$}
In four dimensions, $N=1$ Minkowski superspace is described by a set of 
coordinates  
$Z^A \equiv\left(x^{\a \ad},\th^{\a},\bar{\th}^{\ad}\right)$, where 
$x^{\a \ad} \equiv x^{\underline{a}}$ are the four bosonic real coordinates, 
while $\theta^\a$ (and $\bar{\th}^\ad = (\theta^\a)^{\dag}$) are complex 
two--component Weyl fermions. We use conventions of {\em Superspace} 
\cite{superspace}, with $(\psi^\a)^{\dag} = \bar{\psi}^{\ad}$,   
$(\psi_\a)^{\dag} = - \bar{\psi}_{\ad}$.

The supersymmetry algebra
\bea
&&\left\{Q_{\a},\bar{Q}_{\dot{\a}}\right\}=P_{\a\dot{\a}} \non\\
&&\left\{Q_{\a},Q_{\b}\right\}=\left\{\bar{Q}_{\dot{\a}},
\bar{Q}_{\dot{\b}}\right\}=0 \non\\
&&\left[P_{\underline{a}},P_{\underline{b}}\right]=0
\label{algebra}
\ena
with $\bar{Q}_{\ad} = Q_\a^{\dag}$, is realized by
\bea
&&Q_{\a}=i\left(\pa_{\a}-\frac{i}{2}\bar{\th}^{\dot{\a}}
\pa_{\a\dot{\a}}\right)\non\\
&&\bar{Q}_{\dot{\a}}=i\left(\bar{\pa}_{\dot{\a}}-
\frac{i}{2}\th^{\a}\pa_{\a\dot{\a}}\right)\non\\
&&P_{\a\dot{\a}}=i\pa_{\a\dot{\a}}
\label{charges}
\ena
Under supersymmetry transformations a generic superfield $V$ transforms
according to $\d V = -i \left(\e^{\a}Q_{\a}+\bar{\e}^{\dot{\a}}
\bar{Q}_{\dot{\a}}\right) V$. In particular, the action on the superspace 
coordinates defines supertranslations 
\bea
&&\d x^{\underline{b}} \equiv 
-i\left(\e^{\a}Q_{\a}+\bar{\e}^{\dot{\a}}\bar{Q}_{\dot{\a}}\right)
x^{\underline{b}}= -\frac{i}{2}\left(\e^{\b}\bar{\th}^{\dot{\b}}+
\bar{\e}^{\dot{\b}}\th^{\b}\right)\non\\
&&\d\th^{\b} \equiv -i\left(\e^{\a}Q_{\a}+\bar{\e}^{\dot{\a}}
\bar{Q}_{\dot{\a}}\right)\th^{\b}=\e^{\b}\non\\
&&\d\bar{\th}^{\dot{\b}} \equiv -i\left(\e^{\a}Q_{\a}+\bar{\e}^{\dot{\a}}
\bar{Q}_{\dot{\a}}\right)\bar{\th}^{\dot{\b}}=\bar{\e}^{\dot{\b}}
\label{supert}
\ena
Covariant derivatives with respect to (\ref{supert}) are 
$D_A \equiv \left(D_{\a},\bar{D}_{\dot{\a}}, \pa_{\a\dot{\a}}\right)$ 
where
\bea
D_{\a}=\pa_{\a}+\frac{i}{2}\bar{\th}^{\dot{\a}}\pa_{\a\dot{\a}} 
\quad &,& \quad 
\bar{D}_{\dot{\a}} ~=~ \bar{\pa}_{\dot{\a}}+\frac{i}{2}\th^{\a}\pa_{\a\dot{\a}}
\non\\
\pa_{\a\dot{\a}} &=& -i \left\{D_{\a},\bar{D}_{\dot{\a}}\right\} 
\label{covariant}
\eea
Moreover, they satisfy $\{ D_\a , D_\b \} = \{ \bar{D}_\ad , \bar{D}_\bd \}
=0$ and anticommute with the generators of supersymmetry transformations.
 
We can define left and right grassmannian derivatives 
according to the following rules
\bea
&& (\pa_L)_\a \th^\b ~\equiv~ \overrightarrow{\pa}_\a \th^\b ~=~ \d^{~\b}_\a 
~~~~~~\quad ,\quad 
(\bar{\pa}_L)_\ad \bar{\th}^\bd \equiv \vec{\bar{\pa}}_\ad \bar{\th}^\bd 
~=~ \d^{~\bd}_\ad \non\\
&& (\pa_R)_\a \th^\b ~\equiv~ \th^{\b}\overleftarrow{\pa}_{\a} ~=~
-\d_{\a}^{~\b} \quad , \quad
(\bar{\pa}_R)_\ad \thb^\bd ~\equiv~
\bar{\th}^{\dot{\b}}\overleftarrow{\bar{\pa}}_{\dot{\a}} ~=~
-\d_{\dot{\a}}^{~\dot{\b}}
\label{rule1}
\ena
Their action on a generic superfield is defined as 
\beq
\pa_L V ~\equiv~ \overrightarrow{\pa} V \qquad , \qquad 
\pa_R V ~\equiv~ V \overleftarrow{\pa}
\label{rule2}
\eeq
Notice that these definitions hold independently of the nature of the 
superfield V. In particular, in the case of a spinorial superfield $V_\b$
we have $(\pa_R)_A V_\b \equiv V_\b \overleftarrow{\pa}_A$. As a consequence of
the general definitions (\ref{rule1}, \ref{rule2}) we immediately obtain
$\pa_L V = \pa_R V$ for any tensorial superfield, whereas 
$\pa_L V_\b = -\pa_R V_\b$ for any spinorial superfield.  

From the identities
\bea
&& \left(\overrightarrow{\pa}_{\a}\th^{\b}\right)^{\dag} ~=~
-\bar{\th}^{\dot{\b}}\overleftarrow{\bar{\pa}}_{\dot{\a}} \quad , \quad
\left(\vec{\bar{\pa}}_{\dot{\a}}
\bar{\th}^{\dot{\b}}\right)^{\dag} ~=~
-\th^{\b}\overleftarrow{\pa}_{\a} \non\\
&& \left(\overrightarrow{\pa}^{\a}\th^{\b}\right)^{\dag} ~=~
\bar{\th}^{\dot{\b}}\overleftarrow{\bar{\pa}}^{\dot{\a}} ~~~\quad , \quad
\left(\vec{\bar{\pa}}^{\dot{\a}}
\bar{\th}^{\dot{\b}}\right)^{\dag} ~=~
\th^{\b}\overleftarrow{\pa}^{\a} 
\label{rule3}
\eea
the hermitian conjugation rules for left and right derivatives follow
\beq
((\pa_L)_\a)^{\dag} = -(\bar{\pa}_R)_\ad \qquad , \qquad 
(\pa_L^\a)^{\dag} = (\bar{\pa}_R)^\ad
\eeq

We can also introduce left and right bosonic derivatives which are
simply given by
\beq
(\pa_L)_{\a \ad} x^{\b\dot{\b}} ~\equiv~ \pa_{\a \ad} x^{\b \bd}
~=~ \d_{\a}^{~\b}\d_{\dot{\a}}^{~\dot{\b}} \quad , \quad
(\pa_R)_{\a \ad} x^{\b\dot{\b}} ~\equiv~ 
x^{\b\dot{\b}}\overleftarrow{\pa}_{\a\dot{\a}} 
~=~ \d_{\a}^{~\b}\d_{\dot{\a}}^{~\dot{\b}}
\label{rule4}
\eeq
Their action on a superfield $V$ is defined as $\pa_L V \equiv 
\overrightarrow{\pa} V$
and $\pa_R V \equiv V \overleftarrow{\pa} $. 
Therefore, from (\ref{rule4}), it easily follows that  
$(\pa_L)_{\a \ad} V = (\pa_R)_{\a \ad} V$ for {\em any} superfield. 

As a consequence of the previous identities, left and right covariant 
derivatives can be equally defined. 
Left covariant derivatives act on a generic superfield 
from the left as 
\beq
(D_L)_\a V \equiv \overrightarrow{D}_\a V  \qquad , 
\qquad (\bar{D}_L)_\ad V \equiv 
\vec{\bar{D}}_\ad V
\eeq
where $D_\a$ and $\bar{D}_\ad$ are explicitly given in (\ref{covariant}).
Right covariant derivatives are defined as acting from the
right 
\bea
&& (D_R)_\a V ~\equiv~ V \overleftarrow{D}_{\a}
~=~ 
V \left( \overleftarrow{\pa}_{\a}+\frac{i}{2}
\overleftarrow{\pa}_{\a\dot{\a}} \bar{\th}^{\dot{\a}} \right)
\non\\
&& (\bar{D}_R)_\ad V ~\equiv~ V \overleftarrow{\bar{D}}_{\ad}
~=~ 
V \left(
\overleftarrow{\bar{D}}_{\dot{\a}}=\overleftarrow{\bar{\pa}}_{\dot{\a}}
+\frac{i}{2}\overleftarrow{\pa}_{\a\dot{\a}}\th^{\a} \right)
\eea
It is easy to check that $D_L V = D_R V$ on any tensorial superfield,
whereas $D_L V_\b = -D_R V_\b$. Moreover, left and
right derivatives are related by hermitian conjugation $((D_L)_\a)^{\dag} =
-(\bar{D}_R)_\ad$ and $((D_L)^\a)^{\dag} = (\bar{D}_R)^\ad$.

Defining the right momentum operator as 
$(P_R)_{\a \ad} V =- i V \overleftarrow{\pa}_{\a \ad}$, 
it is easy to show that the algebra 
of right derivatives is the standard one  
$\{ D_R , D_R \} = \{ \bar{D}_R , \bar{D}_R \} =0$ and 
$\{ (D_R)_\a , (\bar{D}_R)_\ad \} = (P_R)_{\a \ad}$. 
Moreover
\bea
&&\left[ D_{\a}^R , D_{\b}^L \right] ~=~ 0 ~=~
\left[\overline{D}_{\dot{\a}}^R,\overline{D}_{\dot{\b}}^L\right]  \non\\
&&\left[D_{\a}^R,\overline{D}_{\dot{\a}}^L\right] V 
~=~ (P_R)_{\a \ad} V \non\\
&&\left[D_{\a}^L , \overline{D}_{\dot{\a}}^R \right] V 
~=~ (P_L)_{\a \ad} V 
\label{B}
\ena
for any tensor superfield. 
When the commutators are applied to a spinor $V_\b$,
a minus sign appears on the r.h.s. of the last two identities due 
to the anticommutation of the spinorial derivatives with $V_\b$.

Following the same procedure one can equally define left and right 
supersymmetry generators as $(Q_L)_A V \equiv \vec{Q}_A V$ and 
\bea
&& (Q_R)_\a V ~\equiv~ V \overleftarrow{Q}_{\a} ~=~ 
V \left[ -i \left(\overleftarrow{\pa}_{\a}-\frac{i}{2}
\overleftarrow{\pa}_{\a\dot{\a}} \bar{\th}^{\dot{\a}} \right) \right]
\non\\
&& (\bar{Q}_R)_\ad V ~\equiv~ 
V \overleftarrow{\bar{Q}}_{\dot{\a}} ~=~
V \left[-i \left( \overleftarrow{\bar{\pa}}_{\dot{\a}}
-\frac{i}{2}\overleftarrow{\pa}_{\a\dot{\a}}\th^{\a} \right) \right]
\eea
The algebra of right generators is again given by (\ref{algebra}).  
The algebra of the commutators on a tensor superfield is
\bea
&&\left[ Q_{\a}^R , Q_{\b}^L \right] ~=~ 0 ~=~
\left[\overline{Q}_{\dot{\a}}^R,\overline{Q}_{\dot{\b}}^L\right]  \non\\
&&\left[Q_{\a}^R,\overline{Q}_{\dot{\a}}^L\right] V
~=~ (P_L)_{\a \ad} V \non\\
&&\left[Q_{\a}^L , \overline{Q}_{\dot{\a}}^R \right] V 
~=~ (P_R)_{\a \ad} V 
\ena 
Instead, when the commutators act on spinor objects 
we get a change of sign on the r.h.s. of the last two equalities.  
\vskip 11pt
In euclidean signature
a reality condition on spinors is applicable only in the presence of
extended supersymmetry. In
the simplest case, $N=2$ euclidean superspace,
the two--component Weyl spinors satisfy a symplectic Majorana condition
\beq
(\th^\a_i )^{\ast} ~=~ \th_\a^i ~\equiv~ C^{ij} \, \th^\b_j \, C_{\b \a}
\quad , \quad
(\thb^{\ad, i} )^{\ast} ~=~ \thb_{\ad, i} ~\equiv~
\thb^{\bd, j} \, C_{\bd \ad} \, C_{ji}
\eeq
with $C^{12} = -C_{12} = i$.
The choice of covariant derivatives and supersymmetry charges is the obvious generalization of the $N=1$ Minkowski case.
\section{Superspace conventions in $d=10$}
We describe Minkowski ten dimensional $N=2$ superspace by the coordinates $(x^m,\th^\a,\hat\th^{\hat\b})$. Depending whether one is in type IIA or IIB case, the spinors $\th^\a$ and $\hat\th^{\hat\b}$ have opposite or same chirality.

In ten dimensions with Minkowski signature one can use Dirac matrices $\G^m=\{I\otimes (i\tau_2),\s^\m \otimes \tau_1,\chi \otimes \tau_1\}$, where $m=0,\dots,9$ and $\m=1,\dots,8$. $\tau_i$, $i=1,2,3$ are the Pauli matrices, $\s^\m$ are eight real symmetric $16\times 16$ off-diagonal Dirac matrices in $d=8$ with euclidean signature, while $\chi$ is the real $16\times 16$ diagonal chirality matrix in $d=8$. The chirality matrix in $d=10$ with Minkowski signature is then $I\otimes\tau_3$ and the charge conjugation matrix $C$, satisfying $C\G^m=-(\G^m)^T C$, is numerically equal to $\G^0$. 

If one uses spinors $\Psi^T=(\a_L,\b_R)$ with spinor indices $\a_L^\a$ and $\b_{R,\bd}$, the index structure of the Dirac matrices and the charge conjugation matrix is
\beq
\G^m=\left(\begin{matrix}
0 & (\s^m)^{\a\bd}\\
({\tilde\s}^m)_{\bd\g} & 0
\end{matrix}\right),\qquad C=\left(\begin{matrix}
0 & c_\a^{~~\bd}\\
c^\bd_{~~\g} & 0
\end{matrix}
\right)
\eeq
where $\s^m=\{I,\s^\m,\chi\}$ and ${\tilde\s}^m =\{-I,\s^\m,\chi\}$. The matrices $c_\a^{~~\bd}$ and $c^{\bd}_{~~\g}$ are numerically equal to $I_{16\times 16}$ and $-I_{16\times 16}$, respectively.
The matrices $\g^m$ used in the text are given by $\g^{m\ad\bd}=c^\ad_{~~\b}(\s^m)^{\b\bd}$ and $\g^m_{\a\b}=c_\a^{~~\bd}({\tilde\s}^m)_{\b\bd}$. From now on and in the text dots are omitted. The spinors $\a_L$ and $\b_R$ form inequivalent representations of $SO(9,1)$. Spin indices cannot be raised or lowered with the charge conjugation matrix, but $\a_L^\a c_\a^{~~\bd} \b_{R,\bd}$ is Lorentz invariant.

One finds that the $16\times 16$ symmetric matrices $\g^m_{\a\b}$ and $\g^{m\a\b}$ satisfy 
\beq
\g^m_{\a\b}\g^n{\b\g}+\g^n_{\a\b}\g^{m\b\g}=2\eta^{mn}\d_\a^\g
\eeq
and
\beq
\g_{m(\a\b}\g^m_{\g)\d}=0
\eeq
which makes Fierz rearrangements very easy.

Our conventions for $d=10$ $N=2$ superspace covariant derivatives and 
supersymmetry charges are
\bea
\label{susu}
&&D_{\a} = \pa_{\a} + {1\over 2}(\g^{m}\th)_{\a} \pa_{m}\,, ~~~~~~
Q_{\a} = \pa_{\a} - {1\over 2}(\g^{m}\th)_{\a} \pa_{m}\,, \cr
&&\hat D_{\hat\a} = \pa_{\hat\a} + {1\over 2}(\g^{m}\hat\th)_{\hat \a} \pa_{m}\,, ~~~~~~
\hat Q_{\hat\a} = \pa_{\hat\a} - {1\over 2}(\g^{m}\th)_{\hat\a} \pa_{m}\,,
\ena
which satisfy
\bea
\label{deriv}
&&\left\{D_\a,D_\b\right\}=\g^m_{\a\b}\pa_m\,, \quad\quad
\left\{\hat D_{\hat\a},\hat D_{\hat\b}\right\}=\g^m_{\hat\a\hat\b}\pa_m\,, 
\quad\quad
\left\{D_\a,\hat D_{\hat\b}\right\}=0\cr
&&
\{D_{\a}, Q_{\b} \} =0\,, ~~~~~
\{\hat D_{\hat\a},\hat Q_{\hat \b} \} =0\,.
\ena

\cd
\addcontentsline{toc}{chapter}{Acknowledgements}
\subsection*{Acknowledgements}
\vskip12pt
First of all I would like to thank Silvia Penati so much for her continuous encouragement, help and support in these many years. 
\vskip 12pt
I would like to thank the Department of Nuclear and Theoretical Physics in Pavia and INFN, sezione di Pavia, for financial support. In particular, I am very grateful to Annalisa Marzuoli, Mauro Carfora and Sergio Ratti for their very kind help and encouragement.
\vskip 12pt
I am very grateful to the Physics Department G. Occhialini, University of Milano-Bicocca, for the very kind hospitality throughout these years and for providing funding that supported my participation into some schools and conferences. I would like to thank all the ``baretto"  friends in particular for the funniest lunch breaks ever.
\vskip 12pt
I would like to thank the C. N. Yang Institute for Theoretical Physics at Stony Brook, U.S.A., for the very kind hospitality during my stay within the International Doctorate Programme between the State University of New York at Stony Brook and the University of Pavia. In particular, I would like to thank Martin Ro\v cek for his kind help and support during my stay there, for the many useful discussions and especially for coming to Italy and participating into my thesis defense as a member of the evaluating committee.
\vskip 12pt
I would like to thank Pietro Antonio Grassi, Marc Grisaru, Dietmar Klemm, Olaf Lechtenfeld, Liuba Mazzanti, Silvia Penati  and Alexander Popov for giving me the opportunity to collaborate with them and to learn so much from them.
\vskip 12pt
Very special thanks go to Dietmar Klemm and Marc Grisaru for their very strong encouragement, help and support.
\vskip 12pt
I am especially grateful to Ulf Lindstr\"om for kindly accepting to be the external referee for this thesis, for his very careful reading of it and for the many useful discussions, comments and suggestions.
\vskip 12pt
Finally, I would like to thank my many ex officemates, Claudio Dappiaggi, Valeria Gili, Liuba Mazzanti, Marcello Musso, Alberto Romagnoni and Andrea Sartirana for tolerating my bad temper, loud voice, invasive plants, my mess and very bad habit of locking one of them in particular inside the office. Now I'm alone in my office and missing all of you.

\end{document}